\pdfoutput=1

\documentclass[11pt,letterpaper]{article}
\usepackage{jheppubcb}
\usepackage{amsmath,amssymb,amsfonts,bm}
\usepackage{graphicx,wrapfig}
\usepackage{multirow}
\usepackage{verbatim}
\usepackage{appendix}
\usepackage{rotating}
\usepackage{multirow}
\usepackage[normalem]{ulem}

\allowdisplaybreaks[1]

\renewcommand\afterTocRuleSpace{\medskip}

\numberwithin{equation}{section}
\newcommand{\be}{\begin{equation}}
\newcommand{\ee}{\end{equation}}
\newcommand{\ba}{\begin{align}}
\newcommand{\ea}{\end{align}}
\newcommand{\ben}{\begin{eqnarray}\displaystyle}
\newcommand{\een}{\end{eqnarray}}
\newcommand{\mb}{\mathbf}

\newcommand{\I}{{\rm I}}
\newcommand{\II}{{\rm I\hspace{-1pt}I}}
\newcommand{\III}{{\rm I\hspace{-1pt}I\hspace{-1pt}I}}

\newcommand{\ra}{{\text{\raisebox{.07cm}{$\alpha$}}}}

\newcommand{\SqS}{S^2\!\times_q\!S^1}
\renewcommand{\SS}{S^2\!\times\!S^1}
\newcommand{\RqS}{\R^2\!\times_q\!S^1}
\newcommand{\RS}{\R^2\!\times\!S^1}
\newcommand{\DqS}{D^2\!\times_q\!S^1}
\newcommand{\DS}{D^2\!\times\!S^1}

\newcommand{\wt}{\widetilde}
\newcommand{\wh}{\widehat}

\newcommand{\ol}{\overline}
\newcommand{\ds}{\displaystyle}

\newcommand{\eg}{\emph{e.g.}}
\newcommand{\ie}{\emph{i.e.}}
\newcommand{\cf}{\emph{cf.}}

\renewcommand{\P}{{\mathbb P}}

\newcommand{\Z}{{\mathbb Z}}
\newcommand{\R}{{\mathbb R}}
\newcommand{\C}{{\mathbb C}}
\newcommand{\Q}{{\mathbb Q}}
\newcommand{\cp}{{\mathbb{CP}}}

\renewcommand{\Re}{{\rm Re}}
\renewcommand{\Im}{{\rm Im}}
\newcommand{\Li}{{\rm Li}}
\newcommand{\Tr}{{\rm Tr \,}}
\newcommand{\bs}{\backslash}
\newcommand{\pd}{\partial}

\newcommand{\ve}{\varepsilon}

\newcommand{\CA}{\mathcal{A}}

\newcommand{\CD}{\mathcal{D}}

\newcommand{\CG}{\mathcal{G}}
\newcommand{\CH}{\mathcal{H}}
\newcommand{\CI}{\mathcal{I}}
\newcommand{\CJ}{\mathcal{J}}
\newcommand{\CK}{\mathcal{K}}
\newcommand{\CL}{\mathcal{L}}
\newcommand{\CM}{\mathcal{M}}
\newcommand{\CN}{\mathcal{N}}
\newcommand{\CO}{\mathcal{O}}
\newcommand{\CP}{\mathcal{P}}

\newcommand{\CR}{\mathcal{R}}

\newcommand{\CT}{\mathcal{T}}

\newcommand{\CV}{\mathcal{V}}
\newcommand{\CW}{\mathcal{W}}
\newcommand{\CX}{\mathcal{X}}

\newcommand{\CZ}{\mathcal{Z}}

\newcommand{\cH}{\mathcal{H}}
\newcommand{\cI}{\mathcal{I}}

\newcommand{\cL}{\mathcal{L}}
\newcommand{\cM}{\mathcal{M}}
\newcommand{\cN}{\mathcal{N}}
\newcommand{\cO}{\mathcal{O}}

\newcommand{\cT}{\mathcal{T}}

\newcommand{\cZ}{\mathcal{Z}}

\newcommand\restr[2]{{
  \left.\kern-\nulldelimiterspace 
  #1 
  \vphantom{\big|} 
  \right|_{#2} 
}}

\newcommand{\SLZ}{SL(2,\Z)}
\newcommand{\SLC}{SL(2,\C)}

\newcommand{\PSLC}{PSL(2,\C}

\title{Holomorphic Blocks in Three Dimensions}

\author[1]{Christopher Beem,\!}
\affiliation[1]{Simons Center for Geometry and Physics, Stony Brook University, \\ Stony Brook, NY 11794-3636, USA}
\emailAdd{cbeem@scgp.stonybrook.edu}

\author[2,3]{Tudor Dimofte,\!}
\affiliation[2]{Institute for Advanced Study, Einstein Dr., Princeton, NJ 08540, USA}
\affiliation[3]{Trinity College, Cambridge CB2 1TQ, UK}
\emailAdd{tdd@ias.edu}

\author[4]{Sara Pasquetti}
\affiliation[4]{Department of Mathematics, University of Surrey, Guildford, Surrey, GU2 7XH, UK}
\emailAdd{s.pasquetti@surrey.ac.uk}

\abstract{We decompose sphere partition functions and indices of three-dimensional $\CN=2$ gauge theories into a sum of products involving a universal set of ``holomorphic blocks''. The blocks count BPS states and are in one-to-one correspondence with the theory's massive vacua. We also propose a new, effective technique for calculating the holomorphic blocks, inspired by a reduction to supersymmetric quantum mechanics. The blocks turn out to possess a wealth of surprising properties, such as a Stokes phenomenon that integrates nicely with actions of three-dimensional mirror symmetry. The blocks also have interesting dual interpretations. For theories arising from the compactification of the six-dimensional $(2,0)$ theory on a three-manifold $M$, the blocks belong to a basis of wavefunctions in analytically continued Chern-Simons theory on $M$. For theories engineered on branes in Calabi-Yau geometries, the blocks offer a non-perturbative perspective on open topological string partition functions.}

\begin{document}

\maketitle


\section{Introduction}
\label{sec:intro}

Recent years have seen a resurgence of interest in supersymmetric gauge theories in three dimensions. In large part, new developments have stemmed from the introduction of techniques for formulating such theories on curved manifolds such as spheres or products thereof ($S^3$ and $\SS$) while preserving some fraction of the original supersymmetry \cite{Kapustin-3dloc, HHL, IY-index, FS-curvedSUSY}. Coupling to more general geometries is also possible \cite{DFS-curvedSUSY, KTZ-curved}. Rather than applying the more familiar technique of topological twisting in order to realize the original flat-space superalgebra in the curved-space theory, the supersymmetry algebras of these theories are deformed to accommodate background curvature. The resulting partition functions can be computed via supersymmetric localization in terms of finite-dimensional matrix integrals. These three-dimensional calculations were inspired by similar techniques developed for computing four-dimensional partition functions on $S^4$ \cite{Pestun-S4}.

In this work we study $\cN=2$ superconformal field theories in three dimensions and their massive deformations. For every $U(1)$ subgroup in the flavor symmetry group of an $\CN=2$ SCFT it is possible to turn on a real mass deformation, and we look specifically at those theories for which such deformations alone are sufficient to render all vacua gapped. We also require that the theories preserve a $U(1)_R$ R-symmetry. Examples include (the infrared limits of) $\CN=2$ SQED, SQCD, and more general gauge theories with perturbative or non-perturbative superpotentials preserving $U(1)_R$. The vacua of the mass-deformed theories on $\RS$ will play a central role for us; typically there are finitely many such vacua, indexed by $\alpha$.

We consider these theories coupled to two compact, curved backgrounds: the ellipsoid $S^3_b$ and the twisted product $\SqS$, where the two-sphere is fibered over $S^1$ with holonomy $\log q$. It has been shown in \cite{Kapustin-3dloc, HHL} and \cite{Kim-index, IY-index, KW-index}, respectively, how the corresponding partition functions $\CZ_b$ and $\CI$ can be calculated from UV Lagrangian descriptions of the theories. The ellipsoid partition function depends on real masses $\mu$ which are complexified by the choice of R-charge assignments for fields in the path integral (as well as the real geometric deformation parameter $b$), while $\CI$, a supersymmetric index, depends on fugacities $\zeta$ for $U(1)$ flavor symmetries and the quantized flux $m$ on $S^2$ of background gauge fields coupled to flavor symmetries (as well as the angular momentum fugacity $q$).

It turns out that neither the ellipsoid partition function nor the index is completely fundamental. It was observed in \cite{Pasquetti-fact} that the ellipsoid partition functions of certain $\CN=2$ theories in the class described above can be expressed as sums of products,
\be \label{factSb}
\CZ_b(\mu,b)\,=\, \sum_\alpha B^\ra(x;q) B^\ra(\wt x,\wt q)\, =:\,  \big|\!\big| B^\ra(x;q) \big|\!\big|_S^2\,,
\ee
where, roughly speaking, each ``holomorphic block'' $B^\ra(x;q)$ is the partition function on a twisted product $\RqS$, labelled by a choice of vacuum $\alpha$ for the massive theory at the asymptotic boundary of spatial $\R^2$. Schematically, the holomorphic block is a ``BPS index'',
\be \label{BPSintro}
B^\ra(x;q)\,\sim\, \Tr_{\CH(\R^2;\alpha)} (-1)^R e^{-\beta H} q^{-J+\frac{R}{2}}x^e\,,
\ee
with fugacity $q$ for the angular momentum on $\R^2$ and fugacities $x$ for flavor symmetries.%
\footnote{Here $(-1)^R$ means $\exp(i\pi R)$, where $R$ is generator of $U(1)_R$. It is more familiar for indices to be written with $(-1)^F$ rather than $(-1)^R$, but both define a protected index and the two conventions are formally related by $q^{\frac12}\to -q^{\frac12}$. We will see that the latter is more appropriate for our purposes.}
In \eqref{factSb}, the complexified masses $\mu$ that enter $\CZ_b$ are related to these fugacities according to $x=\exp(2\pi b\mu)$, $\wt x=\exp(2\pi b^{-1}\mu)$; while $q=\exp(2\pi ib^2)$ and $\wt q=\exp(2\pi ib^{-2})$. A similar sum-of-products expansion was predicted for supersymmetric sphere indices in \cite{DGG-index},
\be
\label{factInd}
\CI(m,\zeta;q) = \sum_\alpha B^\ra(x;q) B^\ra(\wt x,\wt q)\, =:\,  \big|\!\big| B^\ra(x;q) \big|\!\big|_{id}^2~,
\vspace{-9pt}
\ee
where the relations to fugacities and fluxes on $S^2$ are $x=q^{\frac m2}\zeta$, $\wt x=q^{\frac m2}\zeta^{-1}$, and $\wt q = q^{-1}$.

In simple examples, it can be observed that the fundamental objects $B^\ra(x;q)$ appearing in both factorizations are \emph{identical}. The only difference in the two products is in the operation used to relate $(x;q)$ and $(\wt x;\wt q)$ when pairing up blocks. A principle aim of the present paper is to substantiate and elucidate the correspondence \eqref{factSb}--\eqref{factInd}. We conjecture that these factorizations, with identical holomorphic blocks, hold for any $\CN=2$ theory of the type described above, \ie\ any theory with sufficient flavor symmetry to render all its vacua massive. Our approach leads us to a new method for computing the holomorphic blocks $B^\ra(x;q)$ for any theory admitting a UV Lagrangian description, which makes this conjecture eminently testable. We take the first steps towards understanding a variety of surprising and remarkable properties of the blocks --- properties which are largely obscured by the simplicity of \eqref{factSb}--\eqref{factInd}.

Even superficial consideration of these relations suggests that they will yield physically significant insights. The supersymmetric index encodes the (index of the) spectrum of BPS operators in a SCFT, whereas the BPS index counts BPS states in a vacuum of the massive theory obtained by deforming away from the superconformal fixed point by relevant operators.%
\footnote{Along this line, some interesting extensions of the ideas presented in this paper were explored recently in \cite{IV-BPS, LV-NP}.} %
This is reminiscent of a similar correspondence for two-dimensional SCFTs \cite{CV-tt*}, and we will see that the connection to this work runs deeper than this basic similarity. Likewise, the ellipsoid partition function is known to encode important information about the R-charge assignments for the fields of the theory at the conformal point \cite{Jafferis-Zmin}, which apparently can also be recovered from an understanding of the BPS states in massive vacua of the deformed theory. Indeed, the use of supersymmetric localization to perform renormalization-group-invariant computations in a weakly coupled ultraviolet theory has been a general theme in work on ellipsoid and sphere index partition functions. Here, we are in some sense observing the reverse; computations in a ``trivial'' infrared theory allow us to recover interesting information about an interacting UV fixed point.

As an interesting corollary, our study of blocks for the three-manifold theories $T_M$ of \cite{DGG} produces the first concrete examples of non-perturbative path integrals in analytically continued Chern-Simons theory along ``exotic'' integration cycles. Namely, it follows from the work of \cite{Wit-anal, Witten-path, Wfiveknots, DGH} that the corresponding blocks should compute the analytically continued $SU(2)$ Chern-Simons path integral defined on integration cycles labeled by irreducible flat $\SLC$ connections $\CA^\alpha$ on $M$.%
\footnote{The extension to higher rank is also possible, but in this paper we assume that $T_M$ comes from wrapping two M5 branes on a hyperbolic knot complement $M$.} %
This is a non-perturbative completion of perturbative partition functions in the background of a flat connection $\CA^\alpha$ \cite{gukov-2003}, which plays a significant role in the physical interpretation of the Volume Conjecture \cite{kashaev-1997, Mur-Mur}. Our results can then be compared to other non-perturbative objects such as colored Jones polynomials.

We now spotlight some of the more interesting features of the holomorphic blocks that are studied in this paper.

\begin{figure}[t!]
\centering
\includegraphics[width=5in]{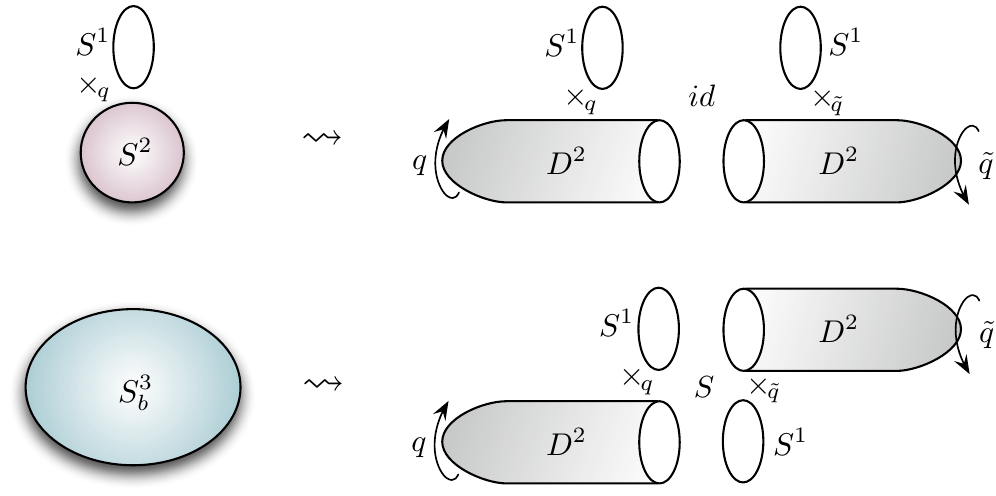}
\caption{Stretching $\SqS$ and $S^3_b$ into a union of semi-infinite $\DqS$ geometries.}
\label{fig:stretch}
\end{figure}

\subsubsection*{A stretch}

In light of the relations \eqref{factSb}--\eqref{factInd}, it is suggestive that both of the spaces $S^3_b$ and $\SqS$ admit Heegaard decompositions as the union of solid tori. In the case of $S^3_b$, the boundaries of the solid tori are identified using the $S$ element of the mapping class group $\SLZ$, together with a reversal of orientation. In the case of $\SqS$ it suffices to use the identity element $id$ together with the same orientation reversal. This accounts for our choice of notation in the norms-squared. Indeed, we even see that the relation between $q$ and $\wt q$ in the two cases corresponds to treating $q=e^{2\pi i\tau}$ as the modular parameter of the boundary of one solid torus, and sending
\be \label{tautransform}
\tau \to \wt \tau= -S\cdot \tau=\frac1\tau\qquad\text{or}\qquad\tau \to \wt\tau = -id\cdot\tau = -\tau~.
\ee
This is a little naive, because supersymmetric partition functions on finite-size solid tori do not obviously correspond to the blocks $B^\ra(x,q)$. Nor do solid tori cut from the $S^3_b$ and $\SqS$ geometries have the same metric, and these constructions are \emph{not} topological; so the solid-torus partition functions coming from the two splittings should not necessarily agree. Nevertheless, the picture is promising.

There is a deformation of the Heegaard splitting that precisely reproduces the factorized forms of \eqref{factSb}--\eqref{factInd}, with the correct relations between parameters $(x,q)$ and $(\wt x,\wt q)$. It is a three-dimensional analogue of the topological/anti-topological fusion setup of Cecotti and Vafa \cite{CV-tt*}, and is also related to recent constructions of Nekrasov and Witten \cite{NekWitten} (see also \cite{DG-Sdual, DGG-index}). 
To describe it, we represent both $S^3_b$ and $\SqS$ as $T^2$ fibrations over an interval with cycles of $T^2$ degenerating smoothly at the ends of the interval, and stretch the interval to infinite length (Figure \ref{fig:stretch}). Topologically, each half takes the form $\DqS$, where $D^2$ is a semi-infinite cigar. The halves are glued together using the appropriate element of the modular group.
On each half, we impose a metric that preserves $U(1)$ rotations of $D^2$ as an isometry and fibers $D^2$ over the remaining circle with a $U(1)$ holonomy $q$. The theory can then be topologically twisted on each half. The resulting geometry, sometimes called a ``Melvin cigar'' (\cf\ \cite{CNV}), will be denoted here by $\DqS$.

\begin{wrapfigure}{r}{2.25in}
\centering
\includegraphics[width=2.0in]{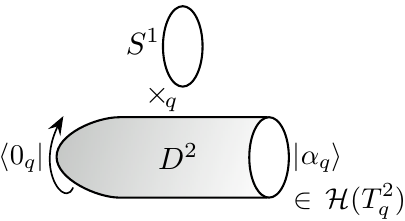}
\caption{A holomorphic block}
\label{fig:cigarstates}
\end{wrapfigure}
We define holomorphic blocks to be the partition function(s) of a theory on $\DqS$. On one hand, the topological twist allows us to deform the geometry to $\RqS$ without changing the partition function, which recovers a BPS index \eqref{BPSintro} that depends on a vacuum $\alpha$. On the other hand, we can interpret the partition function on $\DqS$ as a wavefunction $\langle 0_q|$ in the Hilbert space $\CH(T^2)$ defined in the flat asymptotic region $T^2\times \R$. The infinite Euclidean time evolution in this region projects the wavefunction to the space of exact supersymmetric ground states $|\alpha\rangle$ on $T^2$, which are in one-to-one correspondence with the vacua $\alpha$. It follows that the blocks
\be \label{defBintro}  B^\ra(x;q) := \langle 0_q|\alpha\rangle  \ee
are elements of a discrete and typically finite-dimensional vector space. 

Upon fusing two such semi-infinite geometries with an element $g=S$ or $g=id$ (say) of the modular group, the partition function takes the form
\be  \CZ = \langle 0_q|0_{\tilde q}\rangle  = \sum_\alpha \langle 0_q|\alpha\rangle \langle \alpha|0_{\tilde q}\rangle \sim \sum_\alpha B^\ra(x;q)B^\ra(\wt x;\wt q)\,.
\vspace{-3pt}
\ee
The precise identification of parameters on the two halves, as well as the relation between topological twists, depends on $g$. We will find after a more careful analysis of background field configurations that fusion with the identity precisely reproduces the index identifications \eqref{factInd}, while $S$-fusion reproduces the $S^3_b$ identifications \eqref{factSb}. We will argue that wavefunctions $\langle 0_q|\alpha\rangle$ and $\langle \alpha|0_{\tilde q}\rangle$ on the two sides can be written, in an appropriate sense, in terms of the same holomorphic objects $B^\ra(x;q)$.%
\footnote{Fusion with a general element $g\in SL(2,\Z)$ is expected to produce the partition function of an ellipsoidally deformed lens space. The details are not studied in this paper, but appear in subsequent works \cite{IMY-fact, D-levelk}.}

We have arrived at a stronger, geometric version of the factorization conjecture. The ability to deform $S^3_b$ or $\SqS$ into a union of two copies of the $\DqS$ geometry while leaving the partition functions invariant would imply factorization. The stronger conjecture is that such ``$Q$-exact'' deformations do indeed exist. It is plausible that $Q$-exactness could be established using methods of supersymmetric localization. Several $Q$-exact deformations of three-spheres have already been found \cite{HHL, MartelliSparks}, and they come close to reproducing the stretched $\DqS$ geometries (but no cigar). In two dimensions, an analogous deformation leading to a topological/anti-topological fusion geometry was recently studied in \cite{GomisLee}.

In two and four dimensions, it is a direct consequence of localization that $S^2$ partition functions of theories with $\CN=(2,2)$ supersymmetry \cite{BC-S2, DGFL-S2} and $S^4$ partition functions of theories with $\CN=2$ supersymmetry \cite{Pestun-S4} factorize as a sum (integral) of vortex (instanton) partition functions, respectively:
\be \label{S2S4}
\CZ[S^2] \sim \sum_{\rm vacua} \big|\CZ_{\rm vortex}[\R^2]\big|^2\,,\qquad
 \CZ[S^4] \sim \int_{\rm vacua} \big| \CZ_{\rm inst}[\R^4]\big|^2\,.
\ee
Factorization of the three-dimensional index into holomorphic blocks \eqref{factInd} leads directly to the two-dimensional factorization \eqref{S2S4} of a dimensionally reduced theory in an appropriate limit. As for the analogue in four dimensions, it was observed in \cite{KKL-5dindex} that five-dimensional indices on $S^4\times S^1$ factorize in a way that naturally extends the known $S^4$ factorization.

\subsubsection*{Block integrals}

Our main computational tool is a new integral formula for the blocks, applicable whenever a theory has a UV Lagrangian description as an $\CN=2$ gauge theory. Just like ellipsoid partition functions or indices, blocks are insensitive to renormalization group flow. The formula is motivated by the reduction of the geometry $\DqS$ to supersymmetric quantum mechanics on a half-line.

We begin with the observation that the theory $T_\Delta$ of a free chiral multiplet possesses a single block given by
\be \label{BDintro}
B_\Delta(z;q) = \sum_{n=0}^\infty \frac{z^{-n}}{(1-q^{-1})\cdots(1-q^{-n})}\,.
\ee
where $z$ is the complexified and exponentiated real mass of the chiral. We determine general block integrals to take the schematic form
\be \label{BIintro}
B^\ra(x;q)\;\sim\; \int_{\Gamma^\alpha} \frac{ds}{s}\,\big[\theta(z;q)...\big]B_\Delta(z_1;q)\cdots B_\Delta(z_N;q)\,.
\ee
The integral is over a middle-dimensional cycle $\Gamma^\alpha \subset (\C^*)^r$, where $r$ is the rank of the gauge group. The variables $s\in (\C^*)^r$ are complexified scalars in the gauge multiplets, and each chiral multiplet contributes a factor $B_\Delta(z_i;q)$ to the integrand, where the effective mass $z_i=z_i(s,x)$ may depend on scalars $s$ and non-dynamical real masses $x$. The W-bosons in nonabelian gauge multiplets also contribute factors $B_\Delta(s;q)$ to the denominator. The extra theta-functions $\theta(z;q)$ encode contributions of Chern-Simons and Fayet-Iliopoulos (FI) terms.

This integral mimics the matrix-integral formulas for partition functions $\CZ_b$ and $\CI$ that were derived using localization \cite{Kapustin-3dloc, HHL, IY-index}, yet there is a crucial difference. The universal integrand of \eqref{BIintro} gives rise not just to a single partition function, but to many blocks $B^\ra(x;q)$. The various blocks arise for different choices of integration contour, $\Gamma^\alpha$. Each contour is associated to a critical point of the integrand, which in turn is related to a supersymmetric ground state on $T^2$ --- just as one would expect from quantum mechanics \cite{Witten-Morse, HIV, Witten-path}. Some technical subtleties arising from the nontrivial singularity structure of the functions $B_\Delta(z;q)$ must be dealt with to make this statement more precise and useful.

The integrand of \eqref{BIintro} --- henceforth denoted $\Upsilon(x,s;q)$ --- turns out to be a factorized form of the matrix integrands for ellipsoid partition functions and indices. In fact, this is another way to characterize it. In particular, it will turn out that
\be  \CZ_b = \int_{\R^r} d(\log s)\, \big|\!\big| \Upsilon(x,s;q)\big|\!\big|_S^2\,,\qquad
 \CI = \int_{(S^1)^r} \frac{ds}{2\pi is} \,\big|\!\big| \Upsilon(x,s;q)\big|\!\big|_{id}^2\,.
\ee
Combined with the factorization conjecture, this has the rather beautiful consequence that
\be   \sum_\alpha \bigg|\!\bigg| \int_{\Gamma^\alpha} \frac{ds}{s}\, \Upsilon(x,s;q)\bigg|\!\bigg|_g^2 = \big|\!\big| B(x;q)\big|\!\big|_g^2  =  \int_{\CX[g]} \frac{ds}{s}\, \big|\!\big| \Upsilon(x,s;q)\big|\!\big|_g^2\,,  \ee
for $g=S$ or $g=id$, with $\CX[id]=\R^r$ and $\CX[S] = (S^1)^r$, and with appropriately normalized integration measures. This appears to be a sort of Riemann bilinear relation for the ellipsoid and index partition functions. Physically, this amounts to the statement that fusion of blocks commutes with gauging of symmetries. Fusion also commutes rather trivially with other operations one can perform on SCFTs, such as adding background Chern-Simons levels or superpotentials, which can be combined with the gauging of global symmetries to generate interesting symplectic actions \cite{Witten-SL2, DGG, DGG-index}.

Closely related to blocks and block integrals are a set of $q$-difference equations satisfied by the blocks of a given theory,
\be \label{Wardintro}
\qquad \hat f_i(\hat x,\hat p;q)\cdot B^\ra(x;q)=0\,,\qquad (\hat p\hat x=q\hat x\hat p)\,.
\ee
These equations are a consequence of identities in the algebra of line operators acting at the tip of $\DqS$, as discussed in \cite{DG-Sdual, DGG, DGG-index}.
The identities can be systematically derived for theories with UV Lagrangian descriptions.
The space of blocks can then be described as the vector space of solutions to \eqref{Wardintro} that satisfy certain analytic requirements --- for example that they be meromorphic functions of $q^{\frac12}$ and $x$, with no branch cuts.
The block integral for a theory can be constructed to generate these solutions, much as formal integrals are often used to generate solutions to differential equations (\cf\ \cite{Zeil-AB}) and path integrals manifestly generate solutions to QFT Ward identities. The convergent cycles $\Gamma^\alpha$ are chosen so that the integral solves \eqref{Wardintro}, and a basis of cycles produces a basis of solutions.

When combining conjugate blocks to form the ellipsoid and index partition functions, the number of line-operator identities effectively doubles, %
\be \label{opNintro}
\hat f_i(\hat x,\hat p;q) \cdot \CZ = \hat f_i(\hat {\wt x},\hat {\wt p};\wt q) \cdot \CZ=0\,.
\ee
This is due to of the presence supersymmetric line operators acting at both ends of a stretched geometry (\cf\ \cite{NekWitten}). The requirement that the two sets of operators $(\hat x,\hat p)$ and $(\hat {\wt x},\hat {\wt p})$ commute with each other puts an interesting constraint on the classical relation between $(x,q)$ and $(\wt x,\wt q)$ in a glued geometry, which is indeed satisfied for $S^3_b$ and $\SqS$. Conversely, the observed fact that these partition functions satisfy \eqref{opNintro} for commuting sets of operators strongly suggests that they must factorize as in \eqref{factSb}--\eqref{factInd}. This was the primary motivation behind the prediction of factorization for the index in \cite{DGG-index}.

\subsubsection*{Connections to other topics}

Holomorphic blocks and their fused counterparts have many relations to other constructions in quantum field theory and string theory. We highlight three of them here.

First, as was already pointed out, there is a striking similarity between the gluing of blocks and topological/anti-topological $(tt^*)$ fusion \cite{CV-tt*}. Indeed, the gluing of blocks might be considered a three-dimensional lift of the $tt^*$ setup. The latter construction considers massive $\CN=(2,2)$ theories in two dimensions on a topological two-sphere that has been stretched out into a pair of cigars, $S^2\simeq D^2 \cup D^2$, with (anti-)chiral operators inserted at the north (south) poles. The resulting partition functions obey a set of differential equations --- which determine the partition functions almost completely --- and exhibit properties of special geometry \cite{Str-SG}. The difference equations \eqref{Wardintro} may be thought of as three-dimensional lifts of (part of) the $tt^*$ differential equations. The full three-dimensional story is in some sense richer than in two dimensions, in part because the block geometry can be glued in a variety of topologically distinct ways. We only scratch the surface of the relation between our analysis and the $tt^*$ equations, and there are some notable differences, \eg, the analogous blocks in two dimensions are not generally holomorphic. We expect further investigation of these connections to be fruitful.

Another deep connection --- one which we do not explore extensively in this work --- is to topological string theory. This was pointed out in the original work of \cite{Pasquetti-fact}. Indeed, for a choice of three-dimensional theory that can be engineered in M-theory by wrapping M5 branes on a Lagrangian submanifold of a non-compact Calabi-Yau three-fold, the open A-model partition function in that background is known to compute the BPS index of the theory \cite{GV-I, GV-II, OV, DVV}. Consequently, for theories that arise in such a fashion, we expect 
\be B^\ra(x;q)\,\sim\, \CZ^{\rm top}_{\rm open}(Y;\CL)\,. \label{Btopintro}\ee
The choice of ground state, or vacuum, is usually called a choice of ``phase'' for the brane in the topological string literature. In cases where $\CZ^{\rm top}_{\rm open}$ can be computed, such as for toric branes in local, toric Calabi-Yau threefolds, the relationship can be verified modulo prefactors related to background Chern-Simons couplings (\ie\ Chern-Simons couplings for flavor symmetries), which are crucial in correctly computing $\CZ_b,\,\CI,$ and $B^\ra$.\footnote{For a recent analysis of background Chern-Simons couplings in curved superspace, see \cite{CDFKS, CDFKS-CS}.} Indeed, many of the features of blocks that we encounter here have made an appearance previously in the context of topological string partition functions. For instance, the contour integrals we prescribe for computing holomorphic blocks can be interpreted as non-perturbative completions of the contour integrals that appeared as generalized Fourier transforms of brane wave functions in \cite{adkmv} (see also \cite{KP-wavefunction, CDV-NP}). Furthermore, the line-operator identities that annihilate holomorphic blocks generalize the ``quantum Riemann surface'' which appears in the topological $B$-model on certain geometries \cite{dhsv}. Additionally, a kind of factorization in terms of these open topological string amplitudes has appeared in the context of the open OSV conjecture \cite{ANV, AS-OSV}.

In approaching the problem of blocks from the point of view of gauge theory, we are led to a slightly different perspective on these objects than was natural in the topological string setup. In particular, it is important for the blocks in this paper to be described as holomorphic functions of their parameters in order to make the connection with ellipsoid and index partition functions. Furthermore, the requirement of invariance under large gauge transformations leads to certain differences in the treatment of Chern-Simons terms, including background couplings. It would be extremely interesting to more thoroughly investigate whether these modifications can find a natural interpretation in topological string theory.

Finally, when the theory in question is a three-manifold theory $T_M$ \cite{DGH, DGG}, the blocks correspond to certain non-perturbative path integrals $\CZ_{\rm CS}^\ra(M)$ in analytically continued Chern-Simons theory on $M$. In particular, when the theory arises on a stack of $K$ M5 branes, we expect analytically continued $SU(K)$ Chern-Simons theory. The relation is easiest to understand using the six-dimensional constructions of \cite{Wfiveknots}, which identify independent integration cycles in analytically continued Chern-Simons theory, labelled by a flat $SL(K,\C)$ connection $\CA^\alpha$ on $M$, with BPS indices of the form \eqref{BPSintro}. The Chern-Simons coupling $k$ (complexified and renormalized) is encoded in the twist parameter $q=\exp\big(\frac{2\pi i}{k}\big)$, while boundary conditions for flat connections on the manifold $M$ become flavor fugacities. For example, if $M$ is a knot complement, the eigenvalues of the monodromy of a connection $\CA^\alpha$ at the excised knot are complexified masses.

We can test the correspondence $B^\ra_{T_M}\sim \CZ^\ra_{\rm CS}(M)$ perturbatively by verifying that the ``classical'' asymptotics
\be \label{Volintro}
B^\ra_{T_M}(x;q)\overset{q\to 1}{\sim} \exp\bigg[ \frac{1}{\log q}\CV(M;x;\alpha) +\ldots\bigg]~,
\ee
correctly reproduce the complex volume of a corresponding flat $SL(K,\C)$ connection $\CA^\alpha$ on $M$. In fact, just as in \cite{DGH,DG-Sdual,DGG,DGG-index}, we can argue that this \emph{must} be the case up to a normalization factor because the identities \eqref{Wardintro} for $T_M$ are equivalent to the ``quantum A-polynomial'' equations in Chern-Simons theory \cite{garoufalidis-2004, gukov-2003, Gar-Le}. Both sets of equations almost completely fix the asymptotic expansion \eqref{Volintro} \cite{DGLZ}.

Non-perturbatively, we should compare blocks of a knot-complement theory with colored HOMFLY polynomials $J$ of the knot itself, which are expected to take the form \cite{Wit-anal}
\be\label{HOMintro}
J \sim \sum_\alpha n_\alpha\, \CZ^\ra_{\rm CS}~, 
\vspace{-6pt}
\ee
for appropriate coefficients $n_\alpha$. Unfortunately, we face an important caveat: the theories defined in \cite{DGG} for $K\!=\!2$ only probe \emph{irreducible} $\SLC$ connections on $M$. This is a consistent truncation in analytically continued Chern-Simons theory, but it means that the blocks computed in gauge theory will never correspond to all terms in the sum \eqref{HOMintro}, as they miss reducible flat connections. To circumvent this problem, we take a special limit of knot polynomials, first observed to exist in \cite{GukovZagier, DasbachLin} and later termed ``stabilization'' \cite{Garoufalidis-slopes, GarLe-Nahm}, which seems to project out reducible connections. After taking this limit we find a match in all examples studied.

We should remark that the purpose of relating Chern-Simons path integrals to BPS indices in \cite{Wfiveknots} --- and also in the earlier and very similar approach of \cite{GSV} --- was to provide a physical categorification of knot and three-manifold invariants. Categorification amounts (in part) to replacing an index such as \eqref{BPSintro} with a full Hilbert space of states $\CH(\R^2;\alpha)$ upon which a conserved supercharge acts. In the context of holomorphic blocks, such categorification is likely to lead to new knot homologies --- associated not just to a knot, but to a choice of flat connection in its complement. This is a very interesting subject for future study, and we hope that it will eventually connect to recent work of \cite{FGS-superA, FGS-VC, FGSS-AD}.

\subsubsection*{A jump}

The last major aspect of our work concerns the behavior of $\DqS$ partition functions globally in parameter space. Typically, we will fix $q$ and vary the masses $x$, whereupon we find that holomorphic blocks are subject to Stokes phenomena. That is, the blocks $B^\ra(x;q)$ associated to vacua $\alpha$ in one chamber of parameter space may be related to blocks in a different chamber by a linear transformation,
\be \label{Stokesintro}
\qquad B^\ra \,\longrightarrow\, \sum_\beta M^\alpha_{\;\;\beta} B^\beta\,,\qquad M^\alpha_{\;\;\beta}\in GL(N,\Z)\,.
\vspace{-3pt}
\ee
Such behavior is not too surprising. In the description of blocks \eqref{defBintro} as coming from long cigars, the map between vacua $\alpha$ and supersymmetric ground states $|\alpha\rangle$ can change as parameters are varied. While ground states generically do not mix in a theory with four supercharges, on special loci in parameter space instanton configurations may connect two ground states and lead to a jump such as \eqref{Stokesintro}. Alternatively, this can be described in terms of brane nucleation \cite{HIV}. A similar Stokes phenomenon plays a central role in analytically continued Chern-Simons theory \cite{Wit-anal}. When blocks arise from a finite-dimensional block integral such as \eqref{BIintro}, jumps can be analyzed explicitly using Lefschetz theory for cycles associated to critical points.

The fact that blocks transform as \eqref{Stokesintro} when passing from one chamber to another raises an interesting puzzle. The curved-space partition functions of a theory such as $\CZ_b$ and $\CI$ should not depend on any choice of chamber; yet expressions \eqref{factSb}--\eqref{factInd} do not look invariant under $B^\ra\to (MB)^\ra$. The resolution of the puzzle involves two observations. 

First, we find that blocks $B^\ra(x;q)$ can be expressed as $q$-hypergeometric series that converge both for $|q|<1$ and $|q|>1$, but to two different functions in the two regimes. For example, the free chiral block \eqref{BDintro} takes two different forms 
\be \label{qintro}
B_\Delta(z;q)=\begin{cases} \prod_{n=1}^\infty (1-q^nz^{-1}) & |q|<1 \\[.1cm]
\prod_{n=0}^\infty (1-q^{-n}z^{-1})^{-1} & |q|>1\,,
\end{cases}
\ee
with no analytic continuation across the unit circle. Physically, this arises from a subtlety in our definition of blocks: we use a topologically twisted geometry for $|q|>1$ and an anti-topologically twisted geometry for $|q|<1$. The effect is roughly that bosonic modes contribute in one regime and fermionic modes in the other, switching products from numerator to denominator in expressions such as \eqref{qintro}.

In addition, due to the reflection used in any fusion of $\DqS$ geometries, the twist parameters for the two sides always live on opposite sides of the unit circle. That is, $|\wt q|>1$ whenever $|q|<1$, and vice versa. This is just as we want it for 3d topological/anti-topological fusion. It turns out for the cases that we study that blocks on the two sides of the unit circle have \emph{complementary} transformations at Stokes walls, \eg
\be B(x;q)\,\to\,MB(x;q)\,,\qquad B(\wt x;\wt q)\,\to\, M^{-1\,T} B(\wt x;\wt q)\,,\ee
so that products $\big|\!\big|B(x;q)\big|\!\big|_g^2$ remain invariant. Nevertheless, in every chamber, the blocks at $|q|<1$ and $|q|>1$ agree, in the sense of sharing convergent $q$-hypergeometric series expansions.

We conjecture that this is the case in general. While the presence of conjugate Stokes matrices can be argued directly from the form of block integrals, the statement about sharing series expansions in every chamber is highly nontrivial, and implies very special mathematical properties for the blocks themselves. We will check such behavior in detail for the simplest nontrivial example, the three-dimensional analogue of the $\cp^1$ sigma-model, and discover identities for $q$-Bessel functions that govern the transformations of its blocks.

It is interesting to note that, unlike the index, the physical ellipsoid partition function $\CZ_b$ should be defined for $b^2$ on the positive real axis, implying that $q=e^{2\pi ib^2}$ and $\wt q=e^{2\pi i/b^2}$ are on the unit circle itself. Ellipsoid partition functions appear to have the remarkable property that they can be analytically continued to the cut plane $b^2\in \C\bs\R_{\leq 0}$, and on both the upper and lower half-planes agree with the same product of blocks $\big|\!\big|B(x;q)\big|\!\big|_S^2$. Conversely, we find in examples that products $B^\ra(x;q)B^\ra(\wt x;\wt q)$ for any fixed $\alpha$, with $(x,q)$ and $(\wt x,\wt q)$ identified by the $S$ transformation, can be analytically continued in $b^2$ across the positive real axis, defining a single function on $\C\bs\R_{\leq 0}$.
This surprising property has already been observed for the free-chiral block \eqref{qintro}, in which case the $S$-fusion product is a non-compact quantum dilogarithm \cite{Barnes-QDL, Fad-modular}, \cf\ \cite[Sec. 3.3]{DGLZ}. \vspace{20pt}

The organization of this paper is as follows. In Section \ref{sec:3dintro}, we provide a more careful definition of holomorphic blocks and revisit the geometry of fused $\DqS$ partition functions, aiming to understand the parameters in the products \eqref{factSb}--\eqref{factInd}.
In Section \ref{sec:SQM}, we compactify $\DqS$ to a half-line, and describe aspects of the resulting supersymmetric quantum mechanics. In Section \ref{sec:blockint}, we combine results from quantum mechanics with an understanding of line-operator identities to define block integrals \eqref{BIintro}, and demonstrate how to these integrals can be evaluated in simple examples. This is followed in Section \ref{sec:CP1} by an in-depth study of blocks, Stokes phenomena, and mirror symmetry in the three-dimensional $\cp^1$ sigma model. In Section \ref{sec:CS}, we review the connection to analytically continued Chern-Simons gauge theory in the case of a three-manifold theory $T_M$.


\section{A first look at holomorphic blocks}
\label{sec:3dintro}

The theories under consideration are three-dimensional superconformal field theories with $\cN=2$ supersymmetry and a conserved $U(1)_R$ R-symmetry. However, we will typically work with ultraviolet $\cN=2$ gauge theories that have Lagrangian descriptions and flow to the desired SCFTs in the infrared. The observables we are interested in will be invariant under the flow. Let us therefore review the ingredients that enter into the Lagrangian of such a gauge theory. (For a more complete discussion, see \cite{AHISS}.)

We consider theories whose Lagrangians are written in terms of a set of $r$ gauge multiplets $\{V_a\}$, and chiral matter multiplets $\{\Phi_I\}$, which are the dimensional reductions of the usual $\cN=1$ vector and chiral multiplets in four dimensions. We assume for the moment that gauge symmetry is abelian. In three dimensions, the vector multiplet can be reorganized into a linear multiplet $\Sigma_a = \epsilon^{\alpha\beta}\overline{D}_{\alpha}D_{\beta}V_a$, in terms of which the canonical kinetic Lagrangian takes the following simple form,
\be
\cL_{\text{kinetic}} = \int d^4\theta\left(\,\sum_{a=1}^r\tfrac{1}{e_a^2}\Sigma_a^2+\sum_{I} \Phi_I^{\dagger}\exp\Big(\sum_{a} Q_I^aV_a\Big)\Phi_I\right)~.
\ee
In addition, one may include as an F-term a holomorphic, gauge-invariant superpotential
\be
\cL_{\text{F-term}} = \int d^2\theta\;W(\Phi)+\mathit{h.c.}
\ee
We assume that the superpotential preserves an R-symmetry $U(1)_R$.

The terms introduced so far will preserve some global symmetries. These include symmetries that act manifestly upon the fields in the Lagrangian, as well as ``topological'' $U(1)$ symmetries that act as shifts of the dual photons for any abelian gauge multiplets. Consider a maximal abelian subgroup $\prod_{i=1}^N U(1)_i$ of the full flavor symmetry group.
We can then introduce $N$ non-dynamical background fields $A_\mu^{(i)}$ that couple to the conserved $U(1)_i$ currents, which can be further promoted to background vector superfields $\wh V_i$, with corresponding linear multiplets $\wh \Sigma_i$.
Setting the real scalar components of $\wh\Sigma_i$ to non-zero values $m_i^{\rm 3d}$ turns on real mass deformations of the theory.
Such a deformation for an ordinary flavor symmetry appears in the kinetic terms of the Lagrangian as
\be
\cL_{\text{kinetic}} = \int d^4\theta\left(\Phi^{\dagger} e^{m^{\rm 3d}\theta\bar\theta}\Phi\right)~,
\ee
which, in terms of component fields, leads to mass terms $(m^{\rm 3d})^2|\Phi|^2+i m^{\rm 3d} \epsilon^{\alpha\beta}\bar\psi_{\alpha}\psi_{\beta}\,$. When real mass terms are turned on for topological $U(1)$ symmetries, they appear as Fayet-Iliopoulos (FI) terms for the corresponding dynamical gauge field. In this paper, we collectively denote all real mass parameters as $m_i^{\rm 3d}$, whether they correspond to masses for chirals or to FI terms. In the infrared, they are all on the same footing. We can similarly introduce a non-dynamical background gauge field $A_R$ for the R-symmetry, part of a different supermultiplet; it plays a special role in supersymmetric compactifications on curved spaces.

In three-dimensions one can also include gauge-invariant Chern-Simons interactions. The most general abelian interaction takes the form
\be  
\cL_{\text{CS}} = \int d^4\theta\left(
\tfrac{1}{2}  k_{ab} \Sigma_a V_b +
k_{ia}\widehat\Sigma_a V_i+
\tfrac{1}{2} k_{ij}\widehat\Sigma_i\widehat V_j
\right)\,. 
\ee
The first term is a Chern-Simons interaction for the dynamical abelian gauge fields, while the middle term encodes Fayet-Iliopoulos terms for the dynamical gauge fields, and the last term describes purely background Chern-Simons terms (which are related to the choice of contact terms for conserved current multiplets -- see \cite{CDFKS, CDFKS-CS}). Gauge invariance will sometimes require the inclusion of fractional Chern-Simons terms, the so-called ``parity anomaly'' of three-dimensional gauge theories. This is due to the fact that integrating out charged fermions can shift the effective Chern-Simons matrix according to
\be
(k_{ij})_{\rm eff} = k_{ij}+\tfrac12\sum_{\rm fermions}(q_f)_i(q_f)_j\,{\rm sign}(m_f)~.
\ee
The resulting Chern-Simons levels must be integers. 
It will be important to keep close track of all types of Chern-Simons interactions in order to correctly compute holomorphic blocks for a gauge theory.

Finally, we require theories to have enough flavor symmetry so that real mass deformations completely lift all flat directions in the moduli space (\eg\ all Higgs and Coulomb branches), rendering the theories  massive. More importantly, we demand that after reduction to two dimensions on a circle, the theories at generic values of mass parameters have only discrete, massive vacua. This will be made explicit in Section \ref{sec:bdy-inf}.


\subsection{Cigar compactification}
\label{sec:cigar}

The observable of interest for these gauge theories is the partition function on $\DqS$. Topologically, this geometry is a (noncompact) solid torus with local coordinates $(r,\varphi,\theta)$, where $r\in[0,\infty)$ and $\varphi$, $\theta$ are both periodic with period $2\pi$. The metric is given by
\be\label{cigarmetric}
ds^2 = dr^2 +f(r)^2(d\varphi + \varepsilon \beta d\theta)^2+\beta^2 d\theta^2~,
\ee
where $f(r)\sim r$ near $r=0$ and $f(r)\to\rho$ as $r\to\infty$ (for example, one may take $f(r) = \rho\tanh(r/\rho)$). The cigar parameterized by $(r,\varphi)$ has asymptotic radius $\rho$ and is fibered over the $\theta$-circle so that the cigar rotates by an angle $2\pi\beta\ve$, or alternatively, so that the holomorphic variable $z = r e^{i\varphi}$ is identified around the $\theta$ circle according to $(z,0)\sim(q^{-1}z,2\pi)$, where
\be\label{qdef}
q = e^{2\pi i \varepsilon\beta}=e^{\hbar}~.
\ee

This metric admits no covariantly constant spinors, so in order to preserve supersymmetry we twist the theory. For a generic curved three-manifold, one would need at least $\cN=4$ supersymmetry in three dimensions to define conserved, twisted supercharges. However because the curvature of \eqref{cigarmetric} is valued in $U(1)_E$ (rotations of the tangent space to the cigar fiber), a twisted superalgebra exists for a theory with only $\cN=2$ supersymmetry in three dimensions as long as the theory possesses a $U(1)$ R-symmetry. There are two choices for how to twist the theory, one ``topological'' and one ``anti-topological''. These different choices preserve twisted scalar supercharges $(Q_-,\overline Q_+)$ or $(Q_+,\overline Q_-)$, respectively, where $\pm$ denotes the charge of the operator under $U(1)_E$ (see Appendix \ref{app:3dSUSY} for our conventions). From the perspective of the cigar, this is an $A$-type twist, \cf\ \cite{CV-tt*, Witten-phases}.

In order to implement these partial twists, we introduce a non-trivial profile for some of the non-dynamical background vector fields described above. In particular, for the background field coupling to the R-symmetry of the theory we impose
\be\label{twistbackground}
A_{\mu}^R = A_{0\mu}^R \pm \frac{1}{2}\omega_\mu~,
\ee
where on the right hand side, $\omega_\mu$ represents the $U(1)$-valued spin connection for the metric \eqref{cigarmetric} (its nonvanishing components describe rotations in the tangent bundle to the cigar $D^2$), and $A_0^R$ is a flat connection with holonomy $\exp(i\oint A_R) = e^{\pi i}$ around the non-contractible cycle $S^1_\beta$. The plus sign in \eqref{twistbackground} corresponds to the topological twist, and the minus sign to anti-topological. Note that theories constructed in the UV have no canonical choice of $R$-symmetry in the presence of conserved abelian flavor symmetries. We will usually take the R-symmetry to be such that all fields have integer charges.%
\footnote{This is natural, for example, when the theory in question is viewed as a boundary condition for a four-dimensional $\CN=2$ theory, in which case $U(1)_{R}^{d=3}$ is embedded into $SU(2)_{R}^{d=4}$, \cf\ \cite{DGG-index}. This perspective will play a role in our understanding of line-operator identities for holomorphic blocks.}

Along with the R-symmetry, we are free to couple any conserved flavor current to a line bundle with connection of the form
\be \label{flavorA}
A_{\rm flavor} = A_0 + \kappa\,\omega~,
\ee
where $A_0$ is flat $(dA_0=0)$ and $\kappa$ is any real number. The flat connection $A_0$ is characterized by its holonomy around the non-contractible cycle $S^1_\beta$, which we define to be $e^{2\pi i\vartheta}$, while its holonomy about the contractible cycle is always trivial:
\be
\frac{1}{2\pi}\oint_{S^1_{\beta}}A_0 =:\vartheta~,\qquad \oint_{S^1_\rho}A_0 = 0~.
\ee

\begin{figure}[t!]
\centering
\includegraphics[width=2.9in]{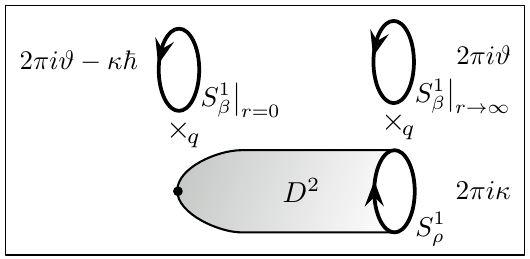}
\caption{Holonomies $i\oint A$ for any connection of the form $A=A_0+\kappa\omega$ around various cycles in the $\DqS$ geometry.}
\label{fig:holonomies}
\end{figure}

It will be useful to record some of the holonomies of the spin connection in this geometry. Since $\DqS$ is not flat, it matters where the holonomies are measured, the most relevant points being at the tip ($r=0$) and in the flat, asymptotic region ($r\to\infty$). We find
\be
\frac{1}{2\pi}\oint_{S^1_{\beta}|_{r=0}}\hspace{-13pt}\omega = -\beta\ve~,\qquad \oint_{S^1_{\beta}|_{r\to\infty}}\hspace{-19pt}\omega=0~,\qquad \frac{1}{2\pi}\oint_{S^1_{\rho}}\omega = 1~.
\ee
where $S^1_\rho$ always refers to the cigar circle in the asymptotic region. Therefore, the holonomies of any connection of the form \eqref{flavorA}, which mixes a flat connection $A_0$ with a multiple of $\omega$, will take the values shown in Figure \ref{fig:holonomies}. In particular, in the case of the connection $A_R$, Figure \ref{fig:holonomies} applies with $\vartheta=\kappa=\pm1/2$. As long as all fields in our theory have integral flavor and R-charge assignments, all holonomies $i\oint A_{\rm flavor}$ and $i\oint A_R$ are invariantly defined modulo $2\pi i$.

Because the geometry is non-compact, the above choices of parameters must be supplemented by a choice of asymptotic boundary conditions at $r\to\infty$. We can describe this choice of boundary conditions in two superficially different but equivalent ways. At fixed $\rho$, the geometry is macroscopically one-dimensional; the whole construction appears as a half-line. Consequently, an appropriate asymptotic boundary condition is to fix the fields to sit in a vacuum of the effective one-dimensional quantum mechanics that results from reduction on an appropriate two-torus. Alternatively, because of the partial twist, the cigar partition function is invariant under changes of the asymptotic radius $\rho$. Thus, we can take the limit $\rho\to\infty$, in which case the geometry becomes approximately $\RqS$, and an appropriate boundary condition is given by a choice of vacuum of the resulting two-dimensional theory. These two descriptions of the boundary conditions are in fact equivalent \cite{NekWitten}.

At large $\rho$, it is natural to describe the resulting partition function as a BPS index, which counts states on the cigar (or, roughly, on $\R^2$, which is the large $\rho$ limit of the cigar geometry) that are annihilated by the two supercharges preserved in the compactification.
We see from holonomies of the various background fields at the \emph{origin} of the cigar that the partition function on $\DqS$ can schematically be written as
\be \label{ZBPS-0}
\Tr_{\CH(D;\alpha)} (-1)^{R} e^{-2\pi \beta H} q^{-J\mp\tfrac{R}{2}}\exp\left({ie\oint_{S^1_\beta|_{r=0}}\hspace{-14pt}A_{\rm flavor}}\right)\,, 
\ee
where $J$ is the generator of $U(1)_E$, $R$ is the generator of the $U(1)$ R-symmetry, and $e=(e_1,...,e_N)$ are the generators (charges) of the abelian flavor symmetries with connections $A_{\rm flavor}=(A_1,...,A_N)$. We have indicated the dependence on the vacuum in which the index is evaluated with the label $\alpha$. The choice of sign in $q^{\mp R/2}$ matches that in \eqref{twistbackground}, and corresponds to topological versus anti-topological twisting.

A more familiar expression for this trace would involve $(-1)^F$ rather than $(-1)^R$. Here, the difference arises from implementing anti-periodic boundary conditions on fermions via a Wilson line for the R-symmetry.
For the purposes of computing a protected index, both $R$ and $F$ are equally good ``fermion numbers'' (the action of all supercharges shifts them by $\pm 1$). Indeed, we can change a $(-1)^R$ index to a $(-1)^F$ index simply by replacing $q^{\frac12}\leftrightarrow -q^{\frac12}$, so the two indices contain identical information.

A more substantial issue is that the Hamiltonian appearing in the above trace is not $Q$-exact for any supercharge.  This is easy to fix, and in the process we learn which variables the index depends on holomorphically. We note that the supersymmetry algebra with $Q=Q_\mp$ is $\{Q,Q^\dagger\}=2(H\mp Z)=:\tfrac 1\pi H_\pm$, where $Z$ is the real central charge (see Appendix \ref{app:3dSUSY}). In the present setting, the central charge of a state with flavor charge $e$ is given simply by $Z=e\cdot m^{\rm 3d}$, with $m^{\rm 3d}=(m^{\rm 3d}_1,...,m^{\rm 3d}_N)$ being the real mass deformations associated to the flavor symmetries. Therefore, we can write
\be 
e^{-2\pi \beta H}\exp\left({ie\oint_{S^1_\beta|_{r=0}}\hspace{-14pt}A_{\rm flavor}}\right) = e^{-\beta H_+}  x_+^{-e} = e^{-\beta H_-}x_-^e \,,
\ee 
where we have introduced the \emph{complexified} fugacities 
\begin{align} \label{defx}
x_\pm &
= \exp(X_\pm) 
= \exp\left(2\pi\beta m^{\rm 3d}\mp i\oint_{S^1_\beta|_{r=0}}\hspace{-14pt}A_{\rm flavor}\right) = \exp\big(2\pi \beta m^{\rm 3d}\mp (2\pi i\vartheta-\kappa\hbar)\big)~.
\end{align}
The logarithmic variables $X_\pm$ can be thought of as two-dimensional twisted masses, rescaled to be dimensionless. They are naturally periodic.
Using this substitution, we can interpret the partition functions on $\DqS$ as indices, with $Q$-exact Hamiltonians:
\be \label{ZBPS}
\begin{array}{@{\qquad}rl@{\qquad}c}
\CZ_{\rm BPS}^\ra(x_+;q)  &=  \Tr_{\CH(D;\alpha)} (-1)^R e^{-\beta H_+} q^{-J-\frac R2} x_+^{\;-e}  & \text{(topological)}~, \\[.1cm]
\CZ_{\ol{\rm BPS}}^\ra(x_-;q)  &=  \Tr_{\CH(D;\alpha)} (-1)^R e^{-\beta H_-} q^{-J+ \tfrac R2} x_-^e & \;\text{(anti-topological)}~.
\end{array} 
\ee
The topological index counts BPS multiplets (those for which $H_+=0$), while the anti-topological index counts anti-BPS multiplets (those for which $H_-=0$). Such indices have been studied extensively in the context of of open topological string amplitudes \cite{GV-I, GV-II, OV, DVV} (\cf, Section \ref{sec:OV}).

So far we have been intentionally ambiguous about the choice of topological versus anti-topological twist on $\DqS$. In defining the holomorphic blocks of a theory, we actually use both. In order for the traces \eqref{ZBPS} to converge and define functions of $x_\pm$ and $q$, it is necessary to analytically continue $q=e^{2\pi i\beta\ve}$ either slightly inside or slightly outside the unit circle. We would certainly like the blocks to make sense as functions. We then define
\be \label{defBBPS}
 B^\ra(x;q) \simeq \begin{cases} \CZ_{\ol{\rm BPS}}^\ra(x;q) & |q|<1 \\
 \CZ_{\rm BPS}^\ra(x;q) & |q|>1\,. \end{cases}
\ee
In each regime, the dependence on $x$ and $q$ is meromorphic. This definition provides a  unification of the topological and anti-topological sectors. Physically, it is clear that the indices \eqref{ZBPS} are closely related: in a CPT-invariant theory, every BPS multiplet contributing to $\CZ_{\rm BPS}^\ra$ has an anti-BPS partner contributing to $\CZ_{\ol{\rm BPS}}^\ra$. Mathematically, we will see in examples (and postulate in general) that  each block $B^\ra(x;q)$ can be written as a single $q$-hypergeometric series that converges both for $|q|<1$ and $|q|>1$, but with no analytic continuation across the unit circle. The inclusion of both sectors in blocks will also be natural in three-dimensional topological/anti-topological fusion.

Now let us say a few words about the finite-$\rho$ description of the geometry. It turns out to be the most relevant description for computing the holomorphic blocks of nontrivial theories, as well as understanding their deeper properties. 
At finite $\rho$, the problem is one of supersymmetric quantum mechanics on the half line $\R_+$ obtained by Kaluza-Klein reduction on the asymptotic two-torus of the cigar geometry. The boundary condition at the tip of the cigar defines a state $\langle 0_q|$ that is annihilated by two supercharges $(Q_-,\ol Q_+)$ or $(Q_+,\ol Q_-)$. The asymptotic boundary condition is not exactly given by a state in the quantum mechanics, but there is a unique state associated to it, defined by propagating inwards from infinity to a finite value of the radial coordinate \cite{Witten-path}. Denoting this state as $|\alpha\rangle$, the block is simply an overlap in the space of supersymmetric ground states of the effective quantum mechanics, $B^\ra \sim \langle 0_q|\alpha\rangle$. More precisely, in order to match \eqref{defBBPS}, we set 
\be \label{defBQM}
B^\ra(x;q) = \begin{cases} \langle 0_{q}|\alpha\rangle_{\text{anti-top}} & |q|<1 \\
\langle\alpha |0_q\rangle_{\text{top}} & |q|>1\,, \end{cases}
\ee
using the anti-topological $\langle 0_q|$ when $|q|<1$ and the conjugated topological state $|0_q\rangle$ when $|q|>1$. Both partition functions have a (local) holomorphic dependence on complexified masses $x$.

\begin{wrapfigure}{r}{2.1in}
\includegraphics[width=1.78in]{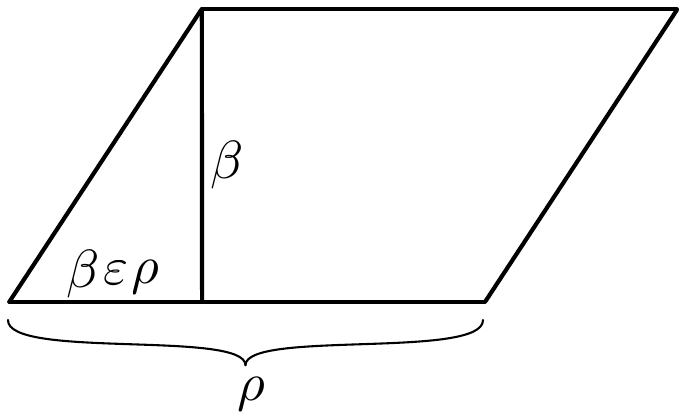}
\caption{The asymptotic torus of $\DqS$.}
\label{fig:T2}
\end{wrapfigure}

Note that the states $|\alpha\rangle$ are supersymmetric ground states of the theory on $T^2$. The presence of holonomies for background gauge fields modify the Hilbert space in which these ground states live, so it is important to keep track of the background fields in the asymptotic geometry. The asymptotic part of $\DqS$ is a product space $\R\times T^2$, where the torus $T^2$ has a flat metric with complex structure parameter $\tau=\ve\beta+i\beta\rho^{-1}$.
The holonomies of the gauge fields around the two cycles of this torus are given by
\be
\frac{1}{2\pi}\oint_{S^1_\beta|_{r\to \infty}}\hspace{-.6cm}A_{\rm flavor} = \vartheta~,\qquad \frac{1}{2\pi}\oint_{S^1_\rho}A_{\rm flavor}=\kappa~.
\ee
The holomorphic blocks are independent of $\rho$ and depend on the dimensionless quantities $q=\exp(2\pi i\beta\ve)=\exp(2\pi i\,\Re\,\tau)$ and $x=\exp X = \exp\big(2\pi\beta m^{\rm 3d}+2\pi i\vartheta+\kappa\hbar\big)$ defined above.

At first glance, the fact that $q$ depends only on $\Re(\tau)$ (though the blocks depend on $q$ holomorphically) may seem peculiar. When we analytically continue in $q$, we will only be analytically continuing in the real part of $\tau$. Notably, such a dependence is familiar in the geometric-Langlands twist of $\cN=4$ super-Yang-Mills theory in four dimensions \cite{Kapustin-Witten}. In that setting, there is a modular complex coupling $\tau$ of the $\cN=4$ Lagrangian, and an affine parameter $t$ (which is actually $\C\P^1$-valued) that parametrizes the combination of scalar supercharges that is promoted to a BRST operator. Topological field theory observables depend only on the combination
\be
\Psi_{\rm GL} = \Re(\tau)+i\,\Im(\tau)\frac{t-t^{-1}}{t+t^{-1}}~,
\ee
called the ``canonical parameter''. When $t=\pm1$, the resulting canonical parameter is just equal to $\Re(\tau)$. We will have more to say about the relation of the present situation to Langlands-twisted SYM in Section \ref{sec:CS} --- we mention it here only to point out that such a dependence on $\Re(\tau)$ may not be all that surprising.


\subsection{Vortices, conformal blocks, and BPS counting}
\label{sec:OV}

BPS indices of the form \eqref{ZBPS} have been encountered frequently in the context of topological string theory as well as in vortex counting. This provides several closely related interpretations of the blocks, which are useful conceptually and sometimes computationally.

Let us first consider the relation to vortex counting. For a gauge theory, we separate the mass parameters into those associated with topological $U(1)$ symmetries (FI parameters) and those associated with ordinary global symmetries that rotate matter fields. We can then place the theory in a background $\RqS$ --- the large-$\rho$ limit of $\DqS$ --- and send $\beta$ (the radius of $S^1$) to zero in such a way that complexified masses \eqref{defx} associated to global symmetries are scaled as
\be x = \exp(\beta m^{\rm 2d})  \label{bmscale}\ee
with $m^{\rm 2d}$ fixed, while complexified FI parameters $x_{\rm FI}$ are kept constant.%
\footnote{It may also be necessary to scale the FI parameters as $x_{\rm FI}\to \beta^cx_{\rm FI}$ for some $c$ in order to obtain a nontrivial $\beta\to 0$ limit, but this is a very different scaling from \eqref{bmscale}.} %
Then the theory reduces to a two-dimensional $\CN=(2,2)$ gauge theory on $\R^2$ with an $\Omega$-deformation (with parameter $\ve$), and the holomorphic blocks reduce to equivariant vortex partition functions \cite{Shadchin-2d, DGH},
\be \label{Btovortex}
B^\ra(x;q)\; \overset{\beta\to 0}{\longrightarrow} \;\CZ^\ra_{\rm vortex}(x_{\rm FI};\ve)\,.
\ee
The field content of the 2d theory is the dimensional reduction of the three-dimensional theory, with all Kaluza-Klein modes discarded. The FI parameters $x_{\rm FI}$ couple to vortex number. The choice of vacuum $\alpha$ descends to a choice of vacuum at the boundary of $\R^2$.

At finite $\beta$, the partition functions on $\RqS$ can be interpreted as a K-theoretic lift of vortex partition functions. This is analogous to the relation between five-dimensional BPS counting and equivariant instanton counting in four-dimensional $\CN=2$ theories \cite{NekSW}. This suggests that for $\hbar=2\pi i\beta\ve$ small (but $\beta$ fixed), holomorphic blocks should have a perturbative expansion
\be \label{BtoWt}
B^\ra(x;q) \sim \exp \Big( \frac1\hbar \wt W(x,s^\alpha;\hbar) \Big)\,, \ee  
where $\wt W$ is an effective twisted superpotential for the effectively two-dimensional theory on $\RqS$, including all Kaluza-Klein modes, in the presence of an $\Omega$-deformation. Such objects were considered in \cite{NS-I, NShatashvili, DG-Sdual}. The twisted superpotential depends on the values $s^\alpha$ of twisted chiral multiplets in a supersymmetric vacuum, a solution (roughly) to $\exp\big(s\frac{\pd}{\pd s}\wt W(x,s;\hbar)\big) = 1$. We will return to this equation in Section \ref{sec:SQM}. In the strict $\ve\to 0$ limit, the superpotential $\wt W(x,s;\hbar=0)$ becomes the undeformed twisted superpotential.

Rather amusingly, the connection to vortices provides a relation between holomorphic blocks and \emph{conformal} blocks in two-dimensional non-supersymmetric CFT. In the extension of the AGT correspondence \cite{AGT} to include half-BPS surface operators, vortex partition functions for the two-dimensional theory on the surface operator are degenerate conformal blocks in Liouville or Toda CFT on an associated Riemann surface \cite{AGGTV, KPW, DGH}. The degenerate conformal blocks are labelled by a discrete choice of operator in the exchange channels denoted by $\alpha$. Then if the same surface operator theory also arose as the reduction of a three-dimensional gauge theory, \eqref{Btovortex} shows that the holomorphic blocks reduce to conformal blocks.

In a related direction, it is well known that five-dimensional BPS indices and four-dimensional instanton partition functions are closely connected to closed topological string amplitudes \cite{GV-I, GV-II}. Similarly, as was mentioned in the Introduction, three-dimensional BPS indices of the form \eqref{ZBPS} and two-dimensional vortex partition functions are related to open topological string amplitudes \cite{OV}. In particular, for theories that can be engineered on M5 branes wrapping a Lagrangian submanifold $\CL$ in a non-compact Calabi-Yau $Y$, the BPS index of the gauge theory counts the number of BPS M2 branes that can end on the M5-branes. Furthermore, this index can be computed by evaluating the open topological string partition function for that geometry $\CZ^{\rm top}_{\rm open}$. The string coupling is encoded in $q=e^{-g_s}$, and both open- and closed-string moduli appear as flavor fugacities $x$. The choice of vacuum is then related to a choice of brane placement.

The topological string partition function can be computed by summing up corrections to the effective action of a two-dimensional $\cN=(2,2)$ theory on $\R^2$ in the presence of a graviphoton background \cite{OV}, leading to an expression of the form
\begin{align} 
B^\ra(x;q)\sim \CZ^{\rm top}_{\rm open} &\;\sim\;  \exp\left[\sum_{J,R,e}\sum_{m=1}^\infty \frac{(-1)^{2J}q^{m(-J-\frac R2+\frac12)}x^{-me}N^e_{J,R}}{m\big(q^{\frac m2}-q^{-\frac m2}\big)}  \right]  \\ 
&\;=\; \prod_{J,R,e} \prod_{n= 0}^\infty \big(1-q^{-J-\tfrac R2-n}x^{-e}\big)^{(-1)^{2J+1}N^e_{J,R}}\,. \label{OV}
\end{align}
Here $N^e_{J,R}$ is the number of BPS M2 branes with given spin, R-charge, and flavor charge. This result has a simple heuristic interpretation in the gauge theory. In three dimensions the central charge of the $\cN=2$ superalgebra is real, so any collection of BPS excitations can potentially form a bound state at threshold. Then the topological string amplitude is counting the single-particle BPS states at a point in moduli space where these bound states can be organized into a Fock space generated by oscillators for the angular momentum modes of quantum fields corresponding to elementary BPS particles \cite{DVV, AY}. The integers $N^e_{J,R}$ describe the number of such quantum fields with given charges, and the angular momentum modes lead to the product over $n$.

There is an important distinction, at least philosophically, to be made between holomorphic blocks and topological string amplitudes. Namely, our definition of holomorphic blocks as supersymmetric gauge theory partition functions on $\DqS$ suggests that they are locally holomorphic functions of their parameters. Topological string amplitudes, on the other hand, are subject to the holomorphic anomaly, and when they are expanded around appropriate large volume points in moduli space they are not necessarily related by analytic continuation. The BPS counting interpretation of holomorphic blocks should then only hold in an appropriate region of parameter space, if ever. We will see an explicit example of this in the context of the free chiral theory of Section \ref{sec:chiralblock}. 


\subsection{Topological/anti-topological fusion in three dimensions}
\label{sec:3dtt*}

Our motivation for studying $\DqS$ partition functions is the conjecture that they form the building blocks for the ellipsoid partition function and supersymmetric index. Let us consider how this comes about.

Two copies of $\DqS$ can be combined naturally to give a three-dimensional analogue of the topological/anti-topological fusion geometry for two dimensional theories with $\cN=(2,2)$ supersymmetry \cite{CV-tt*}. That is to say, they can be ``fused'' as long as the Hilbert spaces defined on their asymptotic boundaries are identical (or more precisely, dual). As in the two-dimensional case, the resulting fused construction does not appear to admit any globally preserved supercharges that annihilate the partition function.%
\footnote{There may nevertheless be non-standard supercharges preserved by this background. Recent work of \cite{GomisLee} has shed light on this issue in two dimensions.} %
Nonetheless, the presence of an infinitely long flat region along which any state must propagate leads to a projection onto the reduced Hilbert space $\cH_0(T^2)$ of supersymmetric ground states of the theory, and consequently the resulting partition function will be quasi-topological, \ie, it will be invariant under all but a finite number of deformations of the $\cN=2$ theory.

The partition function on the fused geometry thus enjoys, by construction, a natural factorization of the form
\be
\cZ_{\rm fused}=\sum_{\alpha,\beta\in\cH_0}n_{\alpha\beta}B^\alpha(x;q)B^{\beta}(\wt x;\wt q)~.
\ee
This is a simple consequence of the fact that only supersymmetric ground states $|\alpha\rangle\in\cH_0$ propagate in the long cylinder connecting the two cigars. The fused partition function is simply the overlap of states generated by the closed of ends of the cigars (after projecting to ground states),
\be\label{fusionoverlap}
\cZ_{\rm fused}=\langle 0_q|0_{\tilde q}\rangle~,
\ee
and inserting a complete set of states spanning $\cH_0$ leads to the factorized form above,
\be\label{fusionschematic}
\cZ_{\rm fused}=\langle0_q|\left(\sum_{\alpha\in\cH_0}|\alpha\rangle\langle\alpha|\right)|0_{\tilde q}\rangle=\sum_{\alpha\in\cH_0}B^{\alpha}(x;q)B^{\alpha}(\wt x;\wt q)~.
\ee
In the ``massive vacuum'' basis for supersymmetric ground states, we expect the intersection matrix $n_{\alpha\beta}$ to be the identity.

Before moving on, some general comments about the relation of this construction to the two-dimensional story of \cite{CV-tt*} are in order. The first and most obvious new ingredient in  three-dimensional fusion is that there are an infinite number of inequivalent constructions of this form, as opposed to the unique two-dimensional topology. For any element $g\in\SLZ$, one can consider copies of $\DqS$ whose boundary tori are related by a $g$-action (along with orientation-reversal), and all of the statements above should go through. The effect of $g$ is indicated in equations \eqref{fusionoverlap}-\eqref{fusionschematic} by the ``tilde'' operation acting on $q$ and the mass parameters $x$. More suggestively, we could write the $g$-fused partition function as
\be \label{gfused}
\cZ_{\rm fused}^{[g]}= \sum_{\alpha\in\cH_0}B^{\alpha}(x;q)B^{\alpha}(\wt x;\wt q) = \big|\!\big| B(x;q)\big|\!\big|_g^2\,.
\ee
Additionally, the equivariant parameter $q$ has no obvious counterpart in the Cecotti-Vafa construction. This equivariance is responsible for the fact that while the asymptotic radius of the cigar (which we call $\rho$) plays a crucial role in the two-dimensional story, it makes no appearance in the definition of three-dimensional holomorphic blocks. Indeed, the two-dimensional ($\beta\to0$) reduction of the holomorphic blocks leads to a two-dimensional partition function for an $\Omega$-deformed theory. The limit of turning off the $\Omega$-deformation, which in general on a non-compact space is a singular limit, should reproduce the traditional $tt^*$ results in the fused setup. Exploring this relation further is left for future work (see also the recent work of \cite{GomisLee, JKLMR}).

The result of three-dimensional fusion is a protected observable of a mass-deformed $\cN=2$ SCFT in three dimensions associated to any lens space topology (the construction here manifestly realizes a genus-one Heegaard splitting of the resulting manifold, which identifies it as a lens space). We conjecture that this observable is equivalent to the more conventional lens space observables that have been defined and computed by supersymmetric localization in recent years, \cf\ \cite{BNY-Lens, Kallen-Seifert, OY-Seifert}. In addition to making the factorization of the ellipsoid partition function observed in \cite{Pasquetti-fact} manifest, this would imply that all other lens space partition functions, such as the supersymmetric index, involve products of the {\it same} holomorphic blocks. 

The admissible pairings between left and right blocks are naturally fixed by the requirement that the two semi-infinite cigars be glued along equivalent tori, and this explains the relations between parameters in equations \eqref{factSb} and \eqref{factInd}. We will be interested in configurations for which an $\SLZ$ action on the torus (acting as usual on $\tau$) induces a modular action combined with a reflection on the parameter $q$. Specifically, if $\tau \mapsto \wt \tau = -\ol{g\cdot \tau}$, we would like $\beta\ve \mapsto \wt\beta\wt\ve = -\ol{g\cdot (\beta\ve)}$ as well.
This is precisely the case in the degeneration limit of the torus, 
\be\label{goodlim}
\beta\ll\rho~,\qquad \tau\to\ve\beta~.
\ee
For a single cigar, this limit has no effect since it amounts to sending $\rho\to\infty$. However, once we start to consider nontrivial fusion geometries, this will be an important constraint.

Notice that although the geometric twist parameters $q=\exp(2\pi i\beta\ve)$ and $\wt q=\exp(2\pi i \wt\beta\wt\ve)$ are related by $\wt\beta\wt\ve = -\ol{g\cdot (\beta\ve)}$, with $g\in\SLZ$ acting as a modular transformation, analytic continuation off of the unit circle will not respect this relation. That is to say that in analytically continuing, the fact that one side is topologically twisted and the other is anti-topologically twisted will lead to an additional complex conjugation in the relationship between $\tau$ and $\bar\tau$. Consequently, after any analytic continuation of $q$ off of the unit circle,
\be   |q|<1\quad \Leftrightarrow \quad |\wt q|>1 \,. \ee
Put differently, a modular transformation alone would preserve the upper half-plane, but a modular transformation combined with a reflection about the origin switches upper and lower half-planes. This dovetails nicely with the definition of blocks $B^\ra(x;q)$ from Section \ref{sec:cigar} as actual functions, using a topological twist outside the unit circle and an anti-topological twist inside the unit circle. In any fused combination $\sum_\alpha B^\ra(x;q)B^\ra(\wt x;\wt q)$, the blocks on the left automatically correspond to an anti-topological twist when the blocks on the right correspond to a topological one, which is just what we need for topological/anti-topological fusion.


\subsubsection{S-fusion}
\label{sec:Sfusion}

We now take a closer look at the fusion geometries that are related to the ellipsoid partition function (S-fusion) and the sphere index (identity-fusion), and relate parameters $(x,q)$ and $(\wt x,\wt q)$ in the two cases.

\begin{figure}[t!]
\centering
\includegraphics[width=6in]{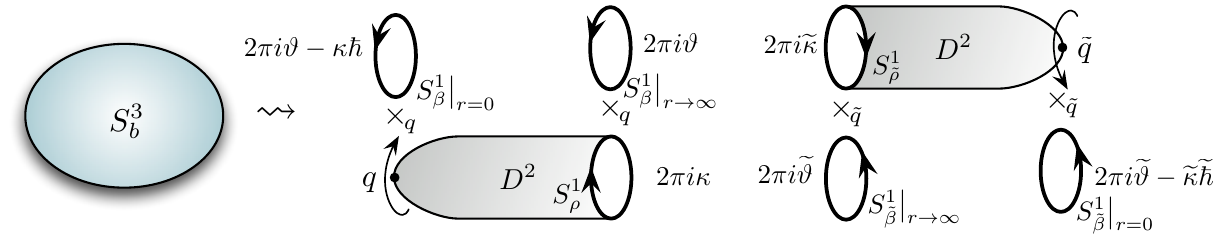}
\caption{Matching holonomies $i\oint A$ of a background gauge field during S-fusion.}
\label{fig:Sfusion}
\end{figure}

If we fuse two blocks whose asymptotic boundaries are related by the element $S\in\SLZ$, as in Figure \ref{fig:Sfusion}, we end up with the topology of the three-sphere. The complex structure $\wt \tau = \wt \beta\wt \ve+i\wt \beta\wt \rho^{-1}$ of the torus on the right is related to that on the left as
\be 
\wt{\tau}=-\overline{\ S\cdot \tau}=\frac{1}{\overline{\tau}}=\frac{\ve+i\rho^{-1}}{\beta(\ve^2+\rho^{-2})}\xrightarrow[\rho\to\infty]{}\frac{1}{\ve\beta}~.
\ee
Thus, in the $\rho\to\infty$ limit, $\wt\beta\wt\ve=(\beta\ve)^{-1}$. Moreover, in this limit, the individual geometric parameters obey
\be \wt \beta = \frac1\ve\,,\qquad \wt \ve = \frac1\beta\,.\ee
For the angular momentum fugacity in the holomorphic blocks, we then find
\be
q = \exp\left(i\oint_{S^1_\beta|_{r=0}}\hspace{-.15in}\omega\hspace{.05in}\right) = e^{2\pi i\beta\ve} = e^\hbar \quad\Rightarrow \quad
 \wt q = \exp\left(i\oint_{S^1_{\tilde\beta}|_{r=0}}\hspace{-.15in}\wt\omega\hspace{.05in} \right) = e^{\tfrac{2\pi i}{\beta\ve}} = e^{-\tfrac{4\pi^2}{\hbar}}\,.
\ee

Now consider the holonomies for a background gauge field that has the form $A=A_0+\kappa\omega$ on the left and $A=\wt A_0+\wt \kappa\wt\omega$ on the right, with $\frac{1}{2\pi}\oint_{S^1_\beta} A = \vartheta$ and $\frac{1}{2\pi}\oint_{S^1_{\tilde\beta}} A = \wt{\vartheta}$ (as in Figure~\ref{fig:holonomies}). The gluing, combining an $S$ transformation and a reflection, requires us to identify
\be \label{Skth}
\wt{\vartheta} = \kappa~,\qquad\wt\kappa = \vartheta~.
\ee
Consequently, if we twist anti-topologically on the left and topologically on the right, the holomorphic variables appearing in the blocks should be
\be x = \exp X\,,\qquad \wt x= \exp\wt X\,,\ee
with
\be \label{ellipsoidrelation}
X = 2\pi\beta m^{\rm 3d} + (2\pi i\vartheta-\kappa\hbar)\,,\qquad \wt X = 2\pi\wt\beta m^{\rm 3d} - (2\pi i\wt{\vartheta}-\wt\kappa\wt\hbar) = \frac{2\pi i}\hbar X\,.\ee
(Note that the relative $\pm$ signs that we must use for $X$ and $\wt X$ come directly from the definitions of the variables $X_\mp$ in \eqref{defx}.)

In addition, the R-symmetry gauge field $A_R$ must have $-\kappa_R=\wt \kappa_R=1/2$ due to the anti-topological/topological twists. The gluing relations \eqref{Skth} then impose $\vartheta_R=-\wt{\vartheta}_R=1/2$. In other words, S-fusion is only consistent if the R-symmetry gauge field has flat component with holonomy $e^{2\pi i\vartheta}=e^{i\pi}=-1$ around the $\beta$-circle on each side. Fortunately, this is exactly how we defined the $\DqS$ partition function in \eqref{twistbackground}. Also recall that as long as all fields have integer R-charges, $\vartheta_R$ and $\wt{\vartheta}_R$ are only defined modulo 1.

We pause here to note that away from the limit \eqref{goodlim}, the combined (S-fused) partition function would indirectly pick up a dependence on $\rho$ and $\wt \rho$, the radii of the cigars. It would be nice to explore the properties of the resulting partition functions and to understand if they constitute a further interesting deformation of the three-sphere partition function.

The relation of holomorphic parameters $(x,q)$ and $(\wt x,\wt q)$ above matches that which emerged in the factorized form of the ellipsoid partition function discovered by \cite{Pasquetti-fact}. In more standard notation, the ellipsoid partition function would depend on $q=\exp(2\pi ib^2)$, $\wt q = \exp(2\pi ib^{-2})$ and $x=\exp(2\pi b\mu)$, $\wt x=\exp(2\pi b^{-1}\mu)$, where $\mu$ are complexified mass parameters relevant to the ellipsoid geometry \cite{HHL}. 

\subsubsection{Identity fusion}
\label{sec:idfusion}

The second fused construction we consider is that with the simplest possible gluing. We choose the element $id\in\SLZ$, which leads to the topology $\SS$.

\begin{figure}[t!]
\centering
\includegraphics[width=6in]{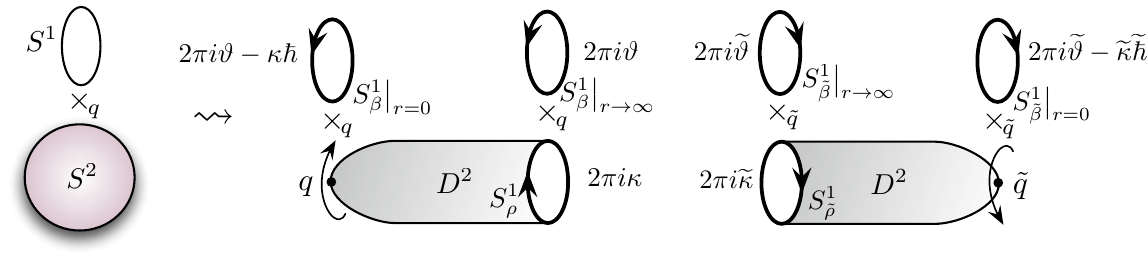}
\caption{Matching holonomies $i\oint A$ of a background gauge field during $id$-fusion.}
\label{fig:Ifusion}
\end{figure}

In this case, the condition for matching the asymptotic tori is simply
\be
\quad\wt\tau=-\overline{\tau}=-\ve\beta+i\beta\rho^{-1}\qquad \Rightarrow\quad \wt \beta \wt \ve = -\beta\ve\,,
\ee
or more precisely
\be \wt \beta = \beta\,,\qquad \wt\ve = - \ve\,,\ee
so that
\be q = e^\hbar\,,\qquad \wt q = e^{-\hbar} = q^{-1}\,.\ee
In this construction, the asymptotic radii $\rho$ and $\wt\rho$ of the cigars play no role, but we usually still work in the limit \eqref{goodlim}. 

The holonomies of a $U(1)$ connection $A_0+\kappa\omega$ on the left and $\wt A_0+\wt \kappa\wt \omega$ on the right must obey
\be \label{relqId}
\vartheta = \wt{\vartheta}\,,\qquad \kappa= -\wt \kappa \quad \text{mod}\;\Z\,,\ee
with the sign in the second equation coming from the reversed orientation in the gluing.
Again, we assume that all fields have integral charges.
Then the fact that $\kappa = -\wt \kappa$ need only be true up to an integer becomes physically relevant: the sum defines a nontrivial magnetic flux of $F=dA$ through $S^2$,
\be -m: = \frac{1}{2\pi}\int_{S^2} F= \kappa +\wt \kappa \,. \ee
Let us then set $\kappa=-\frac{m}2+\kappa_0$, $\wt\kappa=-\frac{m}2-\kappa_0$. If we anti-topologically twist on the left and topologically twist on the right, then the variables in the holomorphic blocks associated to a $U(1)$ flavor symmetry become
\be \label{relxId}
 x = \exp X = q^{\frac m2}\zeta\,,\qquad \wt x= \exp \wt X = q^{\frac m2}\zeta^{-1}\,,\ee
with $\zeta = \exp\big(2\pi i\vartheta-\kappa_0\hbar\big)$.

The connection for the R-symmetry in this geometry has $-\kappa_R=\wt\kappa_R=1/2$, which is right for there to be no net R-flux through $S^2$. In addition, we set $\vartheta_R=\wt{\vartheta}_R = 1/2$. This matches the holonomies of $A_R$ in the S-fusion geometry up to a subtle sign: in S-fusion, we had to have $\vartheta_R=-\wt {\vartheta}_R = 1/2$. The two assignments are equivalent if we are freely allowed to shift $\vartheta_R$ and $\wt {\vartheta}_R$ by integers, \ie\ if all fields in a theory are given integral R-charge assignments. Our ability to write both S-fusion and index-fusion partition functions in terms of \emph{exactly} the same set of holomorphic blocks $B^\ra(x;q)$ seems to rely on this property. 

We conjecture that the fused partition function $\big|\!\big|B(x;q)\big|\!\big|_{id}^2 = \sum_\alpha B^\ra(x;q)B^\ra(\wt x;\wt q)$ is equivalent to the sphere index defined as
\be \label{sphereind}
 \CI(m,\zeta;q) = \Tr_{\CH(S^2;m)}(-1)^R q^{\tfrac R2+J_3}\zeta^e\,, \ee
with $J_3$ and $R$ denoting spin and R-charge in the super-Poincar\'e algebra on (round, untwisted) $S^2\times \R$, $e$ denoting flavor charge as usual, and $m$ denoting the units of magnetic flavor flux through $S^2$. This is the same index defined by \cite{KW-index}, following \cite{KMMR-index, BBMR-index, Kim-index, IY-index, KW-index}, up to the modification $(-1)^F\to (-1)^R$. The expression \eqref{sphereind} is exactly the same index studied by \cite{DGG-index}, as long as fields have integer R-charge assignment.%
\footnote{In \cite{DGG-index}, the naive fermion number $F=2J_3$ was redefined to include additional angular momentum from electric particles in a magnetic monopole background. This has the same effect as replacing $(-1)^F\to (-1)^R$ when states have integer R-charge.} %
The identification of parameters \eqref{relqId}--\eqref{relxId} found for holomorphic blocks is identical to the relation predicted in factorized forms of the index in \cite{DGG-index}.

Note that in writing \eqref{relxId} and obtaining a direct relation to sphere indices, we have tacitly set to zero the real mass parameters $m^{\rm 3d}$ for the flavor symmetries. In the fused geometry, $m^{\rm 3d}$ appears to be an additional free parameter, which could be turned on to further modify \eqref{relxId}. This deformation does not seem to have an analogue for the round $\SqS$ index geometry. Indeed, on $\SqS$, the scalar fields in background gauge multiplets are quantized in units of $\hbar=\log q$, fixed to equal the magnetic flux through $S^2$. (This is actually how the combinations $q^{\frac m2}\zeta^{\pm1}$ arose for the sphere index in \cite{DGG-index}.) A $Q$-exact deformation from round $\SqS$ to two fused copies of $\DqS$ should evidently send quantized masses in the former to vanishing masses in the latter.


\subsection{Difference equations}
\label{sec:diff}

An extremely useful property of partition functions on $\DqS$ is that they are solutions to a system of difference equations, which we now take a moment to explain. The difference equations are a consequence of identities in the algebra of line operators that wrap $S^1$ and act at the tip of the cigar. These supersymmetric line operators are in some sense a three-dimensional lift of the chiral operator insertions that led to $tt^*$ equations in two dimensions. The identities also provide a new perspective on difference equations that arose in the context of open topological string theory \cite{adkmv}. In this paper, they provide a powerful computational tool for analyzing blocks.

The line operators we have in mind were studied extensively in \cite{DG-Sdual, DGG, DGG-index}. They are half-BPS Wilson and 't Hooft lines for the background gauge fields corresponding to the abelian \emph{flavor} symmetries of a theory. For each $U(1)_i$ flavor symmetry (in a maximal torus of the global symmetry group), there is a supersymmetric Wilson line $\hat x_i$ that measures the holonomy of the associated background gauge field, and so acts as multiplication by the complexified mass parameter in \eqref{defx},
\be \label{Wilson} \hat x_i\, B(x;q) = x_i B(x;q)\,, \ee
and there is also an associated 't Hooft line $\hat p_i$ that shifts $x_i\mapsto qx_i$,
\be \label{tHooft} \hat p_i\,B(x;q) = B(x_1,...,x_{i-1},qx_i,...;q)\,. \ee
In terms of the logarithms $X_i$, we have $\hat p_i = \exp(\hbar\,\pd_{X_i})$.
Thus the operators obey $q$-commutation relations
\be \label{opalg}\qquad  \hat p_i\hat x_j = q^{\delta_{ij}}\hat x_j\hat p_i\,;\qquad \hat p_i\hat p_j=\hat p_j\hat p_i\,,\quad \hat x_i\hat x_j=\hat x_j\hat x_i\,. \ee

One nice way to understand these commutation relations is to weakly gauge the flavor symmetries by coupling a three-dimensional theory to an abelian four-dimensional $\CN=2$ theory, thinking of the 3d theory as living on the boundary of a 4d spacetime. In the case of a 3d geometry $\RqS$ (which for this purpose is equivalent to the curved cigar version) the 4d geometry is just $\RqS\times \R_+$. In the bulk, the operators $\hat x$ and $\hat p$ are dynamical Wilson and 't Hooft lines that wrap $S^1$ and can live at any point on $\R_+$. They can move freely along $\R_+$ and act on the boundary, but their ordering along $\R_+$ matters. It is the order in which they can act on the boundary. It was argued in \cite{Ramified, GMNIII} that the OPE of line operators is graded by angular momentum in transverse directions --- \ie\ by the spinning of $\R^2$ --- ultimately implying that as two operators pass each other on $\R_+$ they will $q$-commute.

Alternatively, one may use the AGT correspondence to relate partition functions of certain 3d $\CN=2$ theories to degenerate conformal blocks in Liouville or Toda CFT, as in Section \ref{sec:OV}. In the CFT context, the line-operator identities of 3d $\CN=2$ theory become reinterpreted as standard Ward-Takahashi identities.

When line operators act on the $\DqS$ partition functions of given three-dimensional $\CN=2$ theory, they will obey identities of the form 
\be \hat f_a(\hat x,\hat p;q) \cdot B(x;q) = 0\,, \label{qLSUSY} \ee
where the $f_a$ are \emph{polynomials} in $\hat x_i$, $\hat p_i$, and $q$. There are typically as many operators $\hat f_a$ as there are flavor symmetries, so that the equations \eqref{qLSUSY} completely determine the dependence of $B(x;q)$ on $x$. A more precise statement is that in the ``classical'' commuting limit $q\to 1$, the set of equations
\be \CL_{\rm SUSY}\,:\quad \{f_a(x,p;1)=0\} \ee
cuts out a Lagrangian submanifold $\CL_{\rm SUSY}$ in the space $(\C^*)^{2N}$, where $N$ is the number of flavor symmetries, with respect to a canonical symplectic form $\Omega=\sum_a \frac{dp_i}{p_i}\wedge \frac{dx_i}{x_i}$. We will return to this submanifold in Section \ref{sec:SQM}. The points on $\CL_{\rm SUSY}$ at a fixed value of the $x_i$ --- \ie\ the solutions $p^\alpha_i(x)$ to $f_a(x,p;1)=0$ --- are in one-to-one correspondence with the massive vacua $\alpha$ of the $\CN=2$ theory. In the fully quantum setting, we expect that the holomorphic blocks $B^\ra(x;q)$ provide a complete basis of solutions to the quantized identities \eqref{qLSUSY}. This turns out to be a useful way to characterize the blocks: they are the solutions to the line-operator identities that possess certain analytic properties. We will discover more about the required properties in the following sections.

The identities \eqref{qLSUSY} can be derived systematically for any $\CN=2$ theory that has a UV Lagrangian description. The procedure for doing so was described in \cite{DGG,DGG-index} in the context of ellipsoid partition functions and $\SqS$ indices, but it is entirely local: the line operators and their algebra are localized at points in the geometries that look locally like the tip of $\DqS$. Thus, the systematic procedure applies directly to holomorphic blocks. We will review aspects of the construction in Section \ref{sec:blockint}.

In the case of geometries $S^3_b$ and $\SqS$, there are of course \emph{two} places where line operators can act supersymmetrically, corresponding to opposite tips of cigars in topological/anti-topological fusion. Indeed, these compact partition functions obey two sets of identities,
\be \hat f_a(\hat x,\hat p;q)\cdot \CZ = \hat f_a(\hat {\wt x},\hat {\wt p};\wt q)\cdot \CZ = 0\,, \ee
in two mutually commuting sets of line operators $(\hat x,\hat p)$ and $(\hat {\wt x},\hat {\wt p})$. This was one motivation behind predicting a factorization of the supersymmetric index into blocks in \cite{DGG-index}.
More explicitly, using variables as described in Section \ref{sec:3dtt*} it is known that the Wilson and 't Hooft loops act as
\begin{align} 
S^3_b\,:&\quad  \hat x = e^{2\pi b\mu},\; \hat p = e^{ib\pd_\mu},\,q=e^{2\pi ib^2}\,;\quad \hat{\wt x} = e^{2\pi b^{-1}\mu},\; \hat{\wt p} = e^{ib^{-1}\pd_\mu},\;\wt q=e^{2\pi ib^{-2}}; \notag\\
\SqS\,:&\quad \hat x = q^{\frac m2}\zeta,\; \hat p = \exp({\pd_m+\tfrac\hbar2\pd_{\log\zeta}}),\;q=e^{\hbar}\,; \\
 &\hspace{1.7in} \quad \hat{\wt x}=q^{\frac m2}\zeta^{-1},\; \hat{\wt p} = \exp({\pd_m-\tfrac\hbar2\pd_{\log\zeta}}),\; \wt q=e^{-\hbar}\,, \notag
\end{align}
so that $\hat p\hat x=q\hat x\hat p$ while $\hat{\wt p}\hat {\wt x}=\wt q\hat {\wt x}\hat {\wt p}$.
The multiplicative action of the Wilson loops agrees beautifully with the identification of parameters $(x,q)$ and $(\wt x,\wt q)$ that we found on the two halves of fused geometries in Section \ref{sec:3dtt*}, and provides strong verification for our results there.
In fact, we may observe that once we know the relation between $q$ and $\wt q$ in a fused geometry, the requirement that $q$-shifts commute with multiplication by $\wt x$ (and $\wt q$-shifts commute with multiplication by $x$) fixes the relation between $x$ and $\wt x$ almost entirely.


\subsection{Factorization for the free chiral}
\label{sec:chiralblock}

With the general picture of holomorphic blocks in place, let us consider a simple and fundamental example of factorization: the theory of a free chiral multiplet. This theory  illustrates many of the important properties of holomorphic blocks and fusion, so it is worth introducing it in some detail.

In order to put the theory on curved backgrounds we must specify Chern-Simons terms for the background vector multiplet coupled to the $U(1)$ flavor symmetry. We must further specify R-charge assignments and Chern-Simons contact terms for the R-symmetry vector multiplet. We define the theory $T_\Delta$ as follows,
\be \label{TDdef}
T_\Delta\,:\quad \left\{\begin{array}{l}
\text{chiral fields: $\{\,\phi\,\}$}\qquad
\text{charges:}\;\;\begin{array}{c|c}& \phi \\\hline
    F\, & \,1 \\
    R\, & \,0 \end{array}\qquad
\text{CS matrix:}\;\;\begin{array}{c|cc} & F & R \\\hline
    F\, & -\frac12 & \;\;\,\frac12 \\
    R\, & \;\;\,\frac12 & -\frac12
\end{array}
\end{array}\right\}~,
\ee
where $F$ and $R$ denote the flavor and R-symmetries, respectively. We have encoded the background Chern-Simons couplings both for flavor and R-symmetry gauge fields in a single matrix. For example, there is a Chern-Simons coupling for the flavor symmetry at level $k=-1/2$. The notation $T_\Delta$ is from \cite{DGG}, where this was the theory associated to a single ideal tetrahedron $\Delta$.

Note that the half-integer bare Chern-Simons levels in \eqref{TDdef} cancel the anomaly coming from the fermions in the chiral multiplet $\phi$. For nonzero real mass $m^{\rm 3d}$, they contribute an extra shift by 
\be \label{TDshiftCS}
\Delta k_{ij} = \text{sgn}(m^{\rm 3d})\times\left(\begin{smallmatrix} \;\;\,1/2 & -1/2 \\ -1/2 & \;\;\,1/2\end{smallmatrix}\right)~,
\ee 
to the effective matrix of Chern-Simons levels \cite{AHISS}, changing all the half-integers into integers.

Usually it is only essential to cancel anomalies for a dynamical gauge symmetry. However, a cancellation of anomalies for flavor symmetries becomes important if the flavor symmetries are ever to be weakly gauged --- or if we are to consistently turn on background vevs for flavor gauge fields. This is exactly what we want to do for our partition functions, and indeed it happens that factorization into holomorphic blocks is only possible when all flavor anomalies are cancelled. (This was observed in \cite{Pasquetti-fact} for ellipsoid factorization.)

The ellipsoid partition function for this theory is commonly expressed in terms of variables $(\mu,b)$, where $\mu$ is the mass parameter associated to the flavor symmetry and $b$ is the real deformation parameter for the ellipsoid geometry. They are related to our variables as $X=2\pi b \mu$ and $\hbar=2\pi i b^2$. In \cite{HHL}, it was shown that
\be\label{chiralellipsoid}
\CZ^b_\Delta(X;\hbar) = \exp\Big[\frac{i\pi}{2}\big(\mu-\tfrac i2(b+b^{-1})\big)^2\Big] s_b\big(\tfrac i2(b+b^{-1})-\mu)\,,
\ee
where the function $s_b(x)$ is the non-compact quantum dilogarithm.\footnote{After its introduction in \cite{Barnes-QDL} and rediscovery in \cite{Fad-modular} as a solution of the quantum pentagon identity, the non-compact quantum dilogarithm has appeared with various notations in the literature. The notation ``$s_b$'' adopted here is the one used in \cite{Teschner-TeichMod} and \cite{HHL}. The inverse of this function is called $s_b$ in \cite{TV-6j}. Some of its relevant analytic properties and asymptotics can be found in \cite{DGLZ}.}
Physically, $b$ is real and $\hbar$ is pure imaginary with positive imaginary part, but the partition function \eqref{chiralellipsoid} can be analytically continued to an entire cut plane $\hbar \in \C\bs\{i\R_{<0}\}$. After giving $\hbar$ a nonzero real part, we find that
\be \label{ZDelta}
\CZ^b_\Delta(X;\hbar) = \begin{cases} \ds C^2\prod_{r=0}^\infty \frac{1-q^{r+1}x^{-1}}{1-\wt q^{-r}\wt x^{-1}} & |q|<1 \\ \ds
C^2\prod_{r=0}^\infty \frac{1-\wt q^{r+1}\wt x^{-1}}{1-q^{-r} x^{-1}} & |q|>1\,,
\end{cases}
\ee
where as usual $q=\exp \hbar=\exp 2\pi i b^2\,,$ $\wt q=\exp-\frac{4\pi^2}{\hbar} = \exp 2\pi i b^{-2}$, and $x=\exp X,\, \wt x=\exp \frac{2\pi i}{\hbar}X$. The constant prefactor in the products is $C = \exp\big[\tfrac{-1}{24}(\hbar+\wt \hbar)\big] = \exp\big[\tfrac{-1}{24}\big(\hbar-\tfrac{4\pi^2}{\hbar}\big)\big]$\,.

The sphere index is expressed in terms of variables $(m,\zeta,q)$, as discussed in Section \ref{sec:idfusion}.
It was shown in \cite{DGG-index} (following \cite{KW-index}) that the index, defined only for $|q|<1$, can be written in the form
\be \label{IDelta}
\CI_\Delta(m,\zeta;q) = \prod_{r=0}^\infty \frac{1-q^{r-1}x^{-1}}{1-\wt q^{-r}\wt x^{-1}}\,,
\ee
where $\wt q=q^{-1}$, $x=q^{\frac m2}\zeta$, and $\wt x=q^{\frac m2}\zeta^{-1}$.

As written above, the factorization of the two partition functions is almost obvious. The only nontrivial aspect is that the variables $(x,q)$ appear in the numerators of the products and the dual variables $(\wt x,\wt q)$ in the denominators, or vice versa. Nevertheless, both numerator and denominator can be written in a uniform manner. Let us define the ``tetrahedron block'' as follows:
\be \label{BD}
 B_\Delta(x;q) = (qx^{-1};q)_\infty\,,
\ee
where
\be \label{zqinf}
(z;q)_\infty := \sum_{n=0}^\infty \frac{(-1)^n q^{\tfrac12n(n-1)}z^n}{(q)_n} =
 \begin{cases} \prod_{r=0}^\infty (1-q^rz) & |q|<1 \\[.1cm]
 \prod_{r=0}^\infty (1-q^{-r-1}z)^{-1} & |q|>1\,, \end{cases}
\ee
with
\be (q)_n := (1-q)(1-q^2)\cdots (1-q^n)\,. \ee
The $q$-hypergeometric series defining the function $(z;q)_\infty$ converges for all $|z|<1$ both inside the unit circle $|q|<1$ and outside the unit circle $|q|>1$ to the infinite products indicated on the right side of \eqref{zqinf}. In each regime, the product representation provides an analytic continuation in $z$ to a meromorphic function of $z\in \C$. However, there is no analytic continuation in $q$ between $|q|<1$ and $|q|>1$ --- approaching the unit circle $|q|=1$ from either inside or outside, the function $(z;q)_\infty$ diverges at every rational point (every root of unity).

It is the function $(qx^{-1};q)_\infty$ --- defined piecewise inside and outside the unit circle, but possessing a single $q$-hypergeometric series expansion that makes sense in both regimes --- that we call the tetrahedron block. Due to the reflection used in any fusion operation of two cigars, the parameter $\wt q$ is outside the unit circle whenever $q$ is inside the unit circle, and vice versa. Correspondingly, one half of a fusion geometry is topologically twisted and the other half anti-topologically twisted. Then it is easy to see that the fused partition functions take the simple form
\be \CZ^b_\Delta(X;\hbar) = B_\Delta(x;q)B_\Delta(\wt x;\wt q)
\,,\qquad \CI_\Delta(m,\zeta;q) = B_\Delta(x;q)B_\Delta(\wt x;\wt q)
\,, \ee
with the appropriate definitions of $(x,q,\wt x,\wt q)$ in each case. Quite amazingly, the S-fusion product $\CZ^b_\Delta(X;\hbar) = \big|\!\big|B_\Delta(x;q)\big|\!\big|^2_{S}$\,, is a function that \emph{can} be analytically continued from $\Re\,\hbar<0$ to $\Re\,\hbar>0$ across the positive imaginary axis where $\hbar=2\pi ib^2$ is physical.

The factorization of the ellipsoid partition function in \eqref{ZDelta} only holds modulo the prefactor $C^2$ in \eqref{ZDelta}. This prefactor looks similar to the contribution of a level $\frac{1}{24}$ R-R contact term. 
We will almost always work modulo such R-R contact terms in this paper, in part because it is rather subtle to fix them precisely. One way to (partially) absorb the prefactor in the blocks, if so desired, is to modify
\be \label{BDqinf}
 B_\Delta(x;q) \to (q)_\infty B_\Delta(x;q)\,,
\ee
where we define
\be \label{defqinf}
 (q)_\infty = {(q^{-1})_\infty^{-1}} := \begin{cases} \prod_{r=1}^\infty (1-q^r) & |q|<1 \\
   \prod_{r=1}^\infty (1-q^{-r})^{-1} & |q|>1\,. \end{cases}
\ee
Then $|\!|(q)_\infty|\!|_{S} = -\frac{2\pi}{\hbar}C^2$, while $|\!|(q)_\infty|\!|_{id} = 1$.
Note that R-R contact terms are always invisible in the index (identity fusion) and only appear for the ellipsoid partition function (S-fusion).  

This simple example allows us to investigate the relationship between blocks and BPS-counting partition functions. The infinite-product forms \eqref{zqinf} of the block $B_\Delta$ take roughly the form \eqref{OV} that one expects for a BPS index. We need only identify the elementary BPS excitation that generates the Fock space counted by the BPS index. For the free chiral theory \eqref{TDdef} with $m^{\rm 3d}>0$ ($|x|>1$), there
is a single elementary BPS state coming from the chiral field itself. There is also a single elementary anti-BPS state coming from the anti-chiral which is CPT conjugate to $\phi$. We can then see that the block $B_\Delta(x;q)$ matches the expected BPS-counting partition function for $|q|>1$ and anti-BPS counting for $|q|<1$, as described by \eqref{ZBPS}.

For $m^{\rm 3d}<0$, the match is not exact at first sight. In this region, $\phi$ gives rise to an anti-BPS excitation and it is the anti-chiral multiplet $\phi^\dagger$ (with flavor charge $-1$, opposite statistics, and R-symmetry of the multiplet shifted by $-1$) that creates a BPS particle. However, the blocks as defined do not see this distinction -- they are analytically continued across $\Re\, m^{\rm 3d}=0$ (\ie\ across $|x|=1$) without any trouble. The simplicity of this analytic continuation hides an interesting subtlety of blocks. Indeed, for $m^{\rm 3d}<0$, there are effective Chern-Simons terms for background fields remaining at low energy, as can be seen from \eqref{TDshiftCS}. For example, there is a level $-1$ Chern-Simons term for the flavor symmetry. We are then led to attribute the difference between the analytic continuation and the true BPS counting in this regime to these Chern-Simons terms, and we write
\be \label{chiraltheta}
B_\Delta(x;q) = (qx^{-1};q)_\infty = \frac{\theta(-q^{-\frac 12}x;q)}{(x;q)_\infty}~,
\ee
where
\be \theta(z;q) := (-q^{1/2}z;q)_\infty (-q^{1/2}z^{-1};q)_\infty \label{deftheta0}\ee
is a Jacobi theta-function.
The denominator on the right-hand side of \eqref{chiraltheta} is the prediction of BPS counting, and the theta function is the effect of the Chern-Simons contact terms present in this region of parameter space. Remarkably, this is precisely the prescription for including Chern-Simons contact terms that we will be led to by a more formal analysis in Section \ref{sec:theta}.

The successful interpretation of $B_\Delta(x;q)$ as a BPS index is the first confirmation of the conjecture that the factorized pieces of the ellipsoid partition function and sphere index can be interpreted as partition functions on $\DqS$. We may also consider the limit $q\to 1$, or $\hbar\to 0$. The block $B_\Delta(x;q)$ has an asymptotic expansion given to all orders by
\be \label{BDpert}
B_\Delta(x;q) = (qx^{-1};q)_\infty \overset{\hbar\to 0}{\sim} \exp\frac1\hbar\bigg[\sum_{n=0}^\infty \frac{B_n\hbar^n}{n!} \Li_{2-n}(x^{-1})\bigg]\,,
\ee
where $B_n=\big(1,\frac12,\frac16,0,-\frac1{30},...\big)$ is the $n^{\rm th}$ Bernoulli number.
It makes no difference whether the limit $\hbar\to 0$ is taken from inside or outside the unit circle. This series captures the perturbative contributions of a chiral its KK modes to the twisted superpotential \eqref{BtoWt} of $T_\Delta$ compactified to two dimensions on $\RqS$ \cite{NS-I, DG-Sdual}. We will say more about this in Section \ref{sec:SQM}.

We conclude by mentioning the difference equations which arise from line-operator identities for the theory $T_{\Delta}$. These were described in \cite{DGG, DGG-index}, and they correspond to ``quantized Lagrangians'' for Chern-Simons theory on a tetrahedron \cite{Dimofte-QRS}. The difference equations take the form
\be (-1+\hat p+\hat x^{-1}) B_\Delta(x;q) = 0\,,\ee
in other words $B_\Delta(qx;q) = (1-x^{-1})B_\Delta(x;q)$, and it is easy to see from the infinite products that this is satisfied in both regimes $|q|<1$ and $|q|>1$. In the context of topological strings, this difference equation appeared much earlier in \cite{adkmv}, where it was interpreted as a quantization of the B-model curve which is mirror to $\C^3$. Indeed, the theory one obtains on a single toric brane in $\C^3$, in the canonical framing, is a theory $T_\Delta'$ of a free vortex, which is related to $T_\Delta$ by 3d mirror symmetry. We will compute the blocks of $T_\Delta'$ directly in Section~\ref{sec:free-vortex}.


\subsection{Uniqueness of the factorization}
\label{sec:unique}

It is interesting to ask whether the factorization of $T_\Delta$ partition functions found in Section \ref{sec:chiralblock} is unique. Suppose that we are looking for a function $B(x;q)$ such that
\begin{enumerate}

\item $B(x,q)$ is meromorphic in $x\in \C$ as well as in $q\in \C\bs\{|q|=1\}$\,;

\item there is some natural correspondence between the definitions of $B(x,q)$ in the regimes $|q|<1$ and $|q|>1$ --- \eg\ they have the same convergent $q$-hypergeometric series;

\item $B(x;q)$ is annihilated by the difference operator $\hat p+\hat x^{-1}-1$ in both regimes;

\item $\CZ^b_\Delta(X;\hbar) = B(x;q)B(\wt x;\wt q)$ and $\CI_\Delta(m,\zeta;q) = B(x;q)B(\wt x;\wt q)$\,.

\end{enumerate}
From condition (3), it follows that $B(x;q)=c(x;q)B_\Delta(x;q)$ where the prefactor $c(x;q)$ satisfies
\be c(qx;q) = c(x;q)\,,\ee
so that it is just a constant from the perspective of the difference operator. Then from (1) it follows that $c(x;q)$ must be an elliptic function both inside and outside the unit circle. It would be more standard to write $c(x;q)$ in terms of the logarithmic variables $X=\log x$ and $\hbar=\log q$; then ellipticity says that the function is invariant under $X\to X+2\pi i$ and $X\to X+\hbar$, as well as (here) $\hbar \to \hbar+2\pi i$.

Finally, conditions (2) and (4) require $c(x;q)c(\wt x;\wt q)=1$, given an appropriate relation between regimes $|q|<1$ and $|q|>1$.
The most natural way to satisfy this is to require $c(x;q)$ to be an elliptic ratio of \emph{theta functions}, namely
\be \label{cprod}
 c(x;q) = \prod_{i}  \theta\big( (-q^{1/2})^{b_i} x^{a_i};q\big)^{n_i}
\ee
where the product is finite, $\theta$ is the theta-function from \eqref{deftheta0}, and $a_i$, $b_i$ and $n_i$ are integers that satisfy
\be \label{ccond}  \sum_{i} n_ia_i^2 = 0\,,\qquad \sum_i n_i a_i b_i = 0\,,\qquad \text{and sometimes}\;\;  \sum_i n_i b_i^2 = 0\,. \ee
An example of a function that satisfies the first two constraints is $\theta(x^2;q)/\theta(x;q)^4$.

The first two constraints in \eqref{ccond} imply ellipticity. It is an interesting exercise to check that the constraints also cause the product \eqref{cprod} to satisfy condition (4). For example, the modularity of the theta-functions implies that for S-fusion (for the $S^3_b$ `tilde' operation)
\be \label{thetaexp} \theta\big( (-q^{1/2})^{b} x^{a};q\big)\,\theta\big( (-\wt q^{1/2})^{b} \wt x^{a};\wt q\big) = C^{-2} \exp\Big[-\tfrac{1}{2\hbar}\big(aX+b(i\pi+\hbar/2)\big)^2\Big]\,,
\ee
with $C$ as in \eqref{ZDelta}. Then the constraints \eqref{ccond} ensure that $|\!|c(x;q)|\!|_S^2=1$ modulo a power of $C$. We could partially absorb these powers of $C$ by including factors of $(q)_\infty$ in each theta-function. Usually we work modulo such corrections, which correspond to R-R contact terms, in which case we will ignore the third constraint, as it will only modify the product \eqref{thetaexp} by a sign and some power of $C$.

In the case of identity-fusion (for the sphere index), the product $|\!|c(x;q)|\!|_{id}^2$ is identically equal to $1$, and the third constraint is not needed. Again, this is ultimately due to the fact that the index is insensitive to R-R contact terms.

Thus, from a purely mathematical perspective, we have found that the factorization of the $T_\Delta$ partition functions into the blocks $B_\Delta(x;q)$ is unique \emph{up to} multiplication $B_\Delta(x;q)\to c(x;q)B_\Delta(x;q)$ by modular elliptic functions of the form \eqref{cprod}.
Such an ambiguity will persist throughout this paper for all non-perturbative constructions of blocks. These ratios of theta functions may have a nice physical interpretation in term of ``resolving'' Chern-Simons contact terms in a cigar geometry, which we discuss in Section \ref{sec:theta}.

Notice that if we take $\hbar$ to be small, then an elliptic ratio of theta functions $c(x;q)$ has a trivial perturbative expansion:
\be c(x;q) \overset{\hbar\to 0}\sim i^\#C^\#\,,\ee
for some integer powers of $i$ and $C$.
This is accurate to all orders in $\hbar$, and independent of whether $\hbar$ approaches zero from inside or outside the unit circle. The expansion follows by combining the elliptic and modular properties of $c(x;q)$; or more explicitly by observing that each theta function has an asymptotic expansion $\theta(x;q)\overset{\hbar\to 0}\sim C^{-1}\exp\big(-\frac1{2\hbar}X^2\big)$, similar to the S-fusion product \eqref{thetaexp}, so that in an elliptic ratio \eqref{cprod} all nontrivial asymptotics cancel. Therefore, multiplication by $c(x;q)$ introduces a purely non-perturbative ambiguity into blocks, a non-perturbative ambiguity of a very special type.

The statements made here about uniqueness of $B_\Delta(x;q)$ will apply equally well to blocks of any theory with a single vacuum (hence a single holomorphic block).
If there are multiple massive vacua in a theory, leading to multiple blocks $B^\ra(x;q)$, we have the freedom to rescale each $B^\ra(x;q)$ by an elliptic ratio of theta-functions $c(x;q)$, as well as to perform a linear transformation
\be \label{Mtrans1}
 B^\ra(x;q) \to  \begin{cases} \sum_{\beta} M^\alpha{}_\beta B^\beta(x;q) & |q|<1 \\
 \sum_{\beta} (M^{-1T})^\ra{}_\beta B^\beta(x;q) & |q|>1 \end{cases}
\ee
for a constant matrix $M$. Both of these transformations preserve fused products. The piecewise linear transformation \eqref{Mtrans1} will appear naturally as a Stokes phenomenon for blocks.


\section{Blocks from quantum mechanics}
\label{sec:SQM}

We now take a closer look at holomorphic blocks for gauge theories of the general type discussed in Section \ref{sec:3dintro}. Our aim is to formulate the blocks as certain partition functions in supersymmetric quantum mechanics. This approach is closely aligned with our view that factorization of ellipsoid and index partition functions arises from the three-dimensional analogue of topological/anti-topological fusion. The geometry $\DqS$ is a torus fibration over the half-line $\R_+=\{t\in[0,\infty]\}$, with the torus achieving fixed area and complex structure as $t\to\infty$. Macroscopically, the geometry is then one-dimensional, and by Kaluza-Klein reduction on the torus fiber, one may obtain a description of the problem in the language of supersymmetric quantum mechanics. This will provide a natural and intuitive framework for discussing the general form of holomorphic blocks and their properties, and will lead to a very general picture of how holomorphic blocks should be computed. 

We will see that the blocks of a given gauge theory are computable via a finite-dimensional contour integral. We determine the integrand perturbatively to all orders in $\hbar$. The different blocks then arise from different choices of contours, where the contours are  determined by gradient flow with respect to the superpotential of the quantum mechanics. This approach will provide valuable intuition for the behavior of blocks. However, in this analysis we will not determine the exact non-perturbative integrand and integration contours. Rather, we will combine the present results with the constraints imposed by identities for line operators to generate a non-perturbative block integral in Section \ref{sec:blockint}. It would be interesting to find a non-perturbative completion of the path integral derivation here. Such a derivation would be especially desirable for applications of these ideas to holomorphic blocks in more than three dimensions.

An essential property of the theories that we consider is the presence of massive vacua at generic values of mass parameters. We will always assume that we have deformed a theory to such a point in its parameter space. This gives us control over the dimensional reduction, and ensures, \eg, that the quantum-mechanical path integrals that compute the blocks are divergence-free.


\subsection{Kaluza-Klein reduction}

Our first goal is to describe the effective $\CN=4$ quantum mechanics in the bulk of $\R_+$ coming from a reduction on $\DqS$. For this purpose, it suffices to consider the flat, asymptotic region of $\DqS$, which has the form $T^2\times \R$, and reduce on the torus $T^2$. By continuing to work in the limit $\rho\gg\beta$ and at small $\hbar=2\pi i\beta \rho$, the reduction can be performed in two steps: first reducing on the circle $S^1_\beta$ (in fact, on an exactly periodic cycle of $T^2$ that is slightly offset from $S^1_\beta$) to obtain an effective two-dimensional theory on $S^1_\rho\times\R$, and then reducing on $S^1_\rho$ to quantum mechanics. A non-perturbative version of this derivation would require a one-step reduction on the torus fiber at generic values of $\tau$.

The reduction on the first circle yields an effective $\cN=(2,2)$ supersymmetric theory. Its dynamics are just controlled by twisted F-terms.%
\footnote{This is generally true even in the curved part of the cigar. The effective $\cN=(2,2)$ theory there is A-twisted, so only twisted F-terms are relevant; though the computation of the twisted F-terms is no longer so simple.} %
The computation of these terms in the action has been described previously in \cite{NS-I} (see also \cite{DG-Sdual} for some relevant discussion). We briefly review some of its relevant aspects.

In general, when reducing a three-dimensional gauge theory on a circle of radius R, one can include Wilson lines for global symmetries that complexify the real mass parameters,
\be m_i = m_i^{\rm 3d}+ \frac{i}{R}\oint_{S^1} A_i\,,\qquad i=1,...,N\,. \label{m2d} \ee
As already discussed in Section \ref{sec:3dintro}, complex mass parameters $m_i$ are twisted masses and complexified FI terms in the effective two dimensional theory, and they can be treated as scalars in background twisted chiral multiplets $M_i=m_i-i\sqrt{2}\theta^+\overline\lambda_+ -i\sqrt{2}\overline\theta^-\lambda_-+\ldots$~. The real scalars $\sigma^{\rm 3d}_a$ in gauge multiplets are similarly complexified by Wilson lines of the gauge field and become complex scalars in twisted chiral multiplets $\Sigma_a=\sigma_a-i\sqrt{2}\theta^+\overline\psi_+ -i\sqrt{2} \overline\theta^-\psi_-+\sqrt{2}\theta^+\bar\theta^-(D_a-i\star F_a)+\ldots$~, where
\be \sigma_a = \sigma_a^{\rm 3d} + \frac{i}{R}\oint_{S^1} A_a\,,\qquad a=1,...,r\,. \label{s2d} \ee
As long as the abelian symmetries of the theory are compact, invariance under large gauge transformations of the three-dimensional theory will manifest as periodicity of the complex scalars $\sigma_a$ and $m_i$,
\be  \sigma_a \sim \sigma_a + \frac{2\pi i}{R}\,,\qquad m_i \sim m_i+\frac{2\pi i}{R}\,. \label{period} \ee
This periodicity is not a general property of twisted chiral multiplets in $\CN=(2,2)$ theories. The lone exception is the background twisted chiral whose scalar is an FI parameter, whose imaginary part (a theta-angle) is always periodic. Otherwise, this is a special property of two-dimensional effective theories that descend from three-dimensions.

A single chiral multiplet $\phi$ in three dimensions gives rise to an entire tower of Kaluza-Klein modes in two dimensions. If $\phi$ is charged under an overall $U(1)$ symmetry with associated real mass $m_\phi^{\rm 3d}$ (some linear combination of $\sigma_a^{\rm 3d}$ and $m_i^{\rm 3d}$), the KK mode $\phi_n$ with momentum $n$ on the circle will have a twisted mass given by
\be 
m_{\phi_n}=m_\phi + \frac{2\pi in}{R}\,,\qquad n\in \Z\,. \label{KKspec}
\ee
This spectrum of masses is invariant under shifts $m_\phi \to m_\phi+2\pi i/R$, and the effects of the entire tower of KK modes must be included in order to preserve the periodicity described by \eqref{period}.

The twisted superpotential can be a function of the dynamical and background twisted chiral multiplets, $\wt W(\Sigma_a,M_i)$, and receives one-loop quantum corrections from integrating out massive charged chiral multiplets \cite{Witten-phases}. The contributions from an entire KK tower of chirals can be summed to give %
\be  \delta\wt W(M_\phi) = \sum_{n\in \Z} \big( M_\phi+\tfrac{2\pi i n}{R}\big) \big[\log( R M_\phi+2\pi i n)-1\big] \,\simeq\, \frac R4 M_\phi^2 + \frac{1}{R}\Li_2(-e^{-R M_\phi})\,. \label{WKK} \ee
Any three-dimensional chiral multiplet makes a contribution of the form of \eqref{WKK} to the twisted superpotential, with $M_\phi$ the superfield containing $m_\phi$ (a linear combination of $M_i$ and $\Sigma_a$). The other contributions to the twisted superpotential are tree-level Chern-Simons terms. A supersymmetric Chern-Simons interaction with level matrix $k_{ab}$ contributes $\tfrac R2k_{ab}\Sigma_a\Sigma_b$. Generalizing to include mixed gauge-flavor interactions (\emph{a.k.a.} FI terms) $k_{ai}$ and pure background flavor interactions $k_{ij}$, we obtain a total Chern-Simons contribution
\be \tfrac{1}{R}\wt W_{\rm CS}(\Sigma_a,M_i)= \tfrac12 k_{ab}\Sigma_a\Sigma_b+ k_{ai}\Sigma_a M_i+\tfrac12 k_{ij} M_i M_j\,. \label{WCS} \ee

The superpotential described above (or more importantly, the action derived from it) is not invariant under the large gauge transformations \eqref{period}. This is because we have neglected a crucial ingredient \cite{NS-I, NekWitten}. By working with twisted chiral superfields, we have made a change of integration variables in the path integral from the abelian gauge fields $A_a$ to their gauge-invariant field strengths $F_a$ that appear as auxiliary fields in the twisted chiral multiplets. This is acceptable only so long as we also impose quantization of the field strengths, $\int F_a/2\pi\in\Z$. In order to impose this constraint, we introduce an array of delta functions into the path integral for each integral value of $\int F_a/2\pi$ --- the Dirac comb --- via its fourier series,
\be
\sum_{n_a\in\Z}\exp\left[{-2\pi i n_a\int d^2\theta\,\Sigma_a}\right]~.\label{DiracComb}
\ee
Without this term, the failure of the action to be single-valued is visible in the shifts of the first derivative of the superpotential by integer multiples of $2\pi i$. In particular, the dilogarithm function appearing in \eqref{WKK} has multiple sheets labeled by pairs of integers $(b,c)$ on which the values of the dilogarithm function are related to its value on the principle branch by
\be\label{dilogbranch}
\Li_2(-e^{-x})\rightarrow \Li_2(-e^{-x})+2\pi i b (x+i\pi)+4\pi^2 c~,
\ee
while the quadratic terms which implement Chern-Simons interactions are manifestly multi-valued.
The addition of the overall factor in \eqref{DiracComb} then amounts to summing over the sheets of the covering space $\wt\cM$ of the scalar manifold $\cM$ on which the action is single-valued (note the sheets of the dilogarithm on which the superpotential differs by constant factors are already identified, because the constant shifts are killed by the superspace integration). The entire integrand of the path integral is thus single valued on the original target space $\cM$, {\it i.e.}, the space of periodic scalar field values.

\begin{wrapfigure}{r}{2.2in}
\includegraphics[width=2.1in]{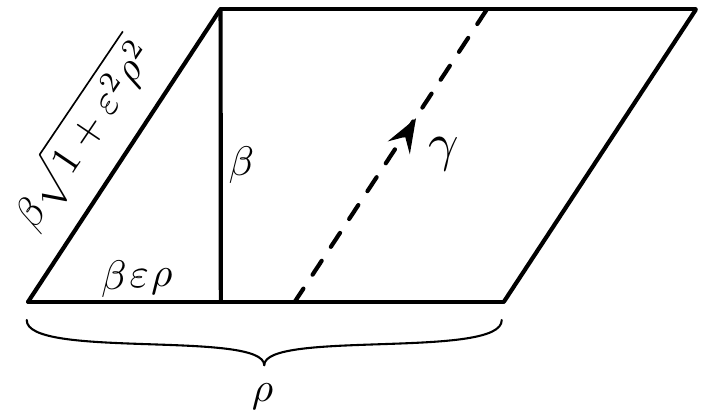}
\caption{The cycle used for the first reduction.}
\label{fig:T2reduce}
\end{wrapfigure}
Now let us return to the setup on $\DqS$, with the asymptotic profile $T^2\times S^1$. We will reduce on the cycle $\gamma$ of $T^2$ shown in Figure \ref{fig:T2reduce}. The fibration in the geometry makes this cycle a slight deformation of the non-contractible circle $S^1_\beta$ at the tip of the cigar. In the limit of fixed, small $\hbar$ and $\rho\to\infty$, the radius of the cycle $\gamma$ in the asymptotic region is given by $R = \beta\sqrt{1+\varepsilon^2\rho^2}\to \tfrac1{2\pi i}\hbar\rho$, whereas for $\hbar=0$ exactly, the radius would just be $\beta$. Once the reduction to two dimensions has been achieved, going down to one dimension is straightforward. This is because in reducing from two dimensions to one dimension, Kaluza-Klein modes on the circle do not correct the twisted F-terms, and so the dimensional reduction can be performed  directly (by demanding that all fields be independent of the periodic direction). The only result is to render the effective twisted superpotential dimensionless via an overall multiplication by the size $2\pi\rho$ of the circle, leading to the quantum-mechanical superpotential
\begin{align}\label{QMSuper}
& W^{\rm QM}(\Sigma_a,M_i) =  2\pi\rho\,\wt W(\Sigma_a,M_i) \notag \\
& \qquad =
\frac i{\hbar} \bigg[\sum_{\phi_i}\left(\tfrac 14 M_{\phi}^2+\Li_2(-e^{-M_{\phi}})\right)+\tfrac 12 k_{ab}\Sigma_a\Sigma_b  +k_{ai}\Sigma_aM_i+ \tfrac 12 k_{ij}M_iM_j\bigg]\,.
\end{align}
We have absorbed factors of $R$ into all the superfields, rendering them dimensionless. From here on out we will work in terms of these renormalized fields, which are cylinder-valued with period $2\pi i$.%
\footnote{We have also implicitly assumed that all chirals $\phi$ have $U(1)_R$ charge equal to one, so that the fermions in these multiplets have R-charge zero. This is relevant for reduction on $\DqS$ where the $U(1)_R$ gauge field has a holonomy $\exp(i\pi)=-1$ around the $S^1_\beta$ circle in the asymptotic region, as we have defined it for holomorphic blocks. For a chiral with general R-charge $R_\phi$, the twisted mass appearing in \eqref{QMSuper} gets modified to $m_\phi \to m_\phi+(R_\phi-1)\pi i$.}

We are left with a one-dimensional $\cN=4$ supersymmetric Landau-Ginzburg model with target space $\cM = (\C^*)^r$ and (up to $\cO(\hbar)$ corrections) superpotential given by \eqref{QMSuper}. The path integral of this theory is somewhat unconventional due to single-valuedness of the action on $\cM$ being achieved through the term \eqref{DiracComb}, which is not easily interpreted as a single contribution to the twisted superpotential. Rather, it is more natural to think of the resulting partition function as one which is formulated on the covering space $\wt\cM$, and then the sum in \eqref{DiracComb} is a sum over deck transformations of the cover. After this summation, the resulting contribution to the path integral will be single-valued on $\cM$.%
\footnote{Note that $\wt\cM$ is not simply the space of fields $\sigma_i$, but is determined by the detailed form of the twisted superpotential and its branching structure.} 
%
This perspective will prove useful for when considering localization for this theory.


\subsection{Vacua and boundary conditions at infinity}
\label{sec:bdy-inf}

To define the path integral of this one-dimensional theory on $\R_+$, we must specify the boundary conditions at $t\to\infty$ and $t=0$. We first consider the asymptotic boundary condition at $t\to\infty$. This amounts to a choice of massive, supersymmetric vacuum for the theory, and we demand that the fields in the path integral asymptotically approach their expectation values in that vacuum. It is crucial that the vacuum have a mass gap --- otherwise massless fluctuations could lead to infrared divergences of the partition function. This is one of many reasons behind our requirement that the original three-dimensional theory be massive at generic points in parameter space. A more precise condition will appear momentarily.

The equations that govern the vacua are given by \cite{NekWitten}
\be\label{multivacua}
\frac{\partial\wt W}{\partial\sigma_a} = 2\pi i n_a~,\qquad n_a\in\Z~.
\ee
This is written more invariantly after passing to single-valued $\C^*$-variables
\be s_a = e^{\sigma_a} \qquad x_i = e^{m_i}\,,\ee
by imposing
\be \exp\bigg(s_a\frac{\pd \wt W}{\pd s_a}\bigg) = 1\,. \label{expvacua} \ee
The left-hand side is a rational function in $(s_a,x_i)$. There are a finite number of distinct solutions to \eqref{expvacua} if and only if the vacua are all massive. We must disallow situations where two roots of \eqref{expvacua} coincide, and, more seriously, situations in which the equations \ref{expvacua} are independent of some $s_a$, leaving them undetermined. This latter possibility indicates that our initial theory did not have enough flavor symmetries to lift its moduli space. Assuming distinct, discrete solutions, we label them $s_a^{(\alpha)}$, with $\alpha$ indexing the vacua.
 
If the sigma model is formulated on $\wt\cM$, where the individual terms in the action are single-valued, then for each term in \eqref{DiracComb} with some fixed integers $\vec{n}$, the boundary condition for vacuum $\alpha$ will impose that the fields approach the image of the vacuum $\alpha$ on the appropriate sheet where \eqref{multivacua} is solved for that value of $n_a$. This is the choice of boundary condition that is naturally invariant under large gauge transformations.

Note that equations \eqref{expvacua} are the same as the equations that govern the supersymmetric vacua of our theory on untwisted $\RS$ \cite{NS-I}. In part, this is just another way of looking at the same construction in the $\rho\to\infty$ limit. 
The reason that we are finding vacuum equations on $\RS$ and not twisted (or Omega-deformed) $\RqS$ is due to our reduction on the cycle $\gamma$ above, which let us ignore the twist. We have in effect demonstrated that vacua on $\RS$ and $\RqS$ are equivalent (given an appropriate redefinition of fields), and this is not too surprising: we would not expect vacua to be charged under the rotations of $\R^2$, \ie\ to have non-trivial spin.

We also recall the construction in \cite{DG-Sdual, DGG} of an auxiliary algebraic variety $\CL_{\rm SUSY}$, the ``supersymmetric parameter space'' of the theory, obtained by adjoining the equations
\be  \exp\bigg(x_i \frac{\pd \wt W}{\pd x_i}\bigg) = p_i\,,\qquad i=1,...,N\, \label{expparam} \ee
to \eqref{expvacua}. These define the effective background FI parameters $p_i$ for the flavor symmetries that would allow supersymmetry to be preserved, were the flavor symmetries to be weakly gauged.  Equations \eqref{expparam}, just like \eqref{expvacua}, are rational in $s_a$ and $x_i$. After eliminating the $s_a$ from the combined system \eqref{expvacua}--\eqref{expparam}, one is left with $N$ polynomial equations that cut out a middle-dimensional algebraic variety in the space $(\C^*)^{2N}$ parameterized by $(x_i,p_i)$
\be \CL_{\rm SUSY}\,:\quad f_i(x,p)=0\,. \label{LSUSY} \ee
By construction, this is a holomorphic Lagrangian variety with respect to the holomorphic symplectic form
\be  \Omega = \sum_{i=1}^N \frac{dx_i}{x_i}\wedge \frac{dp_i}{p_i}\,.\ee
As long as the superpotential is nondegenerate, every solution to \eqref{expvacua} uniquely determines background FI parameters at fixed $x_i$. Therefore, the massive vacua of the effective two-dimensional theory may equally well be characterized as solutions to \eqref{LSUSY} at fixed $x_i$. This is often a more invariant characterization. For example, $\cL_{\rm SUSY}$ is invariant under the action of three-dimensional mirror symmetry (and other infrared dualities) as long as one stays away from massless loci in parameter space. As noted in Section \ref{sec:diff}, the identities for line operators acting on blocks are a quantization of the equations for $\CL_{\rm SUSY}$.


\subsection{Boundary condition at the origin}
\label{sec:bdy-zero}

The boundary condition at $t=0$ is defined by the tip of the cigar in $\DqS$. It is a half-BPS boundary condition supported on all of $\cM$, which was described in \cite{Witten-path}.
The boundary condition assigns to every fixed value of the macroscopic fields $s_a$ a certain weight -- \ie, it can be described as a choice of wavefunction inserted at $t=0$. Wavefunctions in supersymmetric sigma models can be interpreted as differential forms on $\cM$, and the choice of bosonic weight for the wave function needs to supplemented with insertions of fermionic operators to preserve supersymmetry. In this case, the appropriate insertion that preserves supersymmetry corresponds to multiplying the wave function by the holomorphic top form on $\cM$,
\be
\underline{\Omega} = \frac{ds_1}{s_1}\wedge\frac{ds_2}{s_2}\wedge\cdots\wedge\frac{ds_r}{s_r} = d\sigma_1\wedge\cdots\wedge d\sigma_r~.
\ee

The bosonic part of the wavefunction can be determined from its three-dimensional microscopic description. It should just be the holomorphic block for the chiral multiplets of the theory with the gauge fields fixed to background values corresponding to the argument of the wave-function! 
Perturbatively, we can find the wavefunction by summing up contributions to a two-dimensional twisted superpotential in the spinning background $\RqS$ --- \ie\ the tip of the cigar --- in a theory of free chiral fields. The result is 
\be\label{originwavefunction}
\Psi_0(s_a,m_i;\hbar) = \underline{\Omega} \exp\left(\tfrac 1\hbar \wt W_\hbar(s_a,m_i;\hbar)\right)\,,
\ee
with the ``quantum-corrected superpotential'' $\wt W_\hbar$ given (as discussed in Sections \ref{sec:OV} and \ref{sec:chiralblock}) by
\begin{align}
\wt W_\hbar(s_a,m_i;\hbar) &= \tfrac 12 k_{ab}\sigma_a\sigma_b + k_{ai}\sigma_am_i + \tfrac 12 k_{ij}m_im_j + \sum_\phi\big[ \tfrac14 m_\phi^2 + \Li_2(-e^{-m_\phi-\frac\hbar2};\hbar) \big]\,. \label{Wpert}
\end{align}
Here each chiral contributes the perturbative series%
\footnote{Again we are assuming that the chirals have $U(1)_R$ charge one. For general R-charge $R_\phi$, we must shift $m_\phi\to m_\phi+(R_\phi-1)(i\pi+\hbar/2)$. More generally, we can include an effective complex mass $m_R$ coming from the background R-charge gauge field $A_R$ as one of the $m_i$, feeding into the masses $m_\phi$ for chirals or into background Chern-Simons interactions. Its value is fixed to be $m_R = i\pi+\hbar/2$, due to the holonomy of $A_R$ around $S^1_\beta$ at the tip of $\DqS$.}
\be \Li_2(x;\hbar) := \sum_{n=0}^\infty \frac{B_n\,\hbar^n}{n!}\Li_{2-n}(x)\,, \ee
with $B_n$ the Bernoulli numbers $(1,\frac12,\frac16,0,-\frac{1}{30},...)$. We must also include the effect of Chern-Simons interactions for all gauge fields, which are simply quadratic terms in \eqref{Wpert} to all orders in $\hbar$.
This superpotential shifts by non-perturbative terms under large gauge transformations of the background fields --- a feature which will be resolved by the non-perturbative completion of Section \ref{sec:blockint}.


\subsection{Localization of the bulk path integral}

We finally turn to the evaluation of the path integral on $\R_+$. Modulo the subtleties associated to single-valuedness of the superpotential, this is a problem that has been considered many times before --- for a recent discussion, see \cite{Witten-path}. Here we just remind the reader of the relevant facts. The bosonic part of the action for the branch labeled by integers $\vec n$ in \eqref{DiracComb} is given by
\be
I_{\vec n}=\int dt\,d^4\theta\,g_{a\bar b}\Sigma^a\overline{\Sigma^b}+\int dt\,d\theta\,d\bar\theta\, W^{\rm QM}_{\vec n}(\Sigma_a,M_i)+c.c.~,
\ee
where the D-term that contains kinetic terms for the fields is a $Q$-commutator, so the precise form of the K\"ahler metric $g_{a\bar b}$ is irrelevant. The path integral of this theory can be evaluated by localizing to field configurations that are invariant under the action of those supercharges preserved at $t=0$. The result is that the field configurations that contribute to the path integral are precisely those that satisfy gradient flow equations with respect to the potential $\Im\, W^{\rm QM}$ on $\wt \cM$, as a function of the time coordinate $t$. For a choice of asymptotic boundary condition given by a critical point $\alpha$, the path integral localizes onto field configurations that asymptote to $s_a^{(\alpha)}=\exp \sigma_a^{(\alpha)}$ at $t\to\infty$, and that evolve according to
\be
\frac {d\sigma_a}{dt} = g_{a\bar b}\frac{d\, \Im(W^{\rm QM})}{d\ol\sigma_{\bar b}}~.
\ee
This leads to a simple characterization of the state that impinges upon the boundary at $t=0$ --- it is simply the Poincar\'e dual to the downward gradient-flow cycle associated to critical point $\alpha$,
\be
\Psi_{\alpha}(s_a,m_i) \simeq {\rm PD}[\Gamma^\ra]~.
\ee
The partition function is then given by the overlap of this state with the boundary state at $t=0$, leading to the following expression for the holomorphic block:
\be
B^\alpha(x;q)=\cZ_{\rm QM}\simeq\langle 0_q|\alpha_q\rangle = \int_\cM \Psi_{0}\wedge\star\Psi_{\alpha}~.
\ee
Given our identification of the wavefunction $\Psi_{\alpha}$ with the Poincar\'e dual of the cycle $\Gamma^{\ra}$, this simplifies to a contour integral on the gradient-flow cycle itself,
\be\label{qmresult}
B^\alpha(x;w)=\cZ_{\rm QM} \simeq \int_{\Gamma^\ra}\Omega\,\exp\Big(\frac1\hbar\wt W_\hbar(s_a,m_i,\hbar)\Big)~.
\ee
For such an expression to make sense, the cycles $\Gamma^\ra$ must be well matched with the integrand $\exp \frac1\hbar \wt W_\hbar$; in other words, the integrals should converge. This seems quite plausible for small $\hbar$, since the magnitude of the integrand precisely matches the potential for gradient flow in the $\hbar\to 0$ limit: $\Im\,W^{\rm QM}(\sigma_a,m_i) = \Re\big(\frac1\hbar \wt W_\hbar(s_a,m_i;\hbar=0)\big)$\,.

The schematic form of $\cZ_{\rm QM}$ derived here should be correct even when computing holomorphic blocks non-perturbatively (which will be our next goal). This is because the structure of \eqref{qmresult} followed just from general considerations of the behavior of supersymmetric quantum mechanics in this type of setup. However, the definitions of $\Gamma^\ra$ and $\wt W_\hbar$ will certainly be corrected relative to what we found in this section.
The operator insertion $\wt W_\hbar$ requires non-perturbative completion. 
Moreover, the cycles $\Gamma^\ra$ should be true gradient flow cycles of an exact effective superpotential, which agrees with \eqref{QMSuper} only at leading order.
(Although \eqref{QMSuper} was correct to all orders in $\hbar$ given our choice of reduction on the torus in the asymptotic region, there may be a field redefinition between the asymptotic region and the origin that can introduce perturbative $\hbar$ corrections. In addition, the presence of multiple sheets of $\wt M$ must be dealt with, via some manifestly non-perturbative effect.) Nonetheless, the considerations that led to \eqref{qmresult} have allowed us to understand enough of the structure of these contour integrals that we will be able to fix them up in the next section with additional help from line-operator identities.


\section{Block integrals}
\label{sec:blockint}

We have just seen that holomorphic blocks can be expressed as partition functions in an effective supersymmetric quantum mechanics where, given a three-dimensional gauge theory, the dynamical fields in the quantum mechanics are the complexified, exponentiated scalars $\sigma_a$ from the vector multiplets. In this section, we would like to promote those quantum-mechanically motivated contour integrals to a non-perturbative prescription for computing holomorphic blocks.

Given an $\CN=2$ gauge theory with matter, our construction takes the form of a formal contour integral
\be \label{BI-general}
\mathbb B(x;q) = \int_* \frac{ds}{2\pi i s}\, [\text{CS contributions}] \times [\text{matter contributions}]~, 
\ee
which generates solutions to line-operator identities for the theory. The contributions from chiral matter (as well as W-bosons for nonabelian gauge fields) are products of the basic ``tetrahedron block'' $B_\Delta$ given by \eqref{BD}. These are meromorphic functions of the complexified, exponentiated flavor parameters $x_i$ and the $s_a$. The contributions of Chern-Simons terms (dynamical and background) will also be genuine meromorphic functions of the exponentiated $x_i$ and $s_i$, in contrast to the quadratic exponentials that appeared perturbatively.

The integration is over an unspecified middle-dimensional cycle in $(\C^*)^r$, where $r$ is the rank of the gauge group. Indeed, the contour integral over \emph{any} cycle $\Gamma$ will solve the relevant line-operator identities so long as the integral converges, the boundary $\pd\Gamma$ is either empty or at asymptotic infinity, and $\Gamma$ stays sufficiently far away from poles of the integrand. We then propose that all blocks $B^\ra(x;q)$ for a given theory can be obtained by performing the integration over appropriate cycles $\Gamma^\ra$. 

The prescription given by \eqref{BI-general} is a non-perturbative completion of the quantum mechanical integrals described above, so we already know the physical principle by which the correct basis of cycles should be specified. Perturbatively in $\hbar$, the critical points $\alpha$ of the integrand are in one-to-one correspondence with the vacua given by solutions of \eqref{expvacua}. The correspondence can be continued to finite $\hbar$. Then from each critical point $\alpha$ we can define the cycle $\Gamma^\ra$ using downward gradient flow in a neighborhood of the critical point. Far away from the critical points, we will need to adjust the contours to avoid singularities. This is simply a consequence of the fact that we do not have the exact superpotential for the non-perturbative quantum mechanical description. As the parameters $x$ are varied, we expect to see explicit Stokes phenomena whereby the basis cycles are reorganized.

We would like the integral \eqref{BI-general} to define blocks $B^\ra(x;q)$ with the same basic properties as the fundamental chiral block $B_\Delta(x;q)$. Namely:\label{block-criteria}
\begin{enumerate}

\item $\{B^\ra(x;q)\}$ are a set of meromorphic functions of $x\in \C$ and $q\in \C\bs \{|q|=1\}$ with no analytic continuation from $|q|<1$ to $|q|>1$.

\item The perturbative expansions of $B^\ra(x;q)$ in $\hbar$ match on the inside and outside of the unit circle $|q|=1$ for fixed $\alpha$ and $x$. 

\item For each $\alpha$, $B^\ra(x;q)$ can be written as a single $q$-hypergeometric series that converges for $q$ both inside and outside the unit circle.

\item $\{B^\ra(x;q)\}$ form a basis of solutions to the line-operator identities $f_i(\hat x,\hat p;q)B^\ra(x;q)=0$ for the gauge theory.

\item The products $\CZ_b(X;\hbar) = \big|\!\big| B^\ra(x;q)\big|\!\big|_{S}^2$ and $\CI(m,\zeta;q)= \big|\!\big| B^\ra(x;q)\big|\!\big|_{id}^2$ reproduce the $S^3_b$ and $\SqS$ partition functions. Moreover, the $S^3_b$ partition function can be analytically continued from $\hbar <0$ to $\hbar >0$ across the physical half-line $\hbar=2\pi ib^2\in i\,\R_+$.

\end{enumerate}

Some of these properties --- such as (1), (2), and (4) --- follow in a straightforward manner from the construction of the integral. For example, the integrand itself is  a meromorphic function of $x$ and $s$, with no analytic continuation from $|q|<1$ to $|q|>1$ but with the same perturbative expansion and $q$-hypergeometric series in the two regimes. Then demonstrating (1) and (2) is a matter of extending these properties from the integrand to the integrals. Property (3) is still conjectural, though it can be probed in examples where the integrand is evaluated analytically by summing residues. Of course, property (5) is the main conjecture of this paper.

While the blocks defined by the block integral inherit most of the interesting properties of the fundamental tetrahedron block $B_\Delta(x;q)$, they also inherit its main ambiguity: they can be multiplied by elliptic prefactors $c(x;q)$ of the form described in \eqref{cprod}.%
\footnote{We have anticipated that blocks $B^\ra(x;q)$ are naturally associated to a Stokes chamber in parameter space, and transform linearly from one chamber to another. Although \emph{a priori} we could introduce an  ambiguous $c^\alpha(x;q)$ independently for each block $B^\ra(x;q)$, compatibility with the Stokes jumps forces the prefactors to all be equal (at least among any subset of blocks that interact at Stokes walls).} %
This seems to be the price to pay for a non-perturbative completion of this sort. The ambiguity will be most evident in our treatment of Chern-Simons terms in Section \ref{sec:theta}, and might conceivably be resolved with additional physical input. Recall from Section \ref{sec:unique}, however, that the special elliptic functions $c(x;q)$ are invisible to the line-operator identities and drop out of the $S^3_b$ and $\SqS$ partition functions, rendering the ambiguity fairly innocent for many purposes.


\subsection{Assembling Lagrangians and line-operator identities}
\label{sec:Ward-assemble}

Since the principle underlying the block integral \eqref{BI-general} is that it generates solutions to the line-operator identities, introduced in Section \ref{sec:diff}, we will now describe the systematic construction of these  identities. Suppose that the SCFT in question has a UV gauge theory description. We can build up the full gauge-theory Lagrangian by starting with a number of free chirals multiplets and then applying a sequence of elementary moves --- such as adding Chern-Simons levels, gauging flavor symmetries, and adding superpotential terms. The  identities for free chirals are simple, and the idea is to define the elementary moves so that each of them also transforms the identities in a tractable manner.%
\footnote{Such manipulations are closely related to arguments of \cite{Dimofte-QRS} in the context of wavefunctions and difference operators. More generally, they fall under the mathematical theory of holonomic functions, \cf\ \cite{Zeil-holonomic, Zeil-AB}, which we are basically extending to the level of physical gauge theories (following \cite{DGG,DGG-index}).} %
The precise form of transformations for each move can easily be deduced by looking at $S^3_b$ partition functions or indices. We know exactly from \cite{HHL, KW-index} how acting on the Lagrangian of a given theory modifies these partition functions, and also that two copies of the cigar identities must be satisfied by these partition functions (\cf, Section \ref{sec:diff}). Alternatively, the transformations can be derived by direct physical arguments, much along the lines of what was done in \cite{DGG, DGG-index} for theories of class $\CR$. We will generalize the constructions of line-operator identities in \cite{DGG,DGG-index} by allowing for nonabelian gauge groups. However, we will start by reviewing all the steps that are relevant for abelian theories.

To build the Lagrangian for an abelian theory, we first introduce $N$ free chirals $\phi_i$. It is convenient to make sure that our theory has no gauge or flavor anomalies at every stage in the construction --- in particular, we only expect the blocks to be well defined in the absence of anomalies --- so let us start with $N$ copies of the anomaly-free theory $T_\Delta$. The theory $T_\times := T_{\Delta_1}\otimes\cdots \otimes T_{\Delta_{N}}$ has maximal abelian flavor symmetry $U(1)^{N}= U(1)_1\times\cdots \times U(1)_{N}$. The $i$'th flavor symmetry rotates the phase of the chiral $\phi_i$, has a level $-1/2$ Chern-Simons term as dictated by \eqref{TDdef}, and has (say) an associated real mass parameter $x_i$. The operators annihilating the partition function of $T_\times$ on $\DqS$ are simply $N$ copies of the operator for $\cT_\Delta$,
\be \label{prodops}
\qquad \hat f_i^{(\times)} =  \hat p_i + \hat x_i^{-1}-1 \simeq 0\,,\qquad i=1,...,N\,,
\ee
where the Wilson loops $\hat x_i$ act as multiplication and the dual 't Hooft loops $\hat p_i$ act as $q$-shifts, so $\hat p_i \hat x_j=q^{\delta_{ij}}\hat x_j\hat p_i$.

The theory can then be modified arbitrarily by applying the following elementary moves. First, we allow a redefinition of the flavor symmetry by a linear transformation $U\in GL(N,\Q)$, \ie, a redefinition of the basis of $U(1)$'s. Correspondingly, the operators \eqref{prodops} are transformed according to%
\footnote{If $U$ or $U^{-1}$ have non-integer entries, it means that the electric-magnetic charge lattice of the theory is being redefined by a stretch or a squeeze. Correspondingly, the transformation may introduce roots of the electric and magnetic line operators $\hat x_i$ and $\hat p_i$ into the line-operator identities. Depending on the intended physical definition of the theory, it may be desirable (and it is always possible) to eliminate these roots by multiplying the identities on the left by appropriate polynomials in roots of $\hat x_i$ and $\hat p_i$.} %
\be \label{Uxp} \hat x_i \mapsto \prod_j  (\hat x_j)^{U^{-1}_{ij}}\,,\qquad \hat p_i \mapsto \prod_j (\hat p_j)^{U_{ji}}\,.\ee
Similarly, we may redefine the R-symmetry current by adding to it a multiple of the $U(1)_i$ flavor currents. This can equivalently be described as shift of the $U(1)_i$ flavor gauge fields $A_i$, sending $A_i\mapsto A_i+\sigma_i A_R$, for some constants $\sigma_i$. It is easy to see from the $U(1)_R$ holonomies in the $\DqS$ geometry that this will modify the Wilson line operators according to
\be \label{sx} \hat x_i \mapsto (-q^{\frac12})^{\sigma_i}\hat x_i.\ee
(We will always choose $\sigma_i$ to be integers; otherwise one should interpret $(-q^{\frac12})^{\sigma_i}$ as $e^{(i\pi+\frac\hbar2)\sigma_i}$.) A dual transformation is to introduce a mixed flavor/R-symmetry contact term, \ie, a background Chern-Simons interaction $\sim \sum_i\sigma^{(P)}_i \int A_id A_R$. This must act on the line operators as
\be \label{sp} \hat p_i  \mapsto (-q^{\frac12})^{\sigma_i^{(P)}}\hat p_i\,.\ee

More interestingly, we can add Chern-Simons terms for flavor symmetries. The addition of a term $\sum_{ij}\frac12 k_{ij} \int A_idA_j$ with integer level matrix $k_{ij}$, properly supersymmetrized, acts as
\be \label{abCS} \hat x_i\mapsto \hat x_i\,,\qquad  \hat p_i \mapsto q^{-\tfrac12 k_{ii}} \bigg[\prod_j (\hat x_j)^{-k_{ij}}\bigg] \hat p_i\,. \ee
This action can equivalently be described as conjugating all the line operators with the operator $\exp \sum_{ij} \frac{k_{ij}}{2\hbar}\hat X_i\hat X_j$, where $\hat X_i$ are formal logarithms of $\hat x_i$.

Now consider the operation of gauging a flavor symmetry $U(1)_i$. After gauging, shifts in the corresponding parameter $x_i$ will act trivially, since $x_i$ has become dynamical. In the line-operator identities, we must eliminate $\hat x_i$, and then set $\hat p_i \to 1$. The elimination is done by multiplying the polynomial difference operators on the left, and adding and subtracting them --- formally, this is elimination in a \emph{left ideal}. Since we are gauging an abelian symmetry, the three-dimensional theory gains a ``topological'' flavor symmetry $U(1)_J$, coupled to the gauged $U(1)_i$ by an FI term. The total effect of abelian gauging can then be reproduced in two steps. Before gauging, we first introduce a new symmetry $U(1)_J$ coupled to the flavor $U(1)_i$ by a mixed Chern-Simons term, but not to the rest of the theory. This adds new line operators $\hat x_J\,,\hat p_J$ obeying the identity
\be \quad \hat p_J - \hat x_i \simeq 0 \qquad \text{(fixed $i$)}\,,\ee
and shifts $\hat p_i \mapsto \hat x_J^{-1}\hat p_i$. Then we gauge $U(1)_i$ as above, which tells us to eliminate $\hat x_i$ from the line-operator identities and to set $\hat p_i\to 1$. The combined effect on the original operators is the transformation
\be \label{abgauge} \quad \hat x_i \mapsto \hat p_J\,,\qquad \hat p_i \mapsto \hat x_J^{-1}\,, \ee
interchanging Wilson and 't Hooft lines. Indeed, this abelian gauging is equivalent to S-duality in a 4d abelian gauge theory for which the three-dimensional theory under discussion plays the role of a boundary condition. This is the setup discussed in Section \ref{sec:diff}, as well as \cite{Witten-SL2, DGG, DGG-index}.

The final operation is the addition of a gauge-invariant operator $\CO_i$ to the superpotential to break a certain $U(1)_i$ flavor symmetry. The precise form of the operator is unimportant; it may well be a non-perturbative monopole operator, as used in class $\CR$. However, it must have R-charge equal to two so that the $U(1)_R$ R-symmetry of the theory is preserved. Since the cigar partition function is invariant under superpotential deformations, the only effect of adding $\CO_i$ is to fix the value of a corresponding mass parameter $x_i\to 1$. Consequently, the action on line-operator identities is to first eliminate the shift $\hat p_i$ and then to set $\hat x_i=1$. Again, elimination takes place in the left ideal.

By iterating the moves we have just defined, we can construct the Lagrangian for any abelian $\CN=2$ gauge theory, and simultaneously build its line-operator identities. It is interesting to note that all of the nontrivial complexity in the line-operator identities arises from algebraic ``elimination'' steps, such as the elimination of $\hat p_i$ induced by adding a superpotential.

To generalize this program so as to allow for nonabelian gauge groups, we need only modify the gauging rule. Suppose we have a theory with $U(1)^r$ flavor symmetry that is enhanced to a simple nonabelian group $G$. (For this to be the case, the matter content of the theory must fill out complete multiplets of $G$, and the superpotential must be invariant.) We take $\mathbb T=U(1)^r$ to be a maximal torus of $G$. Let $\vec x = (x_1,...,x_r)$ denote the mass parameters associated to this maximal torus (all other parameters are unaffected), and let $\Delta_+$ denote the positive roots of $G$, in a basis corresponding to $\mathbb T$. Then to perform a nonabelian gauging, we first conjugate all the line-operator identities by $ \prod_{\eta\in \Delta_+}\big( \vec x^{\frac12\eta}-\vec x^{-\frac12\eta}\big)$. This is the effect of including W-bosons in the theory. Afterwards we apply the rule above for gauging all the $U(1)$'s in $\mathbb T$, \ie\ we eliminate all $\hat x_i$ ($i=1,...,r$) from the identities and set the conjugate $\hat p_i\to 1$. The validity of this prescription can be verified by looking at $S^3_b$ or $\SqS$ partition functions. 

For example, to gauge an $SU(2)$ flavor symmetry, we identify a single, fixed $U(1)_i$ corresponding to the maximal torus. In a standard normalization of the root $\eta\in\Delta_+$, $x^{\frac12\eta}=x$. Thus we send
\be \label{SU2gauge}
\qquad \hat p_i \mapsto (\hat x_i-\hat x_i^{-1})\hat p_i \frac{1}{\hat x_i-\hat x_i^{-1}} = q^{-1}\frac{1-\hat x_i^2}{1-q^2\hat x_i^2}\hat p_i
\ee
in every identity. Denominators of the form $(1-q\hat x_i)$ can subsequently be removed by multiplying on the left. Then $\hat x_i$ is eliminated completely and $\hat p_i$ is set to $1$. Unlike an abelian gauging \eqref{abgauge}, which preserves the rank of the flavor group, the nonabelian gauging reduces the number of flavor symmetries.

The introduction of dynamical Chern-Simons terms $k\int \Tr(AdA+\frac23 A^3)$ for a nonabelian gauge group simply involves ignoring the cubic $A^3$ part and treating $\Tr(AdA)$ as a sum of abelian Chern-Simons terms for a maximal torus of $G$.  This is exactly how nonabelian Chern-Simons terms contribute to $S^3_b$ and $\SqS$ partition functions --- and more relevantly to the potential of an effective supersymmetric quantum mechanics as in Section \ref{sec:SQM} --- so it must be the case that the action on line-operator identities can be analyzed this way. The abelian Chern-Simons terms can be built up by moves of the form \eqref{abCS} on abelian flavor symmetries before doing a nonabelian gauging.


\subsection{Chern-Simons terms and theta functions}
\label{sec:theta}

We now have two ways of thinking about the block integral \eqref{BI-general}. By considering the reduction on $\DqS$ to supersymmetric quantum mechanics, we expect each matter or gauge multiplet in the Lagrangian to contribute directly to the integrand, after which we integrate over twisted chirals as indicated. Alternatively, we can think of building up both the Lagrangian and the block integral for the theory by a sequence of elementary moves and transformations, as in Section \ref{sec:Ward-assemble}. By starting with a product of blocks $B_\Delta$ for a theory of free chirals, we can transform the mass parameters, multiply by appropriate functions for additional Chern-Simons levels, perform integrations corresponding to gauging, and fix parameters $x_i\to 1$ when symmetries are broken. This allows us to construct the entire block integral by deriving the right transformation rules so that line-operator identities are satisfied at every step.
It will prove useful to keep both perspectives in mind. In either approach, we must find non-perturbative versions of various ingredients and transformations that lead to the desired analytic properties of blocks --- such as being meromorphic functions in the gauge and flavor mass parameters both for $|q|<1$ and $|q|>1$.

For the contribution of a chiral multiplet with background Chern-Simons level $k=-1/2$, the answer was already given in Section \ref{sec:chiralblock}. Any $T_\Delta$ constituent of a larger theory will consequently contribute to the block integral a term given by
\be \label{BD-2}
B_\Delta(x;q) := (qx^{-1};q)_\infty = \sum_{n=0}^\infty \frac{x^n}{(q^{-1})_n} = \begin{cases} \prod_{r=0}^\infty (1-q^{r+1}x^{-1}) & |q|<1~, \\
  \prod_{r=0}^\infty (1-q^{-r}x^{-1})^{-1} & |q|>1~. \end{cases}
\ee
On the one hand, this has the right perturbative expansion (\cf\ \eqref{BDpert}) to match the quantum-mechanics prediction \eqref{Wpert}; on the other, it satisfies the correct line-operator identity $\hat p+\hat x^{-1}-1\simeq 0$.

Now let us consider the addition of a Chern-Simons term at level $+1$ to the theory, for some flavor symmetry with parameter $x$. (This could be a combination of gauge and flavor $U(1)$'s.) In the quantum-mechanics approach, we argued that this simply added a factor
\be \label{logCS} \exp \frac{1}{2\hbar}X^2 = \exp \frac{1}{2\hbar}(\log x)^2\,,  \ee
to the integrand. While the factor \eqref{logCS} does transform line-operator identities the right way, by conjugating
\be \hat p \mapsto \Big(\exp \frac{1}{2\hbar}X^2\Big) \hat p \Big(\exp \frac{-1}{2\hbar}X^2\Big)  = q^{-\frac12}x^{-1}\hat p\,\ee
(compare this with \eqref{abCS}), it is not meromorphic in either $x$ or $q$. A simple mathematical solution is to replace the quadratic exponential \eqref{logCS} with a Jacobi \emph{theta function},
\be \label{CStheta} \exp \frac{1}{2\hbar}X^2\;\leadsto\; \frac{1}{\theta(x;q)}\,, \ee
where, as in \eqref{deftheta0}, we define
\be \label{deftheta}
\theta(x;q) := (-q^{\frac12}x;q)_\infty (-q^{\frac 12}x^{-1};q)_\infty  = \begin{cases} (q)_\infty^{-1}\sum_{n\in \Z}q^{\frac{n^2}{2}}x^n & |q|<1 \\[.1cm]
  (q^{-1})_\infty \Big( \sum_{n\in \Z}q^{-\frac{n^2}2}x^n\Big)^{-1} & |q|>1\,. \end{cases}
\ee
It is easy to see that the theta function in \eqref{CStheta} acts in the expected way on line operators, $\theta(x;q)\,\hat p\,\theta(x;q)^{-1}=q^{-\frac12}x^{-1}\hat p$, but now also has the right analytic properties. In addition, the asymptotic behavior of the theta function,
\be \label{theta-asymp}
\theta(x;q)^{-1} \overset{\hbar\to 0}{\sim} C\exp\left(\frac{1}{2\hbar}X^2\right)\,,
\ee
which terminates at $O(\hbar)$ due to modularity, is correct to reproduce the perturbative Chern-Simons contribution to quantum-mechanics integral, up to a small correction by  the factor $C=\exp\frac{-1}{24}\big(\hbar-\tfrac{4\pi^2}\hbar\big)$\,. The correction term is the same as the one discussed in Sections \ref{sec:chiralblock}--\ref{sec:unique}.\footnote{\label{foot:qinf}
If desired, the correction can be partially absorbed by rescaling $\theta(x;q)\to (q)_\infty\theta(x;q)$, with $(q)_\infty$ as in \eqref{defqinf}, noting that $(q)_\infty \overset{\hbar\to 0}{\sim} (\frac{2\pi i}\hbar)^{1/2} e^{i\pi/4}C$ when $|q|<1$ and $(q)_\infty \overset{\hbar\to 0}{\sim}(\frac{\hbar}{2\pi i})^{1/2}e^{i\pi/4}C$ when $|q|>1$.}
One way to motivate the replacement \eqref{CStheta} is to sum all images of the quadratic exponential \eqref{logCS} under the transformation $X\to X+2\pi i$, thus enforcing periodicity. Then $\sum_{n\in \Z} \exp \frac{1}{2\hbar}(X+2\pi i n)^2 \sim \theta(x;q)^{-1}$ follows from a modular transformation.

We encountered the theta-function above in Section \ref{sec:chiralblock}. There we saw that in order to properly identify the free-chiral block $B_\Delta(x;q)$ with a BPS index, it was important to compute the BPS index in the infrared, and to include a contribution from \emph{effective} Chern-Simons terms induced by massive fermions. This contribution had to take the form of a theta function in order for the index of the chiral to be a continuous function in the mass parameter $x$.

We can offer yet another (related) physical derivation of the replacement \eqref{CStheta} as a consistency condition for holomorphic blocks. Consider a copy of $T_\Delta$ whose free chiral multiplet transforms with charge $+1$ under a $U(1)$ flavor symmetry with parameter $x$, and shift the R-symmetry so that the chiral has R-charge 1. The block of the resulting theory $T_\Delta^1$ is $B_\Delta^1(x;q) = (-q^{\frac12}x^{-1};q)_\infty$. Then let us form a theory $T_{\rm CS}$ made of two copies of $T_\Delta^1$, together with a superpotential coupling
\be \label{supCS} W=\mu\phi_1\phi_2 \ee
between the two chirals $\phi_1$ and $\phi_2$. The superpotential coupling preserves $U(1)_R$ because the product $\phi_1\phi_2$ has R=2. Moreover, the coupling breaks one of the flavor symmetries, with $\phi_1$ and $\phi_2$ having opposite charges $(+1,-1)$ under a single unbroken $U(1)$. Therefore, the block for the combined theory $T_{\rm CS}$ is just
\be \label{blockCS}
B_{\rm CS}(x;q) = B_\Delta^1(x;q) B_\Delta^1(x^{-1};q)= (-q^{\frac12}x^{-1};q)_\infty(-q^{\frac12}x;q)_\infty = \theta(x;q)\,.
\ee

By scaling the coefficient of the superpotential \eqref{supCS}, we can give both $\phi_1$ and $\phi_2$ an arbitrarily large mass, and integrate them out. The block is insensitive to F-terms. Integrating out the fermions in the chiral multiplets leads to shifts of the Chern-Simons level for the unbroken $U(1)$, but the shifts are in opposite directions for $\phi_1$ and $\phi_2$, and cancel out. Nevertheless, an overall Chern-Simons level $k=-1$ is left over from the initial definitions of the theories $T_\Delta^1$. We are ultimately led to associate the block \eqref{blockCS} to the theory of a pure background Chern-Simons term at level $-1$. (An analogous argument at level $+1$ would have led to an inverse theta function as in \eqref{CStheta}.)

This derivation of the contribution of background Chern-Simons terms serves to illustrate an important ambiguity in our prescription. In order to generate a background Chern-Simons term at level $k=4$, we could consider the theory of two chirals described by $T^1_\Delta$ with charges $(+2,-2)$; or four pairs of chirals described by $T^1_\Delta$ with charges $(+1,-1)$. These two situations lead to two different replacement rules for the associated gaussian term,
\be 
\exp \frac{-4}{2\hbar}X^2 \;\leadsto\; \theta(x^2;q)\quad\text{or}\quad \theta(x;q)^4~.
\ee
Either one of these is a reasonable non-perturbative completion of the quadratic exponential. The line-operator identities satisfied by the two theta functions are identical, and the asymptotics only differ by a power of $C$, \ie\ an R--R contact term.

Extrapolating from these simple examples, we can describe a general prescription for Chern-Simons levels. Suppose we have an $N\times N$ Chern-Simons level matrix $k_{ij}$, coupling either gauge or flavor symmetries, as well as a vector $\sigma_i$ of levels for mixed Chern-Simons terms between gauge or flavor symmetries and the R-symmetry. We would like to represent this as a finite product
\be \label{CSreplace}
 \exp\Big[ \frac{1}{2\hbar}\sum_{i,j}k_{ij}X_iX_j +\frac1\hbar \sum_i\sigma_i X_i\big(i\pi+\tfrac\hbar2\big) \Big]\;\leadsto\;
 \prod_{t} \theta\big((-q^{\frac12})^{b_t} x^{a_t};q\big)^{n_t}\,,\ee
where $b_t$ and $n_t$ are integers and $a_t$ are (column) vectors of $N$ integers, such that
\be \label{CSconds}
\sum_t n_t\,a_t(a_t)^T = -k\,,\qquad \sum_t n_t b_t a_t = -\sigma\,.
\ee
The conditions \eqref{CSconds} are a consequence of requiring that the two sides of \eqref{CSreplace} satisfy the same line-operator identities. The same conditions ensure that they have the correct asymptotic expansion as $\hbar\to 0$. The product \eqref{CSreplace} also encodes an R-R Chern Simons coupling at level $k_{RR}=\sum_t n_tb_t^2$, as well as corrections corresponding to the factor $C$ in \eqref{theta-asymp}. As usual, we work modulo such ``constant'' terms.

There are infinitely many ways to choose a finite product \eqref{CSreplace} satisfying \eqref{CSconds}. They correspond to different ways of ``resolving'' Chern-Simons terms via pairs of massive chirals. The physical significance of this for partition functions on $\DqS$ remains unclear, and could benefit from further investigation. In this paper we will treat the choice of non-perturbative resolution as an ambiguity in the block integral.
Note that two different choices of theta functions on the RHS of \eqref{CSreplace} are related by a factor $c(x;q) =  \prod_{t} \theta\big((-q^{\frac12})^{b_t'} x^{a_t'};q\big)^{n_t'}$ where $\sum_t n_t'\,a_t'(a_t')^T=0$ and $\sum_t n_t'b_t'a_t'=0$\,. This is exactly the kind of elliptic function ambiguity discussed in Section \ref{sec:unique}. Recall that such a factor $c(x;q)$ is not only invisible to the line-operator identities (since $\hat p\,c(x;q)=c(x;q)$), but becomes trivial upon fusion, satisfying $\big|\!\big| c(x,q)\big|\!\big|_{S}^2=1$ (modulo powers of $C$) and $\big|\!\big| c(x,q)\big|\!\big|_{id}^2=1$\,.


\subsection{The integrand}
\label{sec:integrand}

We have now compiled all the ingredients necessary to construct the integrand for the block integral \eqref{BI-general}. Combining the observations of the previous two sections leads to the following rules.

Let us consider any $\cN=2$ gauge theory with $U(1)_R$ R-symmetry. Choose a maximal torus $U(1)^N$ for the flavor symmetry group, with associated mass parameters $x_i\in \C^*$, $i=1,...,N$. Also choose a maximal torus $U(1)^r$ for the gauge group, and denote its ``mass parameters'' (\ie\ complexified gauge scalars) $s_i\in \C^*$, $i=N+1,...,N+r$. As a preliminary step, let $T_\Delta^{R_\phi}$ denote the free-chiral theory $T_\Delta$ from Section \ref{sec:chiralblock}, where the R-charge of the (scalar in the) chiral multiplet has been shifted to be $R_\phi$. Explicitly,
\be \label{TDRdef}
T_\Delta^{R_\phi}\,:\quad \left\{\begin{array}{l}
 \text{free chiral $\phi$} \\
 \text{charges:}\quad \begin{array}{c|c}& \phi \\\hline
      F &  +1 \\
      R & \;R_\phi \end{array}\qquad
 \text{CS matrix:}\quad \begin{array}{c|cc} & F & R \\\hline
    F & -\frac12 & \frac12(1-R_\phi) \\
    R & \frac12(1-R_\phi) & -\frac12(1-R_\phi)^2\,.
    \end{array}
 \end{array}\right.
\ee
The contribution of this free chiral constituent of the theory is the block
\be B_\Delta^{(R_\phi)}(y;q) = \big((-q^{\frac12})^{2-R_\phi}y^{-1};q\big)_\infty\,, \ee
where $y$ is the mass parameter of the flavor symmetry. Then we apply the following rules for translating the content of the gauge theory into a block integrand.

\subsubsection*{Chiral matter}

Group every chiral multiplet $\phi$ into a copy of the theory $T_\Delta^{R_\phi}$, where $R_\phi$ is its R-charge. In other words, attach a set of Chern-Simons couplings as in \eqref{TDRdef} to this chiral, and compensate for these couplings (if needed) elsewhere in the Lagrangian. For every such copy of $T_\Delta^{R_\phi}$, add a factor
\be B_\Delta^{(R_\phi)}(y_\phi;q) = \big((-q^{\frac12})^{2-R_\phi}y_\phi^{-1};q\big)_\infty\,, \ee
to the integrand, where $y_\phi$ is the complexified mass of the chiral, a product of $x$'s and $s$'s corresponding to the $U(1)$'s under which it transforms. The grouping of chirals into theories $T_\Delta^{R_\phi}$ ensures that we never encounter anomalous gauge or flavor symmetries.

\subsubsection*{Chern-Simons terms}

After removing the copies of $T_\Delta$, we are left with an $(N+r)\times (N+r)$ integer matrix $k_{ij}$ of levels for (additional) abelianized Chern-Simons couplings. Both gauge and flavor symmetries are included on the same footing in this matrix. We also have an $(N+r)$-dimensional vector $\sigma$ of mixed Chern-Simons couplings between gauge or flavor symmetries and the R-symmetry. Choose a product of theta functions to represent these Chern-Simons terms, as in \eqref{CSreplace}. Namely, introduce a finite product
\be \label{CSreplace-2}
 \text{CS}[k,\sigma;x,s,q] =
 \prod_{t} \theta\big((-q^{\frac12})^{b_t} x^{a_t};q\big)^{n_t}\,,
\ee
where $b_t$ and $n_t$ are integers and $a_t$ are column vectors of $N+r$ integers such that
\be \label{CSconds2} 
\sum_t n_t\,a_t(a_t)^T = -k\,,\qquad \sum_t n_t b_t a_t = -\sigma\,.
\ee
R-R contact terms could also be matched, as discussed above. For example, a Chern-Simons coupling for parameter $x$ at level $+1$ becomes $\theta(x;q)^{-1}$. An FI term that mixes a gauge symmetry with parameter $s$ and a topological flavor symmetry with parameter $x$ is represented as
\be \text{FI:}\quad \frac{\theta(x;q)\theta(s;q)}{\theta(xs;q)}\,.\ee
Such FI terms should be present for all abelian gauge groups (unless the topological flavor symmetries are broken by superpotentials).

\subsubsection*{Nonabelian gauge symmetries}

For every simple nonabelian factor $G$ (of rank $r'$) in the gauge group, identify the subgroup $\mathbb T \simeq U(1)^{r'}\subset U(1)^r$ corresponding to its maximal torus. Suppose that $s_G = (s_1,...,s_{r'})$ are the parameters corresponding to $\mathbb T$, and let $\Delta_+$ be the set of positive roots corresponding to this maximal torus. Then add a factor
\be \label{nagauge-2}
\text{gauge}[G;s,q] =
\prod_{\eta\in\Delta_+} \frac{\theta(s_G^\eta;q)}{(qs_G^\eta;q)_\infty(qs_G^{-\eta};q)_\infty}
\ee
to the block integrand. For example, when $G=SU(2)$, this looks like
\be \label{nagauge-su2}
\text{gauge}[SU(2);s,q] = \frac{\theta(s^2)}{(qs^2;q)_\infty(qs^{-2};q)_\infty}\,.
\ee
Since parameters $s_G$ that are related by a $W(G)$ Weyl group action are equivalent, the domain of integration for the block integral (\ie\ the domain in which we will define convergent cycles) must also be quotiented by $W(G)$. When eventually choosing integration cycles $\Gamma^\ra$, the integral along a cycle that crosses $f$ Weyl-group images of a certain critical point should come with an extra symmetry factor $1/f$.

The perturbative contribution we must reproduce is $\prod_{\eta\in\Delta_+} \big( s_G^{\frac12\eta}-s_G^{-\frac12\eta}\big)$, coming from W-bosons. We know this has roughly the right form by looking at $S^3_b$ or $\SqS$ partition functions, or the action on line operators, \cf\ \eqref{SU2gauge} --- or even more directly, by considering the perturbative contribution to an effective quantum mechanics, as in \cite{NS-I}. However, this expected contribution is not generally meromorphic in $s_G$. To fix it, we write
\be s_G^{\frac12\eta}-s_G^{-\frac12\eta}=s_G^{\frac12\eta}\big(1-s_G^{-\eta}\big) = s_G^{\frac12\eta} \frac{\theta(-q^{\frac12}s_G^\eta;q)}{(qs_G^\eta;q)_\infty(qs_G^{-\eta};q)_\infty}\,,
\ee
and use the theta-function trick from Section \ref{sec:theta} to replace $s_G^{\frac12\eta}$ with the meromorphic function $\theta(s_G^\eta;q)/\theta(-q^{\frac12}s_G^\eta;q)$, which has the same perturbative expansion and difference equation. Then \eqref{nagauge-2} results.

\subsubsection*{Synthesis}

Putting together the three contributions above we obtain the integrand of \eqref{BI-general}. The final step is to integrate over the $r$ parameters $s_i\in \C^*$. We then obtain
\be \label{BI}
\boxed{\mathbb B(x;q) = \int_* \frac{ds}{2\pi is} \prod_G \text{gauge}[G;s,q] \times \text{CS}[k,\sigma;x,s,q]\times \prod_\phi B_\Delta^{(R_\phi)}\big(y_\phi(x,s);q\big)\,.}
\ee
Superpotential couplings play almost no role --- they break the flavor symmetry of a theory and simply restrict (implicitly) the parameters appearing in the integrand. Notice that the integrand, just like the expected blocks, is defined both for $|q|<1$ and $|q|>1$, with no analytic continuation between the regimes.
We predict that after performing the integration on a suitable basis of convergent contours $\Gamma^\ra\subset (\C^*)^r$ this integral will generate blocks that satisfy the five properties outlined in the introduction to this section.

For example, it is almost true by construction that integrals along convergent contours $\Gamma^\ra$ must satisfy the line-operator identities. Although we have given an all-at-once prescription for building \eqref{BI}, we could also have assembled it using iterated elementary moves as in Section \ref{sec:Ward-assemble}: forming a product of $B_\Delta$ blocks, redefining flavor symmetries and shifting R-symmetries, adding Chern-Simons levels, adding nonabelian gauge contributions, gauging by doing integrals, etc. At each step, the expected line-operator identities are obeyed.

Going through this carefully amounts to a proof that the integral \eqref{BI} is annihilated by the correct difference operators for the theory, modulo one important subtlety. We need integration $\int \frac{ds_i}{s_i}$ over some gauge variable $s_i$ to have the effect of eliminating a corresponding operator $\hat s_i$ from the difference equations and trivializing the conjugate shift $\hat p_i\to 1$. This is only true so long as the integration contours $\Gamma^\ra$ used to evaluate \eqref{BI} are invariant under $q$-shifts. That is, if we move an entire contour $\Gamma^\ra$ (multiplicatively) by an amount $q$, we must be able to deform it smoothly back to its original position. This implies that contours must either be closed or end asymptotically at $0$ or $\infty$ in each copy of $\C^*$ in $\cM$. Moreover, contours must stay at least a distance $q$ away from all poles of the integrand. These conditions will play a prominent role when choosing the proper $\Gamma^\ra$.

Let us also comment on the uniqueness of \eqref{BI}. The integrand of the block integral \eqref{BI} has the same ambiguity discussed in Section \ref{sec:unique} and at the end of Section \ref{sec:theta}: it can be multiplied by an elliptic ratio of theta functions, of the form \eqref{CSreplace}, with $\sum_t n_t a_t (a_t)^T = \sum_t n_t b_t a_t=0$ (the ellipticity condition). This ambiguity is inherited from both the choice of Chern-Simons contribution \eqref{CSreplace-2} and the choice of nonabelian gauge contribution \eqref{nagauge-2}. It might be fixed with further physical input, as indicated in Section \ref{sec:theta}. We will just treat it as a mathematical ambiguity.
Since the arguments of the theta functions involve both $x$ and $s$ variables, it is \emph{not} completely clear that upon evaluating \eqref{BI} on a mid-dimensional contour $\Gamma\subset (\C^*)^r$, the ambiguity in the answer will only be an elliptic ratio of theta functions. Nevertheless, it does turn out to be so in examples, and we expect that this will be the case in general.

In one case, it is possible to prove that the ambiguity of the integrand can be promoted directly to the evaluated integral: when the integral along $\Gamma^\ra$ is evaluated by summing residues. Since the entire integrand is formed from $(z;q)_\infty$ functions, its poles typically come in infinite families with spacing $q$, for example at $(s_0,qs_0,q^2s_0,q^3s_0,...)$. As long as this is the case, a factor $c(s,x;q)$ in the integrand consisting of an elliptic ratio of theta functions just contributes the same constant to each residue, since (\eg)
\be c(s_0,x;q) =  c(qs_0,x;q) =  c(q^2s_0,x;q) =  c(q^3s_0,x;q) = ...\,. \ee
This factor then passes outside the integral as (\eg) $c(s_0,x;q)$, which is still an elliptic ratio of theta-functions.

In practice, it is convenient after building the block integrand \eqref{BI} to multiply by an elliptic ratio of the theta functions to reduce the number (and complexity) of theta functions making an appearance. For example, we can substitute $\theta(z;q)^4\leadsto \theta(z^2;q)$, $\theta(y;q)^2\theta(z;q)^2\leadsto \theta(yz;q)\theta(y/z;q)$, $\theta(z;q)\theta((-q^{\frac12})^az;q)/\theta((-q^{\frac12})^bz;q)\leadsto \theta((-q^{\frac12})^{a-b}z;q)$, etc. We will make use of such simplifications in the examples below.


\subsubsection{Fusion commutes with integration}

When computed by supersymmetric localization, both the ellipsoid partition function and the sphere index reduce to integrals that are superficially analogous to the block integral \eqref{BI}. In fact, if we denote the integrand of \eqref{BI} as $\Upsilon(s,x;q)$, then we see that
\be \label{integrand-factor-S}
\CZ_b(X;\hbar) = \int_{\R^r}dS\, \big|\!\big| \Upsilon(s,x;q)  \big|\!\big|_S^2 = \int_{\R^r}dS \, \Upsilon(s,x;q)\Upsilon(\wt s,\wt x;\wt q) \,,\ee
where $s=\exp S$, $\wt s=\exp \frac{2\pi i}{\hbar}S$, etc., and the integration is done on a fixed, canonical cycle: the real slice $\R^n \subset \C^n$. Similarly, the index can be written as
\be \label{integrand-factor-id}
\CI(m,\zeta;q) = \sum_{n\in \Z^r} \oint_{(S^1)^r} \frac{d\sigma}{2\pi i\sigma}  \big|\!\big| \Upsilon(s,x;q)  \big|\!\big|_{id}^2 = \sum_{n\in \Z^r} \oint_{(S^1)^r} \frac{d\sigma}{2\pi i\sigma} \, \Upsilon(s,x;q)\Upsilon(\wt s,\wt x;\wt q) \,,\ee
with the ``identity fusion'' conjugation on the parameters, and $s=q^{\frac{n}{2}}\sigma$, $s=q^{\frac{n}{2}}\sigma^{-1}$. Now the integration is done on $r$ copies of the unit circle in $(\C^*)^r$, another fixed, canonical cycle.

In both expression for $\big|\!\big| \Upsilon(s,x;q)  \big|\!\big|^2$, the ambiguity related to the choice of theta functions completely disappears, modulo the factors of $C = \exp\big[\frac{-1}{24}(\hbar+\wt\hbar)\big]$ for $S$-fusion. In particular, note that
\be \big|\!\big|\theta\big((-q^{\frac12})^b x^a;q\big)\big|\!\big|_{S}^2 = i^{\#}C^{\#} \exp\Big[-\frac{1}{2\hbar}\Big((a\cdot X)^2+(i\pi+\tfrac\hbar2)b(a\cdot X)\Big)\Big]\,, \ee
while
\be \big|\!\big|\theta\big((-q^{\frac12})^b x^a;q\big)\big|\!\big|_{id}^2 = (-q^{\frac12})^{-(a\cdot m)b}\zeta^{-(a\cdot m)a}\,, \ee
so the RHS only depends on the the quantities $aa^T$ and $ba$, and is independent of the precise choice of theta functions on the LHS. These are the correct contributions of (\eg) Chern-Simons terms to the ellipsoid partition function and index, respectively. (Here `$x$' could denote both gauge and flavor parameters.)

Our main conjecture amounts to the statement that if we evaluate the block integral on appropriate integration cycles $\Gamma^\ra$ to find holomorphic blocks,
\be B^\ra(x;q) = \int_{\Gamma^\ra} \frac{ds}{2\pi is} \Upsilon(x,s;q)\,, \ee
then the ellipsoid partition function and index are sums of products of blocks,
\be \CZ_b(X;\hbar)=\big|\!\big|B(x;q)\big|\!\big|_{S}^2 := \sum_\alpha \big|\!\big|B^\ra(x;q)\big|\!\big|_{S}^2\,,\qquad \CI(m,\zeta;q) = \big|\!\big|B(x;q)\big|\!\big|_{id}^2:=\sum_\alpha \big|\!\big|B^\ra(x;q)\big|\!\big|_{id}^2\,. \notag\ee
Putting this together with \eqref{integrand-factor-S} and \eqref{integrand-factor-id} yields a rather beautiful result:
\be \label{comm-S}
\boxed{\int_{\R^r}dS\, \big|\!\big| \Upsilon(s,x;q)  \big|\!\big|_S^2 = \sum_\alpha \bigg|\!\bigg| \int_{\Gamma^\ra} \frac{ds}{2\pi is} \Upsilon(x,s;q)\bigg|\!\bigg|_S^2\,,}\ee
\be \label{comm-id}
\boxed{\sum_{n\in \Z^r} \oint_{(S^1)^r} \frac{d\sigma}{2\pi i\sigma}  \big|\!\big| \Upsilon(s,x;q)  \big|\!\big|_{id}^2 = \sum_\alpha \bigg|\!\bigg| \int_{\Gamma^\ra} \frac{ds}{2\pi is} \Upsilon(x,s;q)\bigg|\!\bigg|_{id}^2\,.}\ee
In other words, fusion commutes with integration!

Mathematically, identities of a similar flavor to the Riemann bilinear relations which describe period integrals on K\"ahler manifolds. Indeed, factorization via Riemann bilinear relations played an important role recently in describing the analytic continuation of complex Chern-Simons theory \cite{Wit-anal}, as well as in the two-dimensional version of topological/anti-topological fusion \cite{CV-tt*}. In \eqref{comm-S}--\eqref{comm-id}, there is a notable deviation from the usual picture in that the left hand side is an integral over a contour of the same dimension as the individual blocks on the right hand side. Nonetheless, the similarity is suggestive.


\subsection{Examples}
\label{sec:simple-ex}

In this section we consider several simple, illustrative examples of block integrals. We look at three gauge theories with gauge groups of rank one, so that the corresponding block integral is one-dimensional. These theories all possess a unique vacuum on $\RS$, and have only one holomorphic block.
The convergent integration cycles for the single block are essentially unique and are easy to identify. In addition, each theory is dual by three-dimensional mirror symmetry to a second theory that consists only of free chirals or free chirals with a superpotential. This allows us to check explicitly that the block integral gives a sensible answer. Since factorization for theories of chirals without gauge interactions is essentially automatic, successfully matching the blocks of the mirror for these theories amounts to a verification of our main conjecture.

These examples will demonstrate one of the most important and surprising properties of the block integrals. Since the integrand of the block integral \eqref{BI} is built entirely from functions $(z;q)_\infty$ (defined in \eqref{zqinf}), it represents two different analytic functions, one for $|q|<1$ and one for $|q|>1$. In each regime we choose an appropriate convergent contour $\Gamma_<$ and $\Gamma_>$, allowing us to calculate blocks both at $|q|<1$ and $|q|>1$. We then find that the expressions for the integrated blocks in the two regimes are related by sharing a $q$-hypergeometric series expansion, verifying property (3) from the introduction to this section.


\subsubsection{The free vortex}
\label{sec:free-vortex}

The first theory is mirror to the tetrahedron theory $T_\Delta$ \eqref{TDdef}. It is summarized as follows:
\be \label{STDelta}
T_\Delta':\quad \left\{ \begin{array}{ll}
  \text{$U(1)$ gauge theory (gauge scalar $s$), coupled to a chiral $\phi$;} \\
  \text{has topological $U(1)_J$ flavor symmetry with parameter (FI term) $x$.} \\
  \text{charges:}\quad \begin{array}{c|cc} & \phi & v \\\hline
    G & 1 & 0 \\
    F & 0 & 1 \\
    R & 0 & 0 \end{array}
  \qquad \text{CS levels:} \quad\begin{array}{c|ccc} & G & F & R \\\hline
     G & \frac12 & 1 & -\frac12 \\
     F & 1 & 0 & 0 \\
     R & -\frac12 & 0 & -\frac12   \end{array}\,.
  \end{array} \right.
 \ee
Here $G$ denotes the gauge symmetry, $F$ the topological $U(1)_J$ symmetry, and $R$ the R-symmetry.
This is the supersymmetric generalization of the abelian Higgs model. At low energies, it is simply the theory of free supersymmetric vortices, created by the monopole operator $v$ --- the ``magnetic'' dual of the chiral in $T_\Delta$.
In \cite{DGG}, the theory $T_\Delta'$ was denoted $\sigma ST\circ T_\Delta$, corresponding geometrically to a tetrahedron that has been rotated through an angle of $2\pi/3$.

The rules of the previous section dictate that the block integral for this theory is given by
\be
\label{STint} \mathbb B_\Delta'(x;q) = \int_* \frac{ds}{2\pi is}\,\frac{\theta(-q^{-\frac12}x;q)}{\theta(-q^{-\frac12}sx;q)}\,B_\Delta(s;q)\,,
\ee
with $B_\Delta(s;q)=(qs^{-1};q)_\infty$, and the theta functions encode the Chern-Simons couplings, including the FI term that couples the topological and gauge symmetries.

Let us denote by the integrand of \eqref{STint} by $\Upsilon_\Delta'(s,x;q)$. For small $\hbar$, it has the leading perturbative expansion
\be \Upsilon_\Delta'(s,x;q) \overset{\hbar\to 0}{\sim}
 \exp\frac{1}{\hbar}\bigg[-\frac12(\log (-x))^2+\frac12(\log (-sx))^2+\Li_2(s^{-1})\bigg] = \exp \frac1\hbar \wt W(x,s)\,,
\ee
where $\wt W(x,s)$ is the superpotential for the associated SQM in Section \ref{sec:SQM}.%
\footnote{In Section \ref{sec:SQM} we were careful in distinguishing various superpotentials and their scalings in 1d and 2d. Here we are less careful.
The potential denoted $\tilde W/\hbar$ here equals $-iW^{\rm QM}$ from Equation \eqref{QMSuper}, and also equals $\tilde W_\hbar(s,m,\hbar=0)/\hbar$ from \eqref{Wpert}. The variables $X$ and $S$ are the scaled versions of $m$ and $\sigma$ there, with period~$2\pi i$.} %
This superpotential has a unique critical point at $s^{(1)}(x)=1-x^{-1}$ which represents the unique vacuum of the theory $T_\Delta'$. After substituting this solution into Equation \eqref{expparam}, we find the supersymmetric parameter space, written in terms of $p=\exp(x\,\pd \wt W/\pd x)$, to be given by $p=1-x^{-1}$. The line-operator identity for the theory is the quantization of this constraint equation,
\be \label{WardST} \hat p+\hat x^{-1}-1 \simeq 0\,. \ee
This is the same as the identity for the mirror theory $T_\Delta$. One can then check that the operator \eqref{WardST} formally annihilates the integral \eqref{STint} so long as the contour of integration is invariant under shifts by $\hbar$.

\begin{figure}[htb]
\centering
\includegraphics[width=6in]{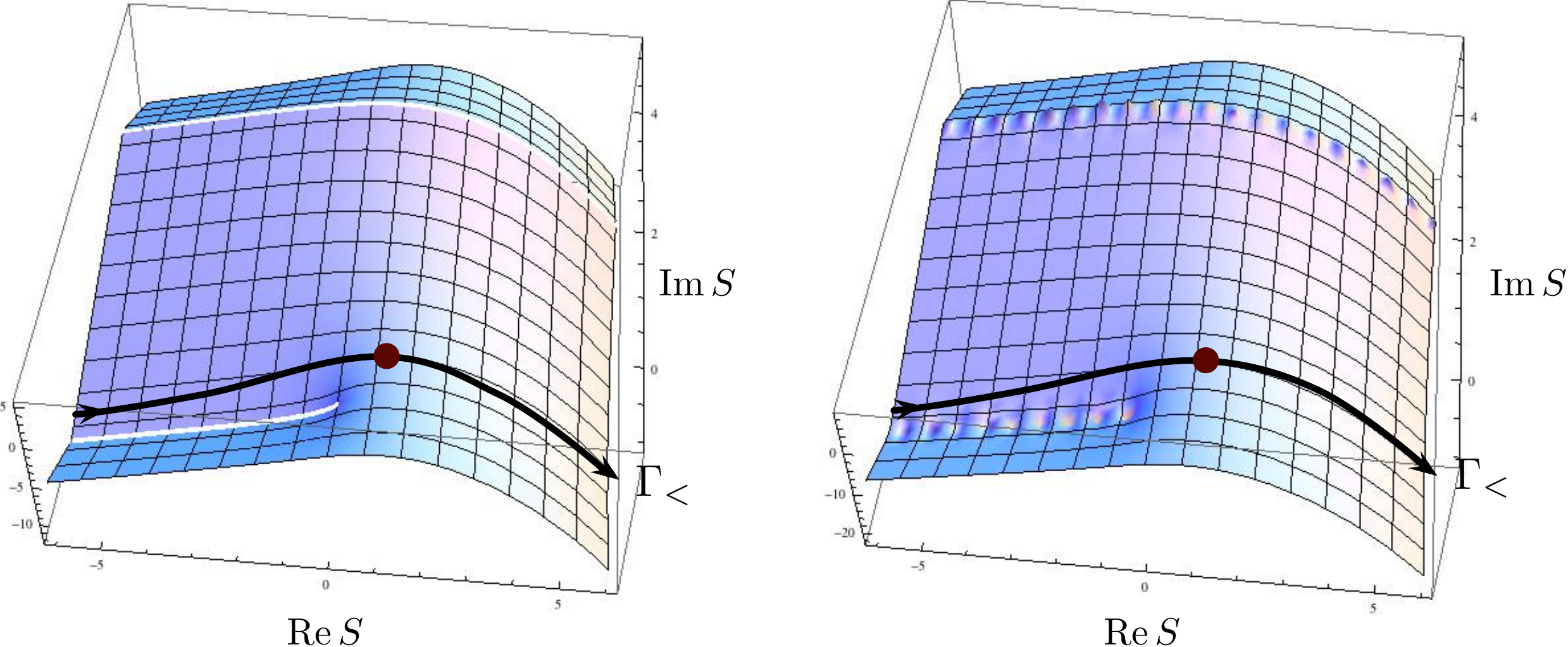}
\caption{The potential $\Re(\wt W(x,s)/\hbar)$ (left) and $\log|\Upsilon_\Delta'|$ (right) at fixed $x=e^{-1+\frac34 \pi i}$ and $\hbar=-\pi/5$. The  $\Im\,S$ direction is periodic. We have indicated the classical critical point and the cycle~$\Gamma_<$.}
\label{fig:ST-}
\end{figure}

Let us now consider the structure of the integral \eqref{STint}. We first take $\hbar$ real and negative (so $q$ is real and inside the unit circle). The potential $\Re\big(\frac1\hbar\wt W(x,s)\big)$ is depicted on the left of Figure \ref{fig:ST-}. On the right, we show the fully quantum-corrected integrand $\log|\Upsilon(x,s;q)|$. These plots appear as functions of the periodic variable $S = \log s$ at fixed $X = \log x$. The integrand has a full line of poles at $s=q^nx^{-1}$ ($S=-X+n\hbar$), $n\in \Z$, coming from the function $\theta(-q^{-\frac12}sx;q)$ (the FI term). There is also a half-line of zeroes at $s=q^n$ ($S=n\hbar$), $n>0$, associated with the chiral matter contribution $B_\Delta(s;x) = (qs^{-1};q)_\infty$. These families of zeroes and poles coalesce into branch cuts of $\wt W$ in the $\hbar\to 0$ limit.

An integration cycle $\Gamma_<$ is also shown in the figure. It is determined uniquely by the requirements that 1) it is nontrivial; 2) the integral along $\Gamma_<$ converges; and 3) the cycle is invariant under $q$-shifts. The cycle $\Gamma_<$ can be seen to match the downward gradient flow cycle for $\wt W(s,x)$ in the neighborhood of the saddle point, but away from the saddle point it is extended towards $S=\pm\infty$. Invariance under $q$-shifts implies that it cannot end at any finite zeroes of the integrand --- even if a naive downward flow would terminate there --- and cannot cross the line of poles. Thus, it can only tend asymptotically to infinity.

Numerical integration along $\Gamma_<$ produces
\be
B_\Delta'(x;q)\,=\, \frac{2\pi i}{\hbar}\int_{\Gamma_<}\frac{ds}{2\pi is} \frac{\theta(-q^{-\frac12}x;q)}{\theta(-q^{-\frac12}sx;q)}B_\Delta(s;q)\,=\, (q)_\infty B_\Delta(x;q)\,,  \qquad (|q|<1)\,.
\ee
Thus we recover the block of $T_\Delta$ up to a normalization factor $\frac{\hbar}{2\pi i}(q)_\infty$, which is of the type discussed around \eqref{defqinf} as being related to R-R contact terms.

\begin{figure}[t!]
\centering
\includegraphics[width=6in]{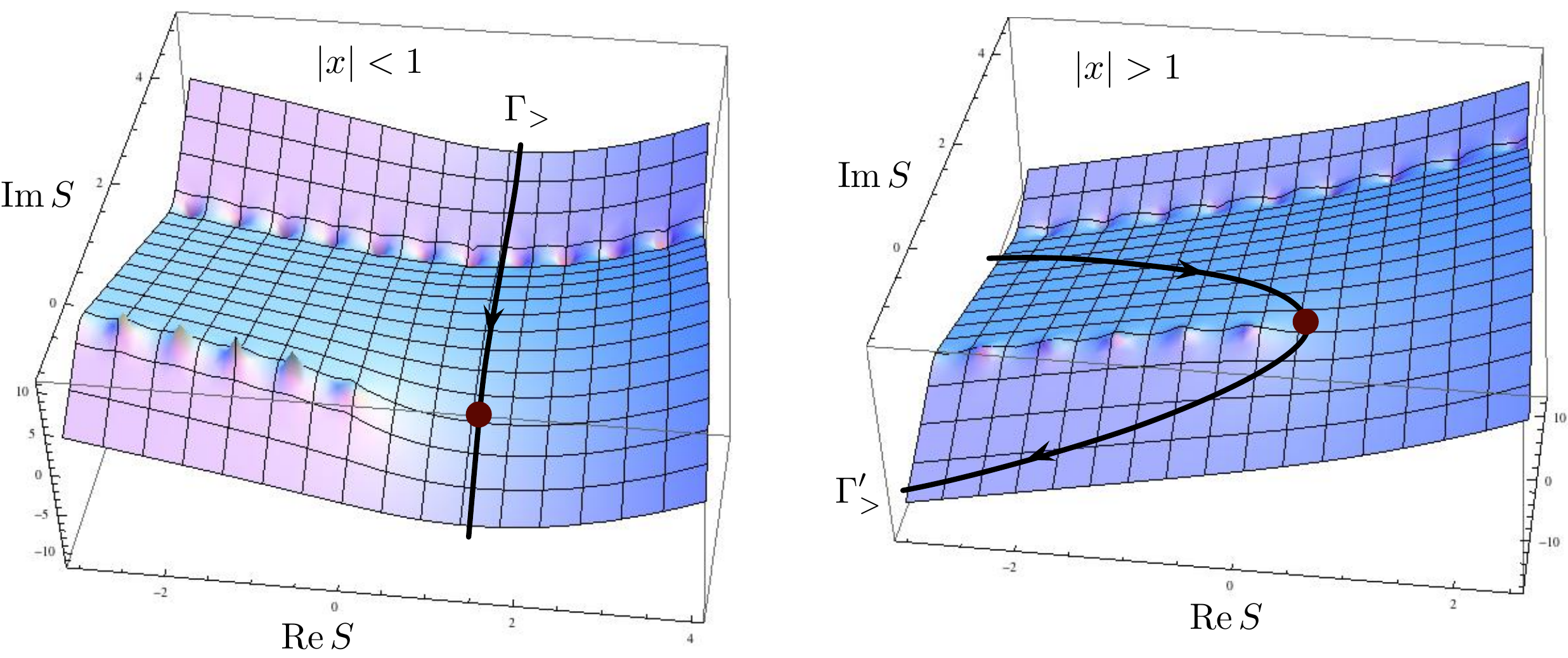}
\caption{Plots of the quantum-corrected potential $\log|\Upsilon_\Delta'|$ at $\hbar=\pi/5$, for $x=e^{-1+\frac34\pi i}$ and $x=e^{2+\frac34\pi i}$.}
\label{fig:ST+}
\end{figure}

Alternatively, we can consider $\hbar$ real and positive ($|q|>1$), and the analysis must be repeated. The theta function $\theta\big(-q^{-\frac12}sx;q\big)^{-1}$ now contributes a line of zeroes to the integrand, whereas the chiral block $B_\Delta(s;q)$ contributes a half-line of poles. Moreover, there are now two regimes to consider for the parameter $x$, $|x|<1$ and $|x|>1$. The integrand is plotted in Figure \ref{fig:ST+}.

When $|x|<1$, the downward-flow cycle in a neighborhood of the critical point is naturally extended to a closed cycle $\Gamma_>$ that is invariant under $q$-shifts. There is no problem crossing the line of zeroes. When $|x|>1$, it is more natural to extend $\Gamma_>$ in the negative $S$ direction, so that it encloses the half-line of poles. These two possibilities are completely equivalent, and topologically the choice of convergent shift-invariant cycle is again unique. The integral can then be evaluated exactly, either by taking the zeroth Fourier coefficient in $z$ (for $|x|<1$) or by summing residues of the enclosed poles at $s=q^{-n}$, $n\in \Z$ (for $|x|>1$). The results agree, giving
\be B_\Delta'(x;q)=\int_{\Gamma_>} \frac{ds}{s} \frac{\theta(-q^{-\frac12}x;q)}{\theta(-q^{-\frac12}sx;q)}B_\Delta(s;q)\,=\, (q^{-1})_\infty^{-1}B_\Delta(x;q)\,,  \qquad (|q|>1)\,.
\ee
Again, we reproduce the block $B_\Delta(x;q)=(qx^{-1};q)_\infty$ of $T_\Delta$, but now in the opposite regime.

The method of summing residues at $|q|>1$ has an interesting physical interpretation: each pole in $s$ at $s=q^{-n}$, $n\geq 0$ corresponds to vortex particles of charge $n$. Taking their residues builds up the block $B_\Delta'(x;q)$, thought of as a BPS index. The authors of \cite{GRR-bootstrap} recently utilized such an interpretation of residues to understand 4d indices in the presence of surface operators, which are the 4d lift of vortices.


\subsubsection{SQED and XYZ}

Our second simple example involves another mirror pair of theories, $N_f=1$ SQED and the XYZ model. Mirror symmetry between these theories, discovered in \cite{AHISS}, was related to ``2--3 moves'' for glued tetrahedra in \cite{DGG}.

The XYZ model is a theory of three $\CN=2$ chiral multiplets coupled by a cubic superpotential. The superpotential preserves a $U(1)_R$ R-symmetry and a $U(1)^2$ flavor symmetry. The theory is summarized as follows:
\be \label{defXYZ}
T_{\rm XYZ}: \left\{ \begin{array}{l}
\text{Chirals $\phi_1,\phi_2,\phi_3$, superpotential $W=\phi_1\phi_2\phi_3$;} \\
\text{$U(1)^2$ flavor symmetry with parameters $x,y$;} \\[.1cm]
\text{charges}:\quad  \begin{array}{c|ccc} & \phi_1&\phi_2&\phi_3 \\\hline
 X & 1 & 0 & -1 \\
 Y & 0 & 1 & -1 \\
 R & 0 & 0 & 2 \end{array} \qquad
 \text{CS levels:}\quad \begin{array}{c|ccc} & X & Y & R \\\hline
   X & 0 & \frac12 & 0 \\
   Y & \frac12 & 0 & 0 \\
   R & 0 & 0 & -\frac12
 \end{array}\,,
\end{array}\right. \ee
where $X$ and $Y$ denote the two flavor symmetries preserved by the superpotential. Note that the R-charges of the chirals are chosen so that the superpotential has $R(W)=2$.

For generic mass parameters $x$, $y$, there is a unique vacuum in which the chirals are set to zero. The block is very easy to write down:
\be B_{\rm XYZ}(x,y;q) = \frac{B_\Delta(x;q)B_\Delta(y;q)B_\Delta(qx^{-1}y^{-1};q)}{\theta(-q^{-\frac12}xy;q)} = \frac{(qx^{-1};q)_\infty(qy^{-1};q)_\infty (xy;q)_\infty}{\theta(-q^{-\frac12}xy;q)}\,.\ee
This factorizes the ellipsoid and index partition functions of the XYZ model in an obvious fashion. We would like to see that it is reproduced by the block integral of the mirror gauge theory.

The mirror, $N_f=1$ SQED, can be defined as
\be \label{defSQED}
T_{\rm SQED}: \left\{ \begin{array}{l}
\text{$U(1)$ gauge theory, two chirals $\varphi_1,\varphi_2$;} \\
\text{$U(1)$ flavor symmetry (parameter $x$);}\\
\text{$U(1)_J$ topological flavor symmetry (parameter $y$);} \\[.1cm]
\text{charges}:\quad  \begin{array}{c|cccc} & \varphi_1&\varphi_2&v_+& v_- \\\hline
 G & 1 & -1 & 0 & 0 \\
 X & 0 & 1 & 0 & -1 \\
 Y & 0 & 0 & 1 & -1 \\
 R & 0 & 0 & 0 & 2 \end{array} \qquad
 \text{CS levels:}\quad \begin{array}{c|cccc} & G & X & Y & R \\\hline
   G & 0 & \frac12 & 1 & -1 \\
   X & \frac12 & -\frac12 & 0 & \frac12 \\
   Y & 1 & 0 & 0 & 0 \\
   R & -1 & \frac12 & 0 & -1
 \end{array}\,,
\end{array}\right. \ee
The axial symmetry is denoted $X$ and the topological symmetry $Y$. This theory has a gauge-invariant meson operator $\varphi_1\varphi_2$ as well as two monopole operators $v_+,v_-$, whose charges we have indicated; they respectively match the three chiral operators $\phi_1,\phi_2,\phi_3$ of the XYZ model. The block integral for SQED, derived via the prescription of Section \ref{sec:integrand}, is
\be \mathbb B_{\rm SQED}(x,y;q) = \int_* \frac{ds}{2\pi is}\, \frac{\theta(-q^{-\frac12}y;q)}{\theta(-q^{-\frac12}sy)}\,B_\Delta(s;q)B_\Delta(xs^{-1};q)\,.\ee
Let us call the integrand $\Upsilon_{\rm SQED}(x,y,s;q)$. Its asymptotic growth at small $\hbar$ is given by
\be \Upsilon_{\rm SQED}(x,y,s;q) \sim \exp\frac1\hbar\left[-\frac12(\log(-y)^2+\frac12(\log(-sy))^2+\Li_2(s^{-1})+\Li_2(x^{-1}s) \right]\,. \ee
The unique critical point of the effective superpotential is at $s^{(1)}\!\! =\! (y-x^{-1})/(y-1)$, corresponding to the unique vacuum of SQED.

\begin{figure}[htb]
\centering
\includegraphics[width=6.1in]{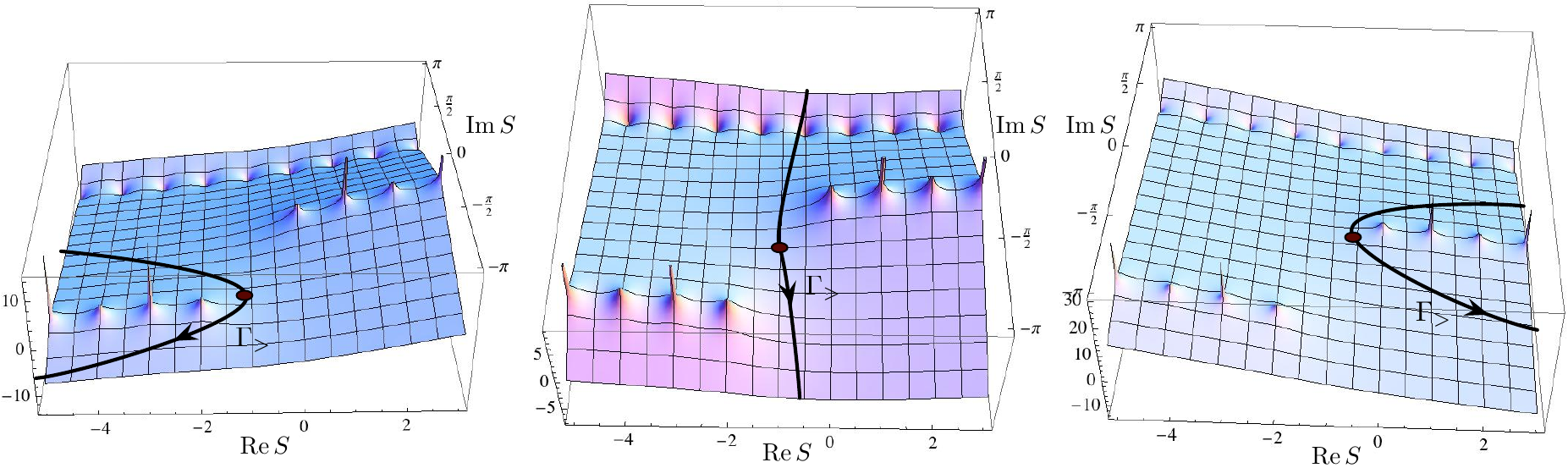}
\caption{The potential $\log|\Upsilon_{\rm SQED}|$ at $\hbar = +1$, $x=\exp(2+\tfrac{2\pi i}{3})$, and three different values (left to right) $y=\exp(-2+\tfrac{4\pi i}{5})$, $y=\exp(-\tfrac12+\tfrac{4\pi i}{5})$, and $y=\exp(2+\tfrac{4\pi i}{5})$. Critical points and integration cycles are shown.}
\label{fig:23+}
\end{figure}

When $\hbar$ is real and positive, the quantum-corrected potential $\log|\Upsilon|$ at finite $q$ is shown in Figure \ref{fig:23+}. Again, we use the periodic, logarithmic variable $S=\log s$ in the figures. The two chiral multiplets of opposite gauge charge produce two half-lines of poles in the integrand, extended in opposite directions. The theta function that encodes the FI term produces a full line of zeroes. There is a unique integration cycle $\Gamma_>$ that is both convergent and invariant under $q$-shifts. Just as in the previous example of the free vortex theory $T_\Delta'$, it is natural to draw the contour in different ways depending on the relative values of $x$ and $y$; but in each case the resulting integration yields the same answer. By summing residues (of either half-line of poles!) or by taking Fourier coefficients, we find
\be  \int_{\Gamma_>} \frac{ds}{2\pi is}\, \Upsilon_{\rm SQED}(x,y,s;q) = (q^{-1})_\infty^{-1} B_{\rm XYZ}(x,y;q)\qquad (|q|>1)\,.\ee

\begin{figure}[htb]
\centering
\includegraphics[width=6.1in]{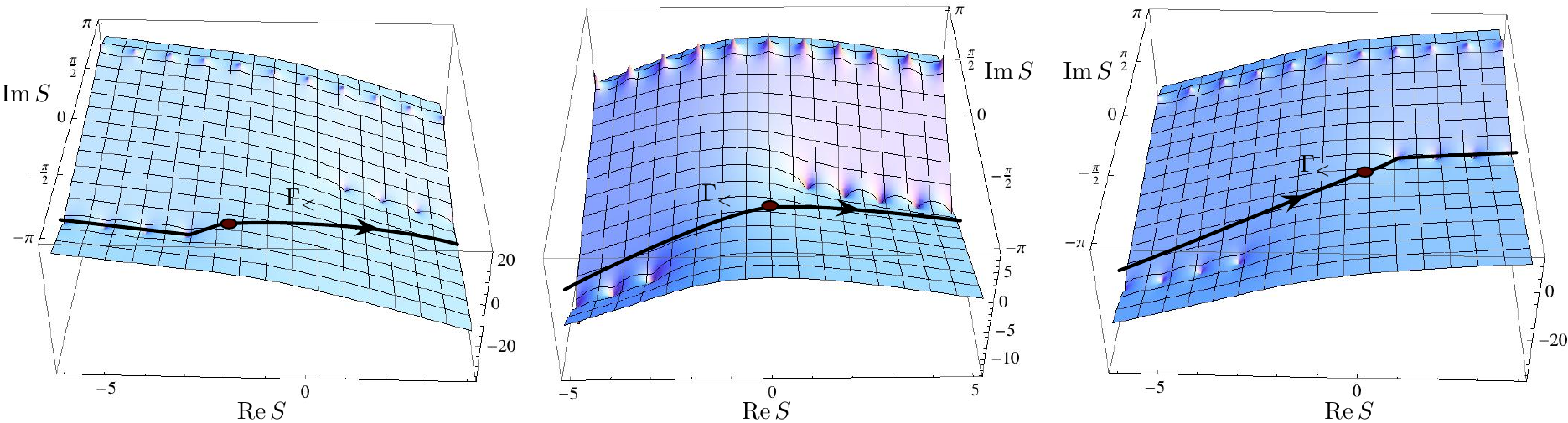}
\caption{The potential $\log|\Upsilon_{\rm SQED}|$ at $\hbar = -1$, $x=\exp(2+\tfrac{2\pi i}{3})$, and three different values (left to right) $y=\exp(-5+\tfrac{4\pi i}{5})$, $y=\exp(-1+\tfrac{4\pi i}{5})$, and $y=\exp(1+\tfrac{4\pi i}{5})$. Critical points and natural integration cycles are shown.}
\label{fig:23-}
\end{figure}
If instead $\hbar$ is real and negative, then poles and zeroes are reversed, and the unique integration cycle $\Gamma_<$ is displayed in Figure \ref{fig:23-}. Integration along $\Gamma_<$ is demonstrably convergent when $1\lesssim |y|^{-1}\lesssim |x|$, as in the center of Figure \ref{fig:23-}. Otherwise, the cycle must be extended along the half-lines of zeroes, which would need to be regularized to produce a convergent answer. In the regime where convergence is apparent, we have verified numerically that
\be \frac{2\pi i}{\hbar}\int_{\Gamma_<} \frac{ds}{2\pi is}\, \Upsilon_{\rm SQED}(x,y,s;q) = (q)_\infty B_{\rm XYZ}(x,y;q)\qquad (|q|<1)\,.\ee
Therefore, $B_{\rm SQED}=B_{\rm XYZ}$ in both regimes, with exactly the same kind of normalization factors that appeared in the previous example.

This example is not entirely independent of the previous one. Indeed, if we send the axial mass $x$ to infinity, SQED reduces to the theory $T_\Delta'$ of a free vortex, while the XYZ model reduces to the theory $T_\Delta$ of a free chiral. This scaling can be performed directly at the level blocks by noting (for example) that $B(x;q)=(qx^{-1};q)_\infty\to 1$ as $|x|\to \infty$.


\subsubsection{The $SU(2)$ appetizer}

Our final rank-one example is a nonabelian gauge theory that appeared in \cite{JY-appetizer} in the context of F-maximization. Let us define the ``appetizer theory'' as
\be \label{defapp}
 T_{\rm app}: \left\{ \begin{array}{l}
\text{$SU(2)$ gauge theory at CS level 1, with adjoint chiral $\phi$\,;} \\
\text{$U(1)$ flavor symmetry (parameter $x$);}\\[.1cm]
\text{charges}:\quad  \begin{array}{c|ccc} & \phi_+ & \phi_0 & \phi_- \\\hline
 G & 2 & 0 & -2 \\
 X & 1 & 1 & 1 \\
 R & 0 & 0 & 0  \end{array} \qquad
 \text{CS levels:}\quad \begin{array}{c|ccc} & G & X & R \\\hline
   G & 2 & 0 & 0 \\
   X & 0 & -\frac32 & \frac32 \\
   R & 0 & \frac32 & -\frac12
 \end{array}\,.
\end{array}\right. \ee
Here we have identified a maximal torus $U(1)\subset SU(2)$, and written charges and effective Chern-Simons levels in terms of this abelianized gauge symmetry, denoted by $G$. We have split the adjoint chiral into its three components. It was conjectured in \cite{JY-appetizer} that $T_{\rm app}$ flows to the theory of a free chiral in the infrared. We have chosen the flavor symmetry and background Chern-Simons terms here so that it flows precisely to a copy of $T_\Delta$ normalized so that the chiral has charge $+2$ under the $U(1)$ flavor symmetry.

The block integral of $T_{\rm app}$ can be constructed following the rules of Section \ref{sec:integrand}:
\be \mathbb B_{\rm app}(x;q) = \int_* \frac{ds}{2\pi is} \frac{B_\Delta(x;q)B_\Delta(s^2x;q) B_\Delta(s^{-2}x;q) }{\theta(s;q)^2(qs^2;q)_\infty(qs^{-2};q)_\infty}\,,\ee
We denote the integrand $\Upsilon_{\rm app}(x,s;q)$. The integrand is symmetric under the Weyl-group action $s\to s^{-1}$. In terms of the logarithmic variable $S=\log s$, a fundamental domain for the Weyl group is given by the strip $0 \leq \Im\, S \leq \pi$, with an identification 
\ben  S& \sim& -S \qquad\qquad (S\in \R)~,\\
S&\sim& 2\pi i-S\qquad (S\in \R+i\pi)~, \een
at the strip's two boundaries. The classical potential in the $\hbar\to 0$ limit is
\be \Upsilon_{\rm app}(x,s;q) \sim \exp\frac{1}{2\hbar} \left[2\log(s)^2+\log(-s^2)^2+2\Li_2(x^{-1})+2\Li_2(x^{-1}s^{-2})+2\Li_2(x^{-1}s^2)\right]\,, \notag \ee
up to a constant, and has a unique critical point in the fundamental domain, corresponding to a solution of $s+s^{-1}=1+x^{-1}$.

\begin{figure}[t!]
\centering
\includegraphics[width=5in]{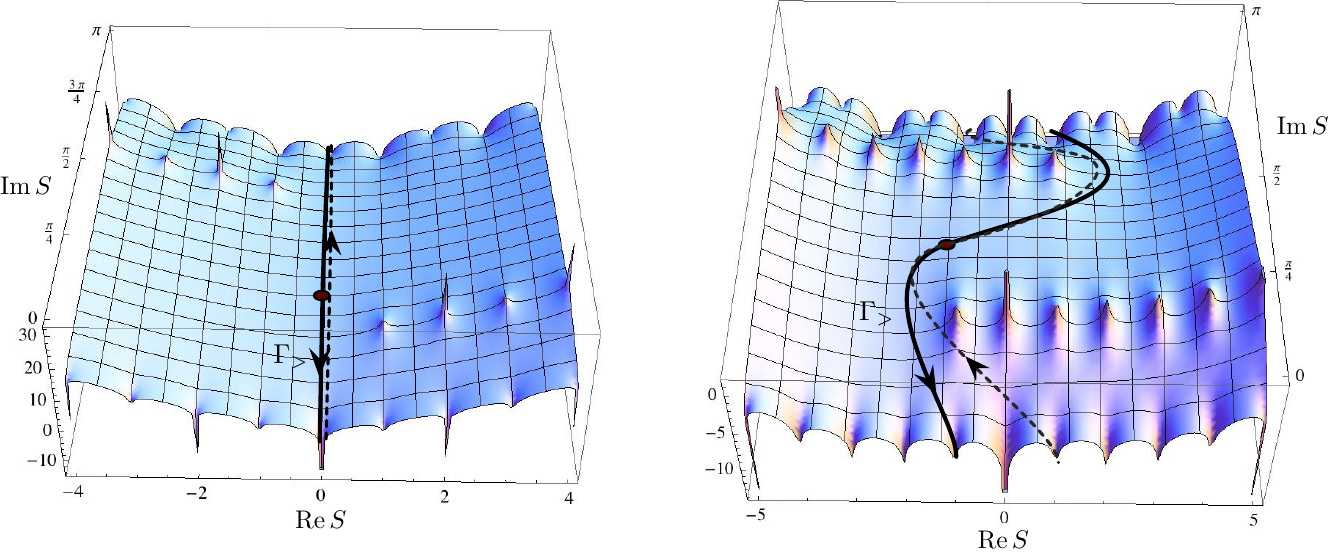}
\caption{Plots of $\log|\Upsilon_{\rm app}(x,s;q)|$ in a fundamental domain, at $\hbar =2$ and two typical values of the flavor parameter $x=ie^{2}$ (left) and $x=ie^{-2}$ (right). That the two segments of $\Gamma_>$ crossing the fundamental domain do not cancel, but rather add constructively, due to the $\Z_2$ orbifolding at the boundaries of the strip. The cycles are deformed slightly to illustrate the doubling.}
\label{fig:SU2+}
\end{figure}

The integrand for $|q|>1$ (with $\hbar$ real) is shown in Figure \ref{fig:SU2+}. The chirals $\phi_\pm$ that are charged under the $U(1)$ in the gauge group contribute half-lines of poles, while the gauge and Chern-Simons factors contribute a half-line of zeroes (of which some are doubled).
Note that the spacing of poles and zeroes is now by multiples of $\hbar/2$ rather than $\hbar$. At any value of $x$ there is one convergent cycle $\Gamma_>$, which forms a closed loop that winds twice around the fundamental domain. Integration along it just picks out the $s^0$ Fourier coefficient of the integrand, and with some work we find that
\be \label{SU2int+}
\frac12 \frac{(q^{-1})_\infty^2}{(q^{-2})_\infty(-q^{-1};q^{-2})_\infty^2} \int_{\Gamma>} \frac{ds}{2\pi is}\Upsilon_{\rm app}(x,s;q) = B_\Delta(x^2;q)\,,\qquad (|q|>1)\,.
\ee
The numerical prefactor $1/2$ appears come from the number of times the cycle $\Gamma_>$ meets images of the critical point.

\begin{figure}[htb]
\centering
\includegraphics[width=2.8in]{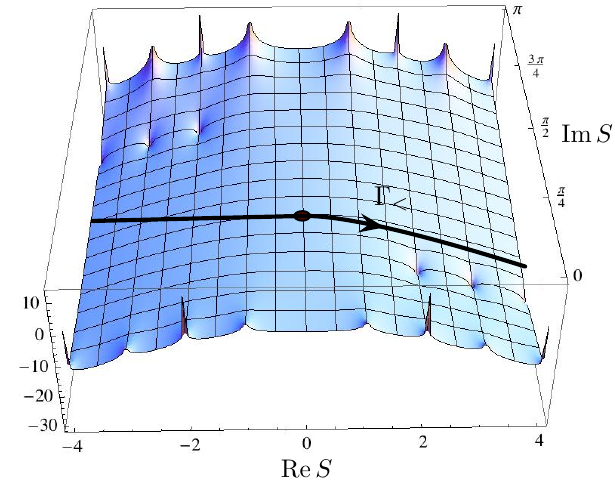}
\caption{Plot of $\log|\Upsilon_{\rm app}(x,s;q)|$ in a fundamental domain, at $\hbar =-2$ and $x=ie^{2}$.}
\label{fig:SU2-}
\end{figure}

For $|q|<1$ and $\hbar$ real, the integrand is shown in Figure \ref{fig:SU2-}. Again there is a unique convergent cycle $\Gamma_<$ in the fundamental domain, at all values of $x$. Numerical integration along $\Gamma_<$ indicates that
\be (\text{const})\frac{2\pi i}{\hbar} \int_{\Gamma_<} \frac{ds}{2\pi is}\Upsilon_{\rm app}(x,s;q) = B_\Delta(x^2;q)\,,\qquad (|q|<1)\ee
for a certain $q$-dependent constant --- presumably a $|q|<1$ version of the prefactor in \eqref{SU2int+}.


\subsection{Defining contours}
\label{sec:contours}

Having understood block integrals for a number of simple examples, we now address the more involved problem of identifying integration cycles $\Gamma^\ra$ for a theory with multiple vacua $\alpha$. We will first make some observations about the problem in general, and then demonstrate these ideas in the context of the $\cp^1$ model in the next section.

For motivation and review, let us recall the simpler case of an integral of the form
\be \label{gen-int} \int_{\Gamma\subset \cM} dS\, \Upsilon(X,S;\hbar)\,, \ee 
where $\CM\simeq \C^r$ and $\Upsilon(X,S;\hbar)$ is a nonvanishing \emph{holomorphic} function of $S\in \C^r$ that depends on additional complex parameters $X$ and $\hbar$. For example, one case that makes frequent appearances in physics is $\Upsilon(X,S;\hbar) = \exp \frac1\hbar f(X,S)$, with $f(X,S)$ a holomorphic function on $\C^r$.
This problem can be considered within the framework of Lefschetz theory,%
\footnote{See, for example, \cite{Moore-Stokes, CV-WC, HIV, Wit-anal, Witten-path} for applications and discussions of Lefschetz theory in physical contexts.} %
which tells us (in part) that there exists a basis $\{\Gamma^\ra\}$ of middle-dimensional integration cycles such that the integral \eqref{gen-int} for any cycle $\Gamma$ on which it converges can be written as an integer linear combination of integrals over the basis contours $\Gamma^\ra$. Since the only non-trivial integrals will be over non-compact contours, the cycles $\Gamma^\ra$ form a basis for the relative homology group
\be \label{simpleHr} H_r(\CM,\CM_\Lambda;\Z)\,,\ee
where $\CM_\Lambda = \big\{S\in \CM\,\big|\,\log|\Upsilon(X,S;\hbar)|\leq \Lambda\}$, for $\Lambda$ sufficiently large and negative. The subset $\CM_\Lambda$ here essentially captures the directions in $\CM$ to which convergent integration may asymptote.

The set $\CM_\Lambda$ depends on the parameters $(X,\hbar)$, although the rank of the homology group \eqref{simpleHr} should not. For any fixed $X$ and $\hbar$ such that $\Upsilon(X,S;\hbar)$ has isolated and nondegenerate critical points $S^{(\alpha)}$, a \emph{distinguished} basis of cycles $\Gamma^\ra$ can be associated to those critical points. This is done by defining $\Gamma^\ra$ to be the set of all points that can be reached by downward gradient flow from $S^{(\alpha)}$, with respect to the potential $\log|\Upsilon(X,S;\hbar)|$ (and the standard K\"ahler metric on $\C^r$). The argument of $\Upsilon$ is constant along such flows, so the flow starting at one critical point $S^{(\alpha_1)}$ can hit another critical point $S^{(\alpha_2)}$ if and only if
\be \label{Stokeswall-gen}
\arg \Upsilon(X,S^{(\alpha_1)};\hbar) = \arg \Upsilon(X,S^{(\alpha_2)};\hbar)\,,\qquad
\ee
and $\log|\Upsilon(X,S^{(\alpha_1)};\hbar)|>\log|\Upsilon(X,S^{(\alpha_2)};\hbar)|$. The condition \eqref{Stokeswall-gen} defines real-codimension-one Stokes walls in parameter space. For generic parameters $(X,\hbar)$, one is far away from these walls, so flows continue indefinitely and all critical-point cycles $\Gamma^\ra$ are well defined. On the other hand, if parameters $(X,\hbar)$ are varied continuously across a Stokes wall, then the natural critical-point basis is shifted,%
\footnote{More generally, the jump is by a multiple of the intersection number between the upward flow from $S^{(\alpha_2)}$ and the downward flow from $S^{(\alpha_1)}$. This played an important role in \cite{HIV}. However, an intersection number greater than one does not occur in the simple case of cycles on $\C^r$, and will not occur in any of the examples studied in this paper.}
\be \label{Stokesjump-gen}
\begin{pmatrix} \Gamma^{\alpha_1} \\ \Gamma^{\alpha_2} \end{pmatrix} \mapsto \begin{pmatrix} \Gamma^{\alpha_1}{}' \\ \Gamma^{\alpha_2}{}' \end{pmatrix} =
 \begin{pmatrix} \Gamma^{\alpha_1}\pm \Gamma^{\alpha_2} \\ \Gamma^{\alpha_2} \end{pmatrix} = \begin{pmatrix} 1 & \pm 1 \\ 0 & 1 \end{pmatrix} \begin{pmatrix} \Gamma^{\alpha_1} \\ \Gamma^{\alpha_2} \end{pmatrix}\,,
\ee
where the sign depends on the relative orientation of the cycles. Two different Stokes walls can intersect on a real-codimension-two locus in $(X,\hbar)$ space. This is a locus where critical points become degenerate. In general, the Stokes walls \eqref{Stokeswall-gen} emanate out from such degenerate loci. Motion in a closed loop around a degenerate locus induces a monodromy that may permute the integration cycles $\Gamma^\ra$.

Unfortunately, the block integrals are \emph{not} of the form \eqref{gen-int}. This is a consequence of the general construction of block integrals, and is particularly clear in the examples of Section \ref{sec:simple-ex}. One mild difference is that the domain of integration for the block integrals is $\CM = (\C/2\pi i\Z)^r$, or a Weyl-group quotient thereof, rather than $\C^r$. (In exponentiated variables, we would say $\CM \simeq (\C^*)^r$.) This simply introduces the possibility of closed homology cycles that encircle the non-trivial one-cycles in $\cM$. More importantly, the integrand $\Upsilon(x,s;q)$, has infinite lines of poles and zeroes, and we have argued that a good integration cycle should never cross the lines of poles. The presence of such a meromorphic integrand makes things more complicated. We will approach this problem in two ways. The first is more intuitive and proves sufficient to compute blocks in all examples encountered here, while the second is somewhat more rigorous (but perhaps slightly less intuitive).

\subsubsection*{Approximate cycles from quantum mechanics}

A lesson we drew from the construction of blocks in supersymmetric quantum mechanics is that there should exist an exact potential $W_{\rm exact}(x,s;q)$ whose only critical points correspond to true vacua of the theory,  which generates gradient flows that serve as exact block cycles $\Gamma^\ra$. While we have not determined this exact potential, we know that perturbatively, at leading order in $\hbar$, it should match $\wt W(X,S)/\hbar$ as defined in \eqref{QMSuper}. (Here, as elsewhere outside of Section \ref{sec:SQM}, we denote $-iW^{\rm QM}$ as $\wt W(X,S)/\hbar$). Consequently, our first approach is to approximate the integration cycles by gradient flow contours of $\wt W(X,S)/\hbar$ while keeping track of the exact, nonperturbative integrand $\Upsilon(s,x;q)$. The asymptotic behavior of the integrand is given at leading order by
\be \label{Ulimit}
\Upsilon(x,s;q) \overset{\hbar\to 0}{\sim} \exp\left[\tfrac1 \hbar \wt W(X,S) \right]\,,
\ee
which agrees with the perturbative potential, particularly around critical points of $\wt W(X,S)$. 

Away from the critical points, we will have to deform the contours by hand in order to make them consistent for block integrals. Notice that as we take $\hbar\to 0$ along a ray of constant phase, the half-lines of zeroes and poles of $\Upsilon$ coalesce into \emph{distinguished} branch cuts for $\wt W(X,S)$. As we saw in the examples of Section \ref{sec:simple-ex}, these distinguished branch cuts are all parallel and have slope $\arg\hbar$.%
\footnote{In the general multi-dimensional setting, the poles and zeroes occur in codimension two, and the distinguished branch cuts are flat walls of codimension one. The ``slope'' of cuts is measured in the space transverse to the zeroes and poles that lie on them. Branch cuts lying in different dimensions may intersect.}
While the true gradient-flow cycles of $\frac1\hbar\wt W(X,S)$ may hit branch cuts (and indeed, as we discussed in Section \ref{sec:SQM}, the quantum mechanical path integral may allow the flow line to cross the branch cuts), crossing a line of poles is disallowed from the perspective of solving line-operator  identities, so we will have to deform the contours away from the branch cuts that correspond to poles.

\begin{figure}[htb]
\centering
\includegraphics[width=5in]{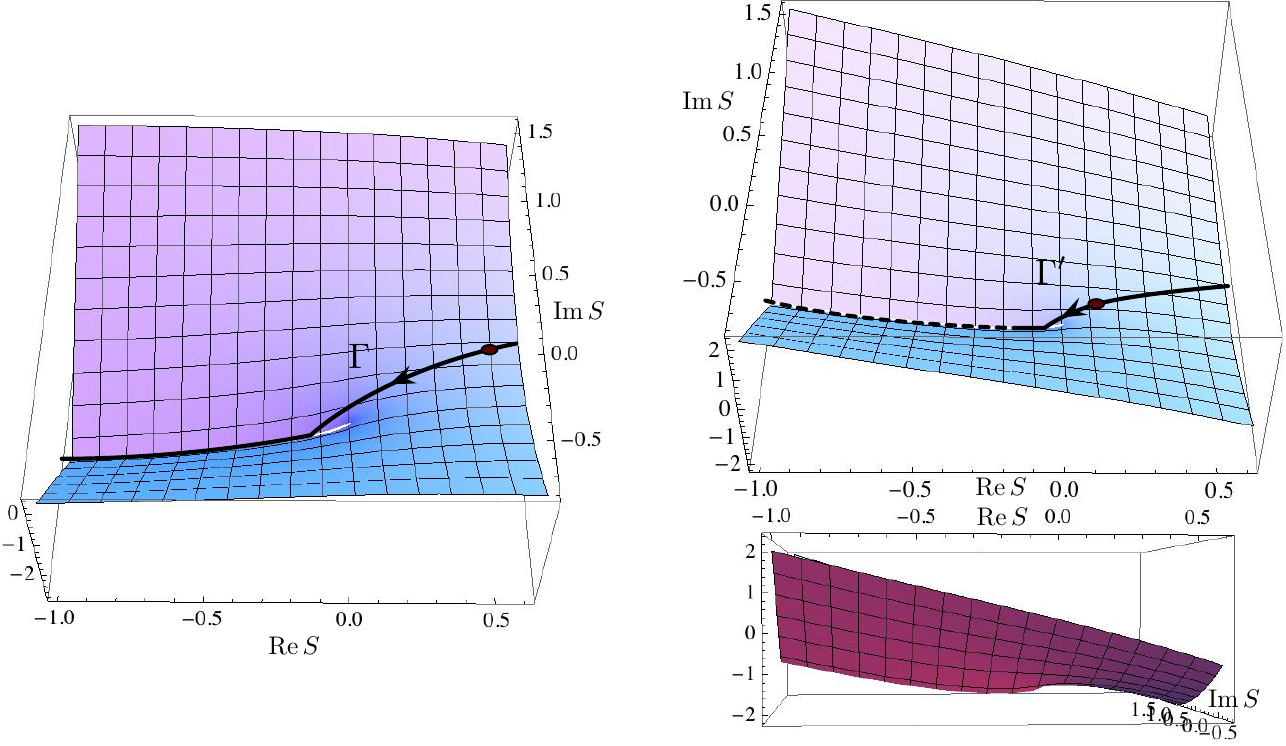}
\caption{Plots of the potential $\Re\big(\frac1\hbar\wt W(x,s)\big)$ for the block integral \eqref{STint} of $T_\Delta'$, showing a cycle that effectively starts to flow upward along a line of zeroes as $x$ is varied. Here $\hbar<0$ and $|x|<1$ on the left while $|x|>1$ on the right. The dashed line is the effective upward flow. The bottom-right view shows the valley corresponding to the line of zeros.}
\label{fig:upzero}
\end{figure}

Alternatively, contours should be allowed to cross or lie on branch cuts corresponding to zeros. For example, we saw in the example of the free vortex theory that a good integration cycle can cross a line of zeros (see Figure \ref{fig:ST+}). For SQED, we found that a good integration cycle should be continued to infinity along lines of zeroes (Figure \ref{fig:23-}).
In some cases, it may even be necessary to allow cycles that continue to infinity along a line of zeros even though it looks like they are flowing \emph{upward} with respect to $\Re\big(\frac1\hbar\wt W(X,S)\big)$. This was the case in SQED on the left or right of Figure \ref{fig:23-} (see also Figure \ref{fig:CPcycles}, page \pageref{fig:CPcycles}). In fact, this bizarre situation can occur in the free-vortex theory as well. In Figure \ref{fig:upzero} we have blown up the region of the $S$-cylinder where the potential of $T_\Delta'$ has a half-line of zeroes at $\hbar<0$ --- it is a close-up of Figure \ref{fig:ST-}. When $|x|<1$ there is a convergent integration cycle that continues to infinity along the line of zeroes, flowing downward. However, when $|x|>1$ the shape of the potential changes and the only candidate integration cycle $\Gamma'$ looks to be flowing upward instead. We expect that blocks should survive the transition from $|x|<1$ to $|x|>1$.  It is conceivable that the integral along $\Gamma'$ might be regularized to converge, since the integrand oscillates very rapidly along the half-line of zeroes.

We can use the approximate cycles $\Gamma^\ra$ defined by the potential $\frac1\hbar\wt W(X,S)$ to study the Stokes phenomenon for blocks. We expect the analysis to hold so long as critical points of $\frac1\hbar\wt W(X,S)$ are sufficiently far away from its branch cuts, so that gradient connecting one critical point to another (at Stokes walls) don't cross any cuts. The approximate location of Stokes walls for a pair of critical-point cycles $(\Gamma^{\alpha_1},\Gamma^{\alpha_2})$ that flow from critical points $(S^{(\alpha_1)},S^{(\alpha_2)})$ is then determined by
\be \label{semiStokes}  \Im\big(\tfrac1\hbar \wt W(X;S^{(\alpha_1)})\big) = \Im\big(\tfrac1\hbar \wt W(X;S^{(\alpha_2)})\big)\,. \ee
The cycle with the greater value of $\Re\big(\tfrac1\hbar \wt W(X;S^{(\alpha)})\big)$ will pick up a copy of the other as parameters are varied across the wall.  Equation \eqref{semiStokes} should be understood on the sheet of $\wt W(X,S)$ (with distinguished branch cuts) that is defined by sending $\hbar\to0$ in $\Upsilon(x,s;q)$ along a ray with constant phase.

This simple analysis can also be used to understand the existence of conjugate Stokes jumps for holomorphic blocks at $|q|<1$ and $|q|>1$. Suppose that we fix two values $\hbar_0,\,\wt \hbar_0$ of the parameter $\hbar$ such that the product $\hbar_0\wt\hbar_0$ is real and negative (for example, $\hbar_0>0$ and $\wt \hbar_0<0$), and only vary the mass parameters $x=\exp X$. It is easy to see that the same sheet of $\wt W$, with the same branch cuts, is relevant for analyzing gradient flows at both values of $\hbar$. Thus the approximate Stokes walls defined by \eqref{semiStokes} occur at the same place in $X$-space. However, because $\Re\big(\wt W(X,S)/\hbar_0)$ and $\Re\big(\wt W(X,S)/\wt \hbar_0\big)$ have opposite sign, the critical-point cycles that jump across a wall will be different:
\be
\begin{pmatrix} \Gamma^{\alpha_1} \\ \Gamma^{\alpha_2} \end{pmatrix} \mapsto \begin{pmatrix} 1 & \pm 1 \\ 0 & 1 \end{pmatrix} \begin{pmatrix} \Gamma^{\alpha_1} \\ \Gamma^{\alpha_2} \end{pmatrix}\quad\text{at $\hbar=\hbar_0$}\quad\Rightarrow\quad \begin{pmatrix} \Gamma^{\alpha_1} \\ \Gamma^{\alpha_2} \end{pmatrix} \mapsto \begin{pmatrix} 1 & 0 \\ \mp1 & 1 \end{pmatrix} \begin{pmatrix} \Gamma^{\alpha_1} \\ \Gamma^{\alpha_2} \end{pmatrix}\quad\text{at $\hbar=\wt\hbar_0$}\,.
\ee
More generally, any Stokes matrices $M,\wt M$ that govern a Stokes phenomenon in $X$-space at ``conjugate'' values of parameters $\hbar_0,\wt\hbar_0$ will satisfy $MM^T=1$.

\subsubsection*{Shift-invariant quantum cycles}

The above description of integration cycles works quite well when $\hbar$ is small, or $q$ is close to $1$. However, it may be useful to define cycles at large $\hbar$ as well. For example, an exact analysis of the blocks $B^\ra(x;q)$, $B^\ra(\wt x,\wt q)$ involved in an S-fusion operation requires simultaneous consideration of cycles at $\hbar$ and $\wt\hbar = -\frac{4\pi^2}{\hbar}$. We therefore provide a second, ``quantum'' prescription for defining integration cycles. We take the non-perturbative integrand $\Upsilon(x,s;q)$ of the block integral (or rather $\log \Upsilon$) to be the potential for gradient flow. We know that $\Upsilon$ cannot be the exact potential for supersymmetric quantum mechanics, because it has too many critical points. Correspondingly, we will see that not all of its convergent critical-point cycles lead to solutions of line-operator identities. These two mismatches ultimately cancel each other out, and can be resolved simultaneously. 

Consider the integrand $\Upsilon(x,s;q)$ of the block integral at finite $q$. It is a meromorphic function of $s$. In addition to the critical points $s=s^{(\alpha)}$ that survive in the limit $q\to 1$ and correspond to vacua on $\RS$, there are an infinite set of ``quantum'' critical points $s=\hat s^{(\beta)}$. They occur in between every two consecutive zeroes or poles on the lines and half-lines that would coalesce into cuts as $q\to 1$. They do not correspond to true vacua of the theory on $\RqS$ (or to SUSY ground states on $T^2 \times \R$) because we know that physical vacua are uncharged under the rotation whose Wilson line implements the $q$ deformation, and so the vacua cannot appear spontaneously when $q\neq 1$.

Now consider downward gradient flows from all the $s^{(\alpha)}$ and $\hat s^{(\beta)}$, with respect to the potential $\log|\Upsilon(x,s;q)|$. The flows define cycles $\Gamma_q^\alpha$ and $\hat\Gamma_q^\beta$, respectively, on which the block integral converges. However, these cycles typically terminate at zeroes of $\Upsilon$ rather than at asymptotic infinity. They are not yet good candidates for cycles on which to compute holomorphic blocks, because they are not invariant under $q$-shifts (as discussed in Section \ref{sec:integrand}), and so do not produce solutions to line-operator identities. Shift-invariant cycles must either be closed or end at asymptotic infinity.

We can solve this problem by taking linear combinations of the cycles $\Gamma_q^\ra$ and $\hat\Gamma_q^\beta$ that are shift invariant. The cycles $\Gamma_q^\ra$ and $\hat\Gamma_q^\beta$ form a countable basis for an abelian group $\bm \Gamma_q$, which is a certain limit of homology groups
\be  H_r(\CM_q,\CM_{q\Lambda};\Z)\,,\ee
as $\Lambda\to -\infty$, where
\be \CM_q = (\C/2\pi i\Z)^r\bs \{\text{poles of $\Upsilon$}\}\,,\qquad \CM_{q\Lambda} = \{S\in (\C/2\pi i\Z)^r\,|\,\Upsilon(x,s;q)|<e^{\Lambda}\}\,.\ee
We say that a (possibly infinite) linear combination of these cycles $\Gamma = \sum_\alpha n_\alpha\Gamma_q^\alpha + \sum_\beta \hat n_\beta\hat \Gamma_q^\beta$  is convergent if $\int_\Gamma dS\,\Upsilon :=\sum_\alpha n_\alpha \int_{\Gamma_q^\alpha} dS\, \Upsilon+ \sum_\beta n_\beta \int_{\hat \Gamma_q^\beta} dS\, \Upsilon$ is finite. We say that such a linear combination is  shift-invariant roughly if a shift by $\pm \hbar$ in any integration variable $S_i$ does not change the integral $\int_\Gamma dS\,\Upsilon$. More precisely, $\Gamma$ is shift-invariant if for every shift by $\pm\hbar$ in a direction $S_i$ there exist two smooth, convergent cycles $\Gamma'$ and $\Gamma''$, with $\Gamma''$ the image of $\Gamma'$ under the shift, such that  $\int_\Gamma dS\,\Upsilon=\int_{\Gamma'}dS\,\Upsilon=\int_{\Gamma''}dS\,\Upsilon$\,. Then we can define the group $\bm \Gamma$ to be the subgroup of $\bm \Gamma_q$ consisting of convergent, shift-invariant elements.

\begin{figure}[htb]
\centering
\includegraphics[width=3.5in]{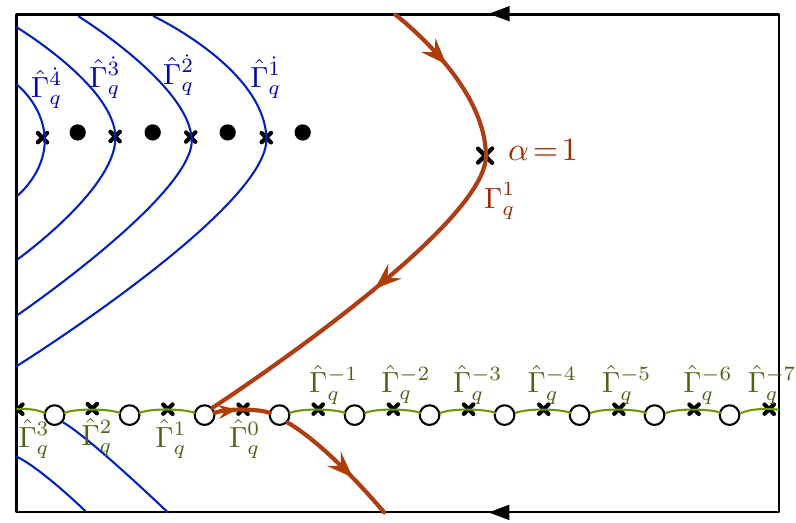}
\caption{A schematic example of gradient-flow cycles for the ``quantum'' potential $\Upsilon$, drawn on a cylinder parameterized by $S\in \C/(2\pi i\Z)$. We have indicated poles by black dots, zeroes by white dots, and critical points by `$\times$'. The convergent, shift-invariant integration cycle is $\Gamma^1 = \hat\Gamma^1_q+\hat\Gamma^0_q$. This picture actually corresponds to the free-vortex theory at $\hbar>0$, \cf\ Figure \ref{fig:ST+}.}
\label{fig:shiftcycles}
\end{figure}

The subgroup $\bm\Gamma$ is the finite-rank group containing the cycles that lead to holomorphic blocks. If we are lucky, every element of $\bm \Gamma$ will contain at least one copy of a cycle $\Gamma^\alpha_q$ corresponding to downward flow from one of the semi-classical critical points $s^\alpha$. (In other words, there will be no convergent shift-invariant cycles formed entirely from quantum cycles $\hat\Gamma^\beta_q$.) This appears to be the case in examples. Then we can identify a distinguished basis of cycles $\Gamma^\ra \in \bm \Gamma$, by defining $\Gamma^\ra$ to be the unique element that contains the minimal positive number of copies of $\Gamma^\alpha_q$ (typically one), and zero copies of all other $\Gamma^{\alpha'}_q$. An example is shown in Figure \ref{fig:shiftcycles}. The basis $\{\Gamma^\ra\}$ will naturally be associated to the vacua of our theory.

Note that this definition of convergent, shift-invariant cycles $\Gamma^\ra$ can naturally include cycles that seemingly flow ``upward'' along a half-line of zeroes, as in Figure \ref{fig:upzero}. We argued previously that these cycles are sometimes necessary if block integrals are to make sense at all values of parameters $x$. They will be included if the infinite sum $\sum_\beta \hat n_\beta\hat \Gamma_q^\beta$ of quantum cycles along a half-line of zeroes can be made to converge.

As we change parameters $x$ and $q$, the integration cycles in $\bm \Gamma_q$ will undergo Stokes phenomena at infinitely many walls, defined by one of the conditions
\begin{subequations} \label{qStokes}
\begin{align} \arg \Upsilon(x,s^{(\alpha_1)};q) &= \arg \Upsilon(x,s^{(\alpha_2)};q)\quad (\text{mod}\;\;2\pi i)\,, \\
\arg \Upsilon(x,\hat s^{(\beta_1)};q) &= \arg \Upsilon(x,\hat s^{(\beta_2)};q)\quad (\text{mod}\;\;2\pi i)\,,  \\
\arg \Upsilon(x, s^{(\alpha)};q) &= \arg \Upsilon(x,\hat s^{(\beta)};q)\quad (\text{mod}\;\;2\pi i)\,.
\end{align}
\end{subequations}
Most of these jumps will just modify the elements of the shift-invariant subgroup $\bm \Gamma$ by a finite number of quantum cycles $\hat \Gamma_q^\beta$, so that the basis for $\bm\Gamma$ associated to vacua of the theory is unchanged. However, at a few, distinguished walls the true basis $\{\Gamma^\ra\}$ of $\bm\Gamma$ jumps. These special jumps are related to the physical Stokes phenomenon. If the classical critical points $s^{(\alpha)}$ are well separated from poles and zeroes, then the physical jumps will occur precisely when a flow from one classical critical point hits another critical point, as in (\ref{qStokes}a). Then the location of the wall will approximate that predicted semi-classically by \eqref{semiStokes}.

In a situation such as S-fusion of blocks, where it is important to keep track simultaneously of parameters $(x=\exp X,q=\exp \hbar)$ and $(\wt x=\exp \frac{2\pi i}{\hbar}X,\wt q = \exp \frac{-4\pi^2}{\hbar})$, it generally does \emph{not} seem that Stokes jumps for blocks $B^\ra(x;q)$ and $B^\ra(\wt x;\wt q)$ can both be analyzed in the semi-classical approximation. The full quantum Stokes analysis outlined here may then be useful.


\section{Case study: the \texorpdfstring{$\cp^1$}{CP1} sigma-model}\label{sec:CP1}

We now consider in detail an example that illustrates the constructions of the previous sections. This will be our main example, and in studying it we will encounter most of the interesting features associated to holomorphic blocks. In particular, we will see how the five proposed properties of holomorphic blocks from Section \ref{sec:blockint} can be satisfied in a theory with multiple vacua, and multiple Stokes chambers. Moreover, in the discussion of connections to Chern-Simons theory in Section \ref{sec:CS}, this example will form the basis for calculations involving the figure-eight knot.

The theory in question (denoted in this section as $T_\I$) can be described in the UV as a gauged linear sigma model, which flows in the IR (in rather subtle ways) to a non-linear sigma model with target $\cp^1$. The UV Lagrangian has two chiral multiplets $\phi_1,\,\phi_2$, transforming with charges $(+1,+1)$ under a dynamical, abelian $U(1)$ gauge symmetry. We denote the scalar in the dynamical vector multiplet as $\sigma^{\rm 3d}$, and its complexification as $S$ or $s=\exp S$. The theory also has flavor symmetry $U(1)_V\times U(1)_J$, as well as a $U(1)_R$ R-symmetry. The vector symmetry $U(1)_V$ rotates the chirals with charges $(+1,-1)$, and has an associated real mass $m^{\rm 3d}$, complexified to $x=\exp X$. The topological symmetry $U(1)_J$ shifts the dual photon, and has an FI parameter $t^{\rm 3d}$, complexified to $y=\exp Y$. For compactification on curved backgrounds, we choose the R-charges of the chirals to be zero. The charges and Chern-Simons levels of the theory are summarized as follows,
\be \label{defcp1}
T_{\I}: \left\{ \begin{array}{l}
\text{Dynamical $U(1)$ gauge theory with two chirals $\phi_1,\phi_2$,} \\
\text{$U(1)_V\times U(1)_J$ symmetry with parameters $x,\,y$; }\\[.1cm]
\text{charges}:\quad  \begin{array}{c|cc} & \phi_1&\phi_2\\\hline
 G & 1 & 1 \\
 X & 1 & -1 \\
 Y & 0 & 0 \\
 R & 0 & 0 \end{array} \qquad
 \text{CS levels:}\quad \begin{array}{c|cccc} & G & X & Y & R \\\hline
   G & 0 & 0 & 1 & 0  \\
   X & 0 & 0 & 0 & 0 \\
   Y & 1 & 0 & 0 & 0 \\
   R & 0 & 0 & 0 & *
 \end{array}\,.
\end{array}\right. \ee

This theory is known to arise in various contexts. It can be engineered on a toric brane in the local Calabi-Yau geometry $\cO(-2)\oplus\cO(0)\rightarrow\P^1$. It also appears on the simplest half-BPS surface operator in 4d $SU(2)$ super Yang-Mills. We will mention some of these interpretations in Section \ref{sec:cpmore}. For now, though, let us start with three-dimensional gauge theory and work our way down to blocks. 


\subsection{Parameter spaces}
\label{sec:cp1param}

\begin{wrapfigure}[15]{r}{2.45in}
\includegraphics[width=2.35in]{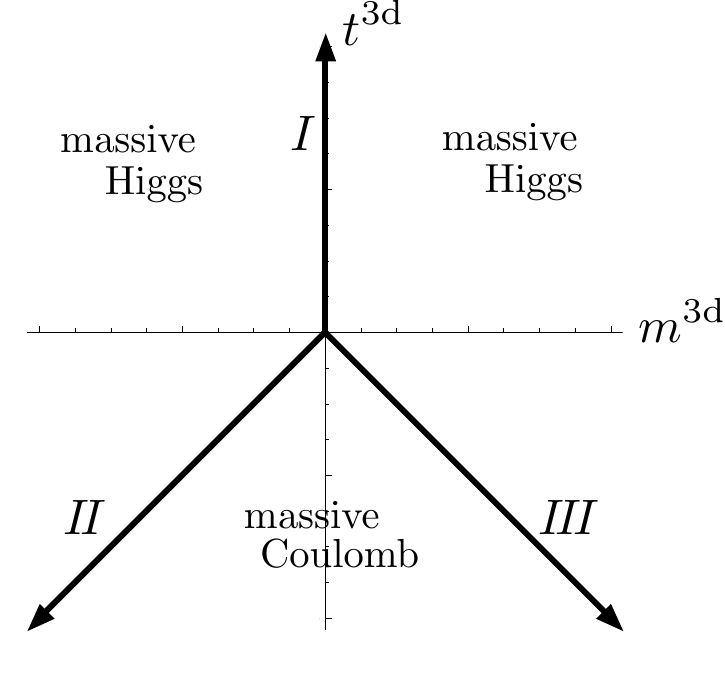}
\caption{Parameter space for 3d $\cp^1$ model.}
\label{fig:CP1param}
\end{wrapfigure}%
The three-dimensional parameter space of $T_\I$ is shown in Figure \ref{fig:CP1param}. It was analyzed carefully in (\eg) \cite{DoreyTongMS}, following \cite{AHISS}. We review some of the relevant details here. For $m^{\rm 3d}=0$ and positive $t^{\rm 3d}>0$ (\ie\ on the ray labelled $\I$), the theory has a $\cp^1$ Higgs branch of vacua, and is well described at low energy as a nonlinear sigma-model with target space a $\cp^1$ of size $\sim t^{\rm 3d}$. For large $t^{\rm 3d}$, the sigma-model is semi-classical. On either side of the ray $\I$, the theory becomes massive, and has two Higgs-branch vacua localized at the poles of $\cp^1$. So far, this is similar to the two-dimensional $\cp^1$ model. The main qualitative difference comes at negative $t^{\rm 3d}$: then there are two $\cp^1$ Coulomb branches of vacua (rays $\II$ and $\III$) at $m^{\rm 3d}=\pm t^{\rm 3d}$. Quantum effects are important in identifying their topology as that of $\cp^1$. In between rays $\II$ and $\III$, $T_\I$ has two massive Coulomb-branch vacua.

There is a $\Z_3$ symmetry in Figure \ref{fig:CP1param}, and this is not a coincidence. The parameter space can be rotated by a third by transforming the mass parameters according to
\be \label{omega}
\big(m^{\rm 3d},t^{\rm 3d}\big)\;\overset{\omega}{\longmapsto}\;\Big(\frac{t^{\rm 3d}-m^{\rm 3d}}{2},\, -\frac{3m^{\rm 3d}+t^{\rm 3d}}{2}\Big)~.
\ee
Upon promoting \eqref{omega} to a transformation of the background vector multiplets in the  $T_\I$ (equivalently, a linear redefinition of the flavor symmetries), one obtains a new theory $T_\II$ that is mirror symmetric%
\footnote{One way to derive the mirror theory $T_\II$ from $T_\I$ is to apply the fundamental $T_\Delta\simeq T_\Delta'$ symmetry of Section \ref{sec:free-vortex} once to the chiral $\phi_1$ and twice to $\phi_2$, and to then integrate out decoupled gauge multiplets. Similarly, to obtain $T_\III$ one may apply the $T_\Delta\simeq T_\Delta'$ symmetry twice to $\phi_1$ and once to $\phi_2$.\label{foot:ST}} %
to $T_\I$. This is also a $\cp^1$ sigma-model, but with modified mass and FI parameters. Consequently, $T_\II$ has a classical $\cp^1$ Higgs branch along ray $\II$. Apply transformation \eqref{omega} again sends 
\be \label{omega2}
\big(m^{\rm 3d},t^{\rm 3d}\big)\;\overset{\omega^2}{\longmapsto}\;\Big(-\frac{m^{\rm 3d}+t^{\rm 3d}}{2},\, \frac{3m^{\rm 3d}-t^{\rm 3d}}{2}\Big),
\ee
and results in a theory $T_\III$ with a $\cp^1$ Higgs branch along ray $\III$. Notice that the transformation has order three, $\omega^3=id$.

Now consider the compactification of $T_\I$ on a circle of radius $\beta$. The mass parameters are complexified by Wilson lines on $S^1$ and we define the conventional dimensionless, single-valued $\C^*$ variables as follows,
\be \label{CP12d}
\begin{array}{rl}
x=e^X\,,\qquad &\ds X: = 2\pi \beta\,m^{\rm 3d}+i\oint_{S^1} A_V\,, \\[.4cm]  y=e^Y\,,\qquad & \ds Y: = 2\pi \beta\,t^{\rm 3d}+i\oint_{S^1} A_J\,,
\end{array}
\ee
where $\Im Y$ is a theta-angle in two dimensions. For the scalar in the two-dimensional vector multiplet we define
\be s = e^S\,,\qquad S := 2\pi\beta\,\sigma^{\rm 3d} +i\oint_{S^1} A\,.\ee
Additionally introducing a Wilson line $i\oint A = i\pi$ for the R-symmetry, the effective twisted superpotential in two dimensions is given by
\be \wt W_\I(S;X,Y) = \tfrac12(S^2+X^2)+S(Y-i\pi)+\Li_2(e^{-S-X})+\Li_2(e^{-S+X})\,. \label{WI} \ee
The term $SY$ is an FI coupling; and we recall that the free chirals with zero R-charge contribute $\tfrac14(S\pm X-i\pi)^2+\Li_2(e^{-S\mp X})$.

The expression in \eqref{WI} has several interpretations. It the effective twisted superpotential of an $\CN=(2,2)$ theory that governs supersymmetric vacua on $\R^2$. It is also the superpotential for an effective $\CN=4$ quantum mechanics on $\R_+$, coming from the reduction of $T_\I$ on $\DqS$ (in terms of Equation \eqref{QMSuper}, we have $-iW^{\rm QM}=\frac1\hbar \wt W_\I$). Moreover, when accompanied with a distinguished choice of branch cuts, $\frac1\hbar \wt W_\I$ describes the leading $\hbar\to 0$ asymptotics of the operator \eqref{originwavefunction} inserted at the origin in quantum mechanics on $\R_+$, or equivalently the leading asymptotics of the integrand of a block integral. 

The mirror symmetry transformation \eqref{omega} should descend to an equivalence of the various compactified versions of $T_\I$, at least in the regions of parameter space where the theory is massive. The extended mirror symmetry transformation should act holomorphically on complex parameters, generalizing \eqref{omega} to
\be \label{omega2d}
(X,Y)\;\overset{\omega}{\longmapsto} \Big(\frac{Y-X}{2},\, -\frac{3X+Y}{2}\Big)\,. \ee
For a precise duality we must supplement \eqref{omega2d} with the addition of Chern-Simons contact terms between between the R-symmetry and vector symmetry, taking the form $\pm i\pi X$ in twisted superpotentials. Then we find
\begin{align} T_\II:\;\; \wt W_\II(S;X,Y) &= \tfrac12S^2-(2X+i\pi)S+X^2-X(Y+i\pi)+\Li_2(e^{-S})+\Li_2(e^{-S+X-Y}) \notag  \\
 &\simeq \wt W_\I\big(S;\tfrac{Y-X}2,-\tfrac{3X+Y}{2}\big)+\frac{i\pi}{2}(Y+X)\,, \label{WII}
\end{align}
\begin{align} T_\III:\;\; \wt W_\III(S;X,Y) &= \tfrac12S^2+(2X-i\pi)S+X^2+X(Y-i\pi)+\Li_2(e^{-S})+\Li_2(e^{-S-X-Y})  \notag  \\
 &\simeq \wt W_\I\big(S;-\tfrac{X+Y}2,\tfrac{3X-Y}{2}\big)+\frac{i\pi}{2}(Y-X)\,, \label{WIII}
\end{align}
where in the relations to $\wt\CW_\I$, indicated by ``$\simeq$,'' we allow shifts of the dynamical field $S$ by multiples of the complex masses $X$ and $Y$.
From the twisted superpotential, we can determine the supersymmetric parameter space $\CL_{\rm SUSY}$ \eqref{LSUSY}. In all three theories $T_\I,\,T_\II,\,T_\III$, it is given by
\be \CL_{\rm SUSY}:\quad \big\{p_y+(y^{-1}-x-x^{-1})+p_y^{-1}=0\,,\;\; p_xp_y-(p_x+p_y)x+1=0\big\}\;\subset\; (\C^*)^4\,.\quad
\label{LSUSYCP1}\ee
The precise match of $\CL_{\rm SUSY}$ among the three theories is a consequence of mirror symmetry, and serves as a verification of the contact terms added to \eqref{WII}--\eqref{WIII}.

Finally, in analyzing Stokes jumps it will be helpful to understand the discriminant locus $\CD$ for these theories. This is the locus in parameter space where the theories become massless, and is the source of Stokes walls. It is important to avoid this locus when defining blocks.%
\footnote{Recall that a mass gap is essential for the effective quantum mechanics on $\R_+$ to be free of infrared divergences. More generally, we only have control over RG flow for compactified theories if we have made them massive; and only in this case can we hope for these calculations to accurately reflect 3d mirror symmetry.} %
For $T_\I$, the vacuum equation is
\be \label{CPvac}
\exp \frac{\pd \wt\CW}{\pd S} = 1\quad\Rightarrow\quad 
 \frac1y=(x^{-1}-s)(1-xs^{-1})\,,
\ee
with solutions
\be \label{CPcrit}
s^{1,2}(x,y) = -\tfrac12\Big[y-x-x^{-1}\pm \sqrt{(y^{-1}-x-x^{-1})^2-4}\Big]\,,
\ee
which has discriminant locus given by
\be \CD:\quad \{y^{-1}=x+x^{-1}\pm 2\} \;\subset\; \C^*\times \C^*\,.\quad\ee
For $T_\II$ and $T_\III$ the vacuum equations look different, but the discriminant locus is the same. Indeed, $\CD$ is invariant under the $\Z_3$ mirror-symmetry action $(x,y)\overset{\omega}{\longmapsto}\big(x^{-\frac12}y^{\frac12},x^{-\frac32}y^{-\frac12}\big)$\,.

\begin{figure}[htb]
\centering
\includegraphics[width=5.7in]{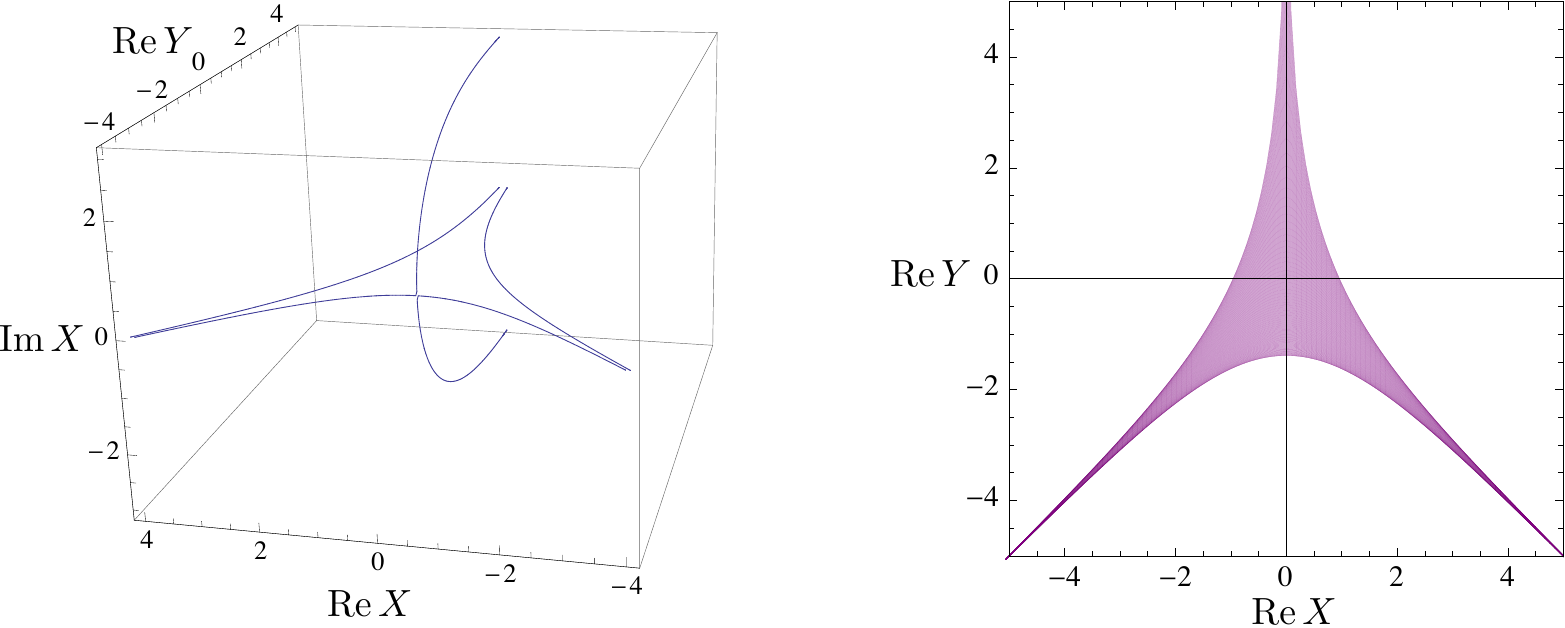}\hspace{.4in}
\caption{Discriminant locus of the $\cp^1$ model on $\RS$: slice at $\Im Y=0$ (left) and projection to $(\Re X,\Re Y)$ (right). The $\Im\,X$ direction is periodic, with period $2\pi$.}
\label{fig:disc}
\end{figure}

We present two visualizations of $\CD$, which has complex codimension one, in Figure \ref{fig:disc}. First we take a three-dimensional slice at $\Im\,Y=0$, which shows some of the branching structure of $\CD$. Then we project all of $\CD$ to the plane parametrized by the real masses $\Re\,X=2\pi\beta m^{\rm 3d}$ and $\Re\,Y=2\pi\beta t^{\rm 3d}$. The latter projection shows the values of $m^{\rm 3d}$ and $t^{\rm 3d}$ for which the compactified theory on $\RS$ could become massless (for some choice of Wilson lines). In the decompactification limit $\beta\to\infty$, the projection reduces to the three rays of Figure \ref{fig:CP1param}.


\subsection{Blocks}
\label{sec:CP1blocks}

Since the $\cp^1$ sigma-model generically has two massive vacua $\alpha=1,2$, there should be two independent holomorphic blocks $B^1(x,y;q),\,B^2(x,y;q)$. We will now compute the two blocks for $T_\I$ using the methods of Section \ref{sec:blockint}: we define a formal integral that solves line-operator  identities, and associate a convergent integration cycle $\Gamma^\alpha$ to each vacuum $\alpha$.

To build $T_\I$ using the iterated construction of Section \ref{sec:Ward-assemble}, we take the following steps.
\begin{enumerate} \setlength{\itemsep}{-.1cm}
\item Tensor together two theories $T_{\Delta_1}\otimes T_{\Delta_2}$ to obtain a theory of two chirals $\phi_1,\,\phi_2$ with $U(1)_1\times U(1)_2$ flavor symmetry and $-1/2$ CS levels for each. The R-charges are $R_{\phi_1}=R_{\phi_2}=0$.
\item Redefine the flavor symmetries as a vector $U(1)_V$ and axial $U(1)_A$, using $U=\frac12\left(\begin{smallmatrix}1 & -1 \\ 1 & 1\end{smallmatrix}\right)\in GL(2)$.
\item There is now minus one unit of Chern-Simons coupling for $U(1)_V$ and for $U(1)_A$. Add one unit of coupling for each to give net zero Chern-Simons levels.
\item Gauge the axial $U(1)_A$, with the addition of an FI term. This produces a new topological flavor symmetry $U(1)_J$.
\item Shift the R-symmetry current by a unit of the topological $U(1)_J$ current, to cancel a Chern-Simons coupling between the $U(1)_R$ connection $A_R$ and the dynamical gauge connection. (The unwanted coupling comes from the full definition of the $T_\Delta$ theories.)
\end{enumerate}

This construction of the theory dictates how to assemble the line-operator identities for the blocks of $T_\I$. We start with canonical identities $\hat p_1+\hat x_1^{-1}-1\simeq 0,\, \hat p_2+\hat x_2^{-1}-1\simeq 0$ for the product $T_{\Delta_1}\otimes T_{\Delta_2}$, and apply the appropriate transformations to obtain the two operators
\be
\frac{1}{\sqrt{\hat p_y \hat x}}\sqrt{\frac{\hat p_x}{-\hat y}}+\frac{1}{\hat p_y\hat x}-1\simeq 0\,,\qquad \sqrt{\frac{\hat x}{\hat p_y}}\frac{1}{\sqrt{-\hat y\hat p_x}}+\frac{\hat x}{\hat p_y}-1\simeq 0\,, \label{WardCP10}
\ee
where ``$\simeq0$'' means ``annihilates blocks.'' Here $\hat x,\,\hat y$ act as multiplication by the complex mass parameters $x=e^X,\,y=e^Y$ \eqref{CP12d}, while $\hat p_x,\,\hat p_y$ are the corresponding shift operators, satisfying $\hat p_x\hat x=q\hat x\hat p_x,\,\hat p_y\hat y=q\hat y\hat p_y$.  By working in the left ideal generated by the two operators in \eqref{WardCP10}, the square roots (which arise from the non-integral transformation of the charge lattice in step (2)) can be eliminated and the operators can be written in the equivalent form
\be \label{WardCP1}
\hat p_y+\Big(\frac{1}{\hat y}-\hat x-\frac{1}{\hat x}\Big)+\frac{1}{\hat p_y} \simeq 0\,,\qquad q^{-\frac12}\hat p_x\hat p_y-\hat x(q^{\frac12}\hat p_x+\hat p_y)+1 \simeq 0\,,
\ee
which makes clear that the line-operator identities are a quantization of the Lagrangian submanifold $\CL_{\rm SUSY}$ \eqref{LSUSYCP1}.

In a similar way, we can follow these steps and build a block integral that formally solves \eqref{WardCP1}. It takes the form
\begin{align} \label{CP1int}
\mathbb{B}_\I(x,y;q) \,&=\, \frac{1}{(q)_\infty}\int_{*} \frac{ds}{2\pi is} \frac{\theta(-q^{-\frac12}y)}{\theta(x)\theta(-q^{-\frac12}sy)} (qs^{-1}x^{-1};q)_\infty (qs^{-1}x;q)_\infty \\ &=\int_{*} \frac{ds}{2\pi is}\Upsilon_\I(x,y,s;q) \,,  \notag
\end{align}
recalling for convenience the definitions
\be (z;q)_\infty := \begin{cases} \prod_{n=0}^\infty (1-q^nx) & |q|<1 \\
 \prod_{n=1}^\infty (1-q^{-n}x)^{-1} & |q|>1 \end{cases}\,, \qquad
 \theta(z)=\theta(z;q) := (-q^{\frac12}z;q)_\infty (-q^{\frac12}z^{-1};q)_\infty\,, \notag
\ee
where $\theta(z;q)$ has been abbreviated to $\theta(z)$. The two products $B_\Delta(sx^{\pm 1};q) = (qs^{-1}x^{\mp 1};q)_\infty$ in the integrand are the contributions of the chirals $\phi_1,\phi_2$, packaged as theories $T_{\Delta_1}$ and $T_{\Delta_2}$, and the remaining theta functions reproduce additional Chern-Simons levels and the FI term. As usual, the choice of theta functions is unique up to multiplication by an elliptic function (Section \ref{sec:theta}). We also include a normalization by $1/(q)_\infty$, defined in \eqref{defqinf}.

To evaluate the blocks \eqref{CP1int}, we need the convergent integration cycles associated to each vacuum. To this end, it helps to analyze the integrand as a function of the cylindrical variable $S=\log s$.  We take $\hbar$ to be real and nonzero, so that $q$ is real and positive, with $|q|<1$ or $|q|>1$. For $|q|<1$ (respectively, $|q|>1$), the integrand has a line of poles (resp., zeroes) along $\Im\,S=\Im\,Y$, coming from the theta-function $\theta(-q^{-\frac12}sy)$ that is associated to an FI term. The poles have spacing $|\hbar|$. There are also two parallel half-lines of zeroes (resp., poles) coming from the contributions of the chirals; the half-lines start at $S = \pm X$ and stretch to $S=-\infty$, with spacing $|\hbar|$. At large $|\Re\,S|$ (close to the ends of the $S$-cylinder), the integrand behaves approximately as $\exp(\tfrac 1{2\hbar} {\rm sign}(\Re\,S)S^2)$, which indicates that naively the convergent integration cycles can and either at $\Re\,S\to \infty$ when $|q|<1$ or at $\Re\,S\to-\infty$ when $|q|>1$.

If we send $\hbar\to 0$ along the real axis, from either the positive or negative side, the integrand has leading asymptotics
\be
\Upsilon_\I(x,y,s;q) \overset{\hbar\to 0}{\sim}\exp\Big[\frac{1}{\hbar}\Big(\frac12(\log x)^2-\frac12(\log (-y))^2+\frac12(\log(-sy))^2 + \Li_2(x^{-1}s^{-1})+\Li_2(xs^{-1})\Big)\Big]\,, \label{hCP1}
\ee
which is equivalent to $\frac1\hbar\wt W_\I(S;X;Y)$ with a distinguished choice of branch cuts. The cuts come from the lines of poles and zeroes of $\Upsilon_\I$ at finite $\hbar$, and have slope zero on the $S$-cylinder; they are the standard cuts for $\log$ and $\Li_2$ as written in \eqref{hCP1}. The two critical points $s^\alpha(x,y)$ of \eqref{hCP1} are located at the solutions to the vacuum equation \eqref{CPcrit}.

\begin{figure}[t!]
\centering
\includegraphics[width=6.1in]{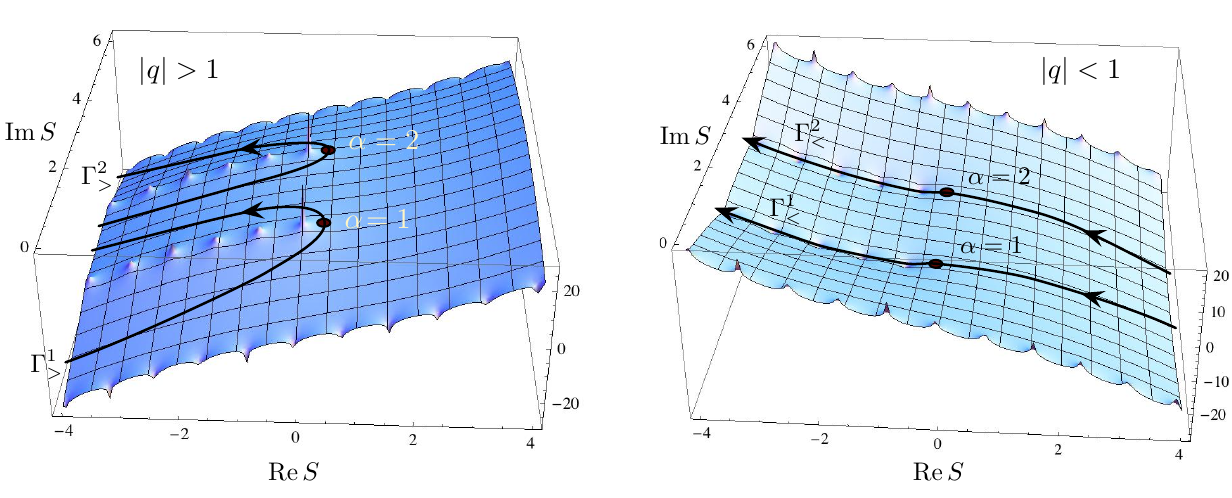}
\caption{Cycles for $T_\I$ blocks at $|q|>1$ (left) and $|q|<1$ (right). We choose $Y=2$, $X=-2\pi i/3$, and $\hbar = \pm \pi/4$. These are plots of the quantum potential $\log|\Upsilon_\I|$.}
\label{fig:CPcycles}
\end{figure}

Now suppose that $|q|>1$. We focus on the classical sigma-model phase of $T_\I$, that is $\Re\,Y\gg 0$ and $\Re\,X \lesssim \Re\,Y$. We will also choose $\Im\,Y\approx 0$ and $\Im X\approx \frac43\pi$ (mod $2\pi$). These imaginary parts (Wilson lines) ensure that we are far away from the discriminant locus in parameter space (Figure \ref{fig:disc}), so that the theory on $\RS$ is massive. Figure \ref{fig:CPcycles} shows a plot of the ``quantum'' potential $\log|\Upsilon_\I|$ at these values of parameters. The critical points are very close to the beginning of the two half-lines of poles, at $S=\pm X$ (mod $2\pi i$). The downward-flow contours in the vicinity of the critical points have an obvious continuation to asymptotic, convergent contours $\Gamma^1_>$ and $\Gamma^2_>$, as indicated. (These could be obtained using either method of Section \ref{sec:contours}.) Moreover, it becomes clear that at this point in parameter space we are not only far from the discriminant locus, but far from any Stokes walls: there is no way that the downward-flow cycle from one critical point could come close to the other critical point.

As in Section \ref{sec:simple-ex}, we largely ignore the line of zeroes coming from $\theta(-q^{-\frac12} ys)$ --- and the fact that the integrand at finite $\hbar$ has an infinite set of ``quantum'' critical points between every pair of zeroes. We demand that the cycles must be well defined in the $\hbar\to 0$ limit (so don't know about ``quantum'' critical points), are asymptotically convergent (so cannot stop on a zero), and resemble downward-flow cycles in a neighborhood of the classical critical points; and this leads to $\Gamma^1_>$ and $\Gamma^2_>$. If we modified the integrand by an elliptic function, more horizontal lines of poles or zeroes could appear in Figure \ref{fig:CPcycles} (or cuts as $\hbar\to 0$), but this would not affect the qualitative features of the cycles.

We can evaluate the integral along $\Gamma^1_>$ and $\Gamma^2_>$ by summing residues, and readily find
\be \label{blocksI+} \quad
\boxed{\begin{array}{rl}
B^1_\I(x,y;q) &:= \ds\int_{\Gamma^1_>}\frac{ds}{2\pi is}\Upsilon_\I = \frac{\theta(-q^{-\frac12}y)}{\theta(x)\theta(-q^{-\frac12}x^{-1}y)}\,\CJ(xy^{-1},x^2;q)\,, \\[.4cm]
B^2_\I(x,y;q) &:= \ds\int_{\Gamma^2_>}\frac{ds}{2\pi is}\Upsilon_\I = \frac{\theta(-q^{-\frac12}y)}{\theta(x)\theta(-q^{-\frac12}xy)}\,\CJ(x^{-1}y^{-1},x^{-2};q)\,,
\end{array}} \quad |q|>1\,.
\ee
Here the function $\CJ$ is related to the so-called Hahn-Exton $q$-Bessel function \cite{Hahn, Exton}, and can be defined by a $q$-hypergeometric series%
\be \label{qBessel}
\CJ(x,y;q) \,:=\, (qy;q)_\infty\sum_{n=0}^\infty \frac{x^n}{(q^{-1})_n(qy;q)_n}\,,\qquad \text{$|q|<1$ or $|q|>1$}\,,
\ee
which converges both for $|q|<1$ and $|q|>1$ and defines a meromorphic function of $x,y\in \C^*$. It is fairly straightforward to check that the functions $B^\ra_\I(x,y;q)$ are solutions to the line-operator identities \eqref{WardCP1}. We will take \eqref{blocksI+}, with the extra normalization factor $(q^{-1})_\infty/2\pi i$, as a definition of the functions $B^\ra_\I(x,y;q)$, fixing the elliptic-function ambiguity. The subscript ``$\I$'' indicates a particular Stokes chamber --- the one where $T_\I$ is a semi-classical sigma-model.

As a physical check, we can take the true classical limit $|y|\to \infty$ of the blocks. This corresponds to a sigma-model with a huge $\cp^1$ target space, with single, light chiral excitations localized at the north and south poles. In this limit, we find
\be B_\I^{1,2}(x,y;q)\,\sim\, (qx^{\pm 2};q)_\infty \ee
up to a theta-function (corresponding to a background Chern-Simons term). We immediately recognize the remaining light chirals, with masses $\sim |2X|$.

What about contours $\Gamma^1_<$ and $\Gamma^2_<$ for $|q|<1$? In the neighborhoods of the critical points, the downward-flow cycles now extend in the horizontal real-$S$ direction, as shown on the right side of Figure \ref{fig:CPcycles}. We are in one of the situations discussed in Section \ref{sec:contours} and shown in Figure \ref{fig:upzero}. Extending toward positive $S$, the choice of cycles is obvious. Extending toward negative $S$, where the potential $\log|\Upsilon_\I|\sim \frac1\hbar \wt W_\I$ typically increases like $\frac{1}{|\hbar|}(\Re\,S)^2$, the most intuitive choice is to extend two independent cycles along half-lines of zeroes. We expect that it is possible to integrate over them with an appropriate regularization.

We do independently expect $B^\ra_\I(x,y;q)$ at $|q|<1$ to be naturally related to $B^\ra_\I(x,y;q)$ at $|q|>1$, \eg\ in the sense of sharing the same $q$-hypergeometric series expansions. We did calculate the $|q|>1$ blocks using a series that converged both for $|q|<1$ and $|q|>1$. Thus, rather than evaluating the integrals along $\Gamma^{1,2}_<$ directly, we will simply define
\be \label{blocksI-}
\begin{array}{rl} B^1_\I(x,y;q) &:= \ds\frac{\theta(-q^{-\frac12}y)}{\theta(x)\theta(-q^{-\frac12}x^{-1}y)}\,\CJ(xy^{-1},x^2;q)\,, \\[.4cm] 
B^2_\I(x,y;q) &:=  \ds\frac{\theta(-q^{-\frac12}y)}{\theta(x)\theta(-q^{-\frac12}xy)}\,\CJ(x^{-1}y^{-1},x^{-2};q)\,,
\end{array} \qquad |q|<1
\ee
as well. We will see that this definition passes a battery of nontrivial tests, and so we predict that if the $\Gamma^{1,2}_<$ integrals were to be regularized and evaluated, they would agree with \eqref{blocksI-}.

There is much hidden in the similarity of the definitions \eqref{blocksI+} and \eqref{blocksI-}. As was the case in the examples and Section \ref{sec:chiralblock}, the blocks cannot be continued from $|q|<1$ to $|q|>1$, and the distinction between the regimes (one topological and one anti-topological) is crucial. We will see now and in Section \ref{sec:CP1Stokes} that the $\CJ$-functions behave very differently at $|q|<1$ and $|q|>1$, and that blocks \eqref{blocksI+}--\eqref{blocksI-} satisfy some extremely restrictive conditions necessary for consistent Stokes phenomena.


\subsubsection{The $q$-Bessel function}
\label{sec:qBessel}

We pause to point out a few surprising properties of the $q$-Bessel function $\CJ(x,y;q)$, some proven and some conjectured. We do not assume that $q$ is real, but we do assume that it is never on the unit circle.

First, for both $|q|<1$ and $|q|>1$, a quick manipulation of the series \eqref{qBessel} shows that
\be \qquad \CJ(x,y;q) = \theta(-q^{\frac12}y) \CJ(xy^{-1},y^{-1};q^{-1})\,,\qquad |q|\neq 1. \label{Jreflect} \ee
When $|q|<1$, we can also re-write $(qy;q)_\infty/(qy;q)_n = (q^{n+1}y;q)_\infty = \sum_{r=0}^\infty \frac{1}{(q^{-1})_r}(q^ny)^r$, leading to
\be \qquad \CJ(x,y;q) = \sum_{n,r=0}^\infty \frac{(-1)^{n+r}q^{\frac12(n+r)(n+r+1)}x^ny^r}{(q)_n(q)_r}\,,\qquad |q|<1\,, \label{symsum} \ee
which shows that the function is symmetric,
\be \qquad \CJ(x,y;q) \,=\,\CJ(y,x;q)\,,\qquad |q|<1\,. \label{Jsym-} \ee
Combining the symmetry with the ``inversion'' \eqref{Jreflect} leads to identities such as
\be \label{Jreflect2} \qquad
\theta(-q^{\frac12}x^{-1}y)\,\CJ(x,y;q) \,=\, \theta(-q^{\frac12}y)\,\CJ(x^{-1},x^{-1}y;q)\,,\qquad |q|>1\,.
\ee
In contrast, when $|q|>1$ the sum \eqref{symsum} never converges and the symmetry \eqref{Jsym-} does not hold! We find, conjecturally, that it is replaced by the relation
\be \label{Jsym+} \qquad
\CJ(x,y;q) - \CJ(y,x;q) \,=\, \frac{\theta(-q^{\frac12}x^{-1})\theta(-q^{\frac12}y)}{\theta(-q^{\frac12}x^{-1}y)}\,\CJ(x,y;q^{-1})\,,\qquad |q|>1\,,
\ee
which we have verified numerically to high precision. This relation ultimately ensures consistent Stokes jumps for the $T_\I$ blocks.

Finally, we experimentally find%
\footnote{We especially thank Don Zagier for pointing this out, and more generally for offering generous and extremely useful guidance in numerical analysis of $q$-series.} an infinite-product expansion
\begin{align} \label{Jprod}
\CJ(x,y;q) &= (qx;q)_\infty(qy;q)_\infty(q^2xy;q)_\infty (q^3x^2y;q)_\infty (q^3xy^2;q)_\infty  \\
 &\quad\times (q^4x^3y;q)_\infty (q^4 x^2 y^2;q)_\infty (q^5x^2y^2;q)_\infty (q^4 xy^3;q)_\infty(q^5x^4y;q)_\infty \cdots, \notag
\end{align}
obtained by treating \eqref{symsum} as a formal series in $q$. The corresponding infinite-product form of the blocks $B^\ra_\I$ of $T_\I$ in the semi-classical sigma-model chamber contains information about Ooguri-Vafa invariants (degeneracies of BPS states), as discussed in Section \ref{sec:OV}.


\subsection{Stokes jumps}
\label{sec:CP1Stokes}

One of the most interesting aspects of holomorphic blocks is their global behavior in parameter space --- namely their Stokes phenomena. In the wavefunction interpretation of the blocks, Stokes walls are locations where there can be tunneling between SUSY ground states $|\alpha\rangle$ in the Hilbert space on $T^2$. We can analyze the structure of Stokes walls quantitatively using the block integral and the methods outlined in Section \ref{sec:contours}. We focus on a region of parameter space where the gradient flows of either the semi-classical potential $\Re\big(\frac1\hbar \wt W_\I(x,y,s;q)\big)$ (with cuts specified by \eqref{hCP1}) or the ``quantum'' potential $\log|\Upsilon_I(x,y,s;q)|$ can connect one critical point to another without passing through branch cuts or lines of poles or zeroes. Then the Stokes walls occur semi-classically when $\Im\big(\frac1\hbar \wt W_\I(x,y,s^{(1)};q)\big)=\Im\big(\frac1\hbar \wt W_\I(x,y,s^{(2)};q)\big)$ at the two critical points $s^{(1)}(x,y)$ and $s^{(2)}(x,y)$. Alternatively, using critical points of the full integrand $\Upsilon_\I(x,y,s;q)$ the walls will be slightly deformed, and are given by $\arg \Upsilon_\I(x,y,s^{(1)};q)=\arg \Upsilon_\I(x,y,s^{(2)};q)$ (mod $2\pi$).

\begin{figure}[t!]
\centering
\hspace{-.3in}\includegraphics[width=5.5in]{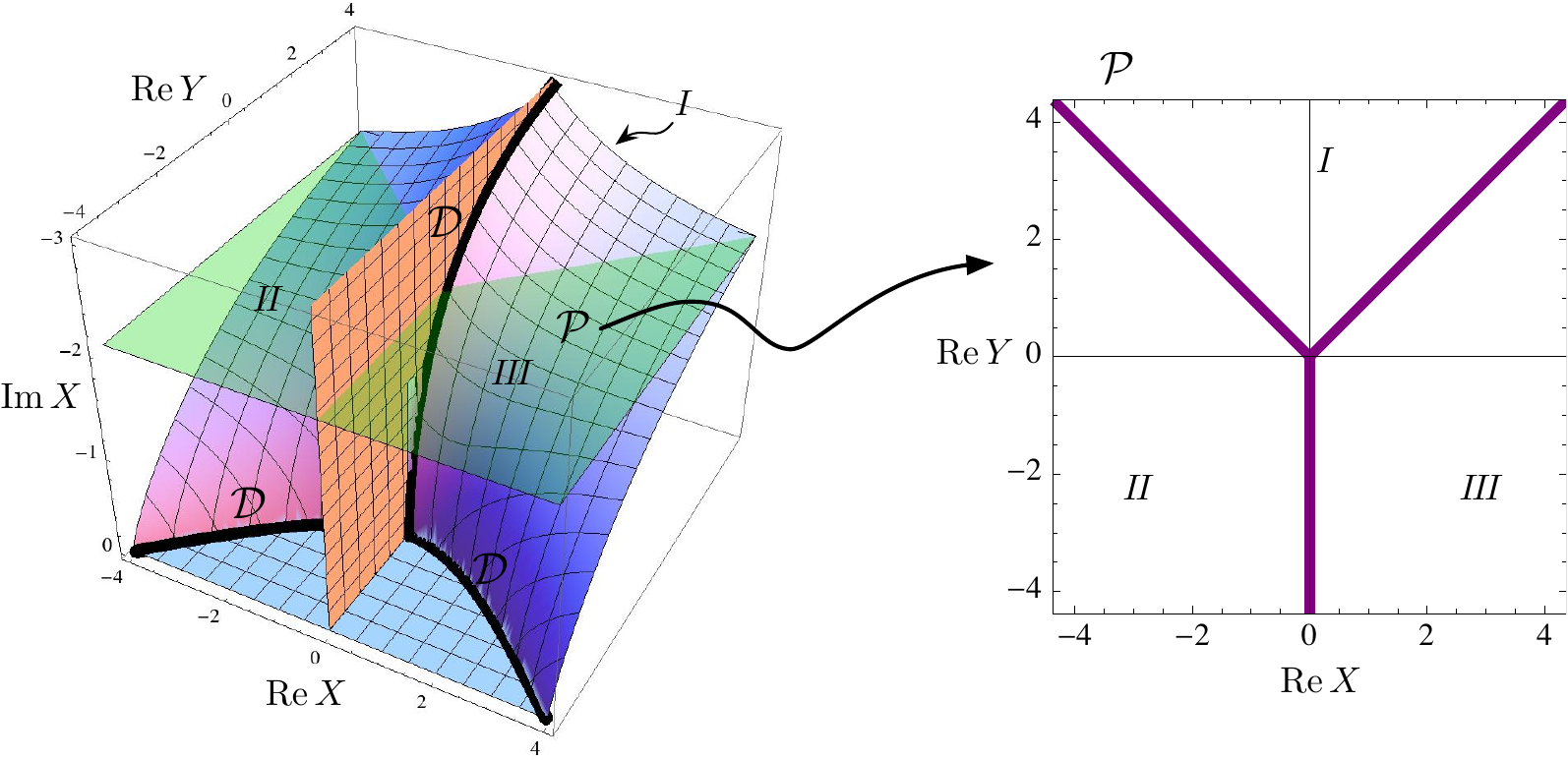}
\caption{Some Stokes walls for the $\cp^1$ sigma-model at $\Im\, Y=0$, emanating from the discriminant locus $\CD$. The plane $\CP$ at $\Im\, X=-2\pi/3$, highlighted in green on the left, is reproduced on the right.}
\label{fig:Stokes}
\end{figure}

We plot the Stokes walls in the region of parameter space with $\Im\,Y=0$ and $\Re\,Y>0$ in Figure \ref{fig:Stokes}, using the semi-classical approximation. We have fixed $\hbar$ to be real and small, but it does not matter whether it is positive or negative. The walls emanate from the discriminant locus $\CD$. Generically, there are three codimension-one walls meeting every branch of $\CD$, in correspondence with the fact that there are two critical points. Indeed, the behavior of any such system with two critical points that are very close to each other (as is the case in a neighborhood of the discriminant locus) is universal. It can be modeled on the Airy integral $\int dS \exp\Big[\frac1\hbar (S^3+aS)\Big]$, where $a$ is a generic parameter in the plane transverse to the discriminant locus. The Airy integral is famously known to have three Stokes walls in the $a$-plane, so this is what we expect to see.

We restrict our analysis to a neighborhood of the real plane $\CP$ parameterized by $\Re\,X$ and $\Re\,Y$ at $\Im X=4\pi/3\equiv -2\pi/3$ (mod $2\pi$) and $\Im Y=0$. This plane has the advantage of being invariant under mirror symmetry transformations \eqref{omega2d}, which will be discussed in Section \ref{sec:CP1mirror}. The discriminant locus $\CD$ intersects the plane $\CP$ at the origin, and three Stokes walls that emanate from it separate the plane into three chambers. Somewhat surprisingly, the walls turn out to be anti-parallel to the initial massless rays in the three-dimensional moduli space of $T_\I$ (Figure \ref{fig:CP1param}). Thus, it makes sense to label the chambers by the Higgs and Coulomb rays $\I$, $\II$, and $\III$.

\begin{figure}[b!]
\centering
\includegraphics[width=4.5in]{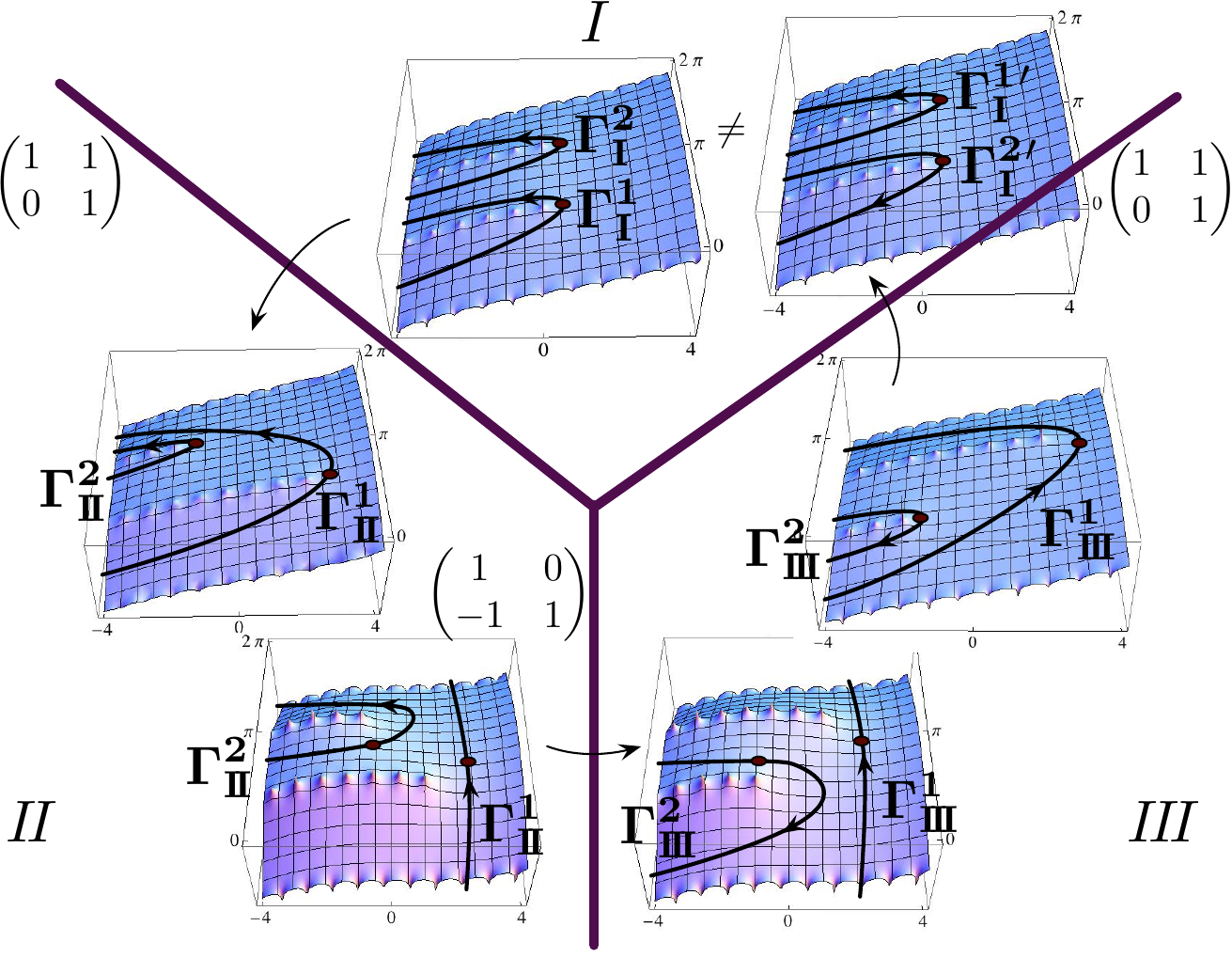}
\caption{The critical-point cycles at $|q|>1$, in various chambers on $\CP$. The plots of $\log|\Upsilon_\I|$ here are drawn for $\hbar=\pi/4$, at points on the circle $|\Re\,X|^2+|\Re\,Y|^2=4$.}
\label{fig:CPchambers}
\end{figure}

We first take $\hbar>0$ and $|q|>1$. In chamber $\I$, where $T_\I$ is approximately a sigma-model onto the $\cp^1$ Higgs branch, we have already found the two blocks $B^{1,2}_\I(x,y;q)$ corresponding to the two massive vacua at the poles of the $\cp^1$ \eqref{blocksI-}. As we vary parameters in the plane to other chambers, the critical points $s^\alpha(x,y)$ start to rotate around each other. Similarly, the two half-lines of poles in the $|q|>1$ integrand (or half-line branch cuts of $h_\I$) slide relative to one another in the $\Re\,S$ direction (their separation in the $\Im\,S$ direction, given by $\Im\,X$, is fixed). The critical points and natural downward-flow cycles associated to them at various values of $(\Re\,X,\,\Re\,Y)$ are sketched in Figure \ref{fig:CPchambers}.

Moving from chamber $\I$ counterclockwise to chamber $\II$, we find that the cycle $\Gamma^1_{I>}$ (we suppress the `$>$' subscript in Figure \ref{fig:CPchambers}) hits the critical point $\alpha=2$, and picks up a copy of $\Gamma^2_{I>}$. Thus
\be \begin{pmatrix} \Gamma^1_{\II>} \\ \Gamma^2_{\II>} \end{pmatrix} 
 = M^{I\to \II}_> \,\begin{pmatrix} \Gamma^1_{I>} \\ \Gamma^2_{I_>} \end{pmatrix}\,,\qquad M^{I\to \II}_> = \begin{pmatrix} 1 & 1 \\ 0 & 1 \end{pmatrix}\,.\ee
Correspondingly, the natural $|q|>1$ blocks associated to critical-point cycles (\ie\ vacua) in chamber $\II$ are not $B^1_\I$ and $B^2_\I$, but rather $(B^1_\II,B^2_\II) = (B^1_\I+B^2_\I,B^2_\I)$.
Continuing on through chamber $\II$, the cycle $\Gamma^1_{\II>}$ at some point appears to become closed (wrapping the cylinder) rather than ending asymptotically at $\Re\,S=-\infty$, but these two choices are homotopic and there is no modification to the blocks. The next Stokes wall is crossed moving into chamber $\III$, where $\Gamma^2_\II$ picks up a copy of $-\Gamma^1_\II$:
\be \begin{pmatrix} \Gamma^1_{\III>} \\ \Gamma^2_{\III>} \end{pmatrix} 
 = M^{\II\to \III}_> \,\begin{pmatrix} \Gamma^1_{\II>} \\ \Gamma^2_{\II_>} \end{pmatrix}\,,\qquad M^{\II\to \III}_> = \begin{pmatrix} 1 & 0 \\ -1 & 1 \end{pmatrix}\,.\ee
And, finally, we cross the third wall to return back to chamber $\I$, finding
\be \begin{pmatrix} \Gamma^1_{I>}{}' \\ \Gamma^2_{I_>}{}' \end{pmatrix}
 = M^{\III\to I}_> \, \begin{pmatrix} \Gamma^1_{\III>} \\ \Gamma^2_{\III>} \end{pmatrix}\,, \qquad
  M^{\III\to I}_>= \begin{pmatrix} 1 & 1 \\ 0 & 1 \end{pmatrix}\,.\ee
Note that the product of matrices
\be \label{CPmonod}
M^{\III\to I}_>M^{\II\to \III}_>M^{I\to \II}_>  = \begin{pmatrix} 0 & 1 \\ -1 & 0 \end{pmatrix}
\ee
is not the identity: there is a monodromy induced by circling around the discriminant locus, so that the cycles in chamber $\I$ have been switched. This is easy to see directly by following the classical critical points $s^\alpha(x,y)$ in \eqref{CPcrit}. As we circle around the discriminant locus, the square root changes sign, taking $s^1(x,y)\leftrightarrow s^2(x,y)$. Physically, we expect the SUSY ground states $|1\rangle$ and $|2\rangle$ in the Hilbert space of $T^2$ to undergo the monodromy \eqref{CPmonod} if one adiabatically varies parameters to circle the discriminant locus.


\subsubsection{Topological and anti-topological regimes}
\label{sec:CP1match}

Now let us consider what happens to the blocks at $|q|<1$. We can do this using the formal integration cycles drawn on the right Figure \ref{fig:CPcycles} for chamber I, by tracing them through the $(\Re\,X,\,\Re\,Y)$ plane, just as we did for $|q|>1$. We can read off Stokes matrices even if we don't know how to compute the integrals directly. However, we already know what the result must be. As discussed in Section \ref{sec:contours}, and as needed for consistency of topological/anti-topological fusion, the (semi-classical) Stokes walls at $|q|<1$ are identical to the ones at $|q|>1$, but the matrices are related by
\be  M_< = (M_>)^{-1\,T}\,. \label{invtran} \ee
Indeed, considering the formal integration cycles leads to precisely these jumps. Given the blocks \eqref{blocksI-} in chamber $\I$, in chambers $\II$ and $\III$ we should then have
\begin{subequations} \label{blocksII-}
\begin{align} \begin{pmatrix} B^1_\II \\ B^2_\II \end{pmatrix} &= (M^{I\to \II}_>)^{-1\,T}\begin{pmatrix} B^1_\I \\ B^2_\I \end{pmatrix} = \begin{pmatrix} B^1_\I \\ B^2_\I-B^1_\I \end{pmatrix}\,, \qquad |q|<1\,,\\[.1cm]
 \begin{pmatrix} B^1_\III \\ B^2_\III \end{pmatrix} &= (M^{\II\to \III}_>M^{I\to \II}_>)^{-1\,T}\begin{pmatrix} B^1_\I \\ B^2_\I \end{pmatrix} = \begin{pmatrix} B^2_\I \\ B^2_\I-B^1_\I \end{pmatrix}\,, \qquad |q|<1\,.\end{align}
\end{subequations}

This brings us to a puzzle, previewed already in the Introduction. When $|q|>1$, the blocks associated to vacua and integration cycles of Figure~\ref{fig:CPchambers} (in contrast to expressions \eqref{blocksII-}), are
\be \label{blocksII+}
\begin{pmatrix} B^1_\II \\ B^2_\II \end{pmatrix}
= \begin{pmatrix} B^1_\I  + B^2_\I \\ B^2_\I \end{pmatrix}\,,\qquad 
 \begin{pmatrix} B^1_\III \\ B^2_\III \end{pmatrix}
= \begin{pmatrix} B^1_\I  + B^2_\I \\ -B^1_\I \end{pmatrix}\,,\qquad
|q|>1\,.
\ee
In the ``classical'' chamber $\I$, we postulated that the $|q|<1$ blocks and the $|q|>1$ blocks should have the same convergent $q$-hypergeometric series expansions, enabling us to derive \eqref{blocksI-} from \eqref{blocksI+}. There is nothing special about chamber $\I$, and this property of being related naturally by a convergent series should hold in all chambers. However, the linear combinations of $|q|<1$ and $|q|>1$ blocks in chambers $\II$ and $\III$ above look very different!

The resolution comes once we remember that --- despite having identical series expansions --- the $|q|<1$ and $|q|>1$ blocks are different analytic functions which obey different identities. For example, let's focus on the blocks $B^1_\II$ in chamber $\II$. For $|q|<1$, we  have $B^1_\II(x,y;q)=B^1_\I(x,y;q)$, with the block given by \eqref{blocksI-}, but we use the symmetry \eqref{Jsym-} to re-write $B^1_\II$ as
\be B^1_\II(x,y;q) = \frac{\theta(-q^{-\frac12}y)}{\theta(x)\theta(-q^{-\frac12}x^{-1}y)}\,\CJ(x^2,xy^{-1};q)\,,\qquad |q|<1\,.\ee
On the other hand, for $|q|>1$, Stokes jumping predicts that the block is given by $B^1_\II(x,y;q)=B^1_\I(x,y;q)+B^2_\I(x,y;q)$, with the RHS given by \eqref{blocksI+}. We can transform the resulting function as follows:
\begin{align}
B^1_\II(x,y;q) &= \frac{\theta(-q^{-\frac12}y)}{\theta(x)\theta(-q^{-\frac12}x^{-1}y)}\,\CJ(xy^{-1},x^2;q)+ \frac{\theta(-q^{-\frac12}y)}{\theta(x)\theta(-q^{-\frac12}xy)}\,\CJ(x^{-1}y^{-1},x^{-2};q) \quad |q|>1  \notag\\[.1cm]
&\hspace{-.75in}\overset{\eqref{Jreflect2}}{=} \frac{\theta(-q^{-\frac12}y)\theta(-q^{\frac12}x^2)}{\theta(x)\theta(-q^{-\frac12}x^{-1}y)\theta(-q^{\frac12}xy)}\CJ(x^{-1}y,xy;q) + \frac{\theta(-q^{-\frac12}y)\theta(-q^{-\frac12}x^2)}{\theta(x)\theta(-q^{-\frac12}xy)\theta(-q^{\frac12}x^{-1}y)}\CJ(xy,x^{-1}y;q) \notag \\
&\hspace{-.65in}= \frac{\theta(-q^{-\frac12}y)\theta(-q^{\frac12}x^2)}{\theta(x)\theta(-q^{-\frac12}x^{-1}y)\theta(-q^{\frac12}xy)}\Big[\CJ(x^{-1}y,xy;q)-\CJ(xy,x^{-1}y;q)\Big]  \label{steptheta} \\
&\hspace{-.75in} \overset{\eqref{Jsym+}}{=} \frac{\theta(-q^{-\frac12}y)}{\theta(x)}\CJ(x^{-1}y,xy;q^{-1}) \notag \\
&\hspace{-.75in} \overset{\eqref{Jsym-}}{=} \frac{\theta(-q^{-\frac12}y)}{\theta(x)}\CJ(xy,x^{-1}y;q^{-1}) \notag \\
&\hspace{-.75in} \overset{\eqref{Jreflect}}{=} \frac{\theta(-q^{-\frac12}y)}{\theta(x)\theta(-q^{-\frac12}x^{-1}y)}\,\CJ(x^2,xy^{-1};q)\,,\notag
\end{align}
where in the middle step \eqref{steptheta} we used $\theta(z)=\theta(z^{-1})$ and $\theta(-q^{-\frac12}z)=-z\,\theta(-q^{\frac12}z)$ to match the theta-prefactors in the second term to those in the first term, up to a sign. Thus, we have arrived at an expression for the $|q|>1$ block in chamber $\II$ that looks identical to the $|q|<1$ block; by the definition of the $\CJ$--function \eqref{qBessel}, they have the same convergent series expansions.

Such manipulations --- taking completely different paths for $|q|<1$ and $|q|>1$ --- work to bring all blocks in all chambers to the same form at $|q|<1$ and $|q|>1$. The result can be summarized as:
\be \label{BIIfinal}
\;\;\;\begin{array}{rl}
 B^1_\II(x,y;q) &= \ds\frac{\theta(-q^{-\frac12}y)}{\theta(x)\theta(-q^{-\frac12}x^{-1}y)}\,\CJ(x^2,xy^{-1};q)\,, \\[.5cm]
 B^2_\II(x,y;q) &= \ds -\frac{\theta(-q^{-\frac12}y)\theta(-q^{\frac12}x^{2})}{\theta(x)\theta(-q^{-\frac12}x^{-1}y)\theta(-q^{\frac12}xy)}\CJ(xy,x^{-1}y;q)\,;
\end{array}\quad |q|<1\;\text{or}\;|q|>1
\ee
\be
\;\;\begin{array}{rl} \label{BIIIfinal}
 B^1_\III(x,y;q) &= \ds\frac{\theta(-q^{-\frac12}y)}{\theta(x)\theta(-q^{-\frac12}xy)}\,\CJ(x^{-2},x^{-1}y^{-1};q)\,, \\[.5cm]
 B^2_\III(x,y;q) &= \ds \frac{\theta(-q^{-\frac12}y)\theta(-q^{-\frac12}x^{2})}{\theta(x)\theta(-q^{-\frac12}xy)\theta(-q^{-\frac12}xy^{-1})}\CJ(x^{-1}y,xy;q)\,.
\end{array}\quad |q|<1\;\text{or}\;|q|>1
\ee


\subsection{Mirror symmetry}
\label{sec:CP1mirror}

Another constraint on the form of holomorphic blocks for $T_\I$ comes from mirror symmetry. The blocks associated to massive vacua in a given chamber of a theory should not depend on the choice of mirror description used to calculate them. In the present case, we have three mirror descriptions $T_\I\simeq T_\II\simeq T_\III$ of the same theory, with the property that each is a semi-classical sigma-model to a $\cp^1$ Higgs branch in a different chamber. Thus, we might expect that theories $T_\II$ and $T_\III$ give us simple descriptions of the blocks $B^\ra_\II$ and $B^\ra_\III$ (respectively), just as $T_\I$ gave an especially simple description of the blocks $B^\ra_\I$ in chamber I.

To be more precise, let's consider again the complexified mirror-symmetry action
\be \label{omegaexp}
(X,Y)\;\overset{\omega}{\longmapsto} \Big(\frac{Y-X}{2},\, -\frac{3X+Y}{2}\Big)\,,\quad\text{or}\quad
(x,y)\;\overset{\omega}{\longmapsto} \Big(\sqrt{\frac yx},\frac{1}{\sqrt{x^3 y}}\Big)\,.
\ee
The theory $T_\II$ is given by applying \eqref{omegaexp} to the flavor symmetries and and parameters of $T_\I$, and shifting some background theta-angles and R-charges. In $(X,Y)$ parameter space, the transformation \eqref{omegaexp} preserves a neighborhood of the plane $\CP$ at $\Im X=4\pi/3$ and $\Im Y=0$, and permutes the chambers of Figure \ref{fig:Stokes} counterclockwise, $\I\to\II\to\III\to\I$. Therefore, by applying \eqref{omegaexp} to the blocks $B^\ra_\I$ associated to vacua in chamber $\I$ (as calculated by $T_\I$, say), we expect to get blocks associated to vacua in chamber $\II$:
\be \begin{pmatrix} B^1_\I \\ B^2_\I \end{pmatrix} \sim \begin{pmatrix} \CJ(xy^{-1},x^2;q) \\ \CJ(x^{-1}y^{-1},x^{-2};q) \end{pmatrix}
\;\overset{\omega}{\longmapsto}\; 
\begin{pmatrix} B^1_\II \\ B^2_\II \end{pmatrix} \sim \begin{pmatrix} \CJ(xy,x^{-1}y;q) \\ \CJ(x^2,xy^{-1};q) \end{pmatrix}\,.
\ee
Similarly, by applying the mirror-symmetry transformation twice, we should get the blocks associated to vacua in chamber $\III$,
\be \qquad\begin{pmatrix} B^1_\I \\ B^2_\I \end{pmatrix} \sim \begin{pmatrix} \CJ(xy^{-1},x^2;q) \\ \CJ(x^{-1}y^{-1},x^{-2};q) \end{pmatrix}
\;\overset{\omega^2}{\longmapsto}\; 
\begin{pmatrix} B^1_\III \\ B^2_\III \end{pmatrix} \sim \begin{pmatrix} \CJ(x^{-2},x^{-1}y^{-1};q) \\ \CJ(x^{-1}y,xy;q) \end{pmatrix}\,.
\ee
This is in beautiful agreement with our calculations in \eqref{BIIfinal}--\eqref{BIIIfinal} above, which used a Stokes analysis and special properties of the $q$-Bessel function $\CJ$. The agreement, of course, is up to a possible relabeling of vacua $1\leftrightarrow 2$, and up to a ratio of theta-functions that encode the extra Chern-Simons contact terms discussed in Section \ref{sec:cp1param}.

There are block integrals for theories $T_\II$ and $T_\III$ that produce expressions \eqref{BIIfinal}--\eqref{BIIIfinal} directly. In deriving them, it is important to be careful about Chern-Simons contact terms. These can be fixed either by matching line-operator identities in candidate mirror theories, or by remembering that the mirror symmetry $T_\I\simeq T_\II \simeq T_\III$ is generated by repeated applications of the simple $T_\Delta\simeq T_\Delta'$ mirror symmetry (\cf\ Footnote \ref{foot:ST}), which was analyzed in Section \ref{sec:simple-ex}. The result for the block integrals is
\begin{subequations}
\begin{align}
\mathbb{B}_\II(x,y;q) \,\,&=\, \int_{*} \frac{ds}{s} \frac{\theta(-q^{-\frac12}y)\theta(-q^{\frac12}x^2)}{\theta(x)\theta(-q^{-\frac12}x^{-1}y)\theta(-q^{-\frac12}sx^{-2})}(qs^{-1};q)_\infty(qs^{-1}xy^{-1};q)_\infty\,, \\
\mathbb{B}_\III(x,y;q) \,\,&=\, \int_{*} \frac{ds}{s} \frac{\theta(-q^{-\frac12}y)\theta(-q^{-\frac12}x^2)}{\theta(x)\theta(-q^{-\frac12}xy)\theta(-q^{-\frac12}sx^2)}(qs^{-1};q)_\infty(q(sxy)^{-1};q)_\infty \,.
\end{align}
\end{subequations}
In chambers $\II$ and $\III$, where the theories $T_\II$ and $T_\III$ become semi-classical, the respective blocks integrals at $|q|>1$ have two well-separated half-lines of poles with critical-point cycles analogous to those in Figure \ref{fig:CPcycles}. Summing up each half-line of residues yields (respectively) the blocks $B^\ra_\II$ in \eqref{BIIfinal} and $B^\ra_\III$ in \eqref{BIIIfinal}.

\subsection{Fusion}
\label{sec:cpfusion}

Now that we have defined holomorphic blocks for the $\cp^1$ sigma-model, let us use them to check our main conjecture: that the ellipsoid partition function and sphere index both decompose into a sum of products of blocks. Exhibiting these decompositions in different mirror frames, in terms of blocks in different Stokes chambers, provides a new way to prove identities among the integral expressions that give ellipsoid partition functions and indices.

The ellipsoid partition function of $T_\I$ is normally expressed in terms of $b^2=\hbar/(2\pi i)$ and complex masses $\mu_x=X/(2\pi b)$, $\mu_y=Y/(2\pi b)$. We also set $\sigma=S/(2\pi b)$. Using the rules described in \cite{HHL}, we find
\begin{align}
 \CZ_b^\I(\mu_x,\mu_y;b) &= \int_\R d\sigma\, e^{-2\pi i\mu_y\sigma} s_b\big(\tfrac{i}2(b+b^{-1})-\mu_x-\sigma\big)s_b\big(\tfrac{i}2(b+b^{-1})+\mu_x-\sigma\big) \\ &= -\frac{2\pi i}{\hbar} C^{-2}\int_\R \frac{dS}{2\pi b} \,\big|\!\big|\Upsilon_\I(x,s;q)\big|\!\big|_S^2\,, \notag
\end{align}
with $C=\exp\big[-\frac1{24}(\hbar+\wt\hbar)\big]$ and $s_b$ and defined in Section \ref{sec:chiralblock}, and with $\Upsilon_\I$ the block integrand from \eqref{CP1int}. The integrand is now a (non-periodic) meromorphic function of $S=2\pi b\sigma\in \C$, with poles at $S = \pm X -\hbar m-2\pi in$, or $\sigma=\pm \mu_x-ibm-ib^{-1}n$, for $n,m\in \Z_{\geq 0}$. The integral can be evaluated by closing the contour in the lower half-plane and summing up  the residues \cite{Pasquetti-fact}. The result is
\begin{align}
\CZ_b^\I(\mu_x,&\mu_y;b) =\notag\\
=&i^{\frac32}C^{-1}e^{\frac{X^2}{\hbar}}\Big( e^{-\frac{1}{\hbar}X(Y-i\pi-\frac\hbar2)}\big|\!\big|\CJ(xy^{-1},x^2;q)\big|\!\big|_S^2+e^{\frac{1}{\hbar}X(Y-i\pi-\frac\hbar2)}\big|\!\big|\CJ(x^{-1}y^{-1},x^{-2};q)\big|\!\big|_S^2\Big) \notag \\
= &i^{\frac32}C^{-3}\big(\big|\!\big|B_\I^1(x,y;q)\big|\!\big|_S^2+\big|\!\big|B_\I^2(x,y;q)\big|\!\big|_S^2\big)\,. \label{cpb1}
\end{align}

Of course, we could also have computed the ellipsoid partition function in a different mirror frame, and we expect to find the same answer. For theories $T_\II$ and $T_\III$, a summation of residues analogous to the one described above produces
\begin{align} 
\CZ_b^\II(\mu_x,\mu_y;b) &= \int_\R d\sigma\, E(\mu_x,\mu_y,\sigma;b)\, s_b\big(\tfrac{i}2(b+b^{-1})-\sigma\big)s_b\big(\tfrac{i}2(b+b^{-1})+\mu_x-\mu_y-\sigma\big) \notag \\
&=i^{\frac32}C^{-3}\big(\big|\!\big|B_\II^1(x,y;q)\big|\!\big|_S^2+\big|\!\big|B_\II^2(x,y;q)\big|\!\big|_S^2\big)\,, \label{cpb2}
\end{align}
\begin{align}
\CZ_b^\III(\mu_x,\mu_y;b) &= \int_\R d\sigma\, E(-\mu_x,\mu_y,\sigma;b)\, s_b\big(\tfrac{i}2(b+b^{-1})-\sigma\big)s_b\big(\tfrac{i}2(b+b^{-1})-\mu_x-\mu_y-\sigma\big) \notag  \\
&=i^{\frac32}C^{-3}\big(\big|\!\big|B_\III^1(x,y;q)\big|\!\big|_S^2+\big|\!\big|B_\III^2(x,y;q)\big|\!\big|_S^2\big)\,, \label{cpb3}
\end{align}
with
\be
E(\mu_x,\mu_y,\sigma;b)=\exp\big[i\pi\big(-\frac32\mu_x^2+\mu_x\mu_y+\frac12\mu_y^2+3\mu_x+\mu_y-\frac{i}{2}(b+b^{-1})(\mu_x+\mu_y)\big)\big]~.
\ee
The blocks appearing here are manifestly the expressions given by  \eqref{BIIfinal}--\eqref{BIIIfinal}. The equivalence of the right-hand sides of \eqref{cpb1}, \eqref{cpb2}, and \eqref{cpb3} then follows immediately from the nontrivial linear identities derived in Section \ref{sec:CP1match}, which show that blocks in different chambers are related by pairs of conjugate Stokes matrices.
For example, when $|q|<1$,
\begin{align} &\big|\!\big|B_\II^1(x,y;q)\big|\!\big|_S^2+\big|\!\big|B_\II^2(x,y;q)\big|\!\big|_S^2  \\ &\hspace{1in}= B_\I^1(x,y;q)\big(B_\I^1(\wt x,\wt y;\wt q)+B_\I^2(\wt x,\wt y;\wt q)\big) + \big(B_\I^2(x,y;q)-B_\I^1(x,y;q)\big)B_\I^2(\wt x,\wt y;\wt q) \notag \\ &\hspace{1in}=\big|\!\big|B_\I^1(x,y;q)\big|\!\big|_S^2+\big|\!\big|B_\I^2(x,y;q)\big|\!\big|_S^2\,. \notag \end{align}
On the other hand, the equivalence of the left-hand sides was studied in \cite{SpirVar-knots} (\cf\ Appendix C therein) using identities for elliptic hypergeometric functions. It appears that Stokes phenomena for holomorphic blocks provides another way to understand such identities.

As for the sphere index of $T_\I$, we follow the formalism of \cite{IY-index, KW-index, DGG-index} to write 
\begin{align} 
\CI(m_x,\zeta_x,m_y,&\zeta_y;q) =\notag\\
&\sum_{n\in \Z}\oint \frac{d\sigma}{2\pi i\sigma} (-q^{-\frac12})^n\zeta_x^{m_x}\zeta_y^n\sigma^{m_y+n}\CI_\Delta(m_x+n,\zeta_x\sigma;q)\CI_\Delta(-m_x+n,\zeta_x^{-1}\sigma;q) \notag\\
=&\sum_{n\in \Z}\oint \frac{d\sigma}{2\pi i\sigma} \big|\!\big| \Upsilon_\I(x,y,s;q) \big|\!\big|_{id}^2\,, \label{cpindex}
\end{align}
where we now identify $x=q^{\frac{m_x}{2}}\zeta_x$, $y=q^{\frac{m_y}{2}}\zeta_y$, $s=q^{\frac n2}\sigma$, and $\wt x=q^{\frac{m_x}{2}}\zeta_x^{-1}$, $\wt y=q^{\frac{m_y}{2}}\zeta_y^{-1}$, $\wt s=q^{\frac n2}\sigma^{-1}$, and we have expressed the integrand in terms of the free-chiral index
$
\CI_\Delta(m,\zeta;q)=\prod_{r=0}^\infty (1-q^{1-\frac m2+r}\zeta^{-1})/(1-q^{-\frac m2+r}\zeta) = \big|\!\big|B_\Delta(x;q)\big|\!\big|_{id}^2~.\notag
$
The integration in $\sigma$ is over the unit circle, and it is assumed that $|q|<1$. For each fixed monopole number $n$, the integrand of \eqref{cpindex} has two families of poles lying on or outside the unit circle. The contour can be deformed to infinity, picking up contributions from all of these poles. It is straightforward to sum up the residues in each family and see that
\be  \CI(m_x,\zeta_x,m_y,\zeta_y;q) = \big|\!\big|B_\I^1(x,y;q)\big|\!\big|_{id}^2+\big|\!\big|B_\I^2(x,y;q)\big|\!\big|_{id}^2\,. \ee
Just as for the ellipsoid partition function, we can express the index in different mirror frames in order to obtain identities among integrals of the form \eqref{cpindex}. Such identities were explored in \cite{KW-index,KSV-index}, and again we observe that blocks provide an interesting window into such relations.

\subsection{Equivariant K-theory, surface operators, and topological strings}
\label{sec:cpmore}

To conclude, we would like to touch upon several other interpretations of the holomorphic blocks for $T_\I$. These include the connections to vortices and topological strings described in Section \ref{sec:OV}. All of the following interpretations were discussed in \cite{DGH} (see also \cite{KPW}).

Let us first consider the dimensional reduction of $T_\I$ to two dimensions. We obtain a classic example of an $\CN=(2,2)$ gauged linear sigma model \cite{Witten-phases}, which in the infrared is described by a nonlinear sigma model with target space $\cp^1$. The FI parameter $Y$, which in two dimensions is complexified by a theta angle, still parameterizes the size of the $\cp^1$ (it is the complexified K\"ahler class). The complex twisted mass $X$ makes the sigma-model classically massive, with two vacua where the bosonic fields are fixed at the north or south poles of $\cp^1$, respectively. For non-zero twisted mass, a map from the worldsheet $\R^2$ to $\cp^1$ will have finite energy only if the asymptotic boundary of $\R^2$ is mapped either to the north or south pole. This effectively compactifies the worldsheet of the IR sigma model: we find a theory of maps from a $\cp^1$ worldsheet to a $\cp^1$ target, with a fixed basepoint. Upon implementing an A-type twist, the theory localizes to \emph{holomorphic} maps of this kind.

The holomorphic blocks of $T_\I$ compute what is known as the equivariant J-function of the moduli space of holomorphic maps from $\cp^1$ to $\cp^1$ \cite{GiventalLee}. Let us recall for convenience that the two blocks associated to vacua in the semi-classical chamber $(\I)$ take the form
\be \label{Jfunction}
 B_\I^{1,2}(x,y;q) \,\sim\, \CJ(x^{\pm 1}y^{-1},x^{\pm 2};q) \,\sim\, \sum_{n=0}^\infty \frac{x^n}{(q^{-1})_n(qx^{\pm 2};q)_n}\,y^{-n}\,, \ee
up to simple theta-function or $(*;q)_\infty$ prefactors. The $n$-th term in the sum is an equivariant K-theoretic character of the moduli space of maps of degree $n$, where $x$ is an equivariant weight for $U(1)$ rotations of the target $\cp^1$ (which is how the vector $U(1)_V$ symmetry acts in the physical theory), and $q$ is an equivariant weight for $U(1)$ rotations of the base. The two choices of vacua correspond to the choice of basepoint for the maps --- whether we map the point at $\infty$ on the base to the north or south poles on the target.

We can also connect the blocks in chamber $\I$ to equivariant vortex counting for the 2d $\CN=(2,2)$ gauged linear sigma model. We take the special limit described in Section \ref{sec:OV}, scaling
\be q = e^{\beta \epsilon}\,,\qquad x=e^{\beta m}\,,\qquad y=-\frac{1}{\beta^{2}y_{\rm FI}}\,;\qquad \beta\to 0\,,\ee
with $y_{\rm FI}$ being the more standard exponentiated FI or vortex-counting parameter in 2d. We have also absorbed factors of $2\pi$ in $\beta$ and $i$ in $\epsilon$ (relative to the rest of this paper). In this limit, the blocks become
\be \label{cp1vortex} B^{1,2}_\I(x,y;q) \,\leadsto\, \sum_{n=0}^\infty \frac{1}{n!\epsilon^n\prod_{j=1}^n (\pm m+j\epsilon)}\,y_{\rm FI}^n\,, \ee
which was the 2d vortex partition function found in \cite{DGH} (see also \cite{Shadchin-2d}).

The reduced 2d theory is also the effective field theory on the simplest possible half-BPS surface operator in 4d pure $SU(2)$ $\CN=2$ super-Yang-Mills theory \cite{Ramified, AGGTV, DGH}. The surface operator can be defined either in the UV (as a gauged linear sigma model) or in the IR (as a $\cp^1$ sigma model). For an appropriate choice of Chern-Simons levels, the $U(1)_V$ symmetry that rotates the $\cp^1$ is enhanced to $SU(2)_V$ and gauged in coupling to the bulk. The vortex partition function \eqref{cp1vortex} can be obtained by placing the coupled 2d-4d system in an Omega background, computing the partition function, and sending the four-dimensional gauge coupling (or rather, the QCD scale) to zero to decouple the bulk physics.

\begin{wrapfigure}{r}{2.3in}
\includegraphics[width=2.2in]{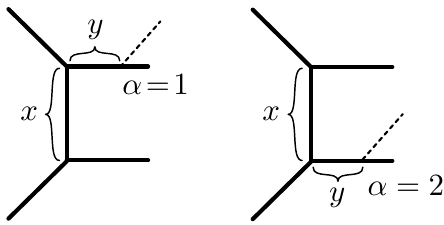}
\caption{Toric geometry for $T_\I$.}
\label{fig:cp1toric}
\end{wrapfigure}

Finally, $T_\I$ can be engineered on an M5 brane that wraps a (toric) Lagrangian cycle in the noncompact Calabi-Yau geometry $\cO(-2)\oplus\cO(0)\to\P^1$. One way to see this connection is to start from a D2-D4-NS5 brane construction of the surface operator described above \cite{HananyHori}, and to perform a chain of string/M-theory dualities. The toric diagram for this setup is shown in Figure \ref{fig:cp1toric}. There are two possible placements for the toric brane, corresponding to the vacua of $T_\I$. The (exponentiated) closed-string modulus which controls the size of the $\P^1$ base corresponds to the mass $x$; while the (exponentiated) open-string modulus corresponds to the FI parameter $y$. The open-string partition functions can readily be computed using the topological vertex \cite{AKMV}, and the answers agree with the blocks \eqref{Jfunction}, again up to theta-functions that encode background Chern-Simons levels. This perspective on holomorphic blocks of $T_\I$ was explored in \cite{Pasquetti-fact}.


\section{Blocks in Chern-Simons theory}
\label{sec:CS}

In this final section, we relate holomorphic blocks to path integrals in analytically continued Chern-Simons theory. For the three-manifold theories to which this relation applies, the dictionary between the two descriptions will shed light on some of the more subtle aspects of holomorphic blocks that we encountered in the previous sections. Moreover, the computability of holomorphic blocks allows us to compute non-perturbative Chern-Simons path integrals that are otherwise out of reach. 

The basic connection between the two comes from M-theory. Namely, consider $K$ M5 branes wrapping a cycle
\be\label{Mgeometry}
M\times D^2\times_q\,S^1\;\;\subset\;\; T^*M \times TN \times_q S^1\,,
\ee
where $D^2$ is a cigar sitting in Taub-NUT space which is locally of the form $TN\simeq T^*D^2$. We let the geometry be a fibration, with $D^2\subset TN$ rotating by an amount $q$ when translated around $S^1$. Then the reduction of the M5-brane theory along $M$ produces an effective three-dimensional $\CN=2$ theory $T_M^{(K)}$ on $\DqS$, as in \cite{DGG}; while the reduction along $\DqS$ produces an analytically continued $SU(K)$ Chern-Simons theory on $M$ \cite{DGH, Wfiveknots}. (More precisely, reduction on the compact torus fiber of $\DqS$ leads to twisted $\CN=4$ Yang-Mills theory on $M\times \R_+$, which provides an analytic continuation of Chern-Simons theory. We will come back to this important detail later.) The renormalized and complexified Chern-Simons level $k$ is related to the twisting of the cigar by
\be  q=\exp (2\pi i\beta\epsilon)=:\exp \frac{2\pi i}{k}\,. \ee

The partition function of $T_M^{(K)}$ on $\DqS$ should be equal to a partition function of Chern-Simons theory on $M$. Both cases require additional input. For $T_M^{(K)}$, we know that we must specify a vacuum $\alpha$ at the infinite end of the cigar, leading to the block $B^\ra_{T_M^{(K)}}$. In Chern-Simons theory we have to choose a convergent integration cycle for the analytically continued path integral --- a middle-dimensional cycle in the space of  $SL(K,\C)$ connections on $M$. A basis of cycles $\{\Gamma_{\rm CS}^\ra\}$ is labelled by \emph{flat} $SL(K,\C)$ connections $\CA^\alpha$ on $M$, or critical points of the Chern-Simons functional \cite{gukov-2003, Wit-anal}. We then expect that
\be  B^\ra_{T_M^{(K)}} =\, \CZ^\ra_{\rm CS}[M]\,, \label{BCS} \ee
where $\CZ^\ra_{\rm CS}$ is the Chern-Simons path integral on $\Gamma^\ra_{\rm CS}$.

There is an important subtlety in identifying the blocks defined in this paper with a Chern-Simons path integral on a cycle defined by the Lefschetz thimbles of \cite{Wit-anal}. Namely, the basis of such Lefschetz thimbles in Chern-Simons theory is \emph{infinite} due to the non-invariance of the analytically continued Chern-Simons functional under large gauge transformations. In general, it was argued in \cite{Wit-anal} that in identifying an integration cycle with a flat connection, one must further specify a lift of the flat connection to the universal cover of the space of connections modulo topologically trivial gauge transformations.

This situation is not dissimilar to what we have encountered in Section \ref{sec:SQM}, where the naive action of the supersymmetric quantum mechanics was multivalued, and required a formulation on an infinite-sheeted cover of the scalar manifold. However, we have seen that three-dimensional gauge theory dictates that we interpret all lifts of a vacuum as images of that vacuum under large gauge transformations which are good symmetries of the theory; hence we are able to compute blocks in terms of exponentiated (single-valued) variables. It is evident that as defined in this paper, holomorphic blocks are related to the four- or six-dimensional realizations of Chern-Simons path integrals with a boundary condition that includes contributions from \emph{all} images of a single flat connection. 

A second indication of this complication is that the individual integration cycles of \cite{Wit-anal} associated to a single image of a flat connection are defined in such a way as to be analytic across $|q|=1$. In fact, they may be adjusted so as to reproduce an $SU(N)$ Chern-Simons partition function for $q$ a root of unity. Holomorphic blocks, on the other hand, are always ill-defined at $|q|=1$, and cannot be analytically continued across the unit circle. This is not so surprising when holomorphic blocks are understood as a (perhaps weighted) sum over images of a single flat connection. In fact, in this sense, holomorphic blocks may be more similar to the $q$-deformation of two-dimensional Yang-Mills theory than to Chern-Simons theory with a compact gauge group, which has the same trouble at $|q|=1$.

In spite of these complications, the general considerations of the first part of this section will apply independent of this detail, and the experimental tests at the end of the section will prove robust enough to evade any of these subtleties.

The relationship between blocks and Chern-Simons path integrals should be very general, applying to any kind of three-manifold $M$, and also to gauge groups besides $SU(K)$. However, it can be studied most concretely if we take $M$ to have non-empty boundary and to admit a (topological) ideal triangulation, \cf\ \cite[Sec 2]{DGG}. For example, we can take $M$ to be a knot complement. From a 6d perspective, the knot is realized as a codimension-two defect; but on $M$ the defect can be regularized to the ideal boundary of a knot complement. Then at low energies the theory $T_M^{(K)}$ (or a certain subsector thereof) is effectively described as one of the ``class-$\CR$'' theories constructed in \cite{DGG}.%
\footnote{In \cite{DGG}, only the case $K=2$ was treated, but it will be shown in upcoming work that general $T_M^{(K)}$ is also in class $\CR$.}
Strikingly, this means that $\CZ^\ra_{\rm CS}[M] = B^\ra_{T_M^{(K)}}$ can be expressed as a finite-dimensional block integral.

In terms of Chern-Simons theory, the reduction of an infinite-dimensional path integral to a finite-dimensional block integral arises by virtue of the triangulation of $M$. Each $B_\Delta(z;q)$ factor in the block integral \eqref{BI} should be interpreted as the Chern-Simons wavefunction on a tetrahedron $\Delta$, obtained by doing an infinite-dimensional path integral with fixed boundary conditions on $\pd \Delta$ parametrized by $z$. Then only finitely many degrees of freedom remain to  be integrated out to glue the tetrahedra together. This type of ``state sum'' or ``state integral'' model has been used fruitfully in the past to construct other kinds of Chern-Simons wavefunctions, \eg\ \cite{TuraevViro, Kashaev-invt}. 

The relation between holomorphic blocks and Chern-Simons path integrals gives an explicit realization of the Stokes phenomena in Chern-Simons theory described by \cite{Wit-anal}. In particular, integration cycles for Chern-Simons path integrals and contours for block integrals are expected to have the same Stokes jumps, so that the latter can be used to understand the former. Another major (potential) advantage of \eqref{BCS} is that it relates categorification in Chern-Simons theory, as discussed in \cite[Sec. 6]{Wfiveknots}, to BPS counting in a comparatively simple three-dimensional gauge theory. One would hope in the future to find ``refined'' versions of blocks $B^\ra_{T_M}\sim \Tr_{\CH(\R^2;\alpha)}(-y)^{2J}q^{\tfrac R2-J}x^e$ that calculate Poincar\'e polynomials for a set of knot homologies labelled by flat connections $\alpha$ on a knot complement.

It is natural to ask if gluing blocks to form ellipsoid partition functions $\CZ_b[\CT]$ or $\SqS$ indices $\CI[\CT]$ for $T_M^{(K)}$ has an analogue in Chern-Simons theory. To some extent, the answer is already known from \cite{DGG,DGG-index}: $\CI[\CT]$ is a physical, non-analytically-continued $SL(K,\C)$ Chern-Simons partition function (with holomorphic and anti-holomorphic contributions); while $\CZ_b[\CT]$ matches the ``state integral'' for Chern-Simons theory defined in \cite{hikami-2006, DGLZ, KashAnd} and reformulated in \cite{Dimofte-QRS}. The latter is an analytically continued $SL(K,\R)$ Chern-Simons partition function on $M$ with special, Teichm\"uller-like boundary conditions at $\pd M$ (\cf\ \cite{Verlinde-TeichLiouv}). Both of these correspondences arise indirectly through comparison of defining properties of the relevant partition functions.

Below, we review some aspects of analytic continuation in Chern-Simons theory and the expected relations to three-dimensional theories arising from six-dimensional constructions. The main goal is to obtain a complete dictionary of parameters on the two sides. Otherwise, there are already long and beautiful expositions of these subjects in the literature \cite{gukov-2003, DGLZ, Wit-anal, Witten-path, Wfiveknots}.

We then consider some explicit examples that test \eqref{BCS} in the case of $K=2$. The experimentally minded reader may wish to skip directly to these examples. We compute the blocks for trefoil ($\mb{3_1}$), figure-eight ($\mb{4_1}$), and $\mb{5_2}$ knot complement theories, and check the expected correspondence between blocks and flat connections by comparing the leading asymptotics of blocks to volumes of flat connections. Finally, we test a new, conjectural, non-perturbative relation between blocks of a knot-complement theory and the so-called ``stabilization limit'' of the colored Jones polynomial for the knot.


\subsection{Chern-Simons theory and analytic continuation}
\label{sec:CSanal}

We begin our review of analytic continuation in Chern-Simons theory and its realization via $\CN=4$ super-Yang-Mills theory on a four-dimensional half-space. We collect results of \cite{gukov-2003, DGLZ, Wit-anal, Witten-path, Wfiveknots} (see also \cite{VCReview}), where this subject is discussed in much greater detail.

For concreteness, we consider Chern-Simons theory with real, compact gauge group $SU(2)$, which has complexification $\SLC$. We also focus on simple three-manifolds with the topology of $S^3$ and an embedded knot $\CK\subset S^3$. The knot is given an orientation and a framing, which is a choice of section for the unit-normal bundle of $\CK$ in $S^3$, or equivalently a prescription for how to deform the knot to a parallel copy of itself. We will always choose the canonical framing in which a knot and its parallel in $S^3$ have zero linking number.

There are two equivalent ways to associate compact $SU(2)$ partition functions to the pair $(\CK,S^3)$. First, one can perform the Chern-Simons path integral over $SU(2)$ connections $A$ on $S^3$, with the insertion of a Wilson-loop operator $W_{\CK,N}(A)$ in the $N$-dimensional irreducible representation of $SU(2)$: 
\be \CZ_{\rm CS}(S^3,\CK;k,N) = \int \CD A\, e^{i\frac{\bar k}{4\pi}I_{\rm CS}(S^3;A)} \,W_{\CK,N}(A)\,, \label{CSdim1} \ee
where $I_{\rm CS}(M;A)$ is the Chern-Simons functional. The answer depends on integers $k,N$, where $k=\bar k+2\,{\rm sgn}(\bar k)$ is the renormalized level of the theory. Alternatively, one can excise the knot, forming the knot complement $M=S^3\bs \CK$, and perform the path integral with a specified singularity at the knot,
\be \hspace{.5in} A \,\sim\, \frac{i\pi N}{k}\begin{pmatrix} 1 & 0 \\ 0 & -1\end{pmatrix} d\theta\text{+ regular}\qquad \text{(modulo conjugation)}\,, \ee
where $\theta$ is an angular coordinate in the plane perpendicular to $\CK$ at a given point. This means that connections have a monodromy with eigenvalues $x^{\pm1/2}$ or $-x^{\pm1/2}$,
\be x\,:=\, e^{\tfrac{2\pi i N}{k}}\,, \ee
around an infinitesimally small loop linking the knot. Then
\be \CZ_{\rm CS}(M=S^3\bs \CK;k,x) = \int_{\text{fixed $x$}} \hspace{-.3in} \CD A\,e^{i\frac{\bar k}{4\pi}I_{\rm CS}(M;A)}\,. \label{CScodim2} \ee
The equivalence of \eqref{CSdim1} and \eqref{CScodim2} is discussed in \cite{EMSS, Witten-IAS, Beasley-WL}. Using either definition, the path integral is a polynomial in $q^\frac12 = \exp\frac{i\pi}{k}$ for every fixed $N$,
\be \CZ_{\rm CS}(S^3,\CK;k,N) \,=\, \CZ_{\rm CS}(M=S^3\bs \CK;k,x) \,\simeq\, J_N(\CK;q) \,=:\, J(x;q)\,, \ee
up to an overall normalization by the empty $S^3$ partition function $\CZ_{\rm CS}(S^3;k)$.
Sometimes it makes sense to think of these partition functions as functions of two variables $x=q^{N}$ and $q$, as indicated by the notation $J(x;q)$. The set of polynomials $J_N(\CK;q)$ coincides%
\footnote{In the mathematics literature, the polynomials $J_N(\CK;k)$ are usually normalized by dividing by the polynomial $J_N(U,k) = (q^{\frac N2}-q^{-\frac N2})/(q^{\frac 12}-q^{-\frac 12})=(x^{\frac12}-x^{-\frac12})/(q^{\frac 12}-q^{-\frac 12})$ of the unknot $U\subset S^3$. This normalization is a little unnatural physically, and we will not use it.} %
with the ``colored Jones polynomials'' of $\CK$ \cite{Jones, Turaev-YB, Wit-Jones}.

The basic idea of analytic continuation is to consider $k$ and $N$, or $x$ and $q$, to be complex numbers, and to  promote $J(x;q)$ to a locally analytic function. While there is no unique way to do this, a natural prescription is suggested by physical path integrals. The two realizations \eqref{CSdim1} (Wilson lines) and \eqref{CScodim2} (monodromy defects) of $J(x;q)$ lead to slightly different continuations, and it is the latter that interests us, because it is ultimately related to the knot-complement theories $T_M$ of \cite{DGG}.

In terms of the path integral \eqref{CScodim2}, analytic continuation arises from formulating the integral over real $SU(2)$ connections $A$ as an integral over a real, middle-dimensional cycle $\Gamma_\R$ in the space of complexified $\SLC$ connections $\CA$. Just as in finite-dimensional complex analysis, the integral over $\Gamma_\R$ may be dominated by contributions from several \emph{complex} critical points of the (now holomorphic) functional $I_{\rm CS}(\CA)$. These critical points are flat $\SLC$ connections $\CA^\ra$ on the knot complement $M=S^3\bs \CK$, with fixed holonomy eigenvalues $\pm x^{\pm \frac12}$ around a small loop linking the knot.%
\footnote{\label{foot:lift}As was stated in the introduction to this section, the actual critical points argued by \cite{Wit-anal} to be relevant when $k,N\notin \Z$ are not just flat connections $\alpha$ in the conventional sense (counted modulo full $SL(2,\C)$ gauge transformations), but flat connections modulo gauge transformations continuously connected to the identity. For a knot complement in $S^3$, every standard flat connection gives rise to a family of $\Z\times \Z$ critical points. The contributions of different critical points in the same family to a sum such as \eqref{Jsum} simply differ by $\exp(2\pi iak+2\pi ib N)$, for integers $a,b\in \Z$. Then, practically speaking, one still expects a formula of the form \eqref{Jsum} to hold with $\alpha$ labeling standard flat connections, so long as the coefficients $n_\alpha$ are allowed to contain sums of factors $\exp(2\pi iak+2\pi ib N)$ when $x$ and $q$ are not roots of unity. As was also discussed in the introduction, the label $\alpha$ for holomorphic blocks is a \emph{standard} flat connection, so that is what we focus on here.} %
Non-perturbatively, one can actually decompose $\Gamma_\R$ into a basis of convergent integration cycles $\Gamma_\R=\sum_\alpha n_\alpha \Gamma_{\rm CS}^\ra$, with each $\Gamma_{\rm CS}^\ra$ defined by downward gradient flow from a critical point $\CA^\ra$ with respect to the real part of $\frac{i\bar k}{4\pi}I_{\rm CS}(\CA)$. One can then try to expand the original partition function as
\be J(x;q) = \sum_{\alpha} n_\alpha \CZ_{\rm CS}^\ra(x;q)\,, \label{Jsum}\ee
where each $\CZ_{\rm CS}^\ra(x;q)$ is the complexified path integral over a fixed integration cycle $\Gamma_{\rm CS}^\ra$. In this context, it makes sense to promote $x$ and $q$ to generic complex variables, and to define the $\CZ^\ra_{\rm CS}(x;q)$ as locally analytic functions. As $x$ and $q$ are varied in the complex plane, the $\CZ^\ra_{\rm CS}(x;q)$ may interact with each other via Stokes phenomena.

The number of flat connections with fixed boundary conditions on a knot complement $M=S^3\bs \CK$ is usually finite. Moreover, the flat connections can be characterized as solutions%
\footnote{More precisely, the A-polynomial parameterizes flat connections on the torus boundary of a knot complement that can be extended to flat connections in the bulk. Thus every flat connection on $M$ maps to a solution to \eqref{Apoly}, and typically the correspondence is one-to-one, though precisely when this is true is not known.} %
at fixed $x$ of a two-variable polynomial
\be A_M(x,p) = 0\,,\qquad x,p\in \C^*\,, \label{Apoly} \ee
called the A-polynomial of the knot \cite{cooper-1994}. This equation relates the square of the holonomy eigenvalue $x$ of a flat connection around a small loop linking $\CK$ with the holonomy eigenvalue $-p$ around a parallel copy
of the knot in $M$ (where the parallel copy is chosen according to the knot's framing). The notation here is chosen to match that of three-dimensional gauge theories $T_M$; in the knot theory literature, the holonomy variables are usually called $m$ and $\ell$ (for ``meridian'' and ``longitude''), with
\be x\,\leftrightarrow\, m^2\,,\qquad p\,\leftrightarrow\, -\ell\,.\ee
The A-polynomial always takes the form $A_M(x,p)=(p+1)A_M^{\rm irr}(x,p)$, with a canonical component $(p+1)$ corresponding to ``abelian'' flat connections that take values in a maximal torus of $\SLC$ (\ie\ that are reducible).

Quantum mechanically, the partition functions $\CZ_{\rm CS}^\ra(x;q)$ should be solutions to an equation
\be \hat A_M(\hat x,\hat p;q)\, \CZ_{\rm CS}^\ra(x;q) = 0\,, \label{Ahat} \ee
where $\hat A_M(\hat x,\hat p;q)$ is a polynomial in operators $\hat x,\hat p$ and $q$ (with $\hat p\hat x=q\,\hat x \hat p$), that reduces to $A_M(x,p)$ in the classical limit $q\to 1$. This is the quantum version of the classical constraint \eqref{Apoly}. In fact, more than this has been conjectured to be true. Namely, it is expected that there exists a quantization $\hat A_M^{\rm irr}(\hat x,\hat p;q)$ of just the irreducible A-polynomial $A_M^{\rm irr}(x,p)$ that annihilates $\CZ^\ra_{\rm CS}(x;q)$ for all flat connections $\CA^\ra$ \emph{except} the abelian one,
\be \hat A_M^{\rm irr}(\hat x,\hat p;q)\, \CZ_{\rm CS}^\ra(x;q) = 0\,,\qquad \alpha\neq {\rm abelian}\,, \label{Airr} \ee
\cf\ \cite{garoufalidis-2004, Dimofte-QRS}. The irreducible A-polynomial has been systematically quantized in \cite{Dimofte-QRS}. This suggests that there exists a consistent truncation of analytically continued Chern-Simons theory to a sector containing just irreducible flat connections. This conjecture is strengthened by the analysis of \cite{Wit-anal}, which demonstrates that the abelian integral can never contribute to other $\CZ_{\rm CS}^\ra(x;q)$ via a Stokes phenomenon.

In the classical limit $k=\frac{2\pi i}{\hbar}\to\infty$, any term $\CZ_{\rm CS}^\ra(x;q)$ is dominated by the critical point that defines it, so
\be \label{volCS}
\CZ_{\rm CS}^\ra \overset{q\to 1}{\sim} \exp \frac1\hbar {\CV}(\CA^\ra)~,\ee
where the ``volume'' $\CV(\CA^\ra)$ of a flat connection is defined by evaluating the classical, holomorphic Chern-Simons functional $-\frac{1}{2}I_{\rm CS}(\CA^\alpha)$. For a flat connection corresponding to a hyperbolic metric on the knot complement $M$ (with a deformation of the cusp at $\CK$ parametrized by $x$), the hyperbolic volume ${\rm Vol}(M;x)$ and Chern-Simons invariant ${\rm CS}(M;x)$ are related to the holomorphic volume by
\be \label{vols}
{\CV}(\CA^\ra;x) = i({\rm Vol}(M;x)+i{\rm CS}(M;x))+i\log|p^\alpha|\arg x+i\pi \log x\,,
\ee
where $p^\alpha(x)$ is the associated solution to the classical A-polynomial $A(p^\alpha,x)=0$. Note that only $\CV(\CA^\ra;x)$ is (locally) a holomorphic function of $x$. On the other hand, the quantity ${\rm Vol}(M;x)+i{\rm CS}(M;x)$ is non-holomorphic but globally well defined up to integer multiples of $4\pi^2$. The LHS has more severe global ambiguities due to branch cuts, ultimately related to the extra factors discussed in Footnote \ref{foot:lift}.


\subsection{To six dimensions and back}
\label{sec:6d}

We now describe the chain of relations that connect Chern-Simons theory in three dimensions to the $(2,0)$ theory in six dimensions, following \cite{Wfiveknots}. We alternate between the purely field theoretic perspective and brane constructions. Our goal is simply to understand the map of parameters between blocks (and their fused products) and analytically continued Chern-Simons theory, though this requires some technology.

\subsubsection*{Chern-Simons from four dimensions}

The first step is a lift to four dimensions, extending Chern-Simons theory on $M$ to $\CN=4$ SYM with the same (compact) gauge group, on a half-space $V=M\times \R_+$. The lift provides a natural and physically meaningful analytic continuation of Chern-Simons theory. The four-dimensional theory must be topologically twisted in order to preserve some supersymmetry on $V$. The correct choice for this application is the geometric Langlands (GL) twist of \cite{Kapustin-Witten}, which breaks the R-symmetry group $SO(6)_R\to SO(3)_R\times SO(3)_R$ and replaces the Lorentz group $SO(3)_E$ of $M$ with the diagonal embedding in $SO(3)_E\times SO(3)_R$. This twist complexifies the gauge connection $A_\mu$ on $M$ with three scalars $\phi_\mu$ that now transform as a one-form, 
\be \label{Acx}
A_\mu\;\to\; \CA_\mu = A_\mu+w\,\phi_\mu\,,
\ee
for some $w$ with $\Im\,w\neq 0$. The twisted theory in the bulk of $V=M\times \R_+$ then localizes to field configurations that obey a flow equation in the ``time'' coordinate on $\R_+$. Namely, all fields except $\CA_\mu$ can be taken to vanish, and this complex connection along $M$ obeys gradient flow with respect to the holomorphic Chern-Simons functional
\be \label{gradCS}
\frac{d}{ds} \CA_\mu = -*_M\frac\delta{\delta \overline{\CA}_\mu}\left[ \frac{i\Psi^{\!\vee}}{4\pi} I_{\rm CS}(M;\CA)\right]\,,
\ee
where $s\in [0,\infty)$ is the coordinate on $\R_+$. This is the same gradient flow in the space of complex connections on $M$ that defined Chern-Simons integration cycles, provided that the parameter $\Psi^{\!\vee}$ appearing here equals $-k$.

One must further specify boundary conditions for the four-dimensional path integral. At the infinite end, asking for finite energy requires $\CA_\mu$ to be at a stationary point of the flow \eqref{gradCS}. Therefore, fields must approach a fixed complex flat connection $\CA_\mu^\ra$ on $M$. The set of possible values for $\CA_\mu$ at $s=0$ then becomes the set of solutions to gradient flow starting from $\CA_\mu^\ra$ at $s=\infty$; this is the cycle $\Gamma^\ra_{\rm CS}$ itself.

At the origin, the appropriate boundary condition is a modified Neumann boundary condition that allows free oscillations of $\CA$. It is well known that a theta-term $\theta^\vee\!\int_V \Tr\,F^2$ in Yang-Mills theory induces a Chern-Simons coupling $\theta^\vee I_{\rm CS}(M;A)=\theta^\vee\!\int_M\Tr(AdA+\frac23 A^3)$ at a boundary $M=\pd V$, and this relationship gets complexified in the twisted $\CN=4$ theory. To be more precise, we recall that GL-twisted $\CN=4$ SYM has two free parameters%
\footnote{We decorate all the parameters here with a ``$\vee$'' in anticipation of an S-duality that appears below.}%
: the complex gauge coupling $\tau^{\!\vee} = \frac{\theta^{\!\vee}}{2\pi}+\frac{4\pi i}{g^{\!\vee}{}^2}$ and the complex twisting parameter $t^\vee\in \cp^1$. (The GL twist preserves two scalar supercharges and their conjugates, and $t^\vee$ parametrizes the projective linear combination of the two charges being used as a BRST operator.) However, all physical correlation functions in the theory depend on only one complex ``canonical parameter''
\be
\Psi^{\!\vee} = \frac{\theta^{\!\vee}}{2\pi}+\frac{4\pi i}{g^{\!\vee}{}^2}\frac{t^\vee-t^\vee{}^{-1}}{t^\vee+t^\vee{}^{-1}}\,.
\ee
The Neumann boundary condition at the origin breaks half of the supersymmetry, relating $t^\vee{}^2=\ol\tau^{\!\vee}/\tau^{\!\vee}$, which fixes the canonical parameter,
\be \label{NPsi}
\text{Neumann:}\quad\Psi^{\!\vee}=\frac{|\tau^{\!\vee}|^2}{\Re\,\tau^{\!\vee}}=\frac{\theta^\vee}{\pi}~,\qquad
\ee
to be real. Moreover, the complexification parameter $w$ in \eqref{Acx} becomes pure imaginary, $w=-i\,\Im\,\tau^{\!\vee}/|\tau^{\!\vee}|$. The correct supersymmetric coupling at $s=0$ then becomes $-\frac{\Psi^{\!\vee}}{4\pi}I_{\rm CS}(M;\CA)$, and the path integral on $M\times \R_+$ simply reduces to an integral over connections on $M$ at $s=0$,
\be \label{SYMCS}
\CZ_{\rm SYM}^\ra(M\times \R_+;\Psi^{\!\vee}) = \int_{\Gamma^\ra_{\rm CS}} D\CA\,\exp\bigg(-\frac{\Psi^{\!\vee}}{4\pi}I_{\rm CS}(M;\CA)\bigg) = \CZ_{\rm CS}^\ra(M;q)\,,
\ee
with $q = e^{-2\pi i/\Psi^{\!\vee}}$.

In this formula, it is clear that $-\Psi^{\!\vee}$ plays the role%
\footnote{In the above formula, we have $\Psi^{\!\vee}=-k$, where $k$ is a renormalized Chern-Simons level; whereas in the standard Chern-Simons path integrals \eqref{CSdim1}--\eqref{CScodim2}, the coupling constant is the unrenormalized $\bar k$. This discrepancy is addressed in \cite{Wfiveknots}, and is related to a change in the path-integral measure.} %
of the Chern-Simons level $k$, but there is no requirement that $\Psi^{\!\vee}$ be an integer. Since $M$ is identified as the boundary of a specific four-manifold $V$, it is not necessary to quantize the level. Indeed, even though $\Psi^{\!\vee}$ is real in \eqref{NPsi}, one can easily analytically continue \eqref{gradCS} and \eqref{SYMCS} to any $\Psi^{\!\vee}\in \C^*$. We emphasize that this is \emph{not} analytic continuation in the coupling $\tau^{\!\vee}$ of SYM, but rather in $\Psi^{\!\vee}=\theta^\vee/\pi$, since the latter is what the twisted theory depends on holomorphically. 

So far the discussion has applied to a closed three-manifold $M$. In order to study Chern-Simons theory on a knot complement $M=S^3\bs \CK$, one must introduce a surface operator along $S=\CK\times \R_+$ in the half-space geometry $V=S^3\times \R_+$.%
\footnote{This preserves the codimension of the monodromy defect at $\CK$ in Chern-Simons theory, which is appropriate for connecting with knot-complement theories $T_M$. An inequivalent way to analytically continue Chern-Simons theory with 4d SYM is to put a Wilson loop operator along $\CK$ at $s=0$ in $S^3\times \R_+$, preserving the dimension of the defect. This case was studied in \cite{Wfiveknots, GW-Jones}.} %
The surface operator preserves the same supersymmetry as the boundary condition at $s=0$. Let us for simplicity fix the gauge group to be $G^{\!\vee}=SU(2)$. Then the simplest surface operator is characterized by four real parameters $(\bm\alpha^{\!\vee},\bm\beta^{\!\vee},\bm\gamma^{\!\vee},\bm\eta^{\!\vee})\in \R/(2\pi\Z)\times\R\times \R\times \R/(2\pi \Z)$ \cite{GW-branes}.
The first three parametrize a singularity of the complexified gauge connection,
\be \CA \sim \frac i2(\bm\alpha^{\!\vee}-w\bm\gamma^{\!\vee})\begin{pmatrix} 1 & 0 \\ 0 & -1 \end{pmatrix}d\theta 
+ \frac{iw\bm\beta^{\vee}}{2}\begin{pmatrix} 1 & 0 \\ 0 & -1 \end{pmatrix}\frac{dr}{r}
+\text{(less singular)}
\ee
in the plane perpendicular to the surface operator; whereas $\eta^{\!\vee}$ is a 2d theta-angle, coupling to two-dimensional instanton number $\int_S F$ in the path integral. (At the surface operator itself, the gauge group is broken from $SU(2)$ to $U(1)$, so $\int_S F$ makes sense.) We see that the holonomy of $\CA$ around a small loop linking the surface operator (and hence the knot $\CK$) has squared eigenvalues
\be x^{\pm 1} = e^{\pm2\pi i(\bm\alpha^{\!\vee}-w\bm\gamma^{\!\vee})}\,.\ee
Moreover, in the case of Neumann boundary conditions at $s=0$, $\bm\eta^{\!\vee}$ must vanish.
The SYM path integral on $S^3\times \R_+$ in the presence of the surface operator  should now evaluate to
\be \label{SYMCSx}
\CZ_{\rm SYM}^\ra(S^3\times \R_+;\Psi^{\!\vee};\bm\alpha^{\!\vee},\bm\gamma^{\!\vee}) = \CZ_{\rm CS}^\ra(S^3\bs \CK;x;q)\,,
\ee
agreeing with the analytic continuation of Chern-Simons theory on a knot complement \eqref{CScodim2}. The partition function involves a choice of flat connection $\CA^\ra$ on $S^3\bs\CK$, depends holomorphically on $x$ and $q$, and is independent of $\bm\beta^{\vee}$.

There is a convenient brane construction of this system in type IIB string theory. One considers two semi-infinite D3 branes wrapping $V$ in the geometry $T^* M\times \R\times\R^3$ that end on an NS5 brane that wraps $M$ and sits at the origin of $\R_+\subset\R$. Codimension-two defects along $\CK\times \R_+$ can then be engineered by including further intersecting branes.

\subsubsection*{Four dimensional S-duality}

$\cN=4$ SYM on $V$ has a useful S-dual description that replaces the Neumann boundary condition at $s=0$ with a Dirichlet-like boundary condition. In terms of type IIB string theory, S-duality maps the NS5-D3 brane system to a D5-D3 brane system, \ie, a stack of semi-infinite D3 branes wrapping $V=M\times \R_+$ and ending on a D5 brane. In the field theory, the D5-D3 boundary condition --- sometimes called a Nahm pole boundary condition --- effectively freezes out the degrees of freedom in the complex connection $\CA$ at $s=0$, and rather than appearing directly as a Chern-Simons path integral the 4d partition function takes the form of an instanton-counting expansion,
\be \label{CSinst}
\CZ_{\rm SYM}^\ra(M\times \R_+;x;q) = \sum_{a,b} n_{a,b}^\alpha q^a x^b\,.\ee
Here $n_{a,b}$ is the (signed) number of solutions to certain instanton equations with 4d instanton number $a\sim \int_V \Tr\,F^2$ and two-dimensional instanton number $b\sim\int_S F$. The numbers $a$ and $b$ may be fractional, as discussed in \cite{Wfiveknots}; in the present case it turns out that $a\in \Z/2$ and $b\in \Z$.

As before, the S-dual description must be supplemented with a choice of flat complex connection at infinity on $\R_+$. There is a one-to-one correspondence between flat connections in the S-dual description and the original description, so we will continue using `$\alpha$' to denote flat connections in this dual description. Technically, the S-dual gauge group is $G=SO(3)$ (with complexification $\PSLC$) rather than $G^\vee=SU(2)$, but the distinction is subtle and is not important in this paper. In particular, on a knot complement $M=S^3\bs \CK$, flat $\PSLC$ connections can always be lifted to $\SLC$ connections.

The parameters $(\Psi,\tau,t)$ of the S-dual theory are related to those of the original as
\be  \label{SYMSparams}
\Psi = -\frac{1}{\Psi^{\!\vee}}\,,\qquad \tau = -\frac{1}{\tau^{\!\vee}}\,,\qquad t = \frac{\tau^{\!\vee}}{|\tau^{\!\vee}|}t^\vee\,.\ee
Moreover, in the presence of the Dirichlet boundary condition, we find
\be t=1\,,\qquad \Psi = \Re\, \tau = \theta/2\pi\,.\ee
Similarly, the S-dual surface-operator parameters are
\be (\bm\alpha,\bm\beta,\bm\gamma,\bm\eta) = (\bm\eta^{\!\vee}=0,|\tau^{\!\vee}|\bm\beta^{\!\vee},|\tau^{\!\vee}|\bm\gamma^{\!\vee},-\bm\alpha^{\!\vee})\,. \ee
Therefore, the parameters $q$ and $x$ of the original SYM theory become
\be \label{qxSdual}
q = e^{2\pi i\Psi}\,,\qquad x = e^{2\pi i (\bm\eta+w|\tau|\bm\gamma)}\,,
\ee
with $w = -i\,\Im\,\tau^{\!\vee}/|\tau^{\!\vee}|=-i\,\Im\,\tau/|\tau|$. These are the correct 4d/2d instanton-counting parameters of the S-dual twisted SYM theory, which enter the path integral \eqref{CSinst}.

\subsubsection*{Lift to six dimensions}

This can now be lifted to six dimensions. The type IIB brane construction can be T-dualized to a D6-D4 system in type IIA string theory, consisting of two semi-infinite D4 branes wrapping $M\times \R_+\times S^1_\beta$ in $T^*M\times\R^3\times S^1_\beta$, and ending on the D6 brane. In turn, this can be lifted to a single configuration of two M5 branes in M-theory embedded in the geometry described in \eqref{Mgeometry}:
\be\label{Mgeometry2}
\text{M5's}:\quad M\times D^2\!\times_qS^1\;\;\subset\;\; T^*M \times TN \times_q S^1\,.\qquad
\ee
A codimension-two defect along a knot $\CK\subset M$ can be engineered with additional intersecting M5 branes. To preserve supersymmetry, the additional branes wrap the conormal bundle of $\CK$ in $T^*M$ as well as $D^2\!\times_q\! S^1$.

From a field-theory perspective, the low-energy theory of the two principal M5 branes is the six-dimensional $(2,0)$ SCFT for Lie algebra $A_1$. It is topologically twisted so that the Lorentz group $SO(3)_E\times SO(2)_E$ on $M\times D^2$ is redefined to be the diagonal in the product of itself and the $SO(3)_R\times SO(2)_R$ subgroup of $SO(5)$ R-symmetry. The twist preserves a scalar supercharge $Q$ (and its conjugate) that has charge $+1$ under the unbroken $SO(2)_R$ R-symmetry. In addition, a codimension-two defect along $\CK\times \DqS$ comes with a global $SU(2)_\CK$ flavor symmetry. We denote the integer charge of states under a maximal torus $U(1)_\CK\subset SU(2)_\CK$ as $e$.

To recover the 4d construction from the 6d theory, one compactifies on the asymptotic torus of $\DqS$. The tip of the cigar generates the Nahm pole boundary condition discussed above, and the scalar supercharge $Q$ coincides with that of GL-twisted SYM in the presence of the boundary. The modular parameter of the compactification torus, $\tau = \beta\ve+i\beta\rho^{-1}$ (\cf\ Figure \ref{fig:T2}), becomes the 4d SYM coupling. We already know, however, that neither the 4d nor the 6d partition functions depend on $\tau$ alone. (For example, in six dimensions, nothing can depend on the radius $\rho$ of the topologically-twisted cigar.) On the other hand, the partition functions should depend analytically on the canonical parameter
\be \Psi = \Re\,\tau = {\beta\ve},\,\ee
which is just the geometric holonomy in $\DqS$.

The M5-brane partition function in this geometry takes the form of a BPS index with respect to $Q$ \cite[Sec. 6]{Wfiveknots},
\be \label{6dindex}
\CZ^\ra(M\times \DS;x;q) = \Tr_{\CH(M\times D;\alpha)}(-1)^{2J} e^{-\beta H} q^{-J+\tfrac R2}x^e\,,
\ee
where $R$ is the generator of $SO(2)_R$ and $J\in \frac12\Z$ generates the $SO(2)_E$ rotations of $D^2$. This index should reproduce the 4d instanton partition function \eqref{CSinst}. Indeed, upon compactification, the angular momentum $J$ becomes 4d instanton number $a$ and we see that it consistently couples to the fugacity $q=e^{2\pi i\Psi}$. Similarly, one can argue that the flavor charge $e$ for a codimension-two defect descends to 2d instanton number $b$ on a surface operator, and that the corresponding fugacity $x$ is given by \eqref{qxSdual}. Finally, the BPS partition function \eqref{6dindex} depends on a choice of vacuum $\alpha$ to set the boundary condition at the infinite end of the cigar, fixing the Hilbert space $\CH(M\times D;\alpha)$. This is equivalent to a choice of vacuum in the four-dimensional setup, \ie\ a choice of flat complex connection $\CA^\ra$ on $M$.

\subsubsection*{Back to three dimensions}

The 6d index \eqref{6dindex} is intentionally written in the same form as the BPS partition functions of the three-dimensional $\CN=2$ theories that we have studied throughout this paper. By taking the 6d theory and reducing on $M$, we obtain a three-dimensional theory $T_M$ on $\DqS$, whose holomorphic blocks are given by \eqref{6dindex}. The parameters/charges $q$, $J$, and $R$ of the 6d theory are equivalent to those that appear in the three-dimensional construction. The chain of dualities reviewed here, however, allows us to identify $q$ with the coupling of analytically continued Chern-Simons theory on $M$ itself,
\be q = e^{2\pi i\beta\epsilon}=  e^{2\pi i\Psi} = e^{\tfrac{2\pi i}{k}}\,. \ee
Beautifully, this reaffirms the analytic dependence on $\Re\,\tau=\beta\ve$ that we found for holomorphic blocks back in Section \ref{sec:cigar}.

Similarly, the three-manifold theory for a knot complement $M=S^3\bs \CK$ is expected to have an $SU(2)_\CK$ flavor symmetry. The complexified twisted mass parameter $x$ for its maximal torus $U(1)_\CK\subset SU(2)_\CK$ has now been identified with the eigenvalue-squared ($m^2$) of the holonomy of a connection $\CA$ at the meridian of the excised knot $\CK$ in $M$:
\be x=\exp\bigg(2\pi\beta m^{\rm 3d}_\CK+i\oint_{S^1_\beta}A_\CK\bigg) = e^{2\pi i(\bm\eta+w|\tau|\bm\gamma)} = m^2\,. \ee
where $A_\CK$ is the background gauge connection for $U(1)_\CK$.
Finally, we see that the choice of vacuum $\alpha$ at the infinite end of the cigar in three dimensions is precisely a choice of flat complex connection on $M$, defining a Chern-Simons integration cycle.

In \cite{DGG}, ideal triangulations of manifolds $M$ were used to define simple candidate UV Lagrangians%
\footnote{A direct M-theory derivation of the UV Lagrangians of \cite{DGG} has been proposed in \cite{CCV}.} %
for the low-energy limit of the associated three-manifold theory. It is important to keep two things in mind about the theories $T_M$ of \cite{DGG}. First, they are gauge theories built using rules such as in Section \ref{sec:Ward-assemble}, and when $M$ is a knot complement, they always have a $U(1)_\CK$ flavor symmetry associated to the knot. It was conjectured that this $U(1)_\CK$ can always be enhanced to $SU(2)_\CK$ at a point in the parameter space of $T_M$. For example, all compact partition functions (and, as we shall see, holomorphic blocks) of $T_M$ are invariant under the inversion of the mass parameter $x\to x^{-1}$, as would be the case for an $SU(2)$ symmetry. However, further analysis along the lines of \cite{GKSTW, DG-E7} is necessary to verify a true enhancement.

Second, the theories $T_M$ of \cite{DGG} do not know about all possible flat connections on a knot complement $M$: they only appear to have vacua $\alpha$ corresponding to the \emph{irreducible} flat connections. Mathematically, this arises from the fact that reducible connections on a knot complement are not naturally obtained by gluing together connections on ideal tetrahedra. Physically, however, the interpretation of this statement has not been fully clarified. (One possible scenario, proposed in \cite{DGG}, is that in the low-energy limit of the $(2,0)$ theory on $M\times \RqS$, superselection sectors develop that decouple reducible connections.)
Fortunately, as was discussed in Section \ref{sec:CSanal}, there does appear to be a consistent truncation of analytically continued Chern-Simons theory that also only sees cycles corresponding to irreducible flat connections. It is this truncated Chern-Simons theory that should be compared to class-$\CR$ constructions of $T_M$.


\subsubsection{Remarks on gluing}

It is of interest to find a six-dimensional description of the ellipsoid partition function and sphere index of a knot complement theory. For this purpose, it would be most natural to study the $(2,0)$ theory on $M\times S^3_b$ or $M\times \SqS$, respectively. Unfortunately, we encounter the same problem as in three dimensions: the theories on $S^3_b$ and $\SqS$ do not use a topological twist to preserve supersymmetry, and it is not clear how to implement the necessary SUSY-preserving modifications directly in six dimensions. Note, for example, that it is not even possible to topologically twist the $(2,0)$ theory in a geometry $M\times S^3$; the R-symmetry group $SO(5)_R$ is too small.

Nevertheless, we can consider the six-dimensional analogue of the stretched construction studied in this paper, and find yet another way to identify parameters $(x,q)$ and $(\wt x,\wt q)$ on the two sides.
Taking a stretched geometry (times $M$) as our starting point, we can reduce to four dimensions and study the boundary conditions and/or brane configurations that reproduce factorized partition functions of the form
\be \langle 0_q|g|0_q\rangle = \big|\!\big|B^\ra_M(q;x)\big|\!\big|^2_g = \sum_\alpha B^\ra_M(q;x)B^\ra_M(\wt q;\wt x), \ee
where now $g\in \SLZ$ acts as an S-duality transformation for GL-twisted super-Yang-Mills theory, with both the coupling $\tau$ and the canonical parameter $\Psi$ transforming in the usual way under $\SLZ$.

For the index, the relevant configuration is a stack of two D3 branes wrapping $M$ and filling an interval $I$, with a D5 brane on one side and a $\ol{\rm D5}$ anti-brane on the other. (This is the analogue of the S-dual framework discussed around \eqref{SYMSparams}.) Supersymmetry is only preserved in this system in the limit of infinite interval length. In field theory, the bulk parameters $q$ and $\wt q$ at the two ends are related due to the reversed orientation as $\wt q=q^{-1}$, just as desired. More interestingly, the parameter $w$ that complexifies connections has the opposite sign near the two boundaries. Consequently, the effective 2d instanton-counting parameters at the two ends of are $x=e^{\bm \eta+w|\tau|\bm \gamma}$ and $\wt x= \ol x= e^{\bm \eta-w|\tau|\bm \gamma}$.

We can learn something about integration cycles for the index by dualizing to a system of D3 branes stretched between an NS5 brane and an $\ol{\rm NS5}$ anti-brane, similar to the original construction of Chern-Simons theory via $\cN=4$ SYM. In the field theory, a Chern-Simons coupling $\Psi I_{\rm CS}(\CA)$ is induced at the NS5 end, while a coupling $-\Psi I_{\rm CS}(\ol \CA)$ is induced at the $\ol{\rm NS5}$ end, with $\ol \CA = A - w\phi$. In the bulk of the (still infinite) interval, the theory localizes to gradient flows for $\CA$ --- or, equivalently, gradient flows for $\ol \CA$. In order to have finite energy, the flows must spend an infinite amount of time near a flat connection%
\footnote{Recall that generically there are no flows between different critical points, so a given flow can only choose a single $\CA^\ra$ to approach to in the middle of $I$.} %
$\CA^\ra$ in the middle of the interval. Flowing away from $\CA^\ra$ toward either end of the interval produces conjugate integration cycles for $\CA$ and $\ol \CA$. The partition function then takes the form
\be \label{SL2Cfact}
\CZ_{\rm SYM}(M\times I;x;q) = \sum_\alpha \left(\int_{\Gamma^\ra_{\rm CS}} \CD\CA\, e^{-i \tfrac{\Psi}{4\pi}I_{\rm CS}(\CA)}\right)\left(\int_{\ol \Gamma^\ra_{\rm CS}} \CD\ol\CA\, e^{i \tfrac{\Psi}{4\pi}I_{\rm CS}(\ol \CA)}\right)\,.
\ee
This is precisely the analytic continuation of the full (non-holomorphic) $\SLC$ Chern-Simons theory on a knot complement $M$, with partition function
\be \label{SL2Cfull}
\CZ_{\rm CS}^{\text{SL}(2,\C)}(M;x;q) = \int \CD\CA\,\CD\ol\CA\,e^{i\tfrac{k}{4\pi}I_{\rm CS}(\CA)-i\tfrac{k}{4\pi}I_{\rm CS}(\ol \CA)}\,.
\ee
(Note that $\CA$ and $\ol \CA$ are independent complex fields in \eqref{SL2Cfact}, whereas they are complex conjugates of each other in \eqref{SL2Cfull}.) The connection between the three-dimensional index of $T_M$ and $\SLC$ Chern-Simons theory on $M$ was also discussed in \cite{DGG-index}.

Of course, for a finite interval the NS5-D3-$\ol{\rm NS5}$ system does not preserve supersymmetry --- at least not at weak string coupling. It is unclear whether this is a serious problem in the low-energy effective gauge theory on the branes. For finite string coupling, it is conceivable that the branes can be arranged in a supersymmetry-preserving bound state. This could provide the geometry corresponding to an index on the untwisted $\SqS$, but for the moment it is just speculation. Further study in this direction should prove interesting. 

The story for the ellipsoid $S^3_b$ (or for more general Lens spaces) is similar. To reproduce the partition function of $T_M$ on $S^3_b$, we should look at 4d SYM on an interval with a Neumann boundary condition at one end and a Dirichlet (or Nahm pole) boundary at the other. In terms of branes, this comes from a system of D3 branes on $M\times I$, stretched between a D5 brane and an $\ol{\rm NS5}$ anti-brane, or vice versa: the boundary branes are related by the element $g=S$ of $\SLZ$. For finite interval length, the system \emph{appears} to break supersymmetry by the S-rule of \cite{HananyWitten}. For infinite interval length, supersymmetry is effectively restored, and the parameters $(q,\wt q)$ at the two ends are related by
\be \wt q = e^{-2\pi i S(\Psi)} = e^{-2\pi i\Psi^{\!\vee}} = e^{\tfrac{2\pi i}{\Psi}}\,.\ee
Meanwhile, to understand the surface-operator parameters $x,\wt x$, observe that the presence of both Dirichlet and Neumann boundaries forces $\bm \alpha = \bm\eta = 0$, and sets $\tau = \Psi = \theta/2\pi$, with zero imaginary part. Then $x=e^{|\tau|w\bm\gamma}=e^{|\Psi|w\bm\gamma}$ at one end is transformed to
\be \wt x = e^{|\Psi^{\!\vee}|w\bm \gamma^{\!\vee}} = e^{w\gamma} = x^{1/\Psi} \ee
at the other, for (say) real and positive $\Psi$. These are precisely the expected relations for the ellipsoid partition function.


\subsection{Examples}
\label{sec:asymp}

As examples of the correspondence between holomorphic blocks and Chern-Simons path integrals, we consider three theories $T_M$, for $M$ the complement of the trefoil ($\mb{3_1}$), figure-eight ($\mb{4_1}$), and $\mb{5_2}$ knots in $S^3$.

From the point of view of Chern-Simons theory, the meridian holonomy of a complex connection at the excised knot will have fixed squared-eigenvalues $x^{\pm 1}$, as discussed in Section \ref{sec:CSanal}. Then the three knot complements $S^3\bs \CK$ for $\CK=\mb{3_1},\,\mb{4_1},\,\mb{5_2}$ admit (respectively) one, two, and three irreducible flat $\SLC$ connections $\CA^\ra$. This counting is confirmed by looking at the (irreducible) A-polynomials of the knot complements:
\begin{subequations} \label{Apolys}
\begin{align} A_{\mb{3_1}}(x,p) &= p-x^3\,,\\
 A_{\mb{4_1}}(x,p) &= p+(x^2-x-2-x^{-1}-x^{-2})+p^{-1}\,,\\
 A_{\mb{5_2}}(x,p) &= x^7p^3-x^2(1-x+2x^3+2x^4-x^5)p^2-(1-2x-2x^2+x^4-x^5)p-1
\end{align}
\end{subequations}
Solving $A_\CK(x,p)=0$ at fixed $x$ yields as many solutions $p^\ra(x)$ as there are irreducible flat connections.

The theories $T_M$, as constructed using the rules of \cite{DGG}, will then have one, two, and three holomorphic blocks $B^\ra(x;q)$, respectively. By construction, the blocks will satisfy line-operator identities $\hat A_\CK(\hat x,\hat p;q)\cdot B^\ra(x;q)=0$, where the operator $\hat A_\CK$ is a quantization of the A-polynomial $A(x,p)$. The quantum A-polynomials are given by \cite{Gar-twist, Dimofte-QRS}
\begin{subequations} \label{qApolys}
\begin{align} \hat A_{\mb{3_1}}(\hat x,\hat p;q) &= \hat p-q^{\frac 32} \hat x^3\,,\\
 \hat A_{\mb{4_1}}(\hat x,\hat p;q) &= (q^{-\frac12}\hat x-q^{\frac12}\hat x^{-1})\hat p+(\hat x-\hat x^{-1})(\hat x^{2}-\hat x-q-q^{-1}-\hat x^{-1}+\hat x^{-2}) \notag\\&+ (q^{\frac12}\hat x-q^{-\frac12}\hat x^{-1})\hat p^{-1}\,,\\
 \hat A_{\mb{5_2}}(\hat x,\hat p;q) &= q^{14}(1-q\hat x^2)(1-q^2\hat x^2)\hat x^7\hat p^3 \notag\\&\hspace{-45pt}
 -q^{\frac52}(1-q\hat x^2)(1-q^4\hat x^2)\hat x^2(1-q^2\hat x-q^2(1-q)(1-q^2)\hat x^2+q^4(1+q^3)\hat x^3+2q^7\hat x^4-q^9\hat x^5)\hat p^2 \notag \\&\hspace{-45pt}
 -(1-q^2\hat x^2)(1-q^5\hat x^2)(1-2q\hat x-q(1+q^3)\hat x^2+q^2(1-q)(1-q^2)\hat x^3+q^5\hat x^4-q^6\hat x^5)\hat p \notag \\&\hspace{-45pt}
  -q^{\frac12}(1-q^4\hat x^2)(1-q^5\hat x^2)\,.
\end{align}
\end{subequations}
The blocks $B^\ra(x;q)$ provide a basis of solutions to these line-operator identities with the analytic properties discussed in Section \ref{sec:blockint}.

Before describing the blocks of these theories, we should make a practical remark about the knot-complement theories $T_M$ constructed in \cite{DGG}. In general, defining the complete theory $T_M$ requires a substantial refinement of the ``standard'' or ``minimal'' ideal triangulation for a knot complement. The reason for this is discussed in \cite[Sec. 4.6]{DGG}. Essentially, the minimal triangulation leads to UV theory that does not contain all the chiral operators $\CO_I$ needed to break flavor symmetries and enforce the gluing of tetrahedra. These operators must be added to the superpotential in order for the theory to truly flow to the correct fixed point $T_M$. Nevertheless, for computing quantities like holomorphic blocks that are independent of superpotential deformations, we can use the ``simplified'' version of $T_M$ coming from a minimal triangulation, and just set the parameters for the unwanted flavor symmetry to zero by hand. This is how we will proceed below. The simplified $T_M$ leads to the exact same blocks as the more complicated true theory.

\subsubsection*{Trefoil $\mb{3_1}$}

We start with the trefoil knot complement $M=S^3\bs \mb{3_1}$. The minimal triangulation of $M$ contains two tetrahedra, and is discussed in Appendix \ref{app:knots}. The ``simplified'' theory $T_{\mb{3_1}}$ derived from this triangulation is rather degenerate. At low energies, it is just a dynamical $U(1)$ Chern-Simons theory at level $-1$. There is a topological $U(1)_J$ flavor symmetry whose background multiplet is coupled to the dynamical vector multiplet via an FI term, and a level $+2$ background CS coupling is turned on for $U(1)_J$. The exponentiated, complexified mass of $U(1)_J$, denoted by $x$, corresponds to the squared meridian eigenvalue for the knot complement.

A block integral can be found using the rules of Section \ref{sec:blockint}, leading to
\be \mathbb{B}_{\mb{3_1}}(x;q)= \int_{*} \frac{ds}{2\pi is} \frac{\theta(sx^{-1})}{\theta(x)^3} \overset{s\to sx}{=} \frac{1}{\theta(x)^3}\int_{*} \frac{ds}{2\pi is} \,\theta(s)\,,\ee
where $\theta(x):=\theta(x;q)$ denotes the theta function first defined in Section \ref{sec:chiralblock}.
There is a single block, and since the integration is $x$-independent, it is easy to see that the block is annihilated by the quantum A-polynomial for any choice of integration cycle:
\be \theta(qx)=\frac{1}{q^{\frac12}x}\theta(x)\qquad\Rightarrow\qquad (\hat p-q^{\frac32}\hat x^3)\frac{1}{\theta(x)^3} = \frac{1}{\theta(qx)^3}-\frac{q^{\frac32}x^3}{\theta(x)^3}=0\,.
\ee
Also note that the block is invariant under $x\to x^{-1}$, which reflects the Weyl symmetry for the meridian holonomy, and the fact that in the full theory $T_{\mb{3_1}}$ there may be enhancement $U(1)_J\to SU(2)_\CK$.

The integral can be performed to normalize the block.
For $|q|<1$, the natural convergent contour $\Gamma_<$ is around the girth of the $s$-cylinder, at $|s|=1$ or $\Re\,S=0$. The integral merely picks out the zeroth Fourier coefficient:
\begin{subequations}\label{B31}
\be \qquad B_{\mb{3_1}}(x;q) = (q)_\infty \int_{\Gamma_<} \frac{ds}{2\pi is}\,\frac{\theta(s)}{\theta(x)^3}= \frac{(q)_\infty}{\theta(x;q)^3}\left[\theta(s;q)\right]_{s^0} = \frac{1}{\theta(x;q)^3}\qquad (|q|<1)\,.\ee
For $|q|>1$, the natural convergent contour is parallel to the $\Re\, S$ direction, from one end of the cylinder to the other. The integrand has a full line of poles at $S=i\pi+\hbar(\Z+\frac12)$, which the contour can never cross. Numerical integration gives
\be \qquad B_{\mb{3_1}}(x;q) = \frac{1}{(q^{-1})_\infty}\int_{\Gamma_>}\frac{ds}{\hbar s}\,\frac{\theta(s)}{\theta(x)^3}= \frac{1}{\theta(x;q)^3}\qquad (|q|>1)\,.\ee
\end{subequations}

\subsubsection*{Figure-eight $\mb{4_1}$}

As is discussed in Appendix \ref{app:knots}, one realization of the simplified figure-eight knot theory $T_{\mb{4_1}}$ is identical to the $\cp^1$ sigma-model of Section \ref{sec:CP1}, with the topological flavor symmetry $U(1)_J$ ``broken'' by hand --- so that the complexified mass (\emph{a.k.a.} FI parameter) $y$ is set to one. This is an oversimplified description of $T_{\mb{4_1}}$ because there exists no operator charged under $U(1)_J$ that can be added to a superpotential to break the symmetry naturally. Nevertheless, we can use the description to write down the block integral:
\be\label{B41int}
\mathbb{B}_{\mb{4_1}}(x;q) =\frac{1}{\theta(x)} \int_* \frac{ds}{2\pi is}\,\frac{1}{\theta(-q^{-\frac12}s)}\,(qs^{-1}x^{-1};q)_\infty(qs^{-1}x;q)_\infty\,.
\ee
Note that we have chosen theta-functions judiciously, to allow the $y\to1$ limit to exist.
There are two vacua, and two critical points, corresponding to the two irreducible flat connections on the figure-eight knot complement. We already know how to find convergent contours in various Stokes chambers. In the semiclassical chamber `$\I$' described%
\footnote{This chamber was loosely described as having large positive FI parameter, $|y|\gg 1$. Here, even though we have set $y\to 1$, part of the chamber still survives. This can be seen qualitatively by taking a slice of the plot in Figure \ref{fig:Stokes} at $\Re\,Y=0$.} %
in Section \ref{sec:CP1}, the two blocks are (\cf\ \eqref{blocksI+})
\be \label{B41}
\begin{array}{rl}
B^1_{\mb{4_1}}(x;q) &= \ds \frac{1}{\theta(x)\theta(-q^{\frac12}x)}\CJ(x,x^2;q)\,,\\[.4cm]
B^2_{\mb{4_1}}(x;q) &= \ds \frac{1}{\theta(x)\theta(-q^{-\frac12}x)}\CJ(x^{-1},x^{-2};q)\,,
\end{array}\,\qquad \text{$|q|<1$ or $|q|>1$}\,,
\ee
modulo factors of $(q^{\pm 1})_\infty$.
Note that, using $\theta(x)=\theta(x^{-1})$, the Weyl symmetry $x\to x^{-1}$ is explicitly realized in this basis of blocks. In fact, the vector $U(1)_V$ flavor symmetry of the simplified theory, with mass $x$, can obviously be enhanced to $SU(2)_V$. The blocks \eqref{B41} are solutions to the difference equation (\ref{qApolys}b).

It is straightforward to check that the S-fusion sum of blocks 
\be
\CZ_b[\mb{4_1}](X,\hbar) = \sum_\alpha B^\ra_{\mb{4_1}}(x;q)B^\ra_{\mb{4_1}}(\wt x;\wt q)~,\vspace{-10pt}
\ee 
with $x=\exp(X),\,q=\exp(\hbar),\,\wt x=\exp(\frac{2\pi i}{\hbar}X),\,\wt q=\exp(-\frac{4\pi^2}{\hbar})$, reproduces the output of the ``state integral'' for analytically continued Chern-Simons theory on the figure-eight knot complement, studied in \cite{hikami-2006, DGLZ, Dimofte-QRS}.
 In fact, different representations of the state integral were given in \cite{hikami-2006,DGLZ} and \cite{Dimofte-QRS}, which were proven to be equal by \cite{SpirVar-knots}.
The different representations are explicitly obtained by substituting $Y=\mu_y=0$ in \eqref{cpb1} and \eqref{cpb2} of Section \eqref{sec:cpfusion}. Thus, they are simply associated to different Stokes chambers of $T_{\mb{4_1}}$.
Similarly, the figure-eight index of \cite{DGG-index} is just $\CI[\mb{4_1}](\zeta,m;q)=\sum_\alpha B^\ra_{\mb{4_1}}(x;q)B^\ra_{\mb{4_1}}(\wt x;\wt q)$ with the usual identification $x=q^{\frac m2}\zeta,\,\wt x=q^{\frac m2}\zeta^{-1},\,\wt q=q^{-1}$.

\subsubsection*{Knot $\mb{5_2}$}

The last example, the $\mb{5_2}$ knot complement. The simplified theory $T_{\mb{5_2}}$, obtained in Appendix \ref{app:knots}, is a variant of the three-dimensional $\cp^2$ sigma-model. Namely, it is a dynamical $U(1)$ gauge theory coupled to three chiral multiplets of charge $+1$, but with a level $-\frac12$ Chern-Simons coupling for the gauge field.%
\footnote{The half-integer Chern-Simons level is necessary as usual to avoid an anomaly.} %
A priori, there is a $U(1)_J$ topological symmetry and a $U(1)^2$ flavor symmetry rotating the chirals.
However, the complex FI term on $\DqS$ (the mass for $U(1)_J$) is set to zero by hand, and the only mass we turn on for the chirals corresponds to a $U(1)_V\subset U(1)^2$ that rotates them with charges $(+1,-1,0)$. Its exponentiated, complexified mass is $x$. We also turn on $-2$ units of background Chern-Simons coupling for $U(1)_V$.

Altogether, the block integral becomes
\begin{align} \label{B52int}
\mathbb{B}_{\mb{5_2}}(x;q) &= \int_*\frac{ds}{2\pi is}\,\frac{\theta(x)}{\theta(-q^{-\frac12}s)}\,(qs^{-1};q)_\infty(qs^{-1}x^{-1};q)_\infty(qs^{-1}x;q)_\infty \notag \\ &= \theta(x)\int_* \frac{ds}{2\pi is}\frac{(qs^{-1}x^{-1};q)_\infty(qs^{-1}x;q)_\infty}{(s;q)_\infty}\,.
\end{align}
There are three vacua, and three relevant critical points. In a Stokes chamber where the theory would look like a massive semi-classical $\cp^2$ sigma model on the Higgs branch, the three corresponding blocks are found to be
\be \label{B52}
\begin{array}{rl}
B^1_{\mb{5_2}}(x;q) &= \ds \theta(x)\,\CG(x,x^{-1},1;q)\,, \\[.4cm]
B^2_{\mb{5_2}}(x;q) &= \ds \frac{\theta(x)}{\theta(-q^{\frac12}x)}\,\CG(x,x^2,x;q)\,,\\[.4cm]
B^3_{\mb{5_2}}(x;q) &= \ds \frac{\theta(x)}{\theta(-q^{\frac12}x^{-1})}\,\CG(x^{-1},x^{-2},x^{-1};q)\,,
\end{array}
\ee
modulo factors of $(q^{\pm 1})_\infty$ and $\hbar$, where
\be \CG(x,y,z;q) := (qx;q)_\infty(qy;q)_\infty \sum_{n=0}^\infty \frac{z^n}{(q^{-1})_n(qx;q)_n(qy;q)_n}\,.
\ee
Note that the series converges and the blocks make sense both for $|q|<1$ and $|q|>1$, as usual. Also note the obvious Weyl symmetry $x\to x^{-1}$ in this basis of blocks. The blocks are annihilated by the quantum A-polynomial \eqref{qApolys}.


\subsubsection{Asymptotics}

Let us denote the asymptotic behavior of a knot-complement theory as
\be \label{BCSasymp}
B_\CK^\ra(x;q) \overset{\hbar\to 0}\sim \exp(\tfrac1\hbar\CV^\ra(x)+\ldots)
\ee
for fixed $x$ in a given Stokes chamber. For $\hbar$ real (say), it does not matter whether we approach $\hbar\to 0$ from positive or negative values, using the $|q|<1$ or $|q|>1$ blocks. Since each $B_\CK^\ra(x;q)$ should equal the analytically continued Chern-Simons partition function $\CZ_{\rm CS}^\ra(S^3\bs \CK;x;q)$ (where vacua $\alpha$ are matched with flat connections $\CA^\ra$), then given \eqref{volCS} we should have
\be \CV^\ra(x) = \CV(\CA^\ra(x))\,, \label{volmatch} \ee
so that the leading asymptotics of the blocks match the holomorphic volume of the corresponding flat connection.

Each $\CV^\ra(x)$ in \eqref{BCSasymp} can be evaluated by a saddle-point expansion of the block integral, evaluating the integrand of the block integral at a critical point $s^\ra(x)$ in the $\hbar\to 0$ limit. These critical points are in one-to-one correspondence with solutions $p^\ra(x)$ of the A-polynomial $A_\CK(x,p)=0$, and thus with flat $\SLC$ connections $\CA^\ra$. By construction, as $x$ varies locally, the $\hbar\to0$ limit of the line-operator identity forces $\CV^\ra(x)$ to satisfy the differential equation
\be x\frac{d\CV^\ra(x)}{dx} = \log p^\ra(x)\,,\quad\text{or}\quad \CV^\ra(x) = \int^x \log p^\ra(x') \frac{dx'}{x'} \ee
for an appropriate $p^\ra(x)$. This is precisely the variation of the holomorphic volume of a connection $\CA^\ra$ (\cf\ \cite{NZ, yoshida-1985}). All that remains is to match the absolute asymptotics of $\CV^\ra(x)$ (rather than the variation) at fixed $x$. 

Although we have been unable to fix the absolute normalization of holomorphic blocks, the ambiguity only involves elliptic ratios of theta functions $\theta(\pm q^{\#}x;q)$ and constant terms $(q^{\pm})_\infty$. The leading asymptotic of an elliptic ratio of theta functions vanishes modulo $\frac{\pi^2}{6}$, and $(q^\pm)_\infty\sim \exp( \pm\pi^2/6)$. Consequently, the normalization of our blocks \emph{is} fixed up to elements of $\frac{\pi^2}{6}\Z$. We can thus establish that our blocks correctly reproduce the volumes of flat connections modulo such shifts. (This ambiguity can be compared to an identical one found in computing volumes of $\SLC$ connections using oriented ideal triangulations \cite{Neumann-combinatorics}.) Note that in different regions of parameter space, the asymptotics of a single block can be controlled by different critical points due to Stokes phenomena. Consequently we must always be sure to use a single basis of blocks which is naturally associated to the critical points in a given Stokes chamber. These will correspond to unique choices of flat connections. Below we report on the numerical evaluation of asymptotics of the blocks given above at fixed values of mass parameters.

Of course, this is hardly the first situation in which the asymptotics of a finite-dimensional integral have been compared to volumes of flat connections. The same was done for ``state integrals'' in Chern-Simons theory --- which should equal the ellipsoid partition functions $\CZ_b[T_M]$ of knot-complement theories  \cite{DGG} --- in, \eg, \cite{hikami-2006, DGLZ, DG-quantumNZ}.%
\footnote{Not to mention the construction of many ad-hoc integrals in the mathematics literature that involve integrands symmetric in $\hbar\leftrightarrow -\frac{4\pi^2}{\hbar}$ and produce asymptotics of colored Jones polynomials, starting with \cite{kashaev-1997}.} %
Namely, it was seen that the saddle-point expansion of a $\CZ_b[T_M]$ integral around particular critical points $\alpha$ (in the $\hbar\to0$ limit) agrees with the holomorphic volumes of connections $\CA^\ra$. From the point of view of blocks, this is no surprise. The integrands of block integrals and localized $\CZ_b[T_M]$ integrals are \emph{identical} perturbatively in $\hbar$. Likewise, the $\hbar\to 0$ asymptotics of a product $B^\ra(x;q)B^\ra(\wt x;\wt q)$, which appears in $\CZ_b[T_M]$, are completely determined by the asymptotics of $B^\ra(x;q)$, because $B^\ra(\wt x;\wt q)$ is non-perturbative: it depends on $\wt x = x^{1/\hbar}$ and $\wt q=e^{-4\pi^2/\hbar}$. Thus, modulo (important) technicalities of Stokes phenomena, the saddle-point expansions of $\CZ_b[T_M]$ integrals ought to match the asymptotics of blocks. In principle, we could also compare subleading asymptotics of blocks to perturbative quantum invariants associated to complex flat connections. However, the procedure for doing so is straightforward, and basically identical to that described for state integrals in \cite{DGLZ, DG-quantumNZ}. Since we gain no new insight from the comparison, there is no reason to include it here.

\subsubsection*{Trefoil asymptotics}

Given the asymptotic expansion
\be \label{thetaas}
\qquad\qquad \theta(x;q)\overset{\hbar\to0}{\sim}e^{-\tfrac{1}{2\hbar}(\log x)^2-\tfrac{\pi^2}{6\hbar}+\tfrac\hbar{24}} \qquad (|q|<1\;\;\text{or}\;\;|q|>1)\,,
\ee
with $\hbar$ approaching zero along the real axis from either the positive or negative directions,
we find for both $|q|<1$ and $|q|>1$ that $B_{\mb{3_1}}(x;q)\sim
\exp\left[\frac1\hbar \CV_{\mb{3_1}}(x)+O(1)\right]$, where
\be \CV_{\mb{3_1}}(x) = \frac32(\log x)^2\quad \text{\Big(mod }\frac{\pi^2}{6}\Big)\,.\ee
This matches the known volume of the unique irreducible flat $\SLC$ connection on the trefoil knot complement, as well as asymptotics of Jones polynomials for the trefoil, \cf\ \cite{Rozansky-triv1, Murakami-torus}. In particular, when the eigenvalue $x$ is set to one, the volume ${\rm Vol}(\mb{3_1})=\Im\,\CV_{\mb{3_1}}(x=1)$ vanishes, in agreement with the fact that the trefoil is not a hyperbolic knot (so its ``hyperbolic volume'' is zero).

\subsubsection*{Figure-eight asymptotics}

From the perspective of Chern-Simons theory, it is desirable to compute asymptotics at $x=1$, since this should  correspond to complete hyperbolic structures. However, the theory becomes massless in this limit and the blocks become singular exactly at $x=1$; we look in a neighborhood of $x=1$ instead.

A saddle-point analysis of the blocks integrals \eqref{B41int} predicts
\be \label{Bcrit41}
B^\ra_{\mb{4_1}}(x;q) \overset{\hbar\to 0}{\sim} \frac{1}{\sqrt[4]{(1-x-x^{-1})^2-4}}\exp\left[ \frac1\hbar \CV_{\mb{4_1}}(x;s^\alpha(x))+O(\hbar^0)\right]
\ee
up to an overall multiple of $i$ and $\exp \frac{\pi^2}{6\hbar}$, where
\be \CV_{\mb{4_1}}(x;s) = \frac{\pi^2}{6}+\frac12(\log x)^2+\frac12(\log(-s))^2+\Li_2(s^{-1}x^{-1})+\Li_2(s^{-1}x)\,, \notag \ee
and $s^\alpha(x)$ are the two solutions to
\be \exp \left(s\frac{\pd \CV_{\mb{4_1}}(x;s)}{\pd s}\right) =1\quad\Rightarrow\quad
s^\alpha(x)=-\frac12(1-x-x^{-1})\pm \frac12\sqrt{x^2-2x-1-2x^{-1}+x^{-2}}\,. \notag\ee
Note that the natural branch cuts of $\log$ and $\Li_2$ in $\CV_{\mb{4_1}}$ are those expected from sending $\hbar\to 0$ along the real axis.
Around the point $x=\exp\frac{1+i}{20}$, it can be checked numerically%
\footnote{We thank D. Zagier for some extremely helpful lessons in numerical testing of asymptotics.} %
that \eqref{Bcrit41} are indeed the correct asymptotics of the blocks \eqref{B41} --- both for $|q|<1$ and $|q|>1$, as long as $\hbar$ is approximately real. Note that the subleading one-loop determinant $1/\sqrt[4]{...}$ in \eqref{Bcrit41} is necessary for a reasonable comparison, because anywhere close to $x=1$, for $\hbar$ real, the leading asymptotic term $e^{\CV/\hbar}$ is highly oscillatory rather than exponentially growing or decaying.

Working at the point $x=\exp\frac{1+i}{20}$, it can be determined numerically that
\be \qquad\qquad\CV_{\bm{4_1}}(x;s^\alpha(x)) \approx \begin{cases} -0.0043301+2.0298796\,i  & \alpha = 1 \\ 0.0043301-2.0298796\,i & \alpha= 2 \end{cases}\qquad \Big({\rm mod}\;\;\frac{\pi^2}{6}\Big)~, \ee
precisely matching the expected holomorphic volumes of the ``geometric'' and ``conjugate'' flat $\SLC$ connections on the figure-eight knot complement, deformed by the nontrivial boundary condition $x$ (\cf\ \cite{thurston-1980}). The oscillatory behavior mentioned above is due to the fact that the real volume $|\Im\,\CV_{\bm{4_1}}|$ is much larger than the Chern-Simons invariant $|\Re\,\CV_{\bm{4_1}}|$. Of course, we can even send $x\to 1$  in \eqref{Bcrit41} to get the exact hyperbolic volume and Chern-Simons invariant for the figure-eight knot at the complete hyperbolic structure,
\be \label{Vol41}
i({\rm Vol}(\mb{4_1})+i{\rm CS}(\mb{4_1})) = \CV_{\mb{4_1}}(1,s^1(1)) = \tfrac{\pi^2}{18}-2\,\Li_2(e^{-\frac{i\pi}{3}}) \approx 2.0298832\,i\quad \big({\rm mod}\;\;\tfrac{\pi^2}{6}\big)\,.
\ee
The Chern-Simons invariant vanishes, as expected. We emphasize, however, that the blocks themselves become singular and no longer have an asymptotic expansion governed by \eqref{Vol41} exactly at $x=1$.

\subsubsection*{Knot $\mb{5_2}$ asymptotics}

A saddle point evaluation of the block integrals yields
\be B^\ra_{\mb{5_2}}(x;q) \overset{\hbar\to 0}{\sim} \frac{1}{\sqrt{\CH(x,s^\alpha(x))}} \exp\left[ \frac1\hbar\CV_{\rm{5_2}}(x;s^\alpha(x))+O(\hbar^0)\right]\,,
\ee
up to an overall multiple of $i$ and $\exp\frac{\pi^2}{6\hbar}$, where
\be\CV_{\rm{5_2}}(x,s) = -\frac12(\log x)^2+\Li_2(s^{-1}x^{-1})+\Li_2(s^{-1}x)-\Li_2(s)\,, \ee
\be \CH(x,s) = s^{-1}(x+x^{-1}+1-s^2-2s^{-1})\,, \ee
and $s^\alpha(x)$ are the three solutions to
\be \exp\left(s\frac{\pd \CV_{\mb{5_2}}}{\pd s}\right)=1\quad\Rightarrow\quad (1-s)(1-sx)(1-sx^{-1})=s^2\,.\ee
We find experimentally that these asymptotics hold for the blocks in the semi-classical Stokes chamber \eqref{B52} as long as $|\log|x||\gtrsim 1.5$. Otherwise, different bases of blocks (different Stokes chambers) must be considered.

Again, the variation of the functions $\CV^\ra(x,s^\alpha(x))$ match the expected variation of the holomorphic volume of flat $\SLC$ connections $\CA^\ra$ by construction, so in the limit $x\to 1$, we should recover the well known complex volumes of irreducible flat connections with parabolic meridian holonomy (unit eigenvalues), fixing the normalization of the asymptotics:
\be\qquad\qquad \lim_{x\to 1} \CV_{\mb{5_2}}(x,s^\alpha(x)) \approx \begin{cases} 
1.11345 + 0\,i & \alpha=1\\
0.26574+2.82812\,i & \alpha=2 \\
0.26574-2.82812\,i & \alpha=3
\end{cases}\qquad \Big(\text{mod}\;\;\frac{\pi^2}{6}\Big) \ee
In particular, for the geometric flat connection $(\alpha=2)$, we find the complete hyperbolic volume ${\rm Vol}(\mb{5_2})=2.82812...$, and the Chern-Simons invariant ${\rm CS}(\mb{5_2})=0.26574...$.


\subsection{Stabilization and specialization}
\label{sec:stab}

We would now like to investigate the specialization of holomorphic blocks to  quantized values of mass parameters: $x=q^N$ for integers $N$. The dictionary established in Sections \ref{sec:CSanal}--\ref{sec:6d} between blocks and Chern-Simons theory suggests that for knot-complement theories $T_M$, $M=S^3\bs \CK$, this limit should have something to with the colored Jones polynomials of the knot $\CK$. We will propose one way to make this connection concrete, by relating blocks to the so-called \emph{stabilization} limit(s) of colored Jones polynomials. It requires treating blocks as formal series in $q$ and $q^N$. Subsequently, we will make a brief, intriguing observation about the specialization $x\to q^N$ when blocks are treated as actual functions of $q$ and $N$; namely, in this case, the dependence of blocks both on vacua $\alpha$ and on Stokes chambers appears to vanish. Altogether, this section is experimental in its approach. We hope that our observations will find a deeper theoretical and physical underpinning in future work.

We first define the stabilization limit. Given the sequence of $SU(2)$ colored Jones polynomials $\big\{J_N(\CK;q)\big\}_{N\in \mathbb N}$ for a knot $\CK\subset S^3$, the stabilized limit $\ol J(\CK;x;q)$ is constructed as follows \cite{GukovZagier, DasbachLin, Garoufalidis-slopes, GarLe-Nahm}. Define the \emph{lower degree} $d(N)\in \frac12\Z$ to be the smallest power of $q$ present in $J_N(\CK;q)$. Since the Jones polynomials satisfy a $q$-difference relation, it follows by a general theorem (\cf\ \cite{Garoufalidis-slopes}) that $|d(N)|$ grows quadratically in $N$. In fact, it is often an honest quadratic polynomial in $N$, and this is the only case we will consider here. Then one can consider the limit of $q^{-d(N)}J_N(\CK;q)$ as $N\to \infty$, and generally this converges (``stabilizes'') to a well-defined formal power series in $q^{\frac12}$:
\be \lim_{N\to \infty} q^{-d(N)}J_N(\CK;q)  = j_0(q) = 1+...\quad \in\; \Z[\![q^{\frac12}]\!]\,.\ee
The convergence to the series $j_0(q)$ is linear in $N$; that is, $q^{-d(N)}J_N(\CK;q) = j_0(q)$ modulo a series whose minimal power of $q$ is roughly $N$. Therefore, one may expect that the sequence $q^{-N}\big(q^{-d(N)}J_N(\CK;q) - j_0(q)\big)$ again converges to a well-defined $q$-series limit $j_1(q)$. The process can (potentially) be repeated to define a formal series
\be 
\ol J(\CK;x;q) = \sum_{r=0}^\infty j_r(q)\, x^r\,, \label{Jstab}
\ee
where the $j_n(q)$ are formal Laurent series in $q^{\frac 12}$, such that for every positive integer $A$
\be \lim_{N\to \infty} q^{-A N}\bigg[ q^{-d(N)} J_N(\CK;q)-\sum_{r=0}^A j_r(q)q^{rN}\bigg] = 0 \label{Jstab2}\ee
in the ring of formal $q$-series. If such $\ol J(\CK;x,q)$ exists, it is called the (lower) stabilization, or stable limit, of the colored Jones polynomials.

The stable limit \eqref{Jstab} has been proven to exist for all alternating knots (including the $\mb{3_1},\mb{4_1},\mb{5_2}$ examples here), and conjectured to exist for all knots \cite{GarLe-Nahm}. In addition to the lower stabilization just described, one can also consider an upper stabilization with similar properties. That is, one defines $d_+(N)$ to be the \emph{maximal} power of $N$ present in $J_N(\CK;q)$, and tries to find $\ol J_+(\CK;x^{-1};q^{-1}) =
\sum_{r=0}^\infty j^{(+)}_r(q^{-1})x^{-r}$, with the $j^{(+)}_r(q^{-1})$ Laurent series in $q^{-\frac12}$, such that $\lim_{N\to \infty} q^{A N}\Big[ q^{-d_+(N)} J_N(\CK;q)-\sum_{r=0}^A j_r^{(+)}(q^{-1})q^{-rN}\Big] = 0$ for all positive integers $A$, in the sense of $(q^{-1})$-series. Since the Jones polynomials of a knot $\CK$ and its mirror image $\ol \CK$ are related by $J_N(\CK;q) = J_N(\ol\CK;q^{-1})$, the lower stabilization for $\CK$ is equivalent to the upper stabilization for $\ol \CK$, and vice versa.

Now, recall the relation \eqref{Jsum} between Jones polynomials and partition functions in analytically continued Chern-Simons theory,
\be \label{Jsum2}
J_N(\CK;q)\sim \sum_{\alpha} n_\alpha \CZ_{\rm CS}^\ra(x;q)\,,
\ee
with $x=q^N$. As discussed in Section \ref{sec:CSanal}, this is a sum over all flat $\SLC$ connections on a knot complement $M=S^3\bs \CK$. On the other hand, we expect that the gauge theory $T_M$ defined by \cite{DGG} only has vacua $\alpha$ and blocks $B^\ra_M(x;q)$ corresponding to irreducible (and in particular nonabelian) flat connections. The main reason we are presently interested in the stable limit of Jones polynomials is that it can effectively project out the abelian flat connection (and perhaps others) from the sum \eqref{Jsum2}. Then if indeed $\CZ_{\rm CS}^\ra(x;q)=B^\ra_M(x;q)$ for $\alpha$ irreducible, we should be able to write stabilizations $\ol J(\CK;x;q)$, multiplied by $q^{d(N)}$, 
as sums of blocks.

We expect that a lower (resp. upper) stable limit projects abelian flat connections out of colored Jones polynomials precisely when the lower (resp. upper) degrees $d(N)$ of the polynomials grow quadratically --- that is, $d(N)\sim aN^2+bN+c$ with $a$ nonzero and negative (resp. positive). One motivation for this is as follows. The AJ Conjecture \cite{garoufalidis-2004} predicts that colored Jones polynomials satisfy an inhomogeneous recursion of the form
\be \hat A^{\rm irr}_M(\hat x,\hat p;q) J_N(\CK;q) = R(x;q)\,, \label{Ainhom} \ee
where $\hat A^{\rm irr}_M$ is a quantization of the nonabelian A-polynomial, $\hat x$ acts as multiplication by $x=q^N$, $\hat p$ multiplies by $(-1)$ and sends $N\to N+1$, and the RHS $R(x;q)$ is a fixed polynomial in $x=q^N$ and $q$. The recursion \eqref{Ainhom} implies a homogeneous recursion of the form $\hat A_M(\hat x,\hat p;q)J_N(\CK;q)=0$, where $\hat A_M$ is a quantization of the complete A-polynomial, including the abelian connection. If $\ol J(\CK;x;q)$ is a lower (say) stabilization of $J_N(\CK;q)$, and if $d(N)$ is quadratic, then it is easy to see that \eqref{Ainhom} also implies
\be \label{Astab}
\hat A^{\rm irr}_M(\hat x,\hat p;q) \cdot \left[ q^{d(N)}\ol J(\CK;x;q)\right]=0\,;
\ee
the inhomogeneous term $R(q^N;q)$ disappears because its degree only grows linearly in $N$. But Equation \eqref{Astab} is precisely the recursion that should be obeyed by the nonabelian functions $\CZ_{\rm CS}^{\alpha\neq{\rm abel.}}(x;q)$, suggesting that the stabilization has projected out the abelian flat connection. Similar remarks apply in the case of upper stabilization.

Recall that the holomorphic blocks $B^\ra_M(x;q)$ of the gauge theory $T_M$ also satisfy \eqref{Astab}, with $\hat A_M^{\rm irr}$ interpreted as an element in the algebra of line operators. In fact, the blocks provide a basis of solutions to the line-operator identity, with certain analytic properties. Then it is natural to expect that a stable series $q^{d(N)}\ol J(\CK;x;q)$ can be directly written in terms of blocks.
We wish to test this idea with our three knot examples from Section \ref{sec:asymp}. 

\subsubsection*{Trefoil stabilization}

We start with the (left-handed) trefoil knot. The Jones polynomials are given by the formula \cite{Morton-torus}%
\footnote{We normalize the Jones polynomials so that the unknot $U$ has $J_N(U)=(q^{N/2}-q^{-N/2})/(q^{1/2}-q^{-1/2})$.} %
\be
J_N(\mb{3_1};q) = 
 \frac{q^{\frac32 N^2-1}}{q^{\frac12}-q^{-\frac12}}\sum_{k=-(N-1)/2}^{(N-1)/2}q^{-6k^2-k}\big(q^{6k-1}-1\big)\,.
\ee
The lower degree $d(N)=\frac12 N-1$ is linear, but the upper degree $d_+(N)=\frac32N^2-1$ is quadratic, and it is not too hard to see that the polynomials have a trivial stabilization, stabilize to the upper limit
\be J_N(\mb{3_1};q) \sim \frac{-(q^{-1})_\infty}{q^{\frac12}-q^{-\frac12}} (-1)^N q^{\frac32 N^2-1} \ol J_+(\mb{3_1};q^{-N};q^{-1}) \,,\qquad\quad\ol J_+(\mb{3_1};x^{-1};q^{-1})=1\,.
\ee
In other words, aside from a constant prefactor $-(q^{-1})_\infty/(q^{\frac12}-q^{-\frac12})$ and the quadratic term $q^{d_+(N)}=q^{\frac32 N^2-1}$, the stabilization is trivial!

On the gauge theory side, we found that $T_{\mb{3_1}}$ has a single block
\be B_{\mb{3_1}}(x;q) = \frac{1}{\theta(x;q)^3}\,.\ee
Upon setting $x=q^N$, the theta function simplifies to $\theta(q^N;q)=q^{-\frac{N^2}{2}}(-q^{\frac12};q)_\infty^2$, and thus we find that the block equals the stabilization up to simple $q$-dependent prefactors and a sign $(-1)^N$. The sign is expected: it comes because we have used a different polarization in defining gauge theories $T_M$ than is standard in Chern-Simons theory. (Put differently, our 't Hooft operator $\hat p$ is related to the standard shift operator $\hat \ell$ of quantum A-polynomials by a sign, $\hat p = -\hat \ell$.) The simple $q$-dependent prefactors are also to be expected. Indeed, our construction of blocks only defines them modulo elliptic functions $c(x;q)$. When specializing to $x=q^N$, an elliptic function just becomes a function of $q$, since $c(q^N;q)=c(1;q)$. Specifically, elliptic ratios of theta-functions just become factors like $(\pm q^{\#};q)_\infty$.

\subsubsection*{Figure-eight stabilization}

The Jones polynomials of the figure-eight knot are given by \cite{Habiro-Jones, Masbaum-Jones}
\be J_N(\mb{4_1};q) = \frac{q^{\frac N2}-q^{-\frac N2}}{q^{\frac12}-q^{-\frac12}}\sum_{k=0}^{N-1} q^{-Nk}(q^{N+1};q)_k(q^{N-1};q^{-1})_k\,. \ee
Since the figure-eight knot is isotopic to its mirror image, the polynomials are invariant under $q\to q^{-1}$, and upper and lower stabilizations agree. Taking the lower, we find%
\footnote{At this point, we must thank S. Garoufalidis and D. Zagier for discussing and sharing data on stabilizations with us, including the formula for the figure-eight knot here. This formula led to the initial realization that stabilizations should be connected to blocks.} %
that
\be J_N(\mb{4_1};q) \sim \frac{-(q)_\infty}{q^{\frac12}-q^{-\frac12}}q^{d(N)}\,\ol J(\mb{4_1};q^N;q)\,,\ee
with $d(N) = -N^2+\frac12 N$ and
\be\label{stabJ41}  \ol J(\mb{4_1};x;q) = \sum_{k,s=0}^\infty \frac{q^{-ks}x^{k+2s}}{(q)_k(q)_s}\,.
\ee 
The series \eqref{stabJ41} is very similar to a $q$-Bessel function, aside from the fact that it does not converge as an actual function of $q$ and $x$ for any $x\in\C^*$ and $|q|<1$, due to the large quadratic powers $q^{-ks}$ in the numerator. It only make sense as a formal power series in $x$, whose coefficients are formal Laurent series in $q$. We can reproduce the series by similarly doing formal manipulations on blocks.

For example, consider the second figure-eight block in \eqref{B41}. We re-write
\begin{align} B^2_{\mb{4_1}}(x;q) &\;= \frac{1}{\theta(x)\theta(-q^{-\frac12}x)}\CJ(x^{-1},x^{-2};q) \notag\\
 &\;= \frac{\theta(-q^{\frac12}x^{-2})}{\theta(x)\theta(-q^{-\frac12}x)}\CJ(x,x^2;q^{-1}) = \frac{\theta(-q^{\frac12}x^{-2})}{\theta(x)\theta(-q^{-\frac12}x)}\sum_{k=0}^\infty \frac{(q^{-k-1}x^2;q)_\infty x^k}{(q)_k} \notag\\
 &\text{``$=$''} \frac{\theta(-q^{\frac12}x^{-2})}{\theta(x)\theta(-q^{-\frac12}x)}\sum_{k,s=0}^\infty \frac{q^{-ks}x^{k+2s}}{(q)_k(q)_s}\,,
\end{align}
where the final equality is not true in the sense of functions, but makes sense for formal series. We also substitute quadratic powers of $q$ for the theta-functions: $\theta(x) \to q^{-\frac{N^2}{2}},\; \theta(-q^{-\frac12}x)\to (-1)^N q^{\frac{N(N-1)}{2}}$ and in general
\be \label{thetaspec}
\theta\big((-1)^aq^bx^c;q\big)\,\to\, (-1)^{acN}q^{-\frac{c^2}{2}N^2-bcN}\,.
\ee
This corresponds to the specialization to $x=q^N$, modulo (potentially divergent) factors which are independent of $N$. As noted above, this specialization is independent of elliptic ambiguities. Altogether, we find
\be B^2_{\mb{4_1}}(x;q)\;\to\; (-1)^N q^{-N^2+\frac N2} \sum_{k,s=0}^\infty \frac{q^{-ks}x^{k+2s}}{(q)_k(q)_s}\,, \label{BJformal41} \ee
in agreement with the stabilization up to the same sign correction $(-1)^N$ and $q$-prefactors.

Curiously, we could also have obtained \eqref{BJformal41} by applying formal power-series identities to the \emph{first} figure-eight block in \eqref{B41}. It could also have been obtained by using any of the blocks in the other two Stokes chambers of the theory! When allowing formal identities and forgetting about the functional meaning of the blocks, the dependence on different vacua $\alpha$ and on Stokes chambers disappears. It is not yet clear what this means, or how general a phenomenon it is.

\subsubsection*{Knot $\mb 5_2$ stabilization}

Finally, the colored Jones polynomials of the $\mb{5_2}$ knot also have a lower stable limit with quadratic growth. By using formulas of \cite{Masbaum-Jones}, we find experimentally that
\be J_N(\mb{5_2};q) \sim \frac{(q)_\infty}{q^{\frac12}-q^{-\frac12}}(-1)^N q^{-\frac52N^2+N+1} \,\ol J(\mb{5_2};q^N;q)\,,
\ee
with
\be \ol J(\mb{5_2};x;q) =
\sum_{r,s,k=0}^\infty \frac{q^{-k(k+1)-(r+s)k}x^{r+2(s+k)}}{(q^{-1})_k(q)_r(q)_s}\,. \ee
Just like for the figure-eight knot, this formal series does not converge to an actual function.
To reproduce it, we can take (say) the third block of \eqref{B52}, and manipulate it as
\begin{align} B^3_{\mb{5_2}}(x;q) &\;=\,\frac{\theta(x)}{\theta(-q^{\frac12} x^{-1})}(qx^{-1};q)_\infty(qx^{-2};q)_\infty\sum_{k=0}^\infty \frac{x^{-k}}{(q^{-1})_k(qx^{-1};q)_\infty(qx^{-2};q)_\infty} \notag \\
&\;=\, \theta(x)\theta(-q^{\frac12}x^{-2}) \sum_{k=0}^\infty \frac{q^{-k(k+1)}(q^{-k-1}x;q^{-1})_\infty(q^{-k-1}x^2;q^{-1})_\infty\, x^{2k}}{(q^{-1})_k} \notag \\
&\text{``=''}\, \theta(x)\theta(-q^{\frac12}x^{-2})\,\ol J(\mb{5_2};x;q) \notag \\
& \to q^{-\frac52 N^2+N} \ol J(\mb{5_2};x;q)\,,
\end{align}
where in the penultimate line we formally expanded a series in $x$, and in the last line we specialized theta-functions as in \eqref{thetaspec}. 

\subsubsection*{Specialization}

We have observed experimentally that by using formal manipulations of $(q,x)$-series, the blocks of a knot complement theory $T_M$ reproduce stable limits of Jones polynomials. We also observed, at least in the limited examples here, that the stable limit could be reproduced from a single block rather than a sum as in \eqref{Jsum2}; and sometimes it does not matter \emph{which} block $B^\ra_M(x;q)$ is used in this process.

We have spent much of this paper considering questions for which blocks $B^\ra(x;q)$ should define honest functions of $x$ and $q$. The above examples suggest that it might be interesting to consider directly the specialization $x\to q^N$, for $N\in \Z$, in these honest functions --- without doing any formal manipulations or rearrangements of power series. We have investigated this limit for the $\mb{4_1}$ and $\mb{5_2}$ knots and found yet another curious result.

Let us take the three blocks \eqref{B52} of the $\mb{5_2}$ knot, specialize theta-functions using \eqref{thetaspec} (recall that this specialization is canonical, independent of any extra elliptic-function prefactors in the blocks), and set $x=q^N$. We obtain three $q$-series that are all convergent as functions and equal:
\begin{align} \label{52Neq}
q^{-\frac{N^2}{2}}\CG(q^N,q^{-N},1;q) = (-1)^Nq^{\frac{N}{2}}\CG(q^N,q^{2N},q^N;q) = (-1)^Nq^{-\frac N2}\CG(q^{-N},q^{-2N},q^{-N};q)\,,
\end{align}
for $|q|<1$ and all $N\in\Z$. We tested these identities numerically. When $|q|>1$, all three sums \eqref{52Neq} diverge at $N\in \Z$, but if we take $N$ to be a continuous variable then the ratio of any two sums converges to one as $N$ approaches integers.

We can consider a similar specialization to $x=q^N$ for the figure-eight knot as well, and obtain the same type of result. It is even more interesting to look at the blocks of the original $\cp^1$ sigma-model. Recall that this theory has two mass parameters $x,y$, and that setting $y=1$ recovers the figure-eight blocks. We can then set $x=q^N$ and $y=q^K$, for integers $N$ and $K$, and rewrite any theta-function prefactors again using essentially \eqref{thetaspec}. We find that all blocks, in all chambers investigated in Section \ref{sec:CP1}, become equal. For example, at $|q|<1$ the six blocks written in Section \ref{sec:CP1} specialize to three distinct sums,
\begin{align} &(-1)^Nq^{N^2-KN-\frac N2}\CJ(q^{N-K},q^{2N};q)
 = (-1)^N q^{N^2+KN+\frac N2}\CJ(q^{-N-K},q^{-2N};q) \\
 &\hspace{1in}=(-1)^Kq^{\frac{K^2}{2}-\frac{N^2}{2}+\frac{K}{2}} \CJ(q^{K+N},q^{K-N};q)\,, \notag
\end{align}
which are equal for all $N,K\in \Z$. This equality of specialized blocks in different chambers is not inconsistent with linear (Stokes) transformations of the exact blocks found in Section \ref{sec:CP1}, because we have rewritten theta-function prefactors using \eqref{thetaspec} rather than substituting $(x,y)=(q^N,q^K)$ in their arguments directly.

Based on these observations, one might hypothesize that when specializing the holomorphic blocks of a knot complement theory $T_M$ to quantized $x=q^N$ --- or more generally when specializing the blocks of any $\CN=2$ SCFT to quantized masses $x_i = q^{N_i}$ --- the dependence on flat connection (or vacuum) and Stokes chamber vanishes. One would then be left with a unique specialized block $B_N(q)$ as a well-defined function of $N$ and $q$ inside or outside the unit circle. 
The physical basis for this unification of blocks is still under investigation. It is reminiscent of topological string constructions which are obtained by large $N$ duality, the K\"ahler parameters (which become mass parameters in an effective QFT description) are frequently quantized in units of the string coupling.

\acknowledgments

We would like to thank Mina Aganagic, Francesco Benini, Thomas Dumitrescu, Guido Festuccia, Abhijit Gadde, Davide Gaiotto, Stavros Garoufalidis, Sergei Gukov, Leonardo Rastelli, Shlomo Razamat, Kevin Schaeffer, Nati Seiberg, Cumrun Vafa, Edward Witten, and Don Zagier for helpful comments and discussions. The work of CB is supported in part by DOE grant DE-FG02-92ER-40697. The work of TD is supported by a William D. Loughlin Fellowship at the Institute for Advanced Study, with additional support from DOE grant DE-FG02-90ER-40542.  The work of SP has been partially supported by a Marie Curie Intra-European Fellowship: FP7-PEOPLE-2009-IEF. CB and TD would like to thank the Aspen Center for Physics and the 2012 Simons Workshop in Mathematics and Physics for hospitality during completion of this work. The Aspen Center for Physics is partially supported by the NSF under Grant No. 1066293.

\appendix

\section{Three-dimensional supersymmetry and BPS indices}
\label{app:3dSUSY}

In this appendix, we briefly review some details of the BPS index for a theory with $\cN=2$ supersymmetry in three-dimensions. The $\cN=2$ supersymmetry algebra in three dimensions follows from dimensional reduction of the $\CN=1$ algebra in four dimensions. We adopt the conventions of \cite{WessBagger}, and take the signature to be $(-,+,+,+)$. In four dimensions, the four supercharges are grouped into a pair of two-component Weyl spinors $Q_\alpha,\,\ol Q_{\dot\alpha}$ of opposite chirality obeying $Q_\alpha^\dagger = \ol Q_{\dot\alpha}$ and
\begin{subequations}
\begin{align} &\{ Q_\alpha, \ol Q_{\dot \alpha}\} = 2\sigma_{\alpha\dot\beta}^mP_m\,, \\ &\{Q_\alpha, Q_\beta\} =  \{\ol Q_{\dot \alpha}, \ol Q_{\dot\beta}\} = 0\,,
\end{align}
\end{subequations}
where $m=0,1,2,3$. Here $\sigma^0=I$ and $\sigma^m$ are the Pauli matrices for $m=1,2,3$.

The little group for massive states in four dimensions is $SO(3)_E\simeq SU(2)_E$, under which the supercharges both transform in spin-$\tfrac12$ representations. Specifically, letting $J_3$ be the generator of rotations in the $1\!-\!2$ plane, we have
\be [J_3,Q_1]=\frac12 Q_1\,\quad [J_3,Q_2]=-\frac12 Q_2\,, \qquad [J_3,\ol Q_{\dot 1}]=-\frac12 \ol Q_{\dot 1}\,\quad [J_3,\ol Q_{\dot 2}]=\frac12 \ol Q_{\dot 2}\,. \ee
In addition, there is an R-symmetry $U(1)_R$ with respect to which the supercharges have charge $\pm 1$,
\be [R,Q_\alpha]=Q_\alpha\,,\qquad [R,\ol Q_{\dot \alpha}]=-\ol Q_{\dot \alpha}\,.\ee

It's most convenient to reduce to three dimensions along the $m=3$ direction. The $m=3$ component of the momentum becomes a real central charge, $P_3=Z$. Massive states in three dimensions transform under the little group $SO(2)_E$, whose generator is $J_3$. We can label the spinor indices $\alpha=(+,-)$ and $\dot \alpha=(-,+)$ to indicate helicity:
\be [J_3,Q_\pm]=\pm \frac 12 Q_\pm\,,\qquad [J_3, \ol Q_\pm]=\pm \frac12 \ol Q_\pm\,. \ee
Also, the R-symmetry descends in a trivial manner to a three-dimensional R-symmetry.

We see from the above commutation relations that the combination $J_3+\tfrac 12 R$ commutes with a pair of supercharges $(Q_-,\,\ol Q_+)$, while $J_3-\tfrac 12 R$ commutes with $(Q_+,\,\ol Q_-)$. Each of these pairs are Hermitian conjugates, and we obtain
\be \{Q_-,\ol Q_+\} = 2(P^0-Z)=:H_+\,,\qquad \{Q_+,\ol Q_-\}=2(P^0+Z)=:H_-\,,\ee
while
\be\ Q_\pm ^2 = \ol Q_\pm ^2 = 0\,.\ee
The positive-definiteness of $H_\pm$ leads to the BPS bound $P^0\geq |Z|$.

We can construct two different BPS indices from this algebra,
\be \cI^+(\beta;q) = \Tr e^{-\beta H_+}(-1)^{2J_3}q^{-J_3-\tfrac R2}\,,\qquad\quad
\cI^-(\beta;q) = \Tr e^{-\beta H_-}(-1)^{2J_3}q^{-J_3+\tfrac R2}\,.
\ee
Let us assume that we have regularized the theory so that the spectrum of the operator in the trace is discrete. Then in the case of $\cI^+(\beta;q)$, the only states that contribute are those annihilated by both $Q_-$ and $\ol Q_+$. Otherwise, the contributions from a state $|\psi\rangle$ and (say) $\ol Q_+|\psi\rangle$ will cancel each other out, because $\ol Q_+$ commutes with ${-J_3-\tfrac R2}$ and anti-commutes with $(-1)^{2J_3}$. Therefore, the index $\CI^+(\beta;q)$ only receives contributions from BPS multiplets. Indeed, the Hamiltonian $H_+$ also annihilates all states that contribute, meaning $P^0=Z$, which is the BPS condition. Similarly, the index $\cI^-(\beta;q)$ only receives contributions from \emph{anti}-BPS multiplets, \ie\ those annihilated by $Q_+$ and $\ol Q_-$, and satisfying $P^0=-Z$.

Both indices are independent of $\beta$. Furthermore, it is useful to observe that neither D-term nor F-term (superpotential) deformations of a theory can affect the indices. For example, a superpotential deformation amounts to an insertion of some operators $\int d^2\theta\, \CO$ and $\int d^2\ol \theta\, \ol \CO$ in the indices. These can be written as $\{Q_-,[Q_+,A]\}$ and $\{\ol Q_+,[\ol Q_-,\ol A]\}$, respectively, for an appropriate $A$, and thus vanish inside both indices. An analogous argument shows invariance under D-terms.

Note that instead of $(-1)^{2J_3}$ we could use $(-1)^R$ in $I^\pm(\beta;q)$ to produce indices with the same essential properties. In fact, the simple replacement $q\to -q$ implements this modification. This is the relevant situation for holomorphic blocks. When $R$ is not integer-valued, $(-1)^R$ means $e^{i\pi R}$.


\section{Combinatorics of triangulated knot complements}
\label{app:knots}

In this appendix, we provide combinatorial details for the knot complement examples of Section \ref{sec:CS}. In particular, we derive the \emph{simplified} theories $T_M$  for the trefoil ($\mb{3_1}$), figure-eight ($\mb{4_1}$) and $\mb{5_2}$ knot complements. It was discussed in Section \ref{sec:CS} that these simplified theories --- corresponding to minimal triangulations of the knot complements --- are somewhat degenerate, and are missing operators necessary to break some flavor symmetries. The real masses for these flavor symmetries are set to zero by hand.

Our notation follows \cite{DGG} and \cite{DGG-index}. The logarithm of the squared meridian eigenvalue $(m)$ is called $X=\log m^2 = U$. The logarithm of the longitude eigenvalue $(\ell)$ is called $P=\log(-\ell)=v$.

\subsubsection*{Trefoil $\mb{3_1}$}

The minimal triangulation of the trefoil knot complement has two tetrahedra. Call the logarithmic shape parameters (complexified dihedral angles) of the tetrahedra $Z,Z',Z''$ and $W,W',W''$. The coordinates defining the gluing, obtained from \texttt{SnapPy} \cite{SnapPy}, can be written as
\be \label{31coords}
\begin{array}{rl}
 X &= -Z''+W'' \\[.05cm]
 C_1 &= Z+W \\[.05cm]
 C_2 &= Z+2Z'+2Z''+W+2W'+2W'' \\[.05cm]
 P &= \tfrac12(3 Z+Z'-3Z''-W-W'+3W'')-i\pi-\tfrac\hbar2\,,
\end{array}
\ee
where $C_1$ and $C_2$ are the sums of angles around the two internal edges. The semiclassical gluing constraint is $C_1=C_2=2\pi i+\hbar$. We ignore the redundant constraint $C_2=2\pi i+\hbar$; use the relations $Z+Z'+Z''=W+W'+W''=i\pi+\hbar/2$ to eliminate $Z'$ and $W'$; and define a momentum coordinate $\Gamma=W''$ conjugate to $C_1$. Then the equations are re-written as
\be \label{31aff} \begin{pmatrix} C_1-2\pi i-\hbar \\ X \\ \Gamma \\ P \end{pmatrix} =
\begin{pmatrix} 1 & 1&0&0 \\ 0&0&-1&1 \\ 0&0&0&1 \\ 1&0&-2&2\end{pmatrix}
\begin{pmatrix} Z\\W\\Z''\\W'' \end{pmatrix}+\big(i\pi+\tfrac\hbar2\big)\begin{pmatrix} -2 \\ 0 \\ 0 \\ -1 \end{pmatrix}\,.
\ee
This defines an affine symplectic transformation in the space of shape parameters. The $Sp(4,\Z)$ matrix appearing here, which we can call $g_{\mb{3_1}}$, decomposes into generators as
\be \label{31symp} g_{\mb{3_1}} =\begin{pmatrix} 1 & 1&0&0 \\ 0&0&-1&1 \\ 0&0&0&1 \\ 1&0&-2&2\end{pmatrix}= 
{ \begin{pmatrix} 1 & 0 & 0 & 0 \\
 0 & 1 & 0 & 0 \\
 0 & 0 & 1 & 0 \\
 0 & 2 & 0 & 1 \end{pmatrix}
 \begin{pmatrix}  1 & 0 & 0 & 0 \\
 0 & 0 & 0 & -1 \\
 0 & 0 & 1 & 0 \\
 0 & 1 & 0 & 0 \end{pmatrix}
\begin{pmatrix} 1 & 1 & 0 & 0 \\
 1 & 0 & 0 & 0 \\
 0 & 0 & 0 & 1 \\
 0 & 0 & 1 & -1
 \end{pmatrix} }\,.
\ee

The combinatorial data in \eqref{31coords}--\eqref{31symp} translates directly into a class-$\CR$ construction of the simplified trefoil theory $T_{\mb{3_1}}$. In general, for a triangulation of a knot complement $M$ into $N$ tetrahedra, the prescription of \cite{DGG,DGG-index} dictates that one should  (\cf\ Section \ref{sec:Ward-assemble})
\begin{enumerate}
\item Tensor together $N$ chiral theories $T_\Delta$, obtaining $T_{\Delta_1}\otimes\cdots\otimes T_{\Delta_N}$, with $U(1)^N$ flavor symmetry and a level $-\frac12$ CS coupling for each $U(1)$.
\item Apply the $Sp(2N,\Z)$ symplectic matrix $g$ as in \eqref{31symp} to the product theory. In particular,
\begin{itemize}
\item[2a.] Generators with block type $\left(\begin{smallmatrix} U & 0 \\ 0 & U^{-1\,T}\end{smallmatrix}\right)$, with $U\in GL(N)$, act by linear redefinitions of the $U(1)^N$ flavor group.
\item[2b.] Generators with block type $\left(\begin{smallmatrix} I & 0 \\ B & I \end{smallmatrix}\right)$, with $B$ symmetric, add background CS couplings with a level matrix $k_{ij}=B_{ij}$.
\item[2c.] $S$-type generators containing pieces that look like $\left(\begin{smallmatrix} 0 & -1 \\ 1 & 0 \end{smallmatrix}\right)$ gauge a $U(1)$ with an FI coupling to the background vector multiplet of a new topological $U(1)_J$.
\end{itemize}
\item Affine shifts, as on the RHS of \eqref{31aff}, are relevant for theories on compactified spaces. On $\DqS$, shifts in ``position'' coordinates (the top half of the shift vector) add units of flavor current to the R-current (shifting the Wilson lines of flavor symmetries by $-i\pi -\frac\hbar2$). Shifts in ``momentum'' coordinates (the bottom half of the shift vector) act by adding mixed Chern-Simons contact terms between the background R-symmetry and flavor symmetry fields.

\item Finally, operators $\CO_{C_i}$ must be added to a superpotential to break the flavor symmetry associated to each internal-edge coordinate $C_i$. If these operators exist, they automatically have R-charge $R_\CO=2$ by virtue of the affine shifts in Step 3.
\end{enumerate}

After applying this prescription to the trefoil triangulation and letting the dust settle, we find a dynamical $U(1)$ gauge theory with two chiral multiplets $\phi_z,\phi_w$. The chirals have charges $(+1,-1)$ under the $U(1)_s$ gauge symmetry, and R-charges $(+1,+1)$. An axial flavor symmetry $U(1)_{c_1}$ is broken by a superpotential coupling $\CO_{C_1}=\phi_z\phi_w$. (There is no operator $\CO_{C_2}$ corresponding to the edge $C_2$, which is why this theory is not complete.) There remains a topological $U(1)_x$ flavor symmetry. There are $+2$ units of CS coupling for $U(1)_x$, $-1$ unit for $U(1)_s$, and an FI term coupling $U(1)_x$ and $U(1)_s$; in total this can be encoded in the CS coupling matrix
\be \frac12\begin{pmatrix} V_s & V_x \end{pmatrix} \begin{pmatrix} -1 & 1 \\ 1 & 2 \end{pmatrix}\begin{pmatrix} \Sigma_s \\ \Sigma_x \end{pmatrix} = -\frac12 V_s\Sigma_s+ V_s\Sigma_x+V_x\Sigma_x\,.\ee
Due to the superpotential interaction, we expect to be able to integrate out the chirals at low energies. No (net) anomalous Chern-Simons couplings are generated.

\subsubsection*{Figure-eight $\mb{4_1}$}

The minimal triangulation of the figure-eight knot also has two tetrahedra. Give them shape parameters $Z$ and $W$. The gluing coordinates from \texttt{SnapPy} are
\be
\begin{array}{rl}
C_1 &=W + 2 W'' + Z + 2 Z'' \\[.05cm]
C_2 &=W + 2 W' + Z + 2 Z' \\[.05cm]
X &=-W' + Z'' \\[.05cm]
P &=\frac12(-W - 3 W' + W'' + Z + Z' + Z'')\,.
\end{array}
\ee
We forget the redundant edge $C_2$, eliminate $Z'$ and $W'$ using $Z+Z'+Z''=W+W'+W''=i\pi+\frac\hbar2$, and (arbitrarily) choose a ``momentum'' coordinate $\Gamma_1$ canonically conjugate to $C_1$, to arrive at the affine symplectic transformation 
\be 
\begin{pmatrix} C_1-2\pi i-\hbar\\X\\\Gamma_1\\P \end{pmatrix} \;=\; 
g_{\mb{4_1}}\cdot \begin{pmatrix} Z\\W\\Z''\\W'' \end{pmatrix} + \big(i\pi+\tfrac\hbar2\big)\,\sigma_{\mb{4_1}}\,,
\ee
with
\be \label{symp41}
g_{\mb{4_1}} = {\scriptstyle \begin{pmatrix}1 & 0 & 0 & 0 \\
 0 & 1 & 0 & 0 \\
 0 & 0 & 1 & 0 \\
 0 & 2 & 0 & 1
\end{pmatrix}\!\!\begin{pmatrix}
 1 & 2 & 0 & 0 \\
 0 & 1 & 0 & 0 \\
 0 & 0 & 1 & 0 \\
 0 & 0 & -2 & 1
 \end{pmatrix}\!\!\begin{pmatrix}
 1 & 0 & 0 & 0 \\
 0 & 0 & 0 & -1 \\
 0 & 0 & 1 & 0 \\
 0 & 1 & 0 & 0
 \end{pmatrix}\!\!\begin{pmatrix}
  1 & 0 & 0 & 0 \\
 0 & 1 & 0 & 0 \\
 0 & 0 & 1 & 0 \\
 0 & 1 & 0 & 1
 \end{pmatrix}\!\!\begin{pmatrix}
 1 & -1 & 0 & 0 \\
 0 & -1 & 0 & 0 \\
 0 & 0 & 1 & 0 \\
 0 & 0 & -1 & -1
 \end{pmatrix} }\,,\quad
 \sigma_{\mb{4_1}}= {\scriptstyle \begin{pmatrix} -2\\-1\\-1\\0 \end{pmatrix}}\,.
\ee

The combinatorial data in \eqref{symp41} leads to a simplified theory $T_{\mb{4_1}}$ that is a specialization of the $\cp^1$ sigma-model. Namely, it is a dynamical $U(1)$ gauge theory with no Chern-Simons terms, coupled to two chiral multiplets both of charge $+1$. The R-charge of each chiral is zero. The vector $U(1)_V$ flavor symmetry has an associated complexified mass $x=\exp X$, while the $U(1)_J$ should be broken by operators $\CO_{C_1}$ and $\CO_{C_2}$ in the superpotential. These operators do not exist in the simplified theory, so instead we set the FI term (the mass of $U(1)_J$) to zero by hand. In fact due to an affine shift, the data \eqref{symp41} dictates that we set the FI term not to zero but to $-i\pi-\tfrac\hbar2$ (\ie\ giving a nonzero theta angle), where $i\pi+\hbar/2$ is the Wilson line of the R-symmetry at the tip of $\DqS$.
 
\subsubsection*{Knot $\mb{5_2}$}

The minimal triangulation for the $\mb{5_2}$ knot has three tetrahedra. Give them shape parameters $Z,W,Y$. From \texttt{SnapPy} we find gluing coordinates
\be
\begin{array}{rl}
 C_1&= W + W'' + Y' + 2 Y'' + Z + Z'' \\
 C_2&=W' + W'' + 2 Y + Z' + Z'' \\
 C_3&=W + W' + Y' + Z + Z'\\
 X&=-W' + Y'' + Z''\\
 P&=W' + 2 W'' + Y + Z + Z'-2\pi i-\hbar \,.
\end{array}
\ee
Then, after removing the redundant edge $C_3$, solving for $Z',W',Y'$ using $Z+Z'+Z''=i\pi+\frac\hbar2$ (etc.), and choosing (arbitrarily) conjugate momenta $\Gamma_1$ and $\Gamma_2$ for $C_1$ and $C_2$, we obtain the affine symplectic transformation
\be \begin{pmatrix} C_1-2\pi i-\hbar\\C_2-2\pi i-\hbar\\X \\\Gamma_1\\\Gamma_2\\P\end{pmatrix} = 
\begin{pmatrix} 1 & 1 & -1 & 1 & 1 & 1 \\
 -1 & -1 & 2 & 0 & 0 & 0 \\
 0 & 1 & 0 & 1 & 1 & 1 \\
 0 & 1 & 0 & 3 & 1 & 2 \\
 0 & 1 & 0 & 2 & 1 & 2 \\
 0 & -1 & 1 & -1 & 1 & 0
\end{pmatrix}\begin{pmatrix}Z\\W\\Y\\Z''\\W''\\Y'' \end{pmatrix}+
\big(i\pi+\tfrac\hbar2\big)\begin{pmatrix}-1\\0\\-1\\0\\0\\0\end{pmatrix}\,.
\ee
This can be written more nicely as
\be \vec X = g^*\big[ g_{\mb{5_2}}\cdot \vec Z + \big(i\pi+\tfrac\hbar2\big)\vec\sigma_{\mb{5_2}}\big]\,, \ee
where $\vec X=(C_1-2\pi i-\hbar,C_2-2\pi i-\hbar,X,...)^T$, $\vec Z=(Z,W,Y,Z'',W'',Y'')^T$, and
\be \label{symp52}
g_{\mb{5_2}} = {\scriptstyle
\begin{pmatrix}
1 & 0 & 0 & 0 & 0 & 0 \\
 0 & 1 & 0 & 0 & 0 & 0 \\
 0 & 0 & 1 & 0 & 0 & 0 \\
 -1 & 0 & 0 & 1 & 0 & 0 \\
 0 & 0 & 0 & 0 & 1 & 0 \\
 0 & 0 & 1 & 0 & 0 & 1
\end{pmatrix}\begin{pmatrix}
1 & 0 & 0 & 0 & 0 & 0 \\
 0 & 1 & 0 & 0 & 0 & 0 \\
 0 & 0 & 0 & 0 & 0 & -1 \\
 0 & 0 & 0 & 1 & 0 & 0 \\
 0 & 0 & 0 & 0 & 1 & 0 \\
 0 & 0 & 1 & 0 & 0 & 0
\end{pmatrix}
\begin{pmatrix}
 1 & 0 & 0 & 0 & 0 & 0 \\
 0 & 1 & 0 & 0 & 0 & 0 \\
 0 & 0 & 1 & 0 & 0 & 0 \\
 0 & 0 & 0 & 1 & 0 & 0 \\
 0 & 0 & 0 & 0 & 1 & 0 \\
 0 & 0 & 1 & 0 & 0 & 1
\end{pmatrix}
\begin{pmatrix}
1 & 0 & -1 & 0 & 0 & 0 \\
 1 & 1 & -2 & 0 & 0 & 0 \\
 0 & 0 & 1 & 0 & 0 & 0 \\
 0 & 0 & 0 & 1 & -1 & 0 \\
 0 & 0 & 0 & 0 & 1 & 0 \\
 0 & 0 & 0 & 1 & 1 & 1
\end{pmatrix}}\,,\quad
\sigma_{\mb{5_2}} = {\scriptstyle \begin{pmatrix} 0 \\ 0 \\ 1 \\ 0 \\ * \\ * \end{pmatrix}}\,;
\ee
whereas $g^*\in Sp(6,\Z)$ simply mixes around and adds Chern-Simons levels for flavor symmetries $U(1)_{C_1}$ and $U(1)_{C_2}$ that will be broken at the end of the day --- so it is irrelevant for the calculation.

The gauge theory obtained from the data \eqref{symp52} is a dynamical $U(1)$ gauge theory coupled to three chiral multiplets all of charge $+1$. The R-charge of each chiral is zero. There is a level $-\frac12$ CS coupling for the gauge field, and a level $-2$ CS coupling for a $U(1)$ flavor symmetry (associated with the complex mass $x=e^X$) under which the chirals have charges $(+1,-1,0)$. The internal edges $C_1$ and $C_2$ correspond a $U(1)$ flavor symmetry that rotates the chirals with charges $(0,+1,0)$, and the topological $U(1)_J$. The expected operators $\CO_{C_1}$ and $\CO_{C_2}$ (and $\CO_{C_3}$) needed to break them do not exist, so instead we set the corresponding masses to zero by hand. Due to a previous affine shift, this results in a fixed theta angle $i\pi+\tfrac\hbar2$ (as opposed to zero) for the dynamical gauge field.

\bibliographystyle{JHEP_TD}
\bibliography{toolbox}

\end{document}